\documentclass{aa}  

\pdfoutput=1 

\usepackage{graphicx}
\usepackage{times}
\usepackage{color}
\usepackage{amssymb}
\usepackage{amsmath}
\usepackage{grffile} % FOR EXTENSIONS
\usepackage{caption}
\usepackage{multicol,multirow}
\usepackage{xtab}
\usepackage{lscape}
\usepackage[caption=false]{subfig}
\usepackage{microtype} % to avoid lines going outside margin
\usepackage[rightcaption]{sidecap}
\usepackage{xcolor}
\usepackage{txfonts}
\usepackage{array}
\usepackage{tabularx}
\usepackage{longtable}
\usepackage{footnote}\makesavenoteenv{longtable}
\usepackage{afterpage}
\begin{document}

\newcommand{\sigmahtwo}{$\Sigma_{\rm H_2}$}
\newcommand{\sigmahi}{$\Sigma_{\rm HI}$}
\newcommand{\sigmasfr}{$\Sigma_{\rm SFR}$}
\newcommand{\hi}{H{\sc i}}
\newcommand{\micron}{$\mu$m}
\newcommand{\taudepl}{$\tau_{\rm depl}$}
\newcommand{\msunpc}{${\rm M}_\odot\,{\rm pc}^{-2}$}
\newcommand{\msunyrkpc}{${\rm M}_\odot\,{\rm yr}^{-1}\,{\rm kpc}^{-2}$}

\title{
The resolved scaling relations in DustPedia\thanks{DustPedia is a project funded by the European Union under the heading 
``Exploitation of space science and exploration data''. 
It has the primary goal of exploiting existing data in the 
\textit{Herschel} Space Observatory and Planck Telescope databases. 
}: \\
Zooming in on the local Universe}
\author{Viviana~Casasola$^{1}$, Simone~Bianchi$^{2}$, Laura~Magrini$^{2}$,
Aleksandr~V.~Mosenkov$^{3}$, Francesco~Salvestrini$^{2}$, Maarten~Baes$^{4}$,
Francesco~Calura$^{5}$, Letizia~P.~Cassar\`{a}$^{6}$, Christopher~J.~R.~Clark$^{7}$,
Edvige~Corbelli$^{2}$, Jacopo~Fritz$^{8}$, Fr\'{e}d\'{e}ric~Galliano$^{9}$, Elisabetta~Liuzzo$^{1}$,
Suzanne~Madden$^{9}$, Angelos~Nersesian$^{4, 10}$, Francesca~Pozzi$^{11, 5}$,
Sambit~Roychowdhury$^{12, 13}$, Ivano~Baronchelli$^{1}$, Matteo~Bonato$^{1}$, Carlotta~Gruppioni$^{5}$, and Lara~Pantoni$^{9,14}$ 
}
\institute{
$^{1}$ INAF -- Istituto di Radioastronomia, Via P. Gobetti 101, 40129, Bologna, Italy\\
\email{viviana.casasola@inaf.it}  \\
$^{2}$ INAF -- Osservatorio Astrofisico di Arcetri, Largo E. Fermi 5, 50125, Firenze, Italy\\
$^{3}$ Department of Physics and Astronomy, N283 ESC, Brigham Young University, Provo, UT 84602, USA\\
$^{4}$ Sterrenkundig Observatorium Universiteit Gent, Krijgslaan 281 S9, B-9000 Gent, Belgium \\  
$^{5}$ INAF -- Osservatorio di Astrofisica e Scienza dello Spazio, via Gobetti 93/3, I-40129 Bologna, Italy \\
$^{6}$ INAF -- Istituto di Astrofisica Spaziale e Fisica cosmica,  Via A. Corti 12, 20133, Milano, Italy \\ 
$^{7}$ Space Telescope Science Institute, 3700 San Martin Drive, Baltimore, Maryland, 21218, USA \\
$^{8}$ Instituto de Radioastronom\'\i a y Astrof\'\i sica, UNAM, Campus Morelia, A.P. 3-72, C.P. 58089, Mexico\\
$^{9}$ AIM, CEA, CNRS, Universit\'{e} Paris-Saclay, Universit\'{e} Paris Diderot, Sorbonne Paris Cit\'{e}, 91191 Gif-sur-Yvette, France \\
$^{10}$ National Observatory of Athens, IAASARS, Ioannou Metaxa and Vasileos Pavlou, GR-15236 Athens, Greece \\
$^{11}$ Dipartimento di Fisica e Astronomia, Universit\'{a} of Bologna, via Gobetti 93/2, 40129, Bologna, Italy \\
$^{12}$ ICRAR, University of Western Australia, 35 Stirling Highway, Crawley,  WA 6009, Australia \\
$^{13}$ ARC Centre of Excellence for All-Sky Astrophysics in 3 Dimensions (ASTRO 3D), Australia \\
$^{14}$ Institut d'Astrophysique Spatiale, Universit\'e Paris-Saclay, 91405 Orsay, France
}
\date{Received ; accepted}

\titlerunning{Resolved scaling relations in DustPedia}
\authorrunning{Viviana Casasola et al.}

\abstract
{}
{We perform a homogeneous analysis of an unprecedented set of spatially resolved scaling relations (SRs) 
between interstellar medium (ISM) components, that is to say dust, gas, and gas-phase metallicity,
and other galaxy properties, such as stellar mass (M$_{\rm star}$), total baryonic content,
and star-formation rate (SFR), in a range of physical scales between 0.3 and 3.4~kpc.
We also study some ratios between galaxy components: dust-to-stellar, dust-to-gas, and dust-to-metal ratios.}
{We use a sample of 18 large, spiral, face-on DustPedia galaxies.  
The sample consists of galaxies with spatially resolved dust maps 
corresponding to 15  \textit{Herschel}-SPIRE 500~$\mu$m resolution elements across the optical radius,
with the morphological stage spanning from T~=~2 to 8, M$_{\rm star}$ from  $2 \times 10^9$ to $1 \times 10^{11}$~M$_\odot$, SFR from 0.2 to 13~M$_\odot$~yr$^{-1}$, 
and oxygen abundance from 12~+~log(O/H)~=~8.3 to 8.8.}
{All the SRs are moderate or strong correlations except the dust-\hi\ SR that does not exist or is weak for most galaxies.
The SRs do not have a universal form but each galaxy is characterized by distinct correlations, affected by local processes and galaxy peculiarities. 
The SRs hold, on average, starting from the scale of 0.3~kpc, and if a breaking down scale exists it is below 0.3~kpc.
By evaluating all galaxies together at the common scale of 3.4~kpc, differences due to peculiarities of individual galaxies are cancelled out and 
the corresponding SRs are consistent with those of whole galaxies.
By comparing subgalactic and global scales, the most striking result emerges from the SRs involving ISM components: 
the dust-total gas SR is a good correlation at all scales, 
while the dust-H$_2$ and dust-\hi\ SRs are good correlations at subkiloparsec/kiloparsec and total scales, respectively.
For the other explored SRs, there is a good agreement between small and global scales  
and this may support the picture where the main physical processes regulating the properties and evolution of galaxies occur locally.
In this scenario, our results are consistent with the hypothesis of self-regulation of the star-formation process.
The analysis of subgalactic ratios between galaxy components
shows that they are consistent with those derived for whole galaxies, from low to high redshift, supporting the idea that also these ratios 
could be set by local processes.}
{Our results highlight the heterogeneity of galaxy properties and the importance of resolved studies on local galaxies in the context of galaxy evolution.  
They also provide fundamental observational constraints to theoretical models and updated references for high-redshift studies.}

\keywords{galaxies: ISM, galaxies: evolution, ISM: dust, extinction, ISM: atoms, ISM: molecules, ISM: abundances}

\maketitle
 
%
%________________________________________________________________

\section{Introduction}
\label{sec:intro}
The scaling relations (SRs) between main global galaxy properties are largely used tools to study the internal physics of different galaxy populations 
and their formation and evolutionary histories.  
Among the first studied relationships, there has been the Tully-Fisher relation for spiral galaxies \citep[][]{tully77}, 
an empirical relation between the mass or intrinsic luminosity of a galaxy, and its asymptotic rotation velocity or emission line width.
Another deeply studied correlation is the fundamental plane for elliptical galaxies \citep[][]{djorgovski87,jorgensen96},
a set of bivariate correlations connecting some of the galaxy properties such as the effective radius, average surface brightness, and central velocity dispersion.
Since the early 2000s, several efforts have been devoted to characterize the correlation between the stellar mass (M$_{\rm star}$) and 
star-formation rate (SFR) in galaxies, 
the so-called star-forming main sequence
\citep[SFMS or MS, e.g.,][]{brinchmann04,daddi07,elbaz07,santini09,santini17,peng10,rodighiero14,schreiber15,tomczak16}.
In the meantime, other correlations have been found between M$_{\rm star}$ and 
the average oxygen abundance, the mass-metallicity (MZ) relation 
\citep[e.g.,][]{lequeux79,garnett87,Vila-Costas92,tremonti04,erb06}, 
and between the SFR and the surface density of cold gas in disks, the Kennicutt-Schmidt (KS) star formation (SF) relation 
\citep[][]{schmidt59,schmidt63,kennicutt98a,kennicutt98b}. 
Variations of these correlations have been largely explored, in particular changing the method of derivation of galaxy properties involved
in the studied SRs.
In the recent years, the number of works dedicated to the interstellar medium (ISM) SRs has also grown, with an increasing level of refinement 
such as separating the different components of the ISM 
\citep[e.g.,][]{saintonge11,corbelli12,boselli14,cortese16,catinella18,lamperti19,lisenfeld19,casasola20,delooze20,ginolfi20,hunt20,dunne22,saintonge22}.
Some of the explored SRs (e.g., MS and MZ) have also been confirmed at various redshifts, showing a dependence on the 
cosmological time and thus allowing us to trace the evolution of galaxy properties with time \citep[e.g.,][]{dave11}. 

To intimately understand the fundamental physics underlying the global SRs, it is necessary to look at small spatial scales,
approaching the intrinsic scale of the SF process, that is the size of the giant molecular clouds 
\citep[GMCs, from a few parsecs to a few hundred parsecs,][]{solomon87}.
The GMCs, that is the major reservoirs of molecular gas, are indeed the sites of most SF in our Galaxy and other galaxies,
and their properties set the initial conditions for protostellar collapse. 
Due to the explosion in multiwavelength resolved data, made possible with 
surveys on large samples of nearby galaxies with instruments covering different spectral bands 
(e.g., 
UV: GUViCS, Boselli et al. 2011;  
Optical: MaNGA, Bundy et al. 2015; 
SAMI, Croom et. al. 2012;
VESTIGE, Boselli et al. 2018; 
NIR: S4G, Sheth et al. 2010;
FIR: KINGFISH, Kennicutt et al. 2011; Boselli et. al. 2010; 
CO: HERACLES-IRAM 30 m, Leroy et al. 2009; ALLSMOG: Cicone et al. 2017; 
COMING, Sorai et al. 2019;  
\hi: VLA THINGS survey, Walter et al. 2008; xGASS Catinella et al. 2018) 
and the detailed work of homogenization of preexistent data to collect large amounts of coherent observations
(e.g., DustPedia, Davies et al. 2017), 
it is now possible to investigate the main SRs also on the subgalactic scale. 
In the literature one refers to such subgalactic SRs, typically at kiloparsec scales, as resolved SRs
\citep[e.g.,][]{viaene14,roychowdhury15,barrera16,barrera21,canodiaz16,hsieh17,pan18,canodiaz19,lin19,
vulcani19,enia20,morselli20,ellison21,sanchez21,abdurrouf22,baker22}.
Most of these works focused on the SF relations, that is those between stellar mass surface density ($\Sigma_{\rm star}$), 
molecular gas mass surface density ($\Sigma_{\rm H2}$), and SFR surface density ($\Sigma_{\rm SFR}$).

In this paper, we study an unprecedented set of resolved SRs involving molecular, atomic and total gas, 
dust, gas-phase metallicity, stars, total baryonic content, SFR, and some ratios of these quantities in the sample of 18 nearby, face-on, spiral galaxies
presented by \citeauthor{casasola17}~(\citeyear{casasola17}, hereafter C17) and extracted from 
the DustPedia\footnote{http://dustpedia.astro.noa.gr/} sample \citep[we refer to][for a detailed description 
of the DustPedia project and sample]{davies17}. 
In C17 we have characterized the radial distribution of dust, stars, gas, and SFR of the sample, and the main result 
was that, on average, the dust-mass surface-density exponential scale length is about 1.8 times higher than the stellar scale length.   
Here we study a collection of SRs for individual galaxies in the range of physical scales from 0.3 to 3.4~kpc and for
all the sample galaxies evaluated together at the common scale of 3.4~kpc, providing useful information for those who want to study 
a given galaxy of our sample and/or need mean trends at subgalactic scales.
The galaxy sample offers the possibility to study SRs at different resolved physical scales searching for the definition of universal calibrations 
and/or of breakdown scales, if any.   
The aim of this paper is therefore to give a view that is as complete as possible of the resolved SRs in spiral galaxies of the local Universe.
The paper is organized as follows.
In Sect.~\ref{sec:sample} we outline the sample selection, in Sect.~\ref{sec:dataset} we describe the data used in this work, 
and in Sect.~\ref{sec:treatment} their treatment is provided. 
We present the results in Sects.~\ref{sec:sr} and \ref{sec:ratios}, and we discuss them in Sect.~\ref{sec:discussion}.
In Sect.~\ref{sec:conclusions} we summarize the entire work.
The paper also includes Appendices~\ref{sec:add-sample}, \ref{sec:peculiar}, and \ref{app:othercal} with additional matter.

\section{The galaxy sample}
\label{sec:sample}
We use the galaxy sample of C17 consisting of 18 DustPedia nearby, large, spiral galaxies
with a small (or moderate) disk inclination.
These galaxies have a diameter $D_{25}\geq$~7\farcm8~\footnote{$D_{25}$ is the major axis isophote at which the optical surface brightness 
falls beneath 25~mag~arcsec$^{-2}$ (we also use $r_{25} = D_{25}/2$).} and a Hubble stage T ranging from 2 to 8. 
They have been imaged over their whole extent with both PACS and SPIRE in \textit{Herschel} and 
 they have a submillimeter diameter $D_{\rm{submm}} \geq 9^{\prime}$, corresponding at least to 15 spatial resolution elements in
the SPIRE 500~$\mu$m maps (FWHM = 36\arcsec).
For this latter reason, we define these galaxies as resolved in dust emission.
Sample galaxies have M$_{\rm star}$ from  $2 \times 10^9$ to $1 \times 10^{11}$~M$_\odot$, SFR from 0.2 to 13~M$_\odot$~yr$^{-1}$, 
and oxygen abundance from 12~+~log(O/H)~=~8.3 to 8.8. 
The sample includes galaxies characterized by different peculiarities, such as the presence of bars, of signatures of interaction
with companions, and/or of an active galactic nucleus (AGN).
Tables~\ref{tab:sample} and \ref{tab:masses} summarize the main properties and global masses and SFR of galaxies, respectively. 
Galaxies are sorted in increasing distance order.
We refer to C17 for further details on the galaxy sample and its selection.

\begin{sidewaystable*}
\caption{\label{tab:sample} Main properties of galaxies.}
\centering
\begin{tabular}{lcccccccccccccccccccc}
\hline
\hline
Galaxy & $\alpha_{\rm J2000}$ $^{(1)}$ & $\delta_{\rm J2000}$ $^{(1)}$ & T $^{(1)}$ & RC3 type $^{(1)}$ & $D_{\rm 25}$ $^{(1)}$ 
& Dist. $^{(1)}$ & $i$ $^{(1)}$ 
& Nuclear $^{(1,2)}$ & 12~+~log(O/H)$_{\rm N2}$ $^{(3)}$ & O/H gradient $^{(3)}$ \\
&              &                                    &                                    & & & & & Activity & $r = 0.4r_{25}$    \\
& [$^{\rm h}$ $^{\rm m}$ $^{\rm s}$]  & [$^{\circ}$ $^{\prime}$ $^{\prime\prime}$]  & &   & [$^{\prime}$] 
& [Mpc] & [$^{\circ}$] & & & dex/($r/r_{25}$) \\
\hline
NGC~300                	& 00 54 53.4  & -37 41 03    	& 7 & SA(s)d             & 19.5  	& 2.0 	& 43.0	& -- 		& $8.432\pm0.022$ & $-0.48\pm0.08$ \\
IC~342                    	& 03 46 48.5  & +68 05 47   	& 6 & SAB(rs)cd        & 20.0  	& 3.1 	& 31.0 	& H		& $8.637\pm0.095$ & $-0.29\pm0.13$ \\
NGC~2403               	& 07 36 51.1  & +65 36 03   	& 6 & SAB(s)cd         & 20.0  	& 3.5 	& 62.9   	& L		& $8.329\pm0.026$ & $-0.02\pm0.09$ \\
NGC~3031 (M~81)  	& 09 55 33.1  & +69 03 55   	& 2 & SA(s)ab          	& 21.4    	& 3.7 	& 59.0 	& L/S1.8 	& $8.632\pm0.031$ & $-0.08\pm0.07$ \\
NGC~7793           	& 23 57 49.7  & -32 35 28     	& 8 & SA(s)d             & 10.5    	& 3.8 	& 49.6   	& H     	& $8.412\pm0.031$ & $-0.42\pm0.12$ \\
NGC~4736 (M~94) 	& 12 50 53.0  & +41 07 13   	& 2 & (R)SA(r)ab       & 7.8     	& 5.2 	& 41.4 	& S2/L   	& $8.735\pm0.078$ & $-0.09\pm0.20$ \\
NGC~6946            	& 20 34 52.2  & +60 09 14    	& 6 & SAB(rs)cd    	& 11.5    	& 5.6 	& 32.6  	& S2/H    	& $8.647\pm0.069$ & $-0.40\pm0.15$ \\
NGC~5236 (M~83) 	& 13 37 00.9  & -29 51 57   	& 5 & SAB(s)c          	& 13.5  	& 6.5 	& 24.0 	& H		& $8.735\pm0.005$ & $-0.10\pm0.01$ \\
NGC~3621             	& 11 18 16.5  & -32 48 51     	& 7 & SA(s)d             & 9.8   	& 6.9 	& 64.7 	& H   	& $8.369\pm0.024$ & $-0.18\pm0.03$ \\
NGC~5457 (M~101)	& 14 03 12.6  & +54 20 57		& 6 & SAB(rs)cd	& 24.0	& 7.0 	& 18.0 	& H		& $8.452\pm0.011$ & $-0.52\pm0.02$ \\
NGC~5194 (M~51) 	& 13 29 52.7  & +47 11 43   	& 4 & SA(s)bc pec    & 13.8    	& 7.9 	& 42.0 	& S2		& $8.728\pm0.016$ & $-0.30\pm0.18$~$^{(4)}$ \\
NGC~5055 (M~63) 	& 13 15 49.2  & +42 01 45   	& 4 & SA(rs)bc          & 11.8    	& 8.2 	& 59.0 	& H/L   	& $8.730\pm0.067$ & $-0.22\pm0.17$ \\
NGC~925               	& 02 27 16.5  & +33 34 44   	& 7 & SAB(s)d          	& 10.7  	& 8.6 	& 66.0 	& H  		& $8.338\pm0.014$ & $-0.30\pm0.05$ \\
NGC~628 (M~74)  	& 01 36 41.8  & +15 47 00   	& 5 & SA(s)c            	& 10.0    	& 9.0 	& 7.0 	& H   	& $8.605\pm0.008$ & $-0.39\pm0.00$ \\
NGC~3521             	& 11 05 48.6  & -00 02 09    	& 4 & SAB(rs)bc        & 8.3   	& 12.0 	& 72.7 	& H/L   	& $8.649\pm0.060$ & $-0.10\pm0.20$ \\
NGC~4725            	& 12 50 26.6  & +25 30 03   	& 2 & SAB(r)ab pec   & 9.8     	& 13.6 	& 54.0 	& S2    	& -- & -- \\
NGC~1365            	& 03 33 36.4  & -36 08 25    	& 3 & SB(s)b             & 12.0    	& 17.7 	& 40.0 	& S1.8    	& $8.634\pm0.015$ & $-0.26\pm0.06$ \\
NGC~1097            	& 02 46 19.0  & -30 16 30    	& 3 & SB(s)b             & 10.5    	& 19.6 	& 46.0 	& L     	& $8.803\pm0.062$ & -- \\
\hline
\hline
\end{tabular}
\tablefoot{
$^{(1)}$ See C17 for source references of the collected galaxy properties.
$^{(2)}$ Classification of nuclear activity:  H = H~{\sc ii} nucleus, S = Seyfert nucleus, and L = LINER.
$^{(3)}$ Oxygen abundances at $r = 0.4r_{25}$ and metallicity gradients from the empirical calibration N2 of 
\citet{pettini04}.
$^{(4)}$ Value of the metallicity gradient from \citet{croxall15}.
For details on notes $^{(3)}$ and $^{(4)}$ see Sect.~\ref{sec:metallicity}. 
}
\end{sidewaystable*}

\begin{table*}
\caption{\label{tab:masses} Global masses and SFR of galaxies.}
\centering
\begin{tabular}{lccccccccccccccccccccc}
\hline
\hline
Galaxy 				& log(M$_{\rm star}$)$^{(1)}$ 	& log(M$_{\rm H2}$)$^{(2)}$		& log(M$_{\rm HI})$$^{(3)}$ 		& log(M$_{\rm dust}$)$^{(1)}$ 	&  SFR$^{(1)}$ \\
					& [M$_{\odot}$] 			& [M$_{\odot}$] 				& [M$_{\odot}$]			& [M$_{\odot}$]			& [M$_{\odot}$~yr$^{-1}$] \\
\hline
NGC~300				& $9.34 \pm 0.18$			& --							& $9.28 \pm 0.03$ 		& $6.70 \pm 0.18$		& $0.20 \pm 0.03$ \\
IC~342        			& $10.31 \pm 0.09$			& $9.70 \pm 0.14$ 				& 10.02 				& $7.35 \pm 0.04$ 		& $4.03 \pm 0.26$ \\
NGC~2403			& $9.47 \pm 0.07$ 			& $8.25 \pm 0.14$				& $9.41 \pm 0.02$ 		& $6.64 \pm 0.04$ 		& $0.73 \pm 0.05$ \\
NGC~3031 (M~81)  	  	& $10.65 \pm 0.06$			& $8.12 \pm 0.13$				& $9.55 \pm 0.02$ 		& $7.02 \pm 0.07$ 		& $0.35 \pm 0.13$ \\
NGC~7793           		& $9.29 \pm 0.13$ 			& $8.10 \pm 0.15$				& $8.91 \pm 0.02$		& $6.56 \pm 0.04$		& $0.46 \pm 0.03$ \\
NGC~4736 (M~94) 	 	& $10.39 \pm 0.07$			& $8.61 \pm 0.14$				& $8.55 \pm 0.02$ 		& $6.39 \pm 0.03$ 		& $0.55 \pm 0.19$\\
NGC~6946            		& $10.43 \pm 0.14$			& $9.94 \pm 0.14$				& $9.73 \pm 0.02$ 		& $7.48 \pm 0.08$ 		& $7.07 \pm 0.56$ \\
NGC~5236 (M~83) 	 	& $10.50 \pm 0.11$			& $9.85 \pm 0.14$				& $9.31 \pm 0.02$ 		& $7.30 \pm 0.12$ 		& $6.66 \pm 1.09$  \\
NGC~3621             	  	& $10.31 \pm 0.12$			& $8.34 \pm 0.13$$^{(5)}$			& $9.86 \pm 0.02$ 		& $7.03 \pm 0.05$ 		& $1.42 \pm 0.23$ \\
NGC~5457 (M~101)		& $10.15 \pm 0.06$			& $9.55 \pm 0.13$				& $10.12 \pm 0.02$ 		& $7.67 \pm 0.05$ 		& $4.75 \pm 0.24$ \\
NGC~5194 (M~51) 		& --$^{(4)}$				& $9.55 \pm 0.13$$^{(5)}$			& 9.55				& --$^{(4)}$			& --$^{(4)}$  \\
NGC~5055 (M~63) 		& $10.77 \pm 0.08$ 			& $9.62 \pm 0.14$				& $9.86 \pm 0.02$		& $7.65 \pm 0.08$		& $2.46 \pm 0.74$  \\
NGC~925               		& $9.73 \pm 0.12$ 			&$<8.14$						& $9.72 \pm 0.00$		& $6.97 \pm 0.04$ 		& $1.15 \pm 0.07$ \\
NGC~628 (M~74)  		& $10.15 \pm 0.07$			& $8.91 \pm 0.14$				& $10.01 \pm 0.00$ 		& $7.58 \pm 0.14$ 		& $2.41 \pm 0.45$  \\
NGC~3521             		& $10.93 \pm 0.08$			& $9.53 \pm 0.15$				& $9.93 \pm 0.00$ 		& $7.73 \pm 0.03$ 		& $3.15 \pm 1.26$  \\
NGC~4725            		& $10.87 \pm 0.07$			& $8.74 \pm 0.51$				& $9.60 \pm 0.00$ 		& $7.39 \pm 0.05$ 		& $0.93 \pm 0.11$  \\
NGC~1365            		& $10.92 \pm 0.14$ 			& $10.00 \pm 0.13$				& $9.98 \pm 0.03$		& $8.00 \pm 0.15$ 		& $12.97 \pm 4.97$  \\
NGC~1097            		& $10.99 \pm 0.11$ 			& $9.79 \pm 0.13$				& $9.91 \pm 0.07$		& $7.57 \pm 0.10$ 		& $6.56 \pm 1.05$  \\
\hline
\hline
\end{tabular}
\tablefoot{
$^{(1)}$ M$_{\rm star}$, M$_{\rm dust}$, and SFR values from \citet{bianchi18} and \citet{nersesian19} obtained through the modeling of 
the galaxy SED with CIGALE code \citep[][]{boquien19}.
$^{(2)}$ M$_{\rm H2}$ values from C20 derived within $r_{25}$ under the assumption of a constant $X_{\rm CO}$ \citep[][]{bolatto13}.
$^{(3)}$ M$_{\rm HI}$ values from \citet{devis19}.
$^{(4)}$ NGC~5194 lacks M$_{\rm star}$, M$_{\rm dust}$, and SFR values because it has not been fitted by CIGALE.
$^{(5)}$ M$_{\rm H2}$ values derived from maps used in this work under the same assumptions of C20.}
\end{table*}

\section{The dataset}
\label{sec:dataset}
In this section we present the set of maps used for this study.
Most of these maps come from the dataset of C17 and, when available, new gas maps have been added.
We use mass surface density maps of various galactic components 
(dust, molecular, atomic and total gas, stars, and total baryonic content) and of the SFR.
All maps are corrected for inclination $i$ (see Table~\ref{tab:sample}). 
We focus on the region within the optical disk ($r \leq r_{25}$) for all maps of all galaxies.
This represents a sort of normalization parameter, already used in the literature \citep[e.g.,][]{casasola20,morselli20}.
\subsection{The dust and stellar mass and SFR maps} 
\label{sec:dust}
We use dust mass, stellar mass and SFR surface density maps produced for the analysis performed in C17.
Details on the derivation of these maps
are given in C17.
Here we recall their main steps.
The dust mass surface density has been derived by comparing a modeled SED, convolved with the \textit{Herschel} filter response functions,
with the observed data at each position within a galaxy.
We adopted the optical properties and grain size distributions of the THEMIS\footnote{The 
Heterogeneous Evolution Model for Interstellar Solids, http://www.ias.u-psud.fr/themis/index.html} 
dust model, as described in \citet{jones13} and its updates \citep{kohler14,ysard15,jones17}.

We use the maps of the stellar mass surface density derived 
by means of the 
\textit{Spitzer}-IRAC 3.6 and 4.5~$\mu$m images and following the prescription of \citet{querejeta15}. 
This method has the advantage of isolating
the old stellar light from contaminant emission 
(e.g., hot dust and the 3.3~$\mu$m polycyclic aromatic hydrocarbon feature) in the 3.6 
and 4.5~$\mu$m bands, that is able to significantly contribute to the 3.6~$\mu$m flux 
\citep[see ][]{meidt12}.

We use the SFR surface density maps derived from GALEX-FUV emission corrected for dust extinction using WISE 22~$\mu$m emission
according to the calibration of \citet{bigiel08}, by replacing the 24~$\mu$m intensity with the 22~$\mu$m 
emission.
This calibration is based on the stellar initial mass function of \citet{calzetti07}, consisting of two power laws, with slope of $-1.3$ in the range $0.1-0.5$~M$_\odot$ and slope of $-2.3$ in the range $0.5-120$~M~$_\odot$ 
\citep[see also][]{leitherer99}.
This combined (GALEX-FUV~+~ WISE 22~$\mu$m) SFR calibration is widely used in the literature \citep[e.g.,][]{cortese12,huang15,muraoka19,yajima21}.

The maps of dust and stellar mass surface density and of SFR surface density
are available for the entire sample. 
The uncertainties on these maps take into account uncertainties associated with the methods used for their derivation.

\begin{table*}
\tiny
\caption{\label{tab:maps} 
Gas maps used in this work.}
\centering
\begin{tabular}{lllcllc}
\hline\hline
Galaxy          			& Ref. $^{12}$CO 	& Instrument $^{12}$CO (FWHM)  		& $^{12}$CO line & Ref. \hi\	& Instrument \hi\ (FWHM) & Total gas maps    \\
\hline
NGC~300                         	& 	--				& --							& --				& -- 		& --				&  \\
IC~342                            	&      C17       			& NRO 45m (15\arcsec)            		& (1--0)			& C17     	& VLA (38\arcsec)	&  $\checkmark$\\
NGC~2403                       	&      C17				& IRAM 30~m (11\arcsec)         		& (2--1)			& C17	& VLA (6\arcsec) 	&  $\checkmark$         \\
NGC~3031 (M~81)           	&      C17  $^{(1)}$	         & BIMA (6\arcsec)                     	& (1--0)			& C17	& VLA (6\arcsec) 	&  $^{(6)}$ \\
NGC~7793                        &       --        			& --							& --				& C17	& VLA (6\arcsec)	& \\
NGC~4736 (M~94)          	&      C17       			& IRAM 30~m (11\arcsec)                 	& (2--1) 			& C17 	& VLA (6\arcsec)	&  $\checkmark$\\
NGC~6946                        &      C17       			& IRAM 30~m (11\arcsec)                 	& (2--1)   			& C17	& VLA (6\arcsec)	&  $\checkmark$\\
NGC~5236 (M~83)          	&      This work $^{(2)}$	& NRO 45m (15\arcsec)			& (1--0)   			& C17 	& VLA (6\arcsec)	&  $\checkmark$\\
NGC~3621                        &	This work $^{(3)}$     & ALMA (15\arcsec)				& (2--1) 			& C17 	& VLA (6\arcsec)	& $\checkmark$\\
NGC~5457 (M~101)		&      C17         			& IRAM 30~m (11\arcsec)         		& (2--1)			& C17 	& VLA (6\arcsec) 	&  $\checkmark$        \\
NGC~5194 (M~51)          	&      C17       			& IRAM 30~m (11\arcsec)       		& (2--1)         		& C17	& VLA (6\arcsec) 	& $\checkmark$  \\
NGC~5055 (M~63)           	&      C17     			& IRAM 30~m (11\arcsec)                 	& (2--1)   			& C17	& VLA (6\arcsec)	&  $\checkmark$\\
NGC~925                          &      C17       			& IRAM 30~m (11\arcsec)                 	& (2--1)   			& C17	& VLA (6\arcsec)	&  $\checkmark$\\
NGC~628 (M~74)             	& 	C17       			& IRAM 30~m (11\arcsec)                 	& (2--1) 			& C17	& VLA (6\arcsec)	&  $\checkmark$\\
NGC~3521                        &     This work $^{(4)}$      & NRO 45m (15\arcsec)            	& (1--0) 			& C17	& VLA (6\arcsec)	&  $\checkmark$\\
NGC~4725                        &	C17       			& IRAM 30~m (11\arcsec)                 	& (2--1)			& C17       & WSRT (13\farcs22)	&  $\checkmark$\\
NGC~1365                        &	This work $^{(5)}$	& ALMA (16\farcs5$\times$8\farcs1)	& (1--0) 			& -- 		& --				& \\
NGC~1097                        &      C17         			& ATFN Mopra 22~m (30\arcsec)     	& (1--0)   			& -- 		& --				& \\
\hline
\hline
\end{tabular}
\tablefoot{
$^{(1)}$~NGC~3031 has been observed but not detected (see C17 and the main text of this paper for details).
$^{(2)}$~$^{12}$CO(1--0) map of NGC~5236 from the Nobeyama CO Atlas of Nearby Spiral Galaxies survey \citep[][]{kuno07}.
$^{(3)}$~$^{12}$CO(2--1) map of NGC~3621 from the PHANGS (Physics at High Angular Resolution in Nearby GalaxieS) survey \citep[][]{leroy21b}.
$^{(4)}$~$^{12}$CO(1--0) map of NGC~3521 from the COMING survey \citep[][]{sorai19}.  
$^{(5)}$~$^{12}$CO(1--0) map of NGC~1365 from the ALMA Science Archive.
$^{(6)}$~Total gas map of NGC~3031 has been not produced (see note $^{(1)}$ of this table).
}
\end{table*}

\subsection{The gas mass maps} 
\label{sec:gas}
We use molecular (H$_2$) and atomic (\hi) gas mass surface density maps from C17 with some updates. 
Table~\ref{tab:maps} collects the gas dataset used in this work.
The $\Sigma_{\rm H2}$ maps have been derived by both $^{12}$CO(1--0) and $^{12}$CO(2--1) emission line observations, 
based on available data, under the assumption of optically thick $^{12}$CO emission.
The derived $\Sigma_{\rm H2}$ maps do not take into account the contribution of helium.
We adopt both a constant value for the CO-to-H$_{2}$ conversion factor
($X_{\rm CO} = N({\rm{H_2}})/I_{\rm {CO}}$, where $N({\rm{H_2}})$ is the molecular gas column density 
in cm$^{-2}$ and $I_{\rm {CO}}$ is the CO line intensity in K~km~s$^{-1}$) and 
a metallicity-dependent conversion factor.
We aim at exploring the effect of these two assumptions on $X_{\rm CO}$, 
which provide a conservative range of molecular gas mass estimations, on the spatially resolved SRs.
We adopt the constant value of $X_{\rm CO} = 2.0 \times 10^{20}$ cm$^{-2}$ (K~km~s$^{-1}$)$^{-1}$ with $\pm$30$\%$ uncertainty \citep[e.g.,][]{bolatto13},
corresponding to a CO-to-H$_{2}$ conversion factor expressed in terms of 
$\alpha_{CO}$ ($\alpha_{CO} = {\rm M(H_{2})}/{\rm L_{CO}}$, where ${\rm M(H_{2})}$ is the H$_{2}$ mass in M$_\odot$ and ${\rm L_{CO}}$ 
the CO line luminosity in K~km~s$^{-1}$~pc$^{2}$) of $\alpha_{CO} = 3.2\,{\rm M}_\odot\,{\rm pc}^{-2}$~(K~km~s$^{-1}$)$^{-1}$ 
\citep[][]{narayanan12}. 
We assume the calibration for the metallicity-dependent $X_{\rm CO}$ by \citet{amorin16}:
${\rm log}(\alpha_{\rm CO}) = 0.68 - 1.45[12 + {\rm log(O/H} - 8.7)]$, corresponding to $X_{\rm CO} \propto (Z/Z_\odot)^{-1.5}$.
This calibration provides a value of $\alpha_{\rm CO}$ including the contribution of  helium, so we correct  
the relation of \citet{amorin16} to be consistent with our measurements of H$_2$ mass.
The calibration of \citet{amorin16} is derived combining low-metallicity starburst galaxies with more metal-rich galaxy objects, including the Milky Way and 
Local Volume galaxies from \citet{leroy11}, and is in agreement with other observational determinations \citep[e.g.,][]{genzel12,schruba12} 
and model predictions \citep[e.g.,][]{wolfire10}.    
The calibration of \citet{amorin16} is also consistent with the more recent determination by \citet{hunt20} of a 
$X_{\rm CO} \propto (Z/Z_\odot)^{-1.55}$ in local star-forming galaxies.
The adopted calibrations of $X_{\rm CO}$ by \citet{bolatto13} and \citet{amorin16} are
the same ones used in the study of the global ISM SRs for DustPedia late-type galaxies
of \citet[][hereafter C20]{casasola20}, allowing therefore a consistent comparison of the SRs between small and global scales.  
In the case of the use of $^{12}$CO(2--1) observations, we assumed a $^{12}$CO line ratio $R_{21} = I_{\rm CO(2-1)}/I_{\rm CO(1-0)}$ = 0.7, 
a typical value for nearby spiral galaxies \citep[e.g.,][]{schruba11,leroy13,casasola15,yajima21}.
This value of $R_{21}$ is also consistent within the errors with that recently determined by \citet{denbrok21},
$R_{21} = 0.64 \pm 0.09$.

With respect to the CO data used in C17, we added the following:
\textit{i)} the $^{12}$CO(1--0) map of NGC~5236 obtained with the Nobeyama 45-m radio telescope 
and available thanks to the Nobeyama CO Atlas of Nearby Spiral Galaxies survey \citep[][]{kuno07};
\textit{ii)} the $^{12}$CO(2--1) map of NGC~3621 obtained with ALMA and available thanks to the PHANGS 
survey\footnote{https://sites.google.com/view/phangs/home} \citep[][]{leroy21b}; 
\textit{iii)} the $^{12}$CO(1--0) map of NGC~3521 obtained with the Nobeyama 45-m radio telescope and available 
thanks to the COMING survey \citep[][]{sorai19};
\textit{iv)} the $^{12}$CO(1--0) map for NGC~1365 obtained with ALMA and available in the ALMA Science Archive\footnote{https://almascience.eso.org/asax/}.
In particular, we replace the IRAM--30m $^{12}$CO(2--1) map of NGC~3521 used in C17 with $^{12}$CO(1--0) observations mentioned above.
This substitution is dictated by the fact that the use of the $^{12}$CO(1--0) emission line eliminates the uncertainty on the $R_{21}$ 
line ratio in the H$_2$ mass derivation and the COMING map, differently by that used in C17, covers the entire optical radius of NGC~3521.
As regards NGC~3621, we point out that the PHANGS survey provides $^{12}$CO(2--1) maps convolved to angular resolutions of 2\arcsec, 7\farcs5, 
11\arcsec, and 15\arcsec, in addition to the native resolution of $\sim$1\arcsec.
Since we compare maps at 36\arcsec\ resolution, we use the $^{12}$CO(2--1) PHANGS map of NGC~3621 at 15\arcsec\ resolution.
We use the integrated intensity map obtained by the high completeness ``broad'' masking scheme, 
optimized to include all emission from the galaxy \citep[for details on PHANGS-ALMA Data Processing and Pipeline see][]{leroy21a,leroy21b}. 
The new $^{12}$CO data adopted in this work are treated following the same prescriptions described above and in C17 in order to derive
$\Sigma_{\rm H2}$ maps.
Combining all $^{12}$CO data, $\Sigma_{\rm H2}$ maps are available for 15/18 sample galaxies 
(they would be 16/18 but NGC~3031 is not detected in CO, see Table~\ref{tab:maps}).
The uncertainties on $\Sigma_{\rm H2}$ maps are calculated as the quadrature sum of the uncertainty on the $^{12}$CO flux
and on the $X_{\rm CO}$ conversion factor. 
Under the assumption of the metallicity-dependent $X_{\rm CO}$, uncertainties in $\Sigma_{\rm H2}$ take also into account uncertainties 
on the metallicity maps (see Sect.~\ref{sec:metallicity}) and on the calibration of $X_{\rm CO}$ with the metallicity.

We use maps of the atomic gas mass surface density ($\Sigma\rm_{HI}$) from C17, derived  from \hi\ 21~cm line intensity images
under the assumption of optically thin \hi\ emission.
The produced $\Sigma_{\rm HI}$ maps do not take into account the contribution of the helium.
The $\Sigma_{\rm HI}$ maps are available for 15/18 sample galaxies. 
The uncertainties on $\Sigma_{\rm HI}$ maps take into account uncertainties associated to \hi\ observations.

Combining available data on molecular and atomic gas, we have produced total gas-mass surface-density maps 
summing atomic and molecular gas masses and correcting this sum for the helium contribution by multiplying by a factor of 1.36 
\citep[see][]{schruba11}, $\Sigma_{\rm{tot\,gas}}$ = 1.36~$\times$~($\Sigma\rm_{HI}$~+~$\Sigma\rm_{H2}$).
The $\Sigma_{\rm{tot\,gas}}$ maps are available for 13/18 sample galaxies. 

\subsection{The total baryonic content maps} 
\label{sec:bar}
We produce total baryonic content maps summing stellar, total gas, and dust masses, $\Sigma_{\rm bar} = \Sigma_{\rm star} +  \Sigma_{\rm tot\,gas} + \Sigma_{\rm dust}$.
We stress how complete estimations of $\Sigma_{\rm bar}$ such as the ones presented here have rarely been attempted in the similar studies.
Taking into account the analysis of all galaxies evaluated together at 3.4~kpc scale, for the almost totality of the $\geq$3$\sigma$ pixels  $\Sigma_{\rm star} > \Sigma_{\rm tot\,gas}$
and all $\geq$3$\sigma$ pixels have a mean value $\Sigma_{\rm star}/\Sigma_{\rm tot\,gas} = 11.74 \pm 0.62$.
The $\Sigma_{\rm bar}$ maps are available for 13/18 sample galaxies.

\subsection{The gas-phase metallicity maps} 
\label{sec:metallicity}
We use the oxygen abundance, 12~+~log(O/H), to trace the gas-phase metallicity in our sample of galaxies.
We adopt a solar oxygen abundance of 12 + log(O/H) = $8.69 \pm 0.05$ from \citet{asplund09}.  
Among the various gas-phase metallicity determinations for DustPedia galaxies provided in \citet{devis19}, 
we select the empirical calibration N2 from \citet{pettini04}, based on the ratio of [N{\sc ii}]~$\lambda$6584$\AA$ and H$\alpha$,
since it is available for all sample galaxies, except for NGC~4725 (see Table~\ref{tab:sample}). 
The N2 metallicities at the galactocentric distance $r = 0.4r_{25}$ of our galaxy sample span from 12~+~log(O/H)$_{\rm N2} = 8.3$ to~8.8 
and the metallicity gradients from 0.0 to $-0.5$~dex~$r_{25}^{-1}$.
Azimuthally averaged metallicity maps were derived from these values, and from the galaxy inclinations and position angles used by \citet{devis19}.

We decide to discard the values of metallicity gradients derived from the N2 calibration 
of NGC~5194 and NGC~1097 because they are unreliable and quite unusual for typical nearby spiral galaxies 
\citep[e.g.,][]{magrini11,magrini16,sanchez14,belfiore17,maiolino19,yates21}.
We have checked the values of the slopes and intercepts of the metallicity gradients of these two DustPedia galaxies provided by other calibrations 
(e.g., O3N2--\citeauthor{pettini04}~et~al.~\citeyear{pettini04}, KK04--\citeauthor{kobulnicky04}~et~al.~\citeyear{kobulnicky04}, 
PG16S--\citeauthor{pilyugin16}~et~al.~\citeyear{pilyugin16}) and collected in \citet{devis19}.   
The galaxies NGC~5194 and NGC~1097 have negative, positive, or flat metallicity gradients, 
depending on the adopted strong line ratio and relative calibration.
We refer to \citet{devis19} for the discussion on the comparison between different metallicity calibrations for DustPedia galaxies
\citep[see also e.g.,][for similar comparisons]{maiolino19,yates21}.     
Fortunately, \citet{croxall15} detected the temperature-sensitive auroral lines 
[O {\sc iii}]~$\lambda$4363$\AA$, [N {\sc ii}]~$\lambda$5755$\AA$ and 
[S {\sc iii}]~$\lambda$6312$\AA$ for 28 H{\sc ii} regions of NGC~5194, 
permitting the derivation of the oxygen abundance from the so-called direct method, as part of the CHemical Abundances of Spirals project \citep[CHAOS,][]{berg15}.   
\citet{croxall15} found that the central O/H abundance of NGC~5194 is roughly solar or possibly slightly
super-solar by $10\%-30\%$, depending on the adopted solar oxygen abundance, and this galaxy has a relatively flat metallicity gradient 
with a slope of $-$0.30~dex~$r_{25}^{-1}$ (or $-$0.023~dex~kpc$^{-1}$), consistent with the findings of \citet{bresolin04}.
Since the mean O/H abundance of NGC~5194 derived from the N2 calibration is consistent with the values derived 
from direct measures (see Table~\ref{tab:sample}), we adopt the O/H gradient found by \citeauthor{croxall15}~(\citeyear{croxall15}, 
see their Eq.~(3)).   
To our knowledge, a direct measurement of the oxygen abundance for NGC~1097 is instead not available. 
For this galaxy, we prefer to use the estimate of the mean metallicity.

Although the use of the metallicity gradient instead of the mean metallicity should contribute to refine the results, 
we have to remind the readers of one main caveat in our approach.
We use metallicity maps derived from metallicity gradients instead of ``real'' and complete metallicity maps. 
In this way we have neglected the azimuthal variations that may be present at each galactocentric radius
or local chemical inhomogeneities as present, for example, in H{\sc ii} regions
\citep[e.g.,][]{rosolowsky08,sanders12,clark19}.
However, it is also worth mentioning that in massive spiral galaxies the metallicity variation is dominated by the radial gradient 
while the azimuthal scatter is marginal \citep[e.g.,][]{berg15,croxall16,kreckel19}. 
Therefore, our approach is legitimate and 
it is expected to give a good approximation of the real map.

The metallicity maps are available for 17/18 sample galaxies. 
The uncertainties on the metallicity maps are calculated as the quadrature sum of the uncertainties on the mean metallicity and 
on the metallicity gradient.

\section{Treatment of data}
\label{sec:treatment}

\subsection{Image convolution and resizing}
\label{sec:convolution}
All images are convolved at the resolution of 36\arcsec, the same of the $\Sigma_{\rm{dust}}$ maps. 
This resolution is that of the \textit{Herschel}-SPIRE 500~$\mu$m maps used in the spectral energy distribution (SED) fitting procedures in C17.  
All maps are resampled to a pixel size equal to the adopted dust resolution, 36\arcsec, that is one pixel per beam
\citep[see, e.g.,][for a similar treatment of images]{viaene14,vutisalchavakul14,casasola15,saikia20}.
This choice differs from that adopted in C17 where the maps have been resampled to a pixel size equal to 12\arcsec (1/3~$\times$~FWHM).
Studying pixel-by-pixel SRs, we prefer to assume the pixel size equivalent to the spatial resolution so that
the pixels can be considered as roughly statistically independent, and there should be little correlation among them.
In other words, each pixel is treated as a single datapoint.

We also produced an additional set of maps regridded to the common 3.4~kpc scale, that is the physical scale corresponding to 36\arcsec\ for the 
most distant galaxy in the sample (NGC~1097). 
Consistent with what is discussed above, also in this case the pixel size of all maps is resampled to the angular scale corresponding 
to 3.4~kpc for each galaxy.
The galaxies NGC~3621 and NGC~4736 are excluded from the analysis at 3.4~kpc because the convolution at this resolution produces maps of only 4 pixels size. 

With all collected and properly treated maps, we explore several resolved (pixel-by-pixel) SRs.
These procedures were performed using CASA\footnote{Common Astronomy Software Applications, \citet{McMullin07}.} (versions from 5.1.1 to 5.1.5),
IDL\footnote{http://idlastro.gsfc.nasa.gov/, Landsman (1993).},
IRAM/GILDAS\footnote{http://www.iram.fr/IRAMFR/GILDAS/, Guilloteau \& Lucas (2000).}, and
IRAF\footnote{IRAF is the Image Reduction and Analysis Facility. 
IRAF is written and supported by the National Optical Astronomy Observatories (NOAO) in Tucson, 
Arizona. NOAO is operated by the Association of Universities for Research in Astronomy (AURA), Inc. 
under cooperative agreement with the National Science Foundation.}
software packages.

\subsection{Fitting method}
\label{sec:fit}
After eliminating possible corrupted pixels and pixels with negative values, we use those with $\geq$3$\sigma$ significance within $r_{25}$ of
all galaxy maps in order to study the pixel-by-pixel SRs.
We stress that the $\geq$3$\sigma$ pixels represent $\sim$97--100$\%$ of the pixels within the optical radius of the explored galaxies. 
Given the low number of nondetections within $r_{25}$, their exclusion has no systematic impact on the inferred slopes as a function of the spatial scale, contrary to what may happen
when exploring the entire galactic extension \citep[e.g.,][]{pessa21}.

For a given SR, we fit the data in logarithmic space as follows:
\begin{eqnarray}
{\rm log}(\Sigma{_{y}}) = q + m\,\times\, {\rm log}(\Sigma_{x}),
\label{eq:fit}
\end{eqnarray}

\noindent
where $\Sigma_{x}$ and $\Sigma_{y}$ can be
$\Sigma_{\rm{HI}}$, $\Sigma_{\rm{H2}}$, $\Sigma_{\rm{tot\,gas}}$, $\Sigma_{\rm{dust}}$,  $\Sigma_{\rm{stars}}$, and $\Sigma_{\rm{bar}}$
in units of \msunpc\ and $\Sigma_{\rm{SFR}}$ in units of  \msunyrkpc,  $m$ is the slope and $q$ the intercept
of the linear fit.
For each linear fit we also provide the Pearson correlation coefficient $R$, the standard deviation of residuals (dispersion) $\sigma$,  
and the number of datapoints taken into account.

\begin{figure*}
  \centering 
  \subfloat{\includegraphics[width=0.48\textwidth]{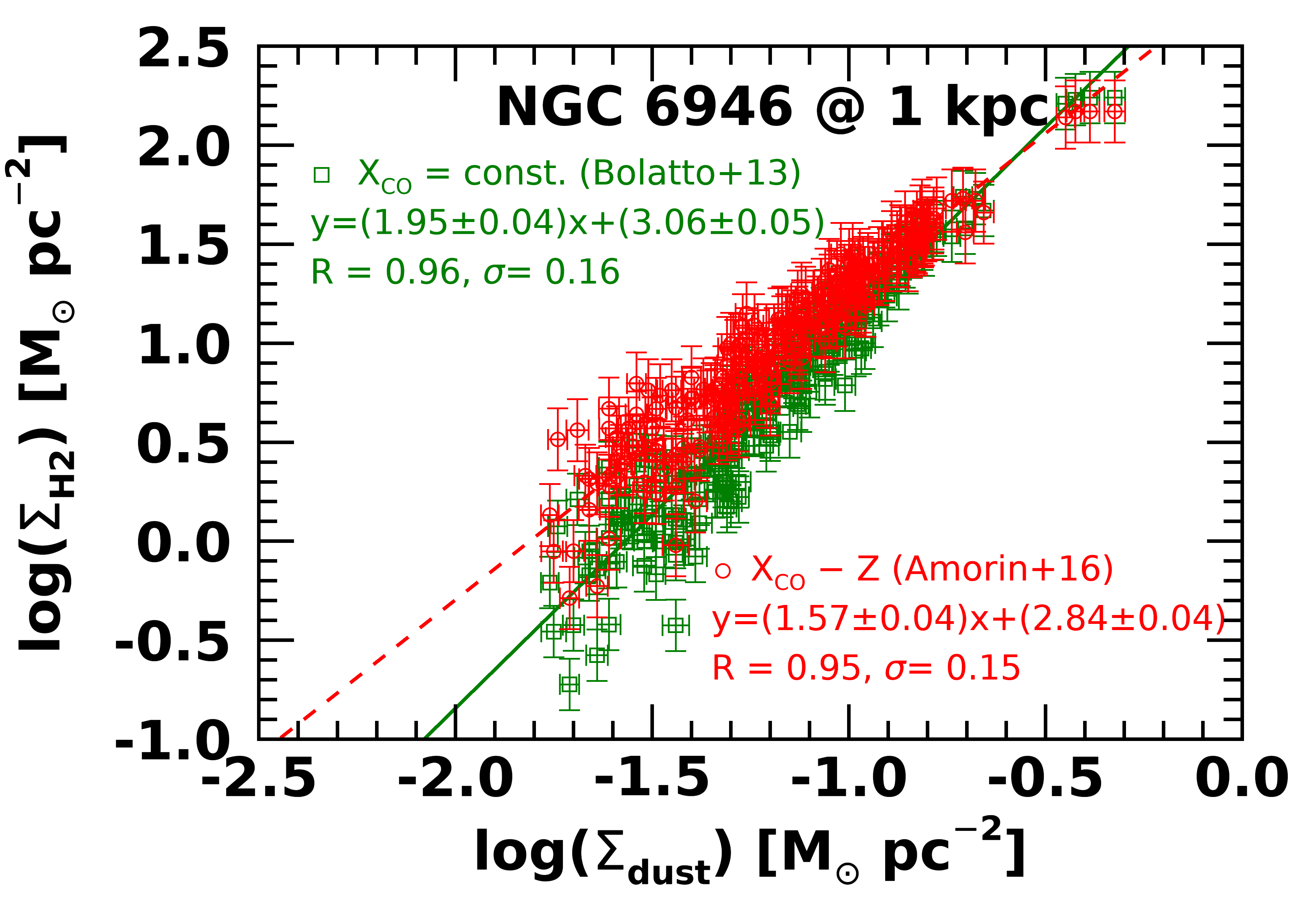}}
  \qquad 
  \subfloat{\includegraphics[width=0.48\textwidth]{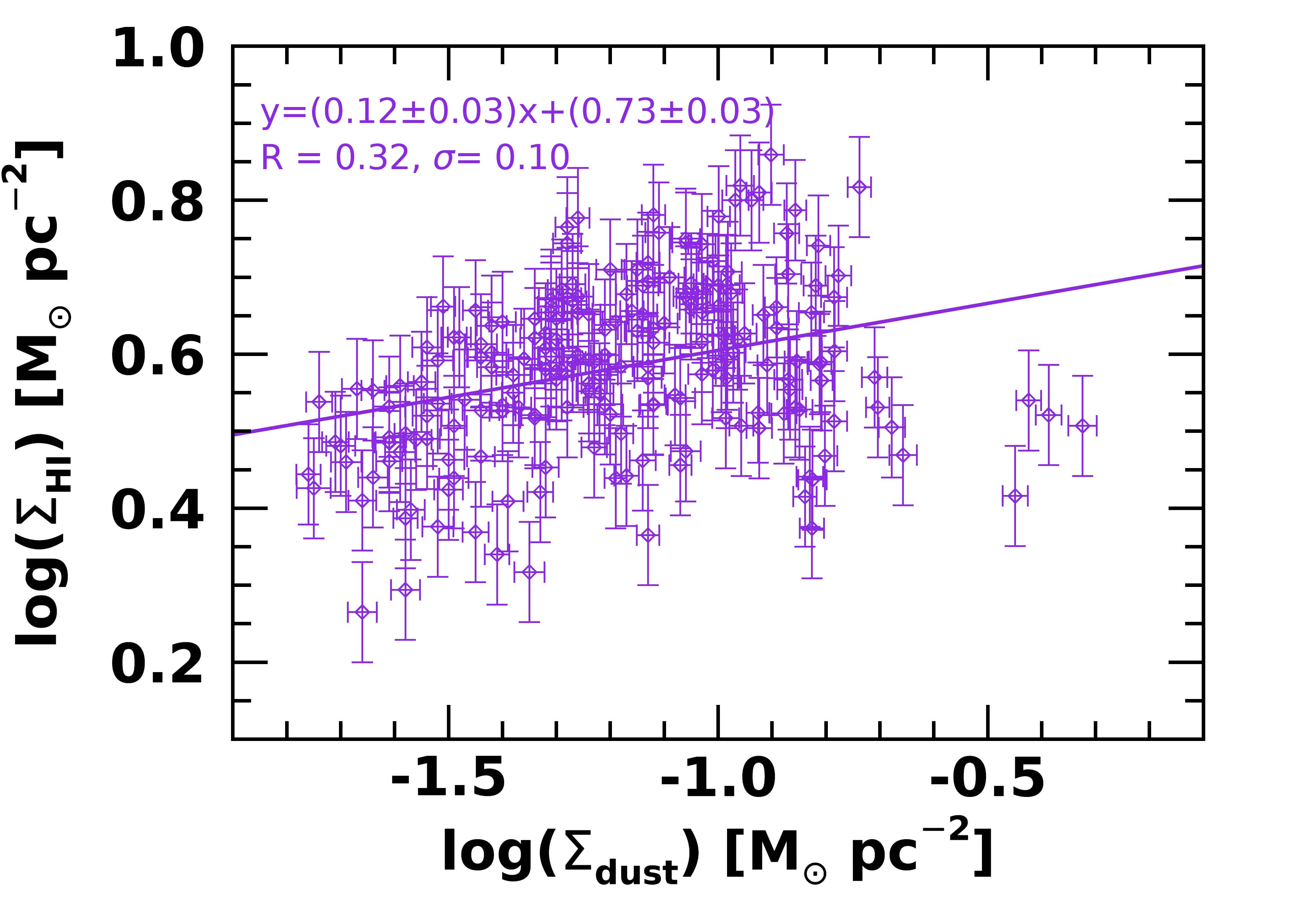}} 
   \qquad 
  \subfloat{\includegraphics[width=0.48\textwidth]{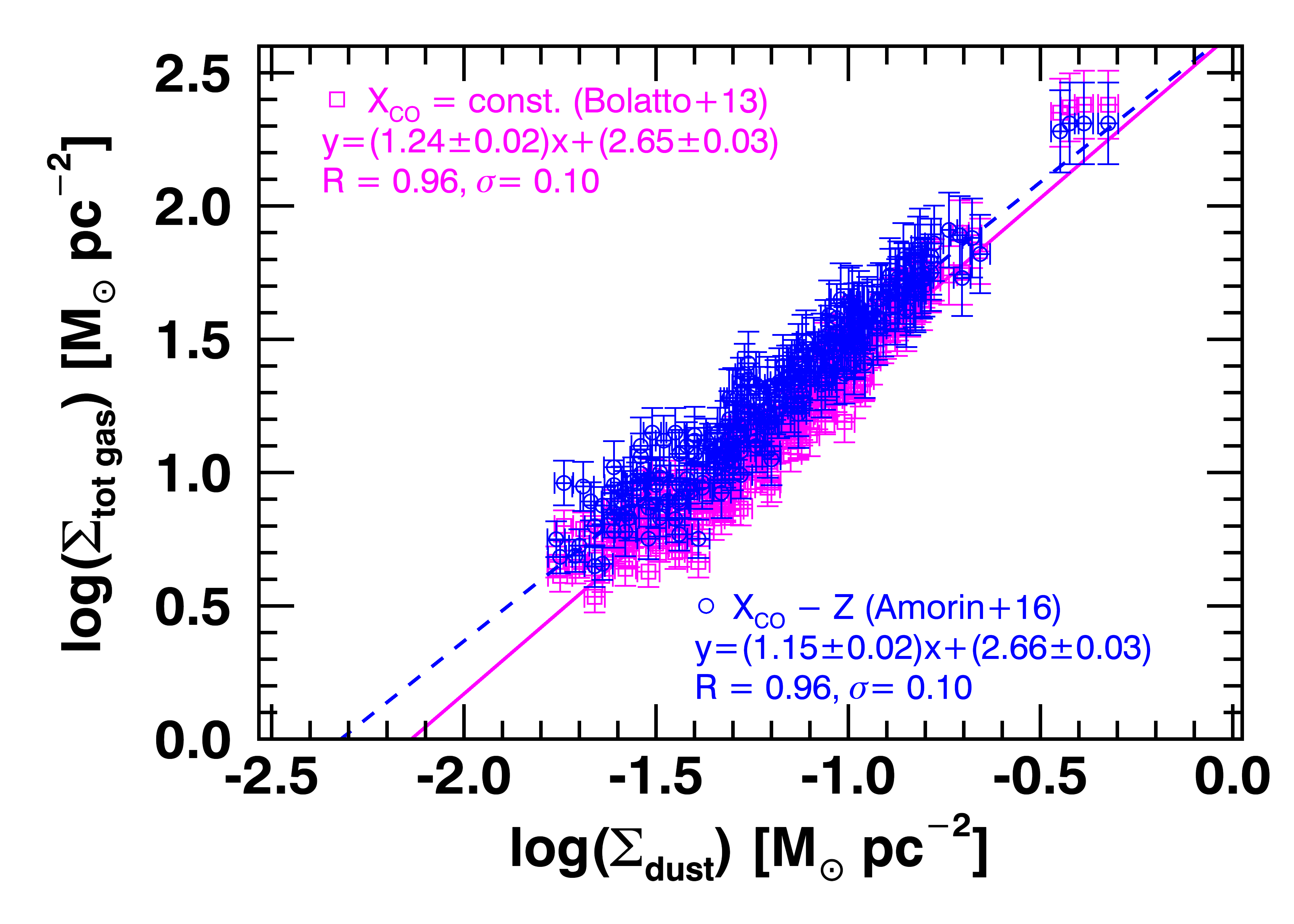}}
  \qquad 
  \subfloat{\includegraphics[width=0.48\textwidth]{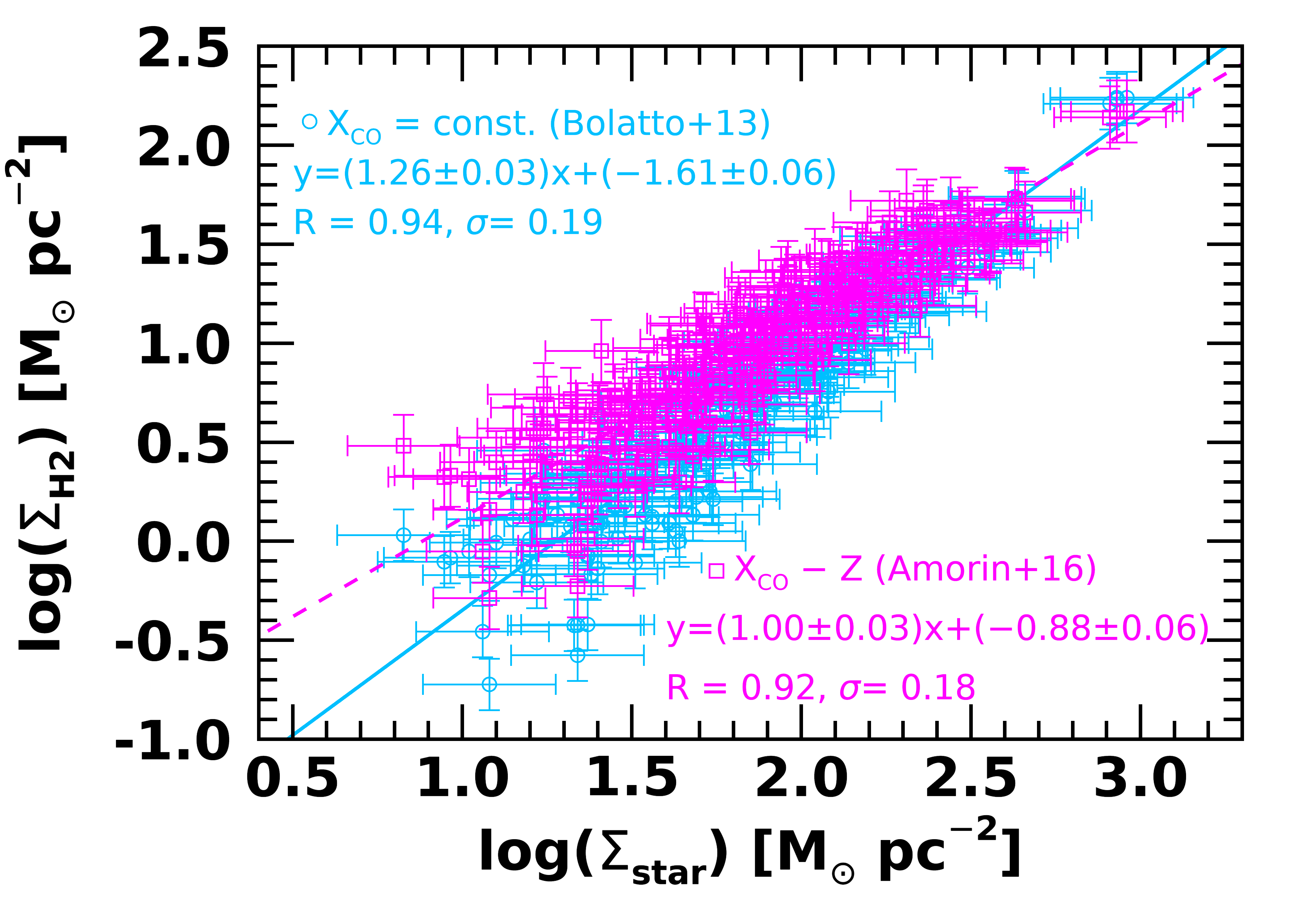}}
   \qquad 
  \subfloat{\includegraphics[width=0.48\textwidth]{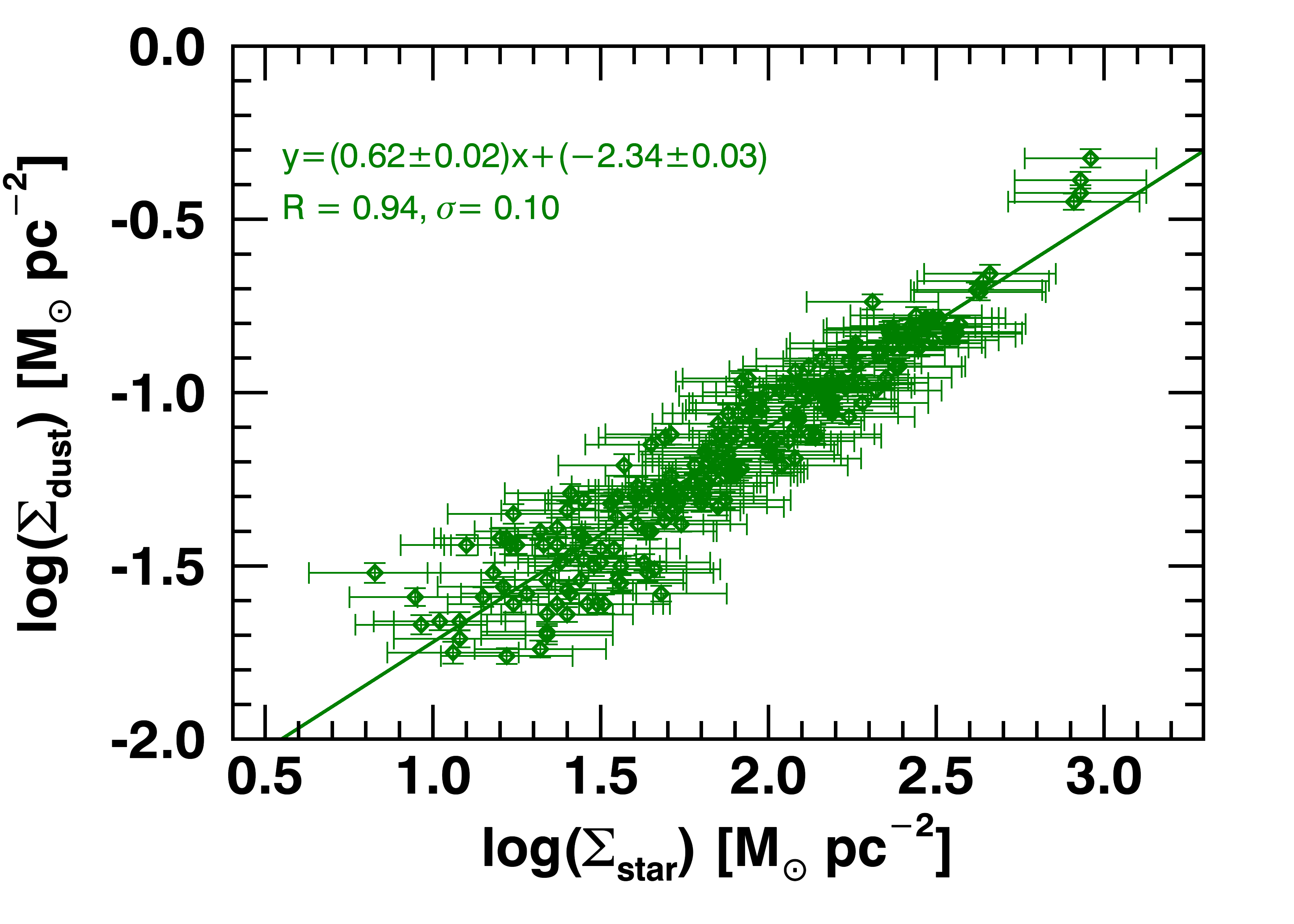}}
   \qquad
  \subfloat{\includegraphics[width=0.48\textwidth]{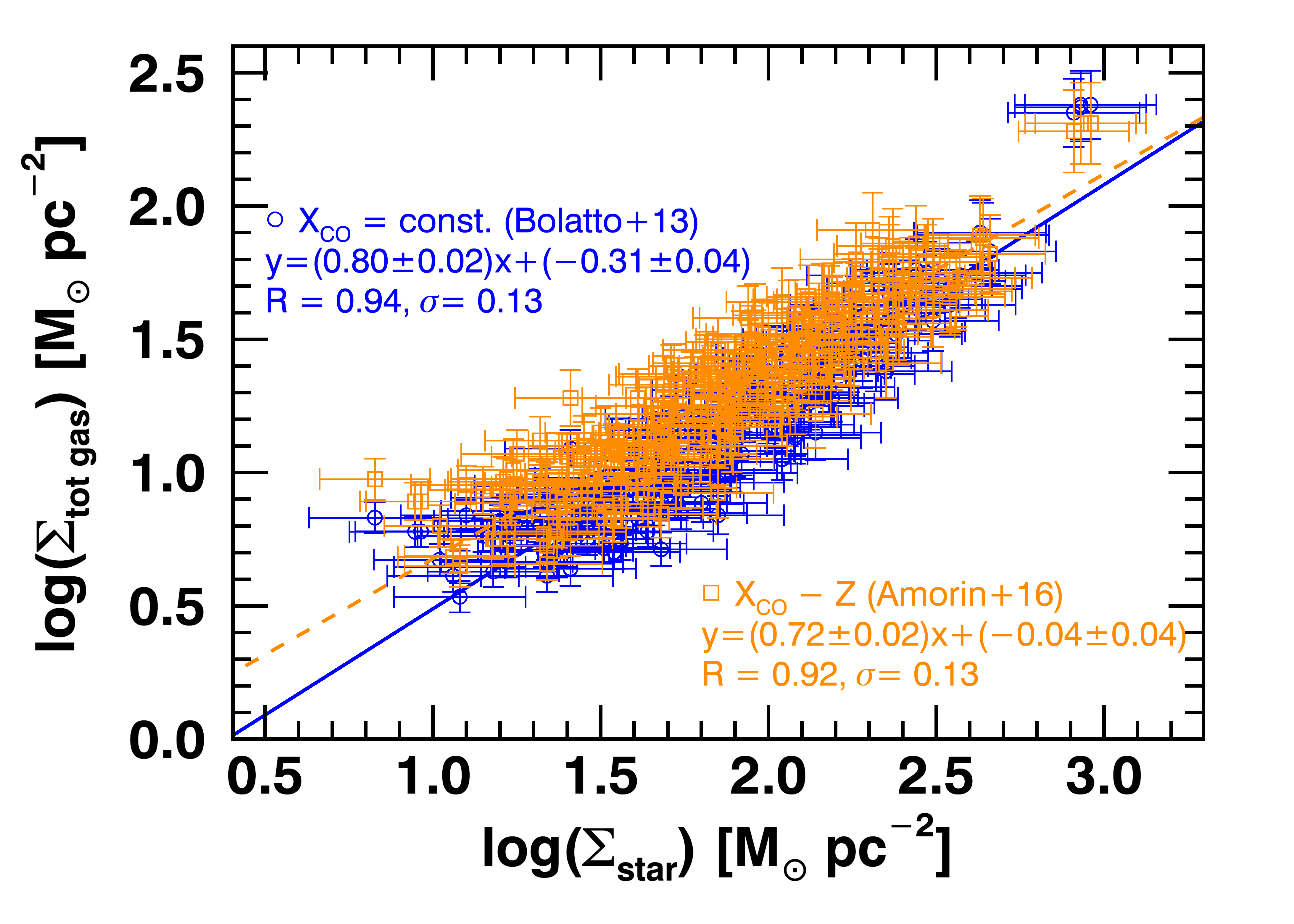}} 
  \qquad
    \subfloat{\includegraphics[width=0.48\textwidth]{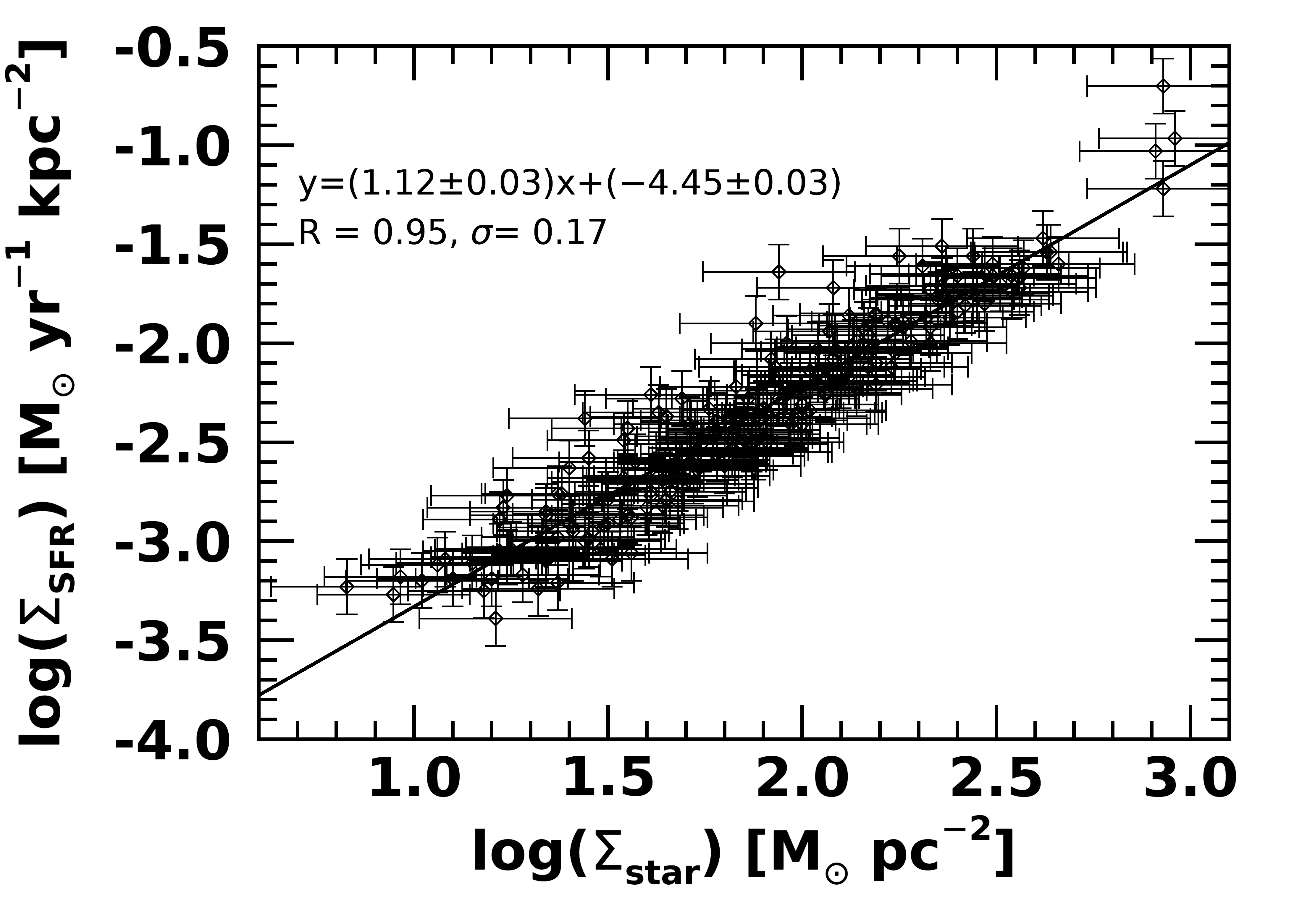}}
  \qquad 
  \subfloat{\includegraphics[width=0.48\textwidth]{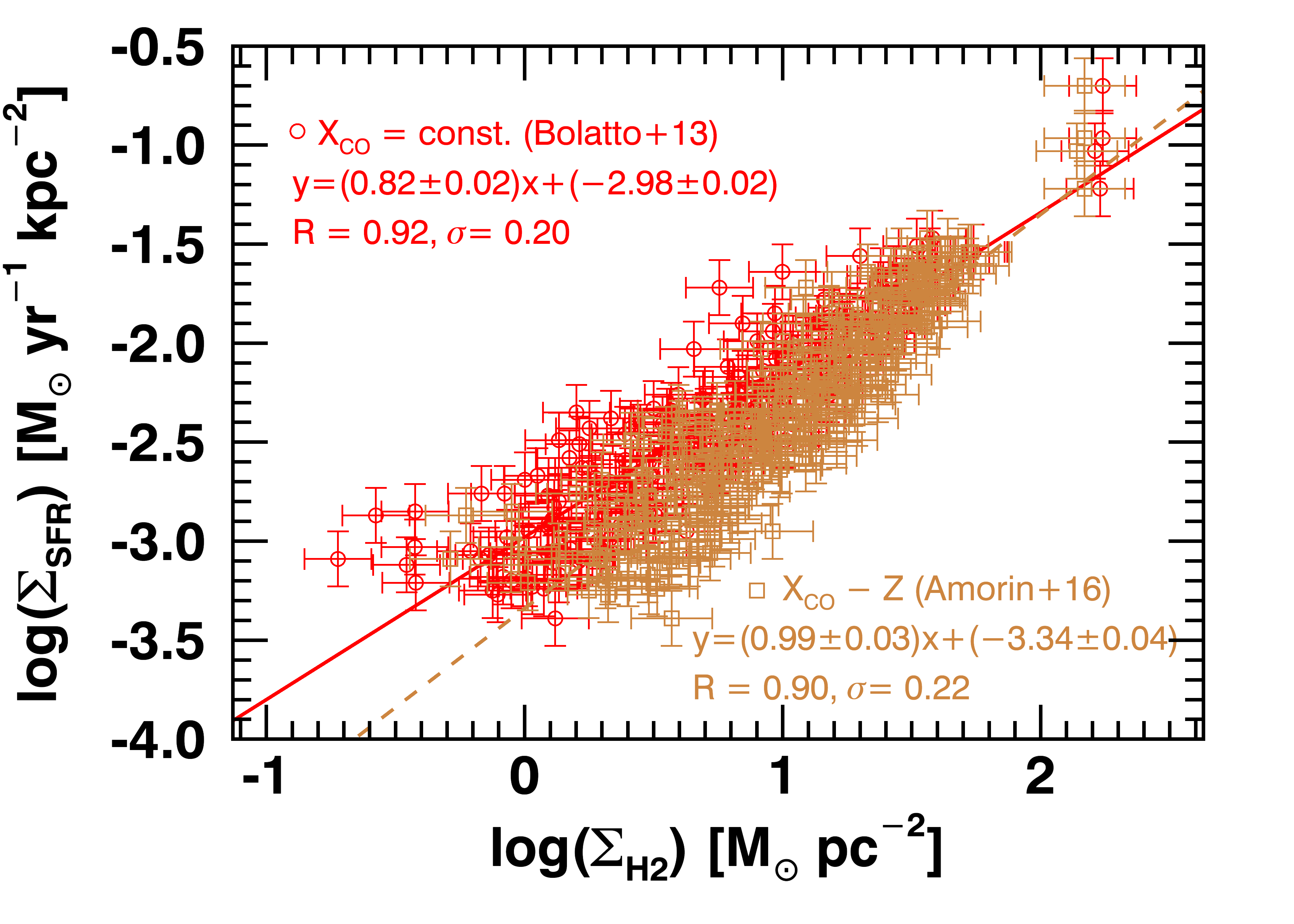}} 
\end{figure*}

\begin{figure*}
  \centering 
  \subfloat{\includegraphics[width=0.48\textwidth]{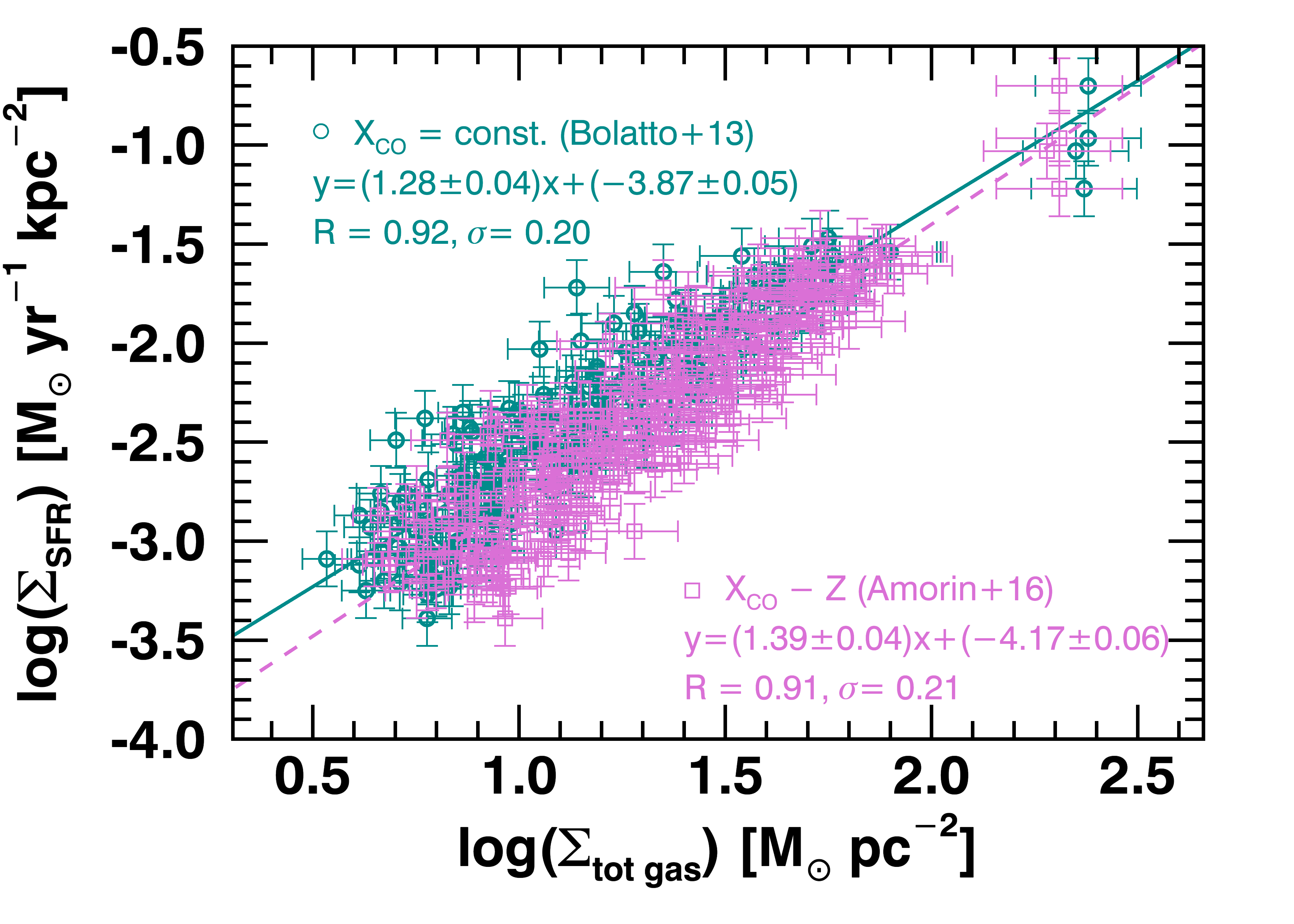}}
  \qquad 
  \subfloat{\includegraphics[width=0.48\textwidth]{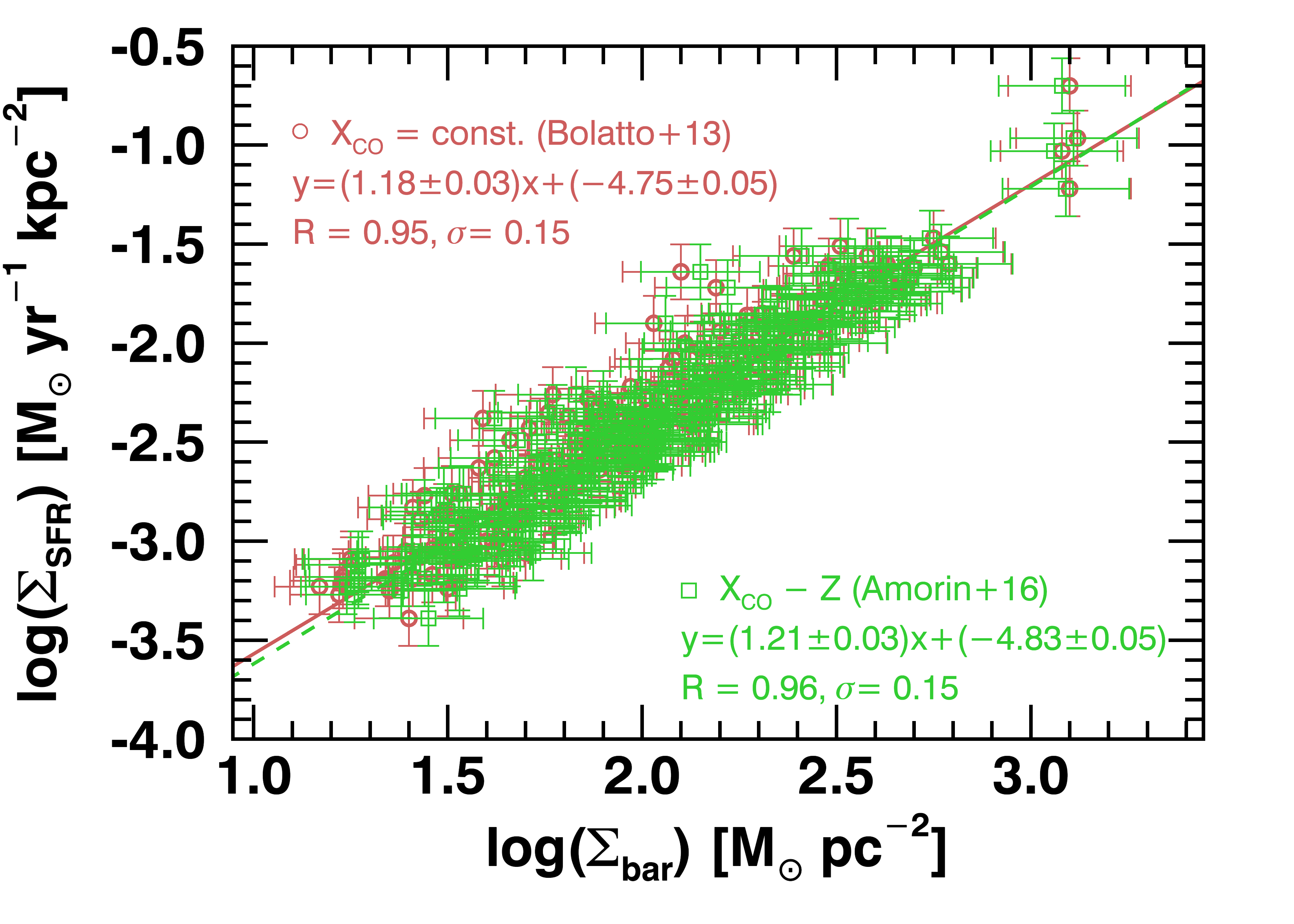}} 
\caption[]{Resolved SRs for the galaxy NGC~6946 within $r_{25}$ at the resolution of 1~kpc, in logarithmic scale.
From left to right:
$\Sigma_{\rm dust}$--$\Sigma_{\rm H2}$, $\Sigma_{\rm dust}$--$\Sigma_{\rm HI}$,  $\Sigma_{\rm dust}$--$\Sigma_{\rm tot\,gas}$,
$\Sigma_{\rm star}$--$\Sigma_{\rm H2}$, $\Sigma_{\rm star}$--$\Sigma_{\rm dust}$, $\Sigma_{\rm star}$--$\Sigma_{\rm tot\,gas}$,
$\Sigma_{\rm star}$--$\Sigma_{\rm SFR}$, $\Sigma_{\rm H2}$--$\Sigma_{\rm SFR}$, $\Sigma_{\rm tot\,gas}$--$\Sigma_{\rm SFR}$, 
$\Sigma_{\rm bar}$--$\Sigma_{\rm SFR}$.
Two sets of data are shown when the H$_2$ gas is involved according to the two assumptions on $X_{\rm CO}$ (see the legends).
The lines are the linear fits to the data: 
continuum line for the constant $X_{\rm CO}$ \citep{bolatto13}, dashed line for the metallicity-dependent $X_{\rm CO}$ \citep{amorin16}.
Fit parameters are also given in the figures.}
\label{fig:n6946}
\end{figure*} 

\begin{figure*}
  \centering 
  \subfloat{\includegraphics[width=0.48\textwidth]{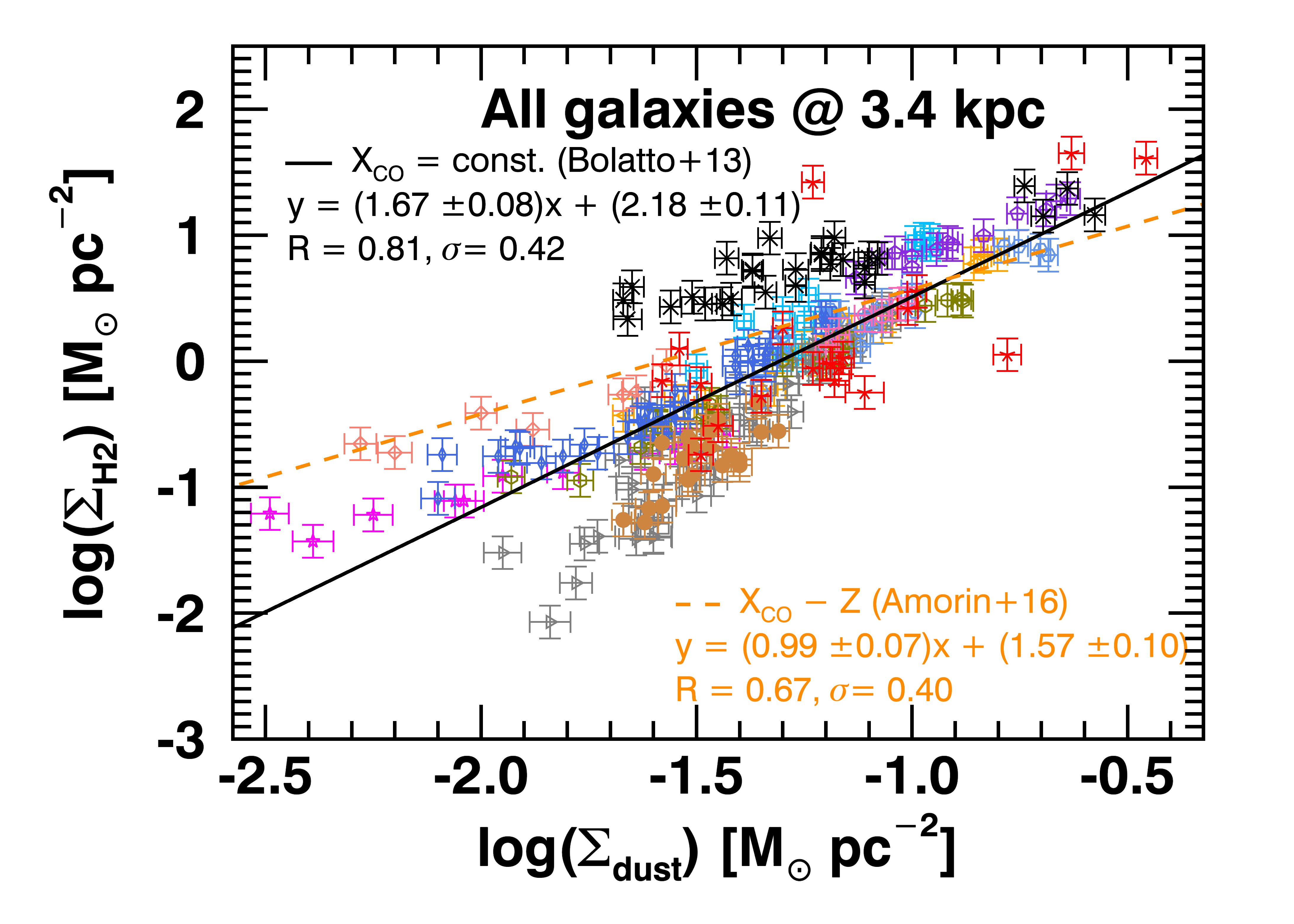}}
  \qquad 
  \subfloat{\includegraphics[width=0.48\textwidth]{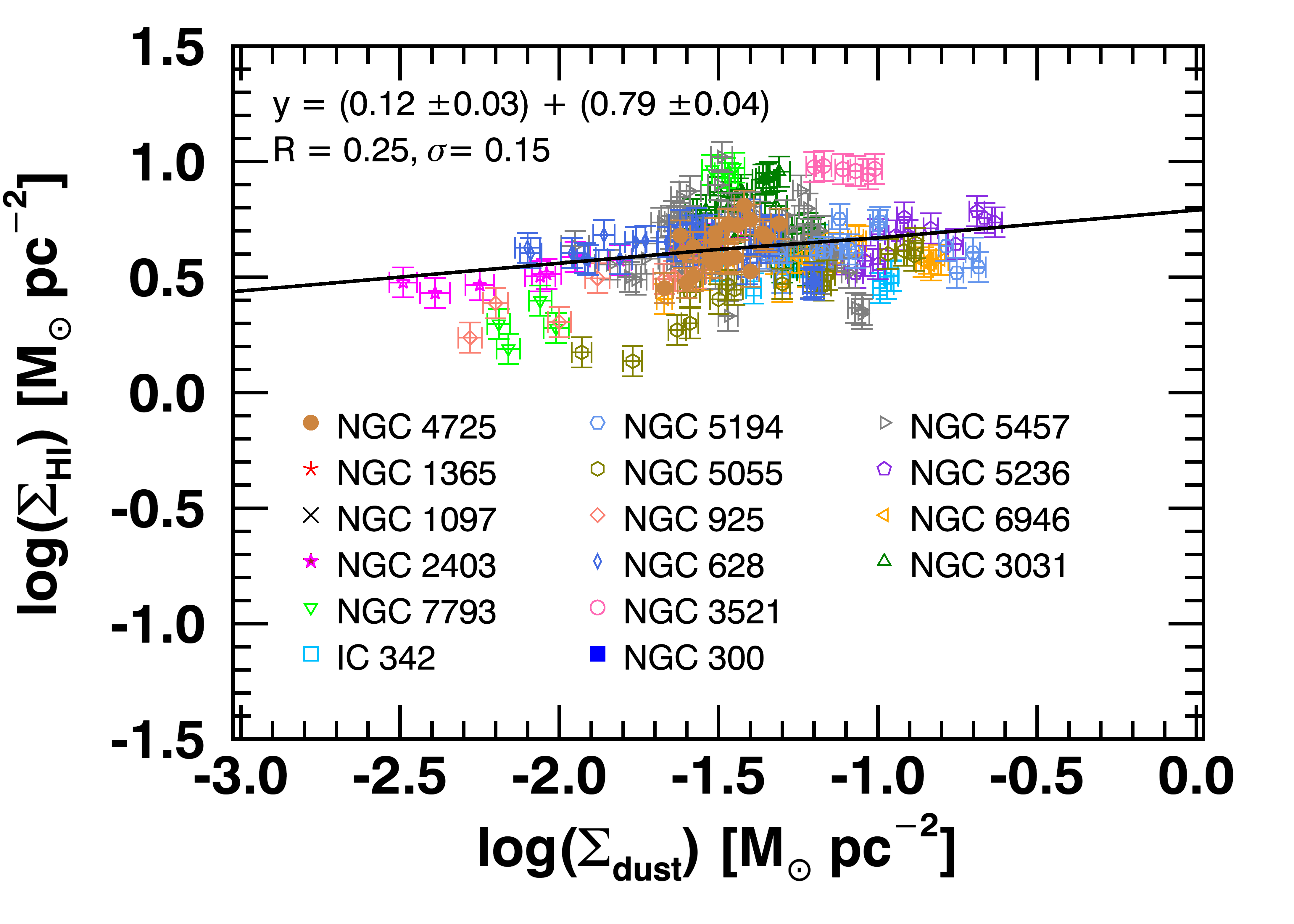}} 
   \qquad 
  \subfloat{\includegraphics[width=0.48\textwidth]{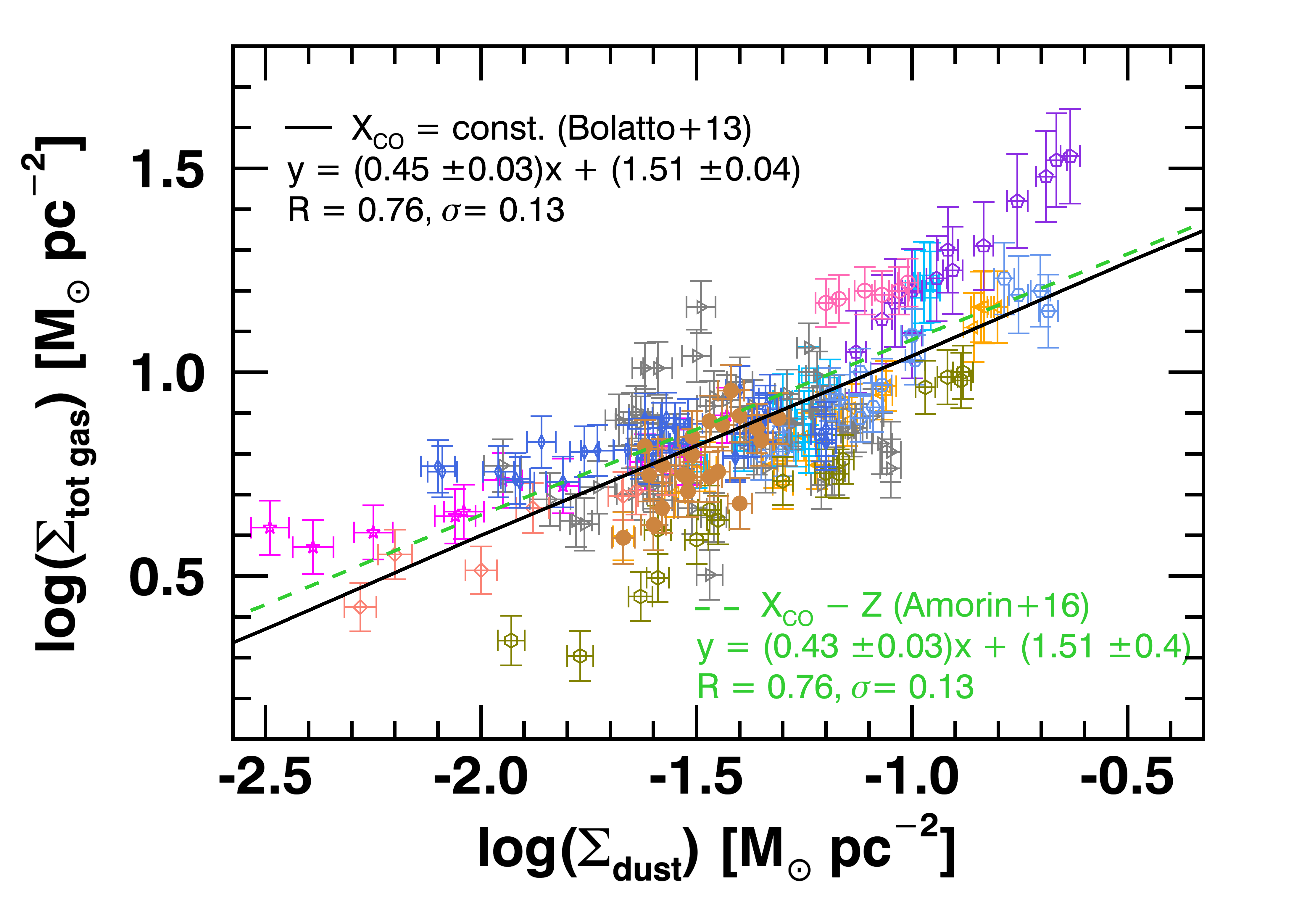}}
  \qquad 
  \subfloat{\includegraphics[width=0.48\textwidth]{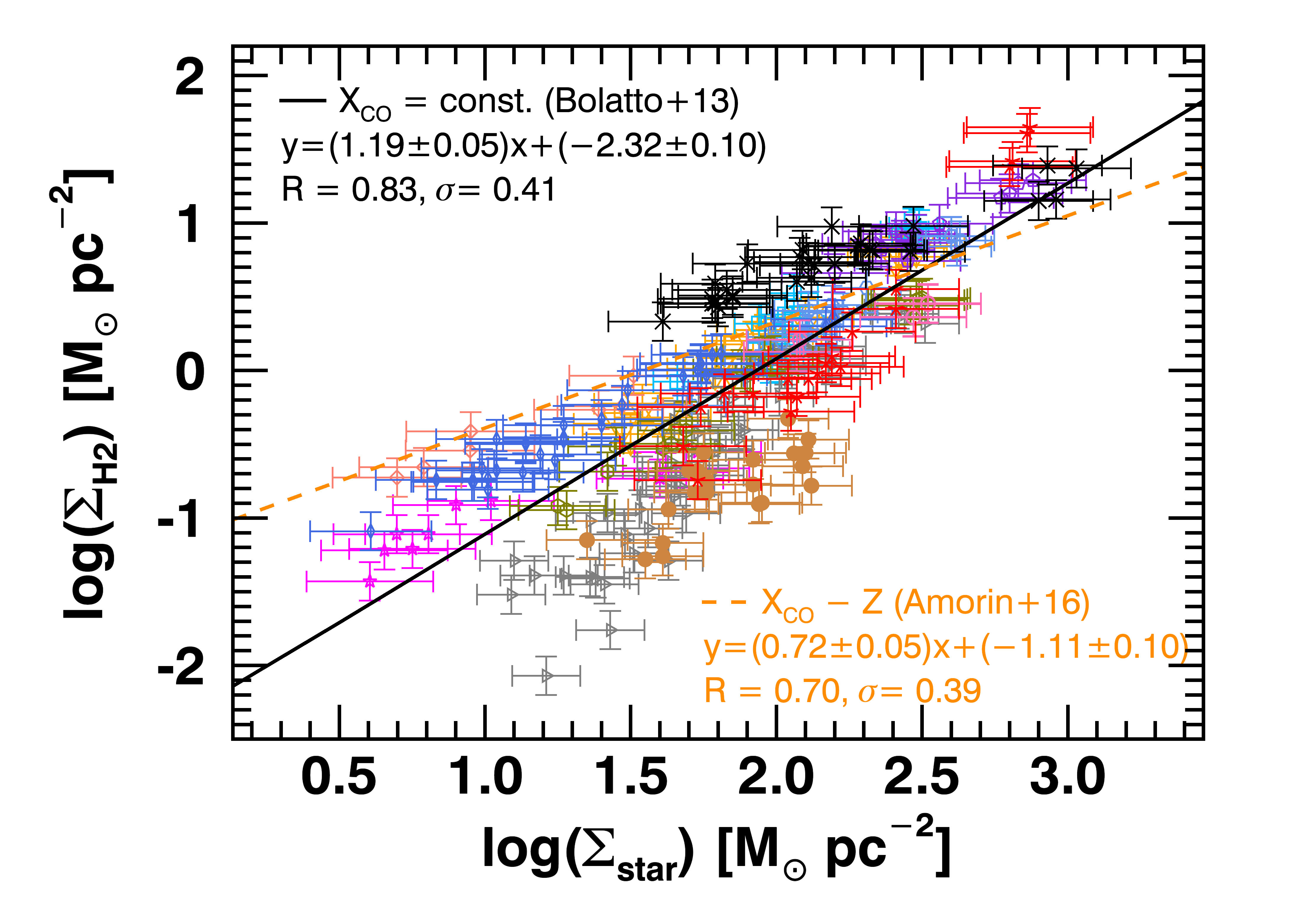}}
   \qquad 
  \subfloat{\includegraphics[width=0.48\textwidth]{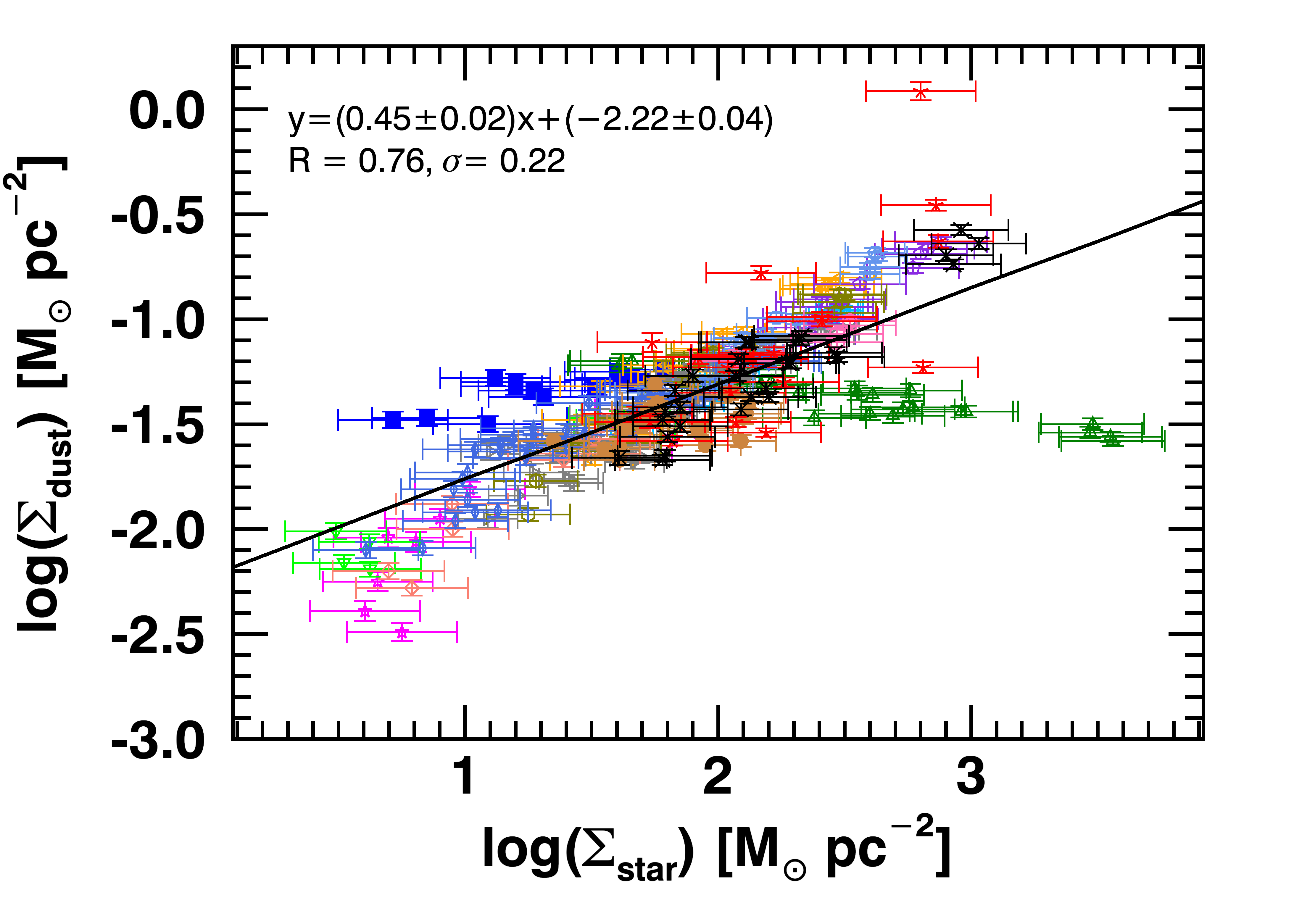}}
   \qquad
  \subfloat{\includegraphics[width=0.48\textwidth]{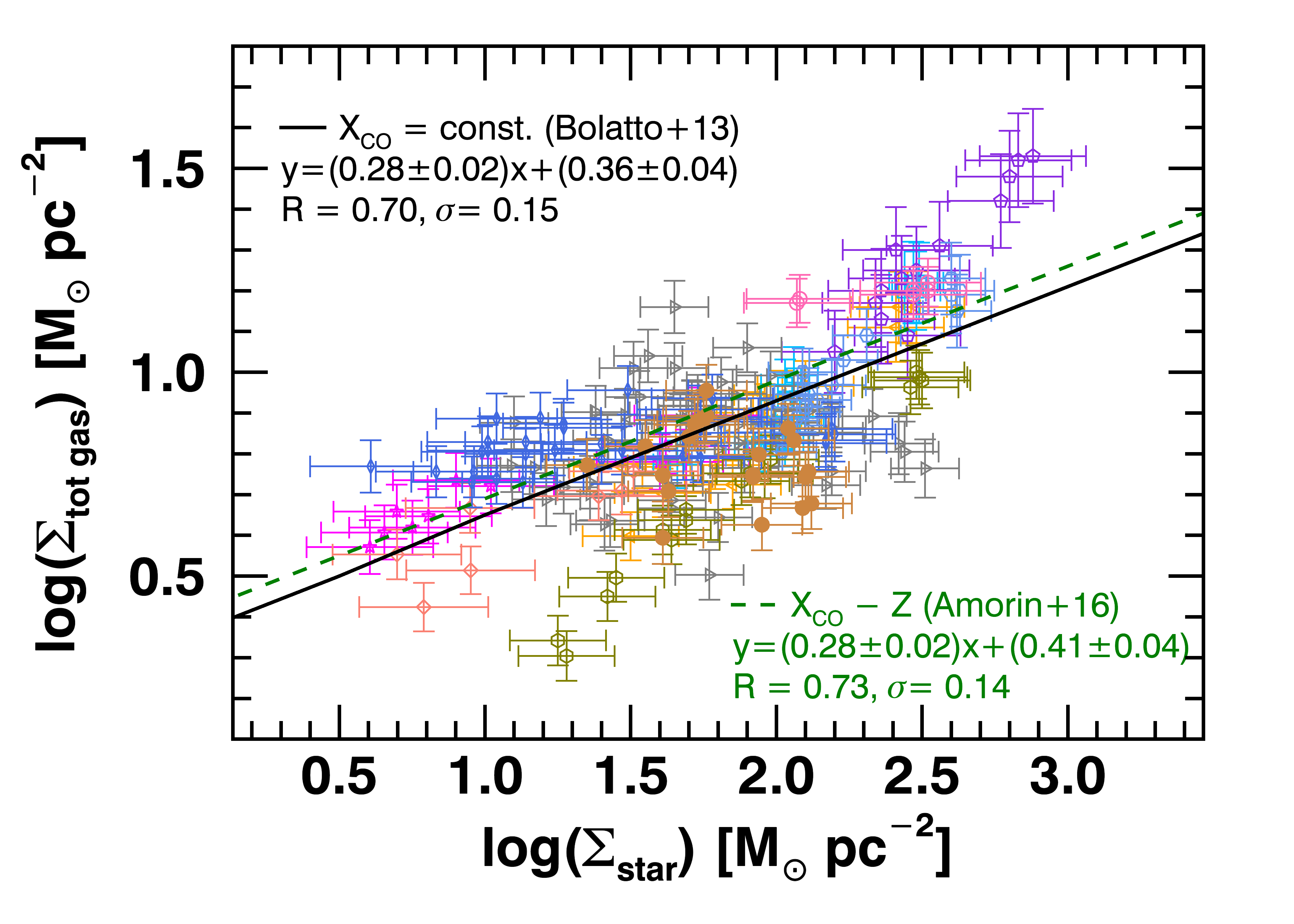}} 
  \qquad
    \subfloat{\includegraphics[width=0.48\textwidth]{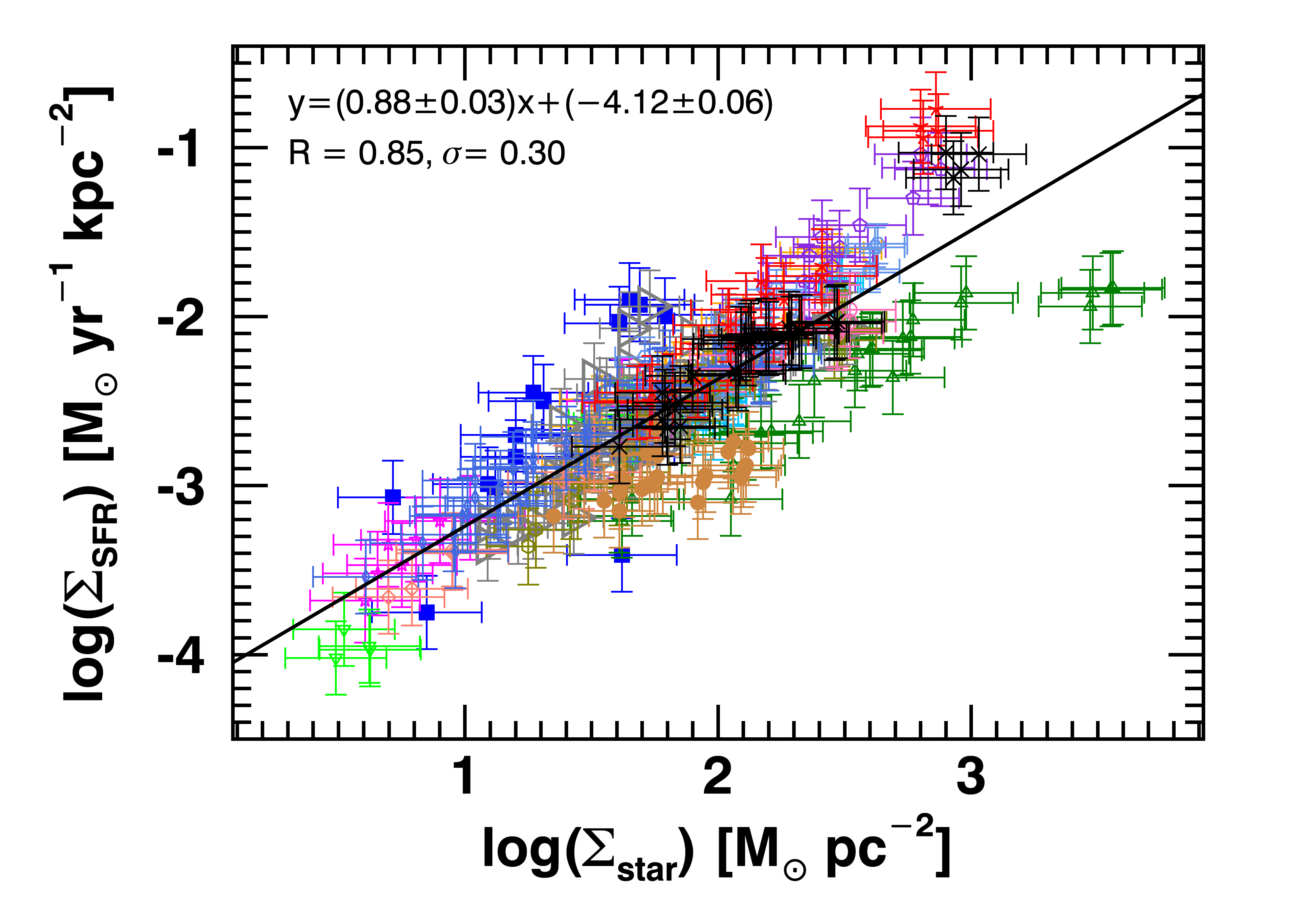}}
  \qquad 
  \subfloat{\includegraphics[width=0.48\textwidth]{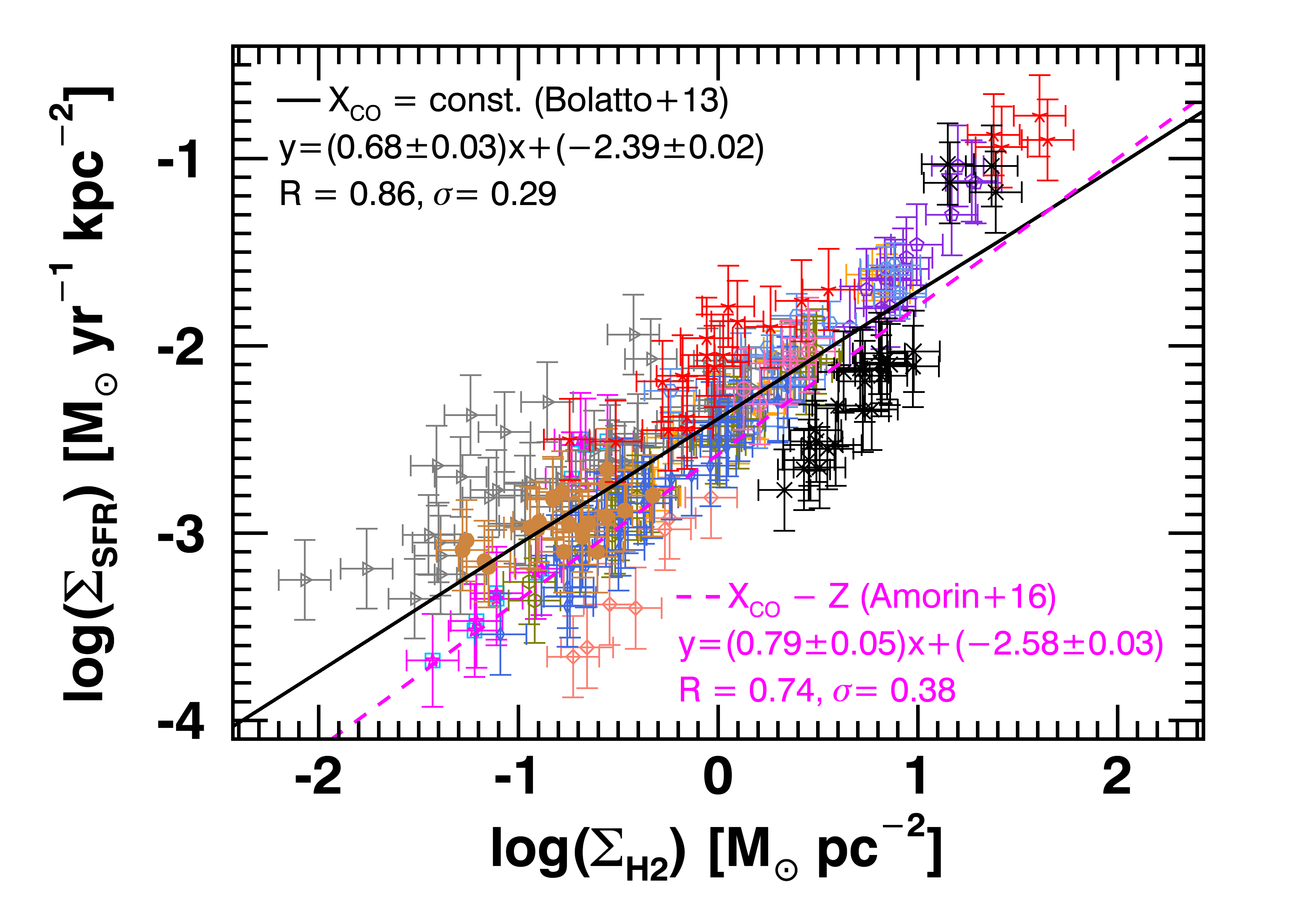}} 
\end{figure*}

\begin{figure*}
  \centering 
  \subfloat{\includegraphics[width=0.48\textwidth]{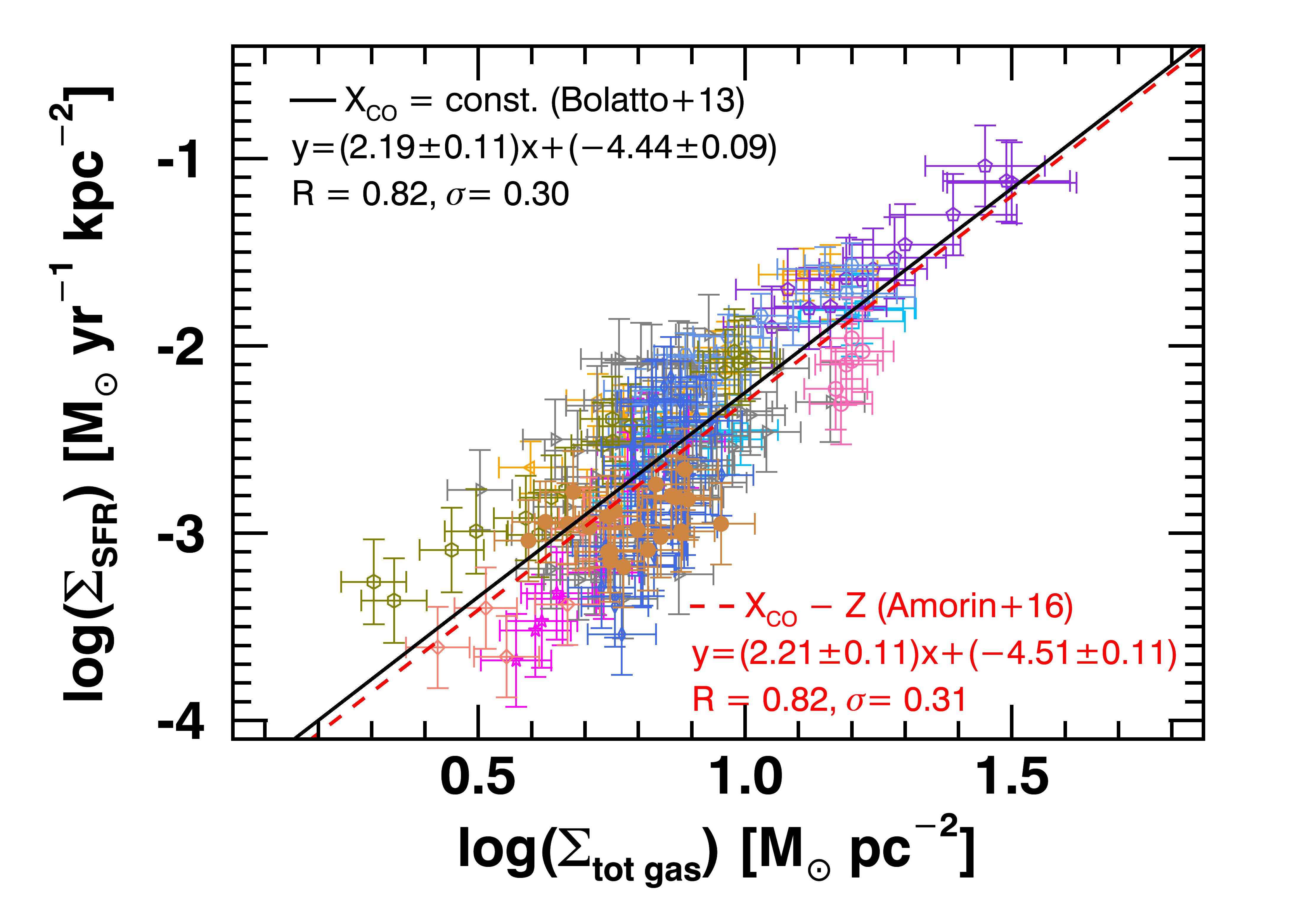}}
  \qquad 
  \subfloat{\includegraphics[width=0.48\textwidth]{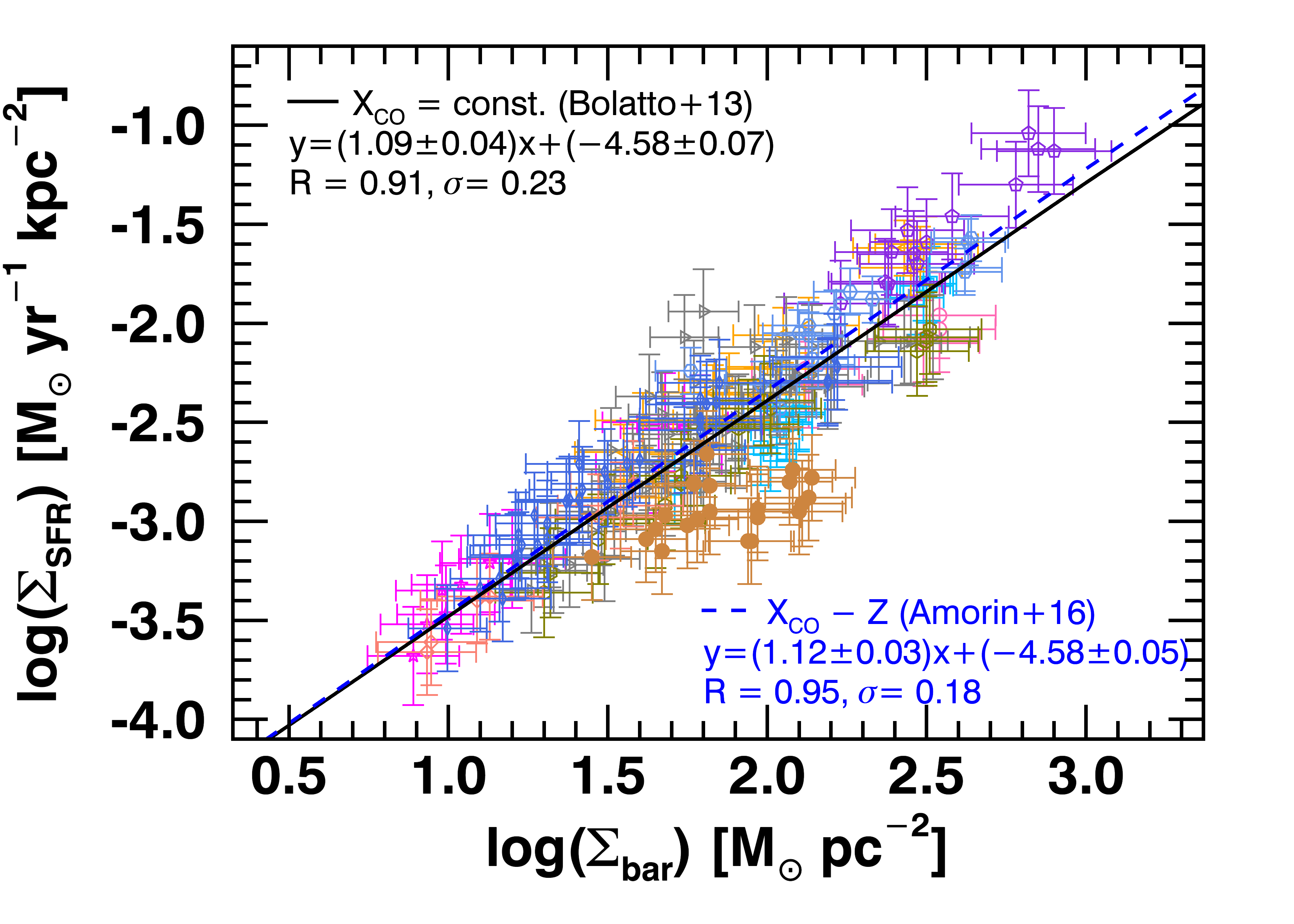}} 
\caption[]{
Same as Fig.~\ref{fig:n6946} for all sample galaxies evaluated together at the common scale of 3.4~kpc.
Only data referring to the assumption of the constant $X_{\rm CO}$ are shown, while the drawn lines refer to the fits
under the two assumptions on $X_{\rm CO}$. 
Different symbols (and colors) represent different galaxies according to the legend shown in the panel displaying the 
$\Sigma_{\rm dust}$--$\Sigma_{\rm HI}$ SR.}
\label{fig:all}
\end{figure*}

\section{The resolved scaling relations}
\label{sec:sr}
In this section we present the results obtained from the study of the pixel-by-pixel SRs for our galaxy sample.
We explore the following SRs: $\Sigma_{\rm dust}$--$\Sigma_{\rm H2}$, 
$\Sigma_{\rm dust}$--$\Sigma_{\rm HI}$, $\Sigma_{\rm dust}$--$\Sigma_{\rm tot\,gas}$ (these three together hereafter called ISM SRs), 
$\Sigma_{\rm star}$--$\Sigma_{\rm H2}$, $\Sigma_{\rm star}$--$\Sigma_{\rm dust}$, $\Sigma_{\rm star}$--$\Sigma_{\rm tot\,gas}$, 
$\Sigma_{\rm star}$--$\Sigma_{\rm SFR}$, $\Sigma_{\rm H2}$--$\Sigma_{\rm SFR}$, $\Sigma_{\rm tot\,gas}$--$\Sigma_{\rm SFR}$,
$\Sigma_{\rm bar}$--$\Sigma_{\rm SFR}$.
When the molecular gas is involved, two prescriptions on $X_{\rm CO}$ are explored. 

\subsection{Results for individual galaxies}
\label{sec:ism-rel-individual}
Figure~\ref{fig:n6946} shows the example of the SRs derived for NGC~6946 at the resolution of 1~kpc.
For this galaxy, all components are, on average, moderately or strongly correlated\footnote{Typically, if ${0<|R|<0.3}$ the correlation is defined weak,
if ${0.3<|R|<0.7}$ the correlation is defined moderate, if ${|R|>0.7}$ the correlation is defined strong.}
within the optical disk, under both assumptions on $X_{\rm  CO}$ when the molecular gas is involved in a given SR.
Only, the $\Sigma_{\rm dust}$--$\Sigma_{\rm HI}$ SR is a weak or moderate correlation.

In Appendix~\ref{sec:add-sample}, Fig.~\ref{fig:add-ism} displays the SRs for the entire sample and 
Table~\ref{tab:fit-sr} collects 
properties of the corresponding linear fits to the data.
The analysis of the full sample shows that each galaxy is characterized by distinct, in terms of slope and $R$ (and $\sigma$), SRs 
on scales between 0.3 and 3.4~kpc.   
Nevertheless, we identify some recurrent behaviors for each SR:
for instance, the $\Sigma_{\rm dust}$--$\Sigma_{\rm H2}$ SR has a superlinear slope for most galaxies.
Details on galaxies deviating from recurrent trends are provided in Appendix~\ref{sec:peculiar}.
All the explored SRs are moderate or strong correlations except the $\Sigma_{\rm dust}$--$\Sigma_{\rm HI}$ SR 
that does not exist or is a weak relationship for most galaxies. 

The results found by assuming the metallicity-dependent $X_{\rm CO}$ are similar, in terms of slopes and $R$,
to those under the prescription of the constant $X_{\rm CO}$.
However, there is evidence, for each galaxy, that the slopes and $R$ values change a little bit as a function of the prescription on $X_{\rm CO}$.  
For example, the slopes and $R$ values of the SRs with $\Sigma_{\rm H2}$ in the $y-$axis 
(e.g., $\Sigma_{\rm dust}$--$\Sigma_{\rm H2}$) 
under the metallicity-dependent $X_{\rm CO}$ prescription are lower than those under the constant  one.
In this type of SRs, the values of slopes typically get lower for steeper gas-phase metallicity gradients 
(e.g., NGC~628), for significantly lower mean metallicities than that of solar (e.g., NGC~925),
or for the combination of the two conditions (e.g., NGC~5457).

We have tested two other prescriptions on $X_{\rm CO}$, one from \citet{bolatto13} and the other one from \citet{madden20}.
Since these tests do no give extremely different results from those obtained 
under the adopted prescriptions, we do not show them for the entire sample.
Details on these tests are described in Appendix~\ref{app:othercal}.

Since we have also all maps convolved at the common scale of 3.4~kpc we are able to check the existence of possible dependencies 
of SRs as a function of scale (see Table~\ref{tab:fit-sr}).
The slopes turn out to be approximately independent on the physical scale, in the range of scales we explore.
In particular, for a given galaxy the slope of a SR at the scale imposed by the dust map and that at 3.4~kpc
are almost always consistent within a factor of $(1-3)$ times the uncertainties.
Overall, there is no evidence of a dependence of $R$ of all SRs on the scale.

\subsection{Results for the sample}
\label{sec:ism-rel-sample}
Figure~\ref{fig:all} shows the results obtained from the study of the pixel-by-pixel SRs for all sample galaxies
evaluated together at 3.4~kpc.
The parameters of these correlations are collected in Table~\ref{tab:fit-sr}.

In general, all the explored SRs are strong correlations and the $\Sigma_{\rm bar}$--$\Sigma_{\rm SFR}$ SR is the strongest one. 
The most striking results concern the ISM SRs.
The $\Sigma_{\rm dust}$--$\Sigma_{\rm HI}$ SR is a weak correlation also for all galaxies evaluated together at 3.4~kpc.
These results differ from those found from the global study of DustPedia late-type galaxies 
(C20), where both dust--\hi\ and dust--total gas are better correlated than dust--H$_2$. 
In particular, comparing the ISM SRs studied for individual galaxies at subkiloparsec/kiloparsec (subkpc/kpc) 
scales (Sect.~\ref{sec:ism-rel-individual}), those for all galaxies at 3.4~kpc (this section), and those at global scales (C20), 
it emerges that the $\Sigma_{\rm dust}$--$\Sigma_{\rm tot\,gas}$ SR is a good correlation at all (from small to global) scales, 
and the strength of the $\Sigma_{\rm dust}$--$\Sigma_{\rm H2}$ and $\Sigma_{\rm dust}$--$\Sigma_{\rm HI}$ SRs 
vary as a function of the sampled scale.
The atomic gas is globally a very good tracer of dust within the optical disk of spiral galaxies, while \hi\ is almost always
unrelated with dust at scales between 0.3 and 3.4~kpc.   
The molecular gas is globally a good tracer of dust within the optical disk of spiral galaxies, but the $\Sigma_{\rm dust}$--$\Sigma_{\rm H2}$ SR
strengthens at subkpc/kpc scales.
For the other SRs there is a good agreement, in terms of slopes and $R,$ between small-scale SRs we explore and integrated SRs present in the literature (see Sect.~\ref{sec:intro} for some references),
supporting a scenario where the main physical processes regulating the properties and evolution of galaxies may locally occur.

\begin{figure*}
\includegraphics[width=0.5\textwidth]{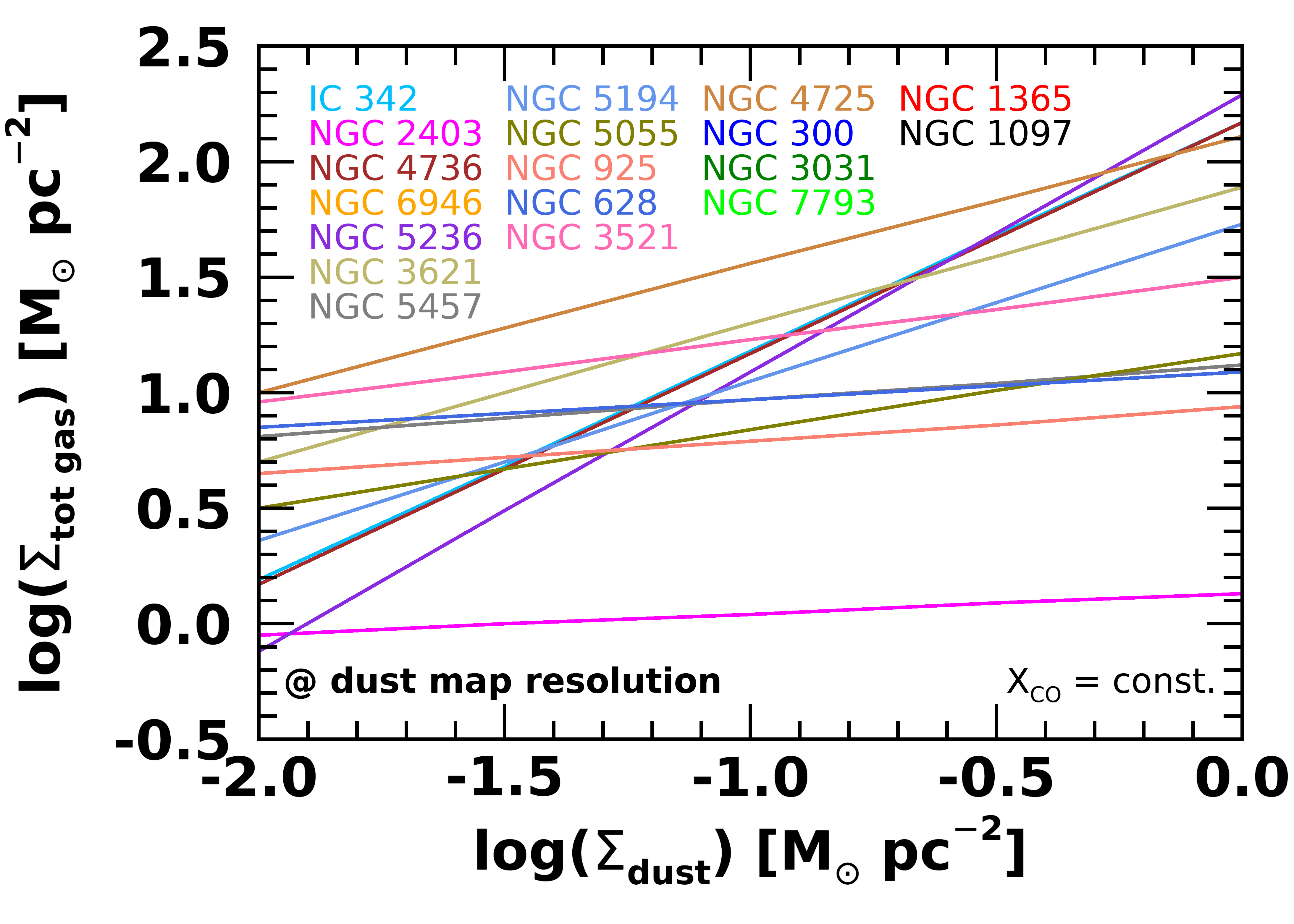}
\includegraphics[width=0.5\textwidth]{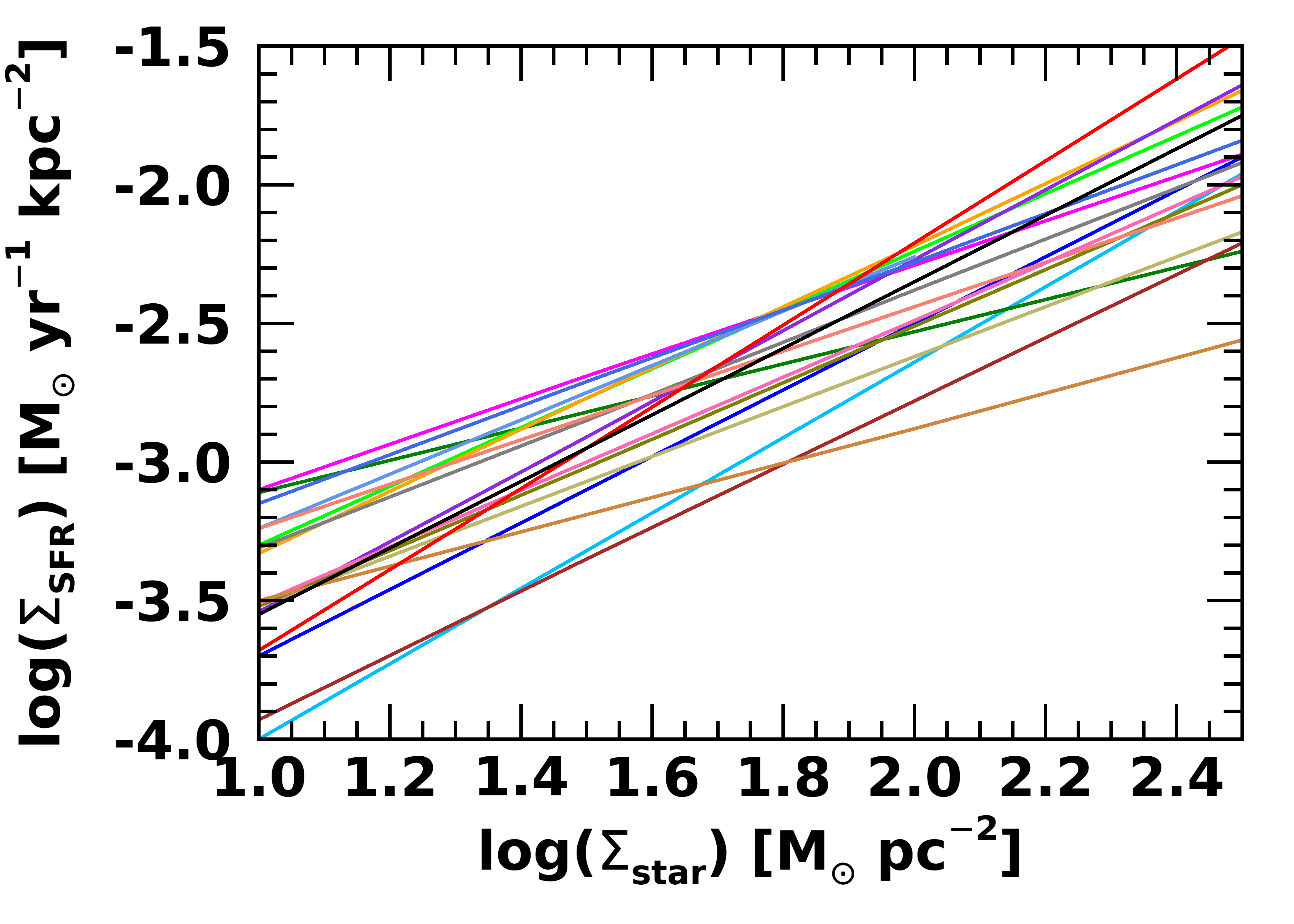}  
\caption{
Linear fits of the dust-total gas (left panel) and SFMS (right panel) SRs 
of sample galaxies individually evaluated at the physical scale imposed by the dust map resolution (36$^{\prime\prime}$), from 0.3 to 3.4~kpc (see 
Sect.~\ref{sec:ism-rel-individual}).
Each colored fit corresponds to a galaxy according to the legend shown in the left panel.
The total gas refers to the assumption of the constant $X_{\rm CO}$ \citep[][]{bolatto13}.
}
\label{fig:slopes-dustres}
\end{figure*}

\begin{figure*}
\includegraphics[width=0.5\textwidth]{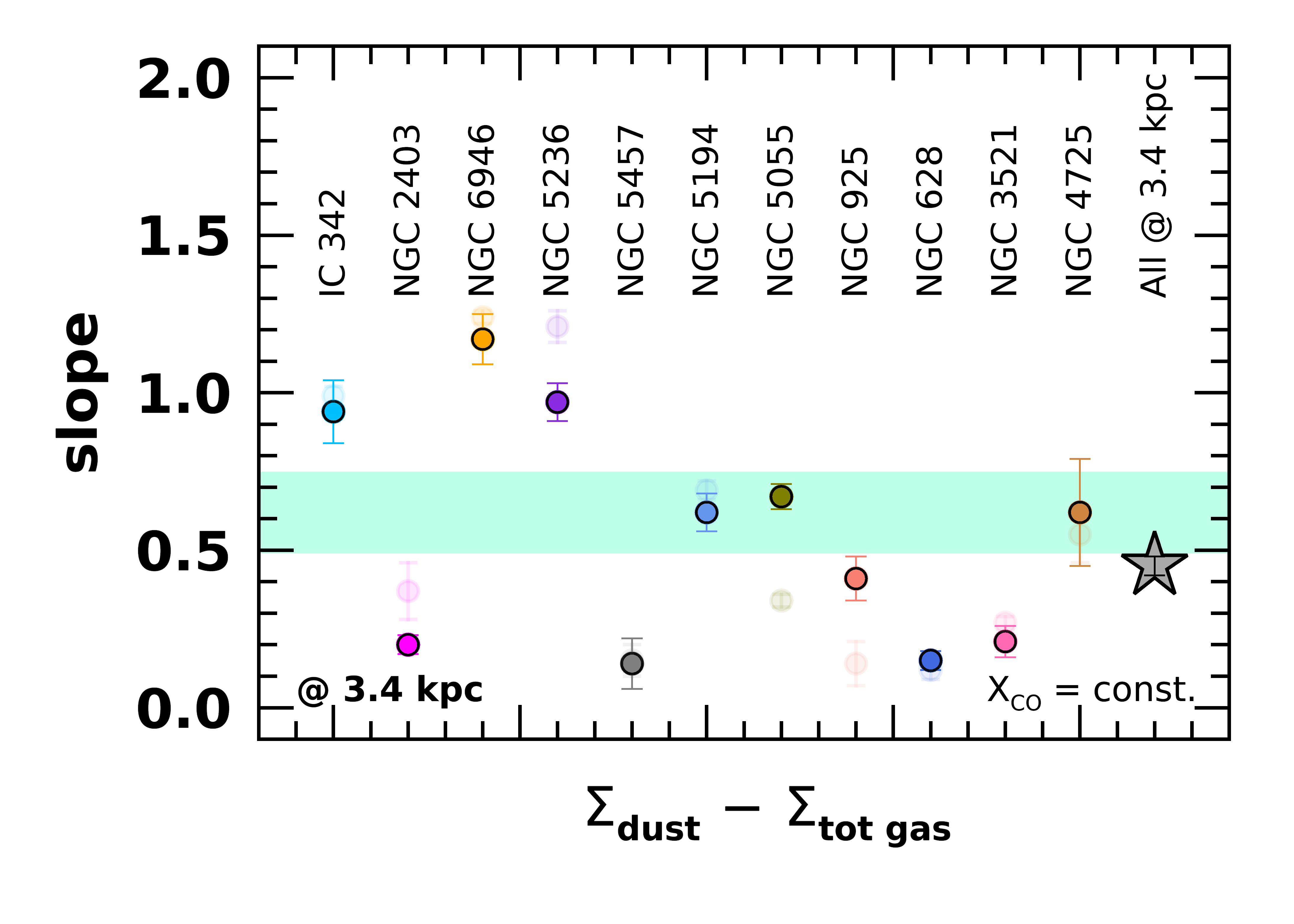}
\includegraphics[width=0.5\textwidth]{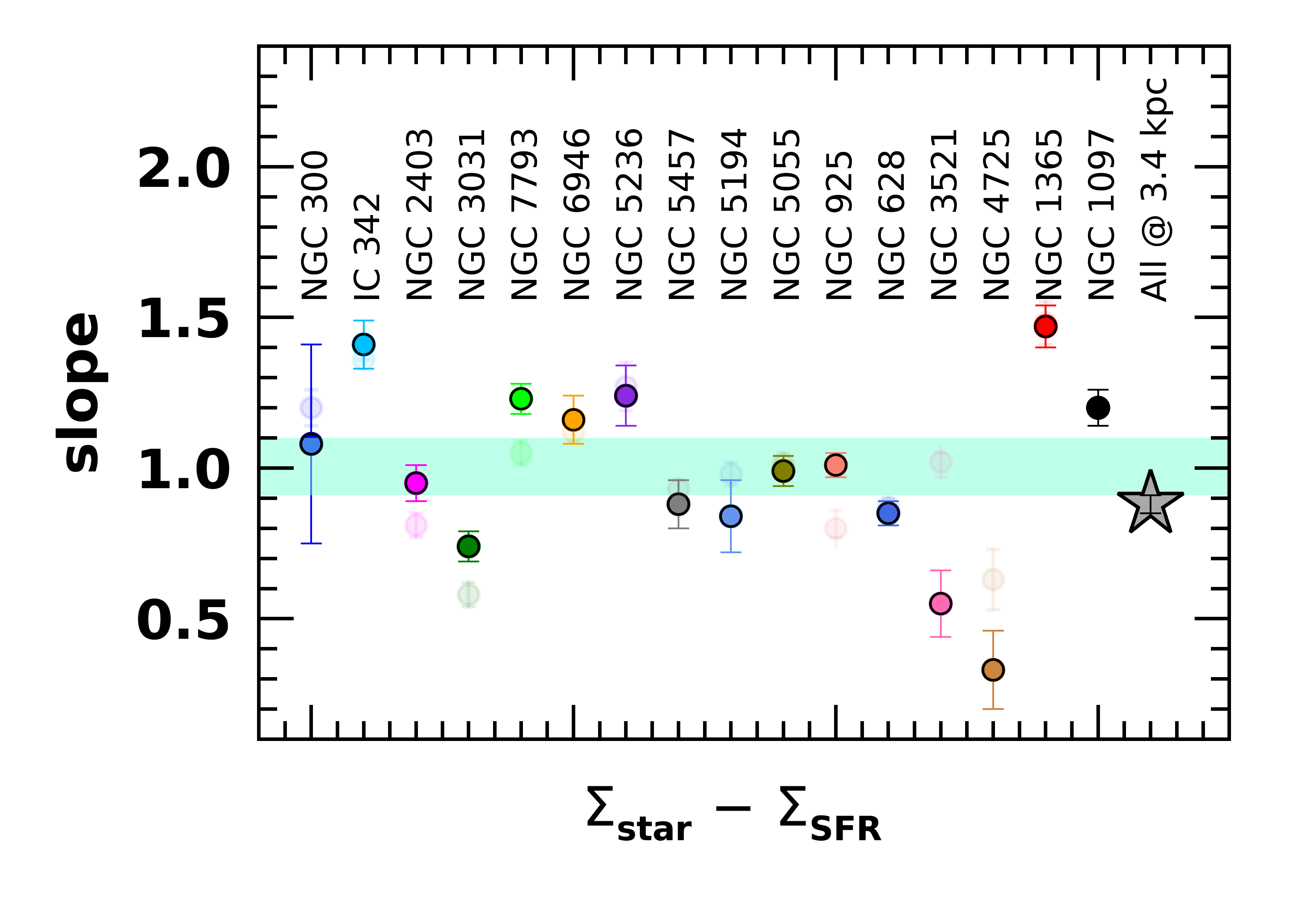}  
\caption{
Slopes of the dust-total gas (left panel) and SFMS (right panel) SRs 
of sample galaxies individually evaluated at 3.4~kpc (see Sect.~\ref{sec:ism-rel-sample}).
Each colored dot corresponds to a galaxy.
The transparent symbols represent the corresponding slopes at the resolution imposed 
by the dust map (see Sect.~\ref{sec:ism-rel-individual}).
The light blue band is the median slope with error 
and the gray star the slope of all galaxies at 3.4~kpc (see Table~\ref{tab:fit-sr}).
The total gas refers to the assumption of the constant $X_{\rm CO}$ \citep[][]{bolatto13}.}
\label{fig:slopes}
\end{figure*}

A short summary of the main results presented in this section is shown in 
Figs.~\ref{fig:slopes-dustres} and \ref{fig:slopes}.
They refer to the dust-total gas and SFMS SRs as an example and the emerging considerations can be extended to all SRs. 
Figure~\ref{fig:slopes-dustres} shows the nonuniversality of the resolved SRs 
of sample galaxies individually evaluated at the scale imposed by the dust map, 
from 0.3 to 3.4~kpc.
Figure~\ref{fig:slopes} displays that also when the slopes of a given SR are evaluated at the common scale of 3.4~kpc there is a variety of results 
but the slope of all galaxies at 3.4~kpc (gray star) is consistent with their median slope (light blue band).
In this figure, transparent symbols represent the slopes at the scale imposed by the dust map and the comparison with the slopes at 3.4~kpc 
shows the approximate independence, for a given galaxy, of the slope of a SR on the physical scale (see Sect.~\ref{sec:ism-rel-individual}).
Table~\ref{tab:fit-sr} collects the median slopes of the SRs at 3.4~kpc.  

\section{Resolved ratios in the local Universe}
\label{sec:ratios}
In this section we study some ratios focusing on ISM components: dust-to-stellar, dust-to-gas, and dust-to-metal ratios.
We explore the behavior of these ratios as a function of various galaxy physical quantities typically used as tracers of the evolutionary stage 
(especially in global analyses).   
In particular, we study the ratios against 12~+~log(O/H), $\Sigma_{\rm star}$, specific SFR 
(the SFR per unit stellar mass) surface density ($\Sigma_{\rm SSFR}$), and gas fraction 
($f_{\rm gas}$ = $\Sigma_{\rm tot\,gas}$/($\Sigma_{\rm tot\,gas}$ + $\Sigma_{\rm star}$)). 
We present the results taking into account all galaxies evaluated together at 3.4~kpc.
Table~\ref{tab:ratio-values} collects mean values of the ratios and Table~\ref{tab:ratios} the properties of the fits to data.

\begin{figure*}
\includegraphics[width=0.5\textwidth]{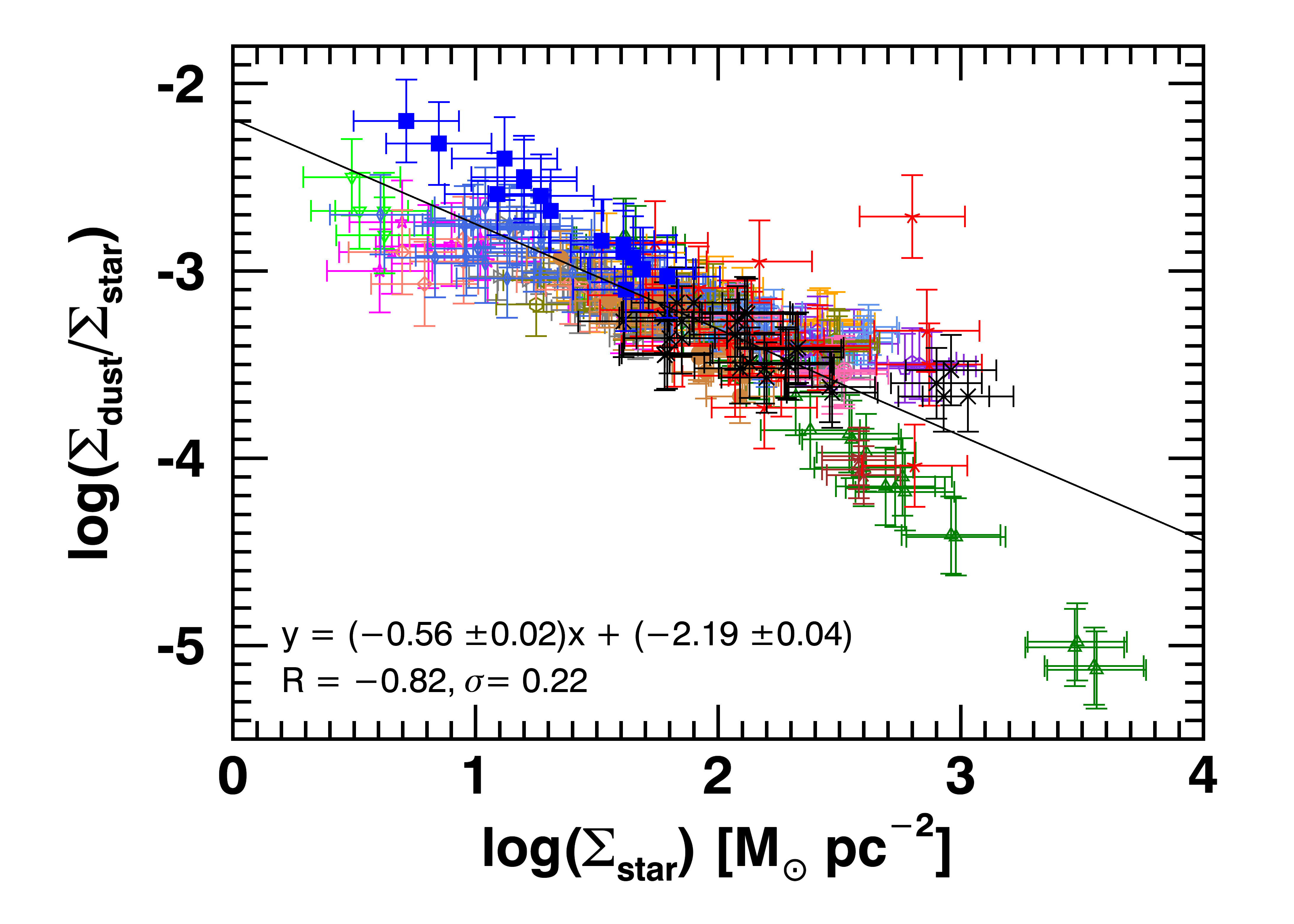}
\includegraphics[width=0.5\textwidth]{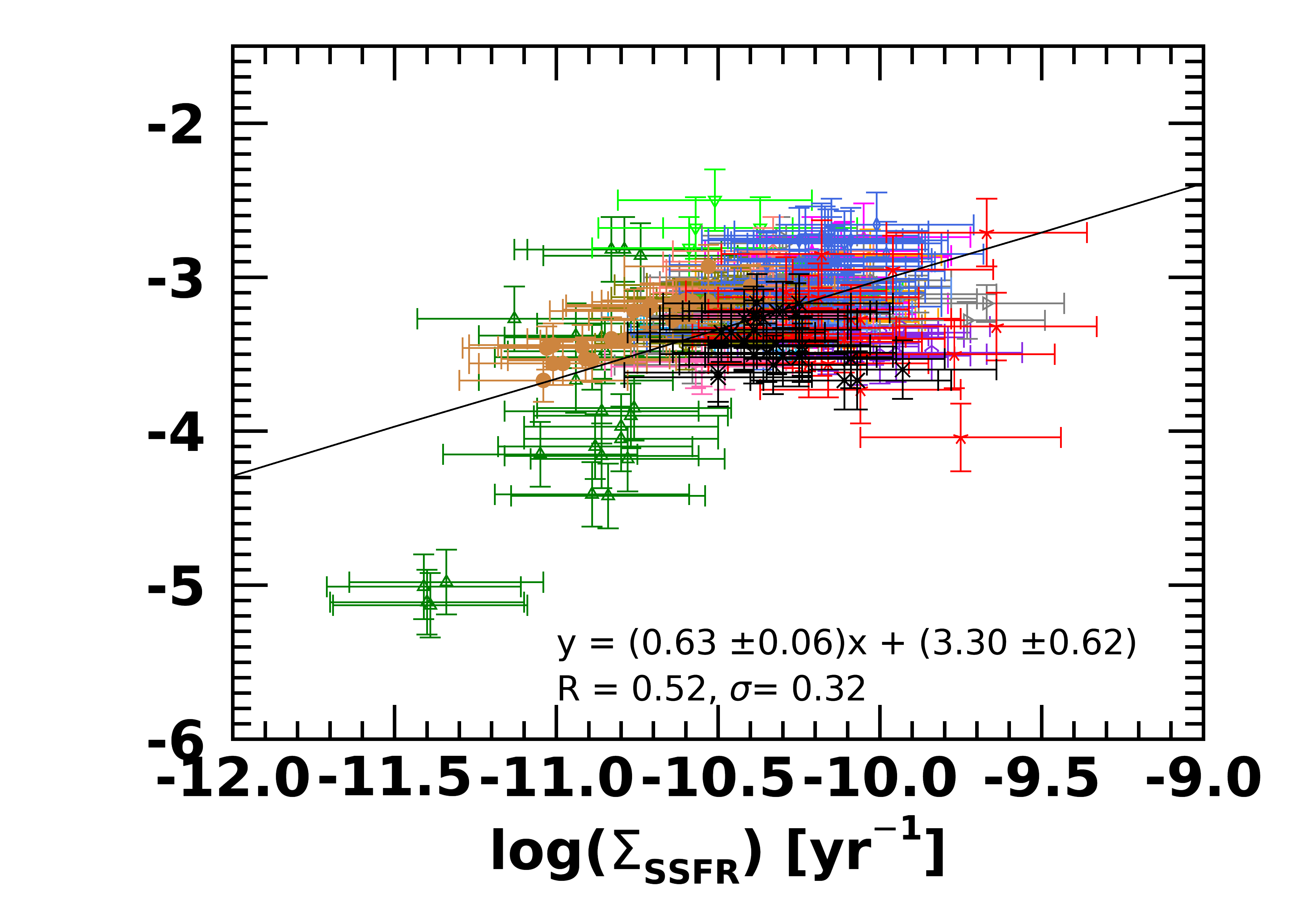}  
\caption{
Logarithm of dust-to-stellar mass ratio as a function of logarithm of stellar mass (left panel) and SSFR surface density (right panel).
The continuum line is the linear fit to the data (see Table~\ref{tab:ratios}).
Symbols are same as Fig.~\ref{fig:all}.
}
\label{fig:mstar-ssfr-dustonstar}
\end{figure*}

\begin{figure}[h]
\includegraphics[width=0.5\textwidth]{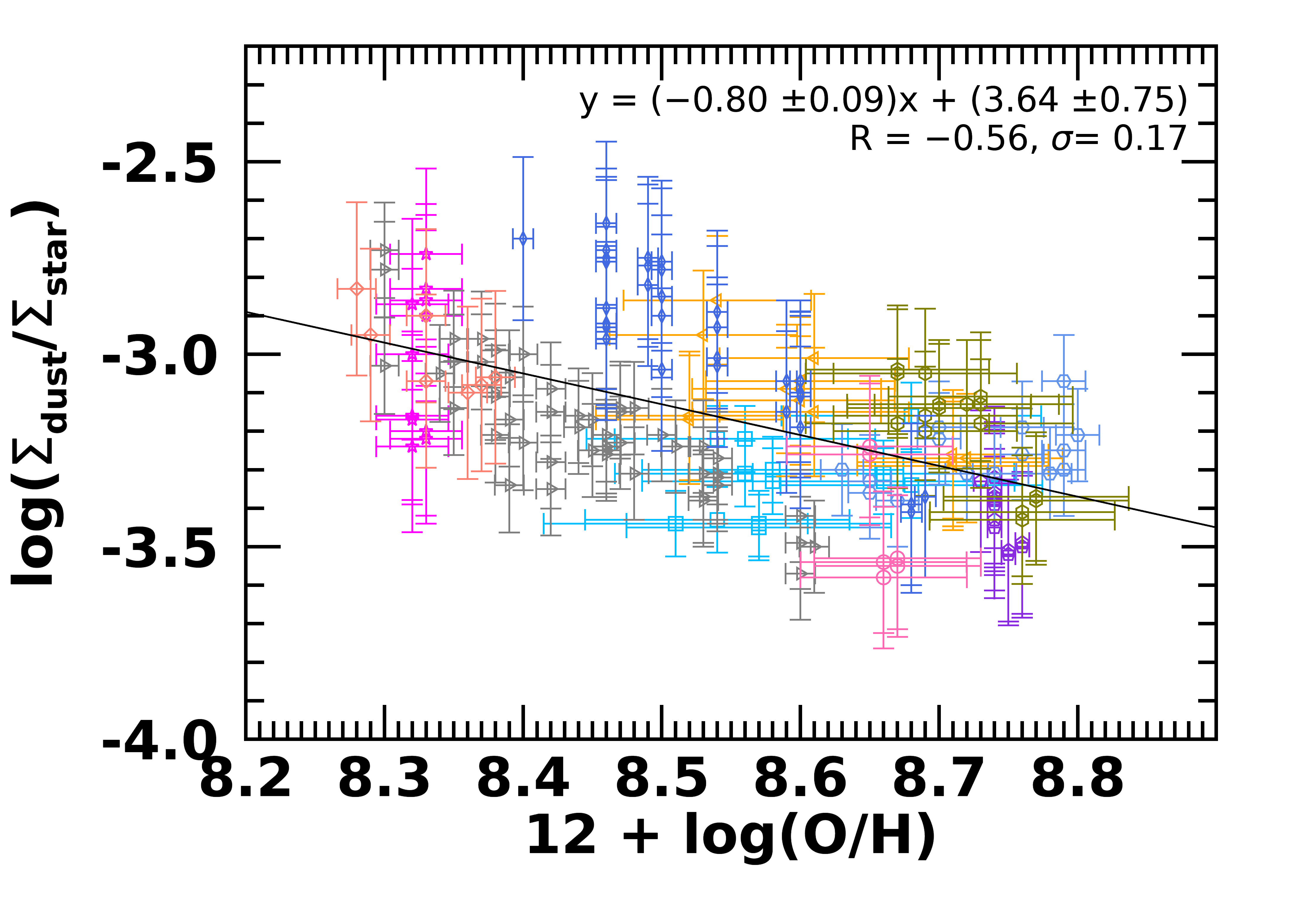}  
\caption{
Same as Fig.~\ref{fig:mstar-ssfr-dustonstar}, but as a function of gas-phase metallicity.}
\label{fig:oh-dustonstar}
\end{figure}

\subsection{Dust-to-stellar mass ratio}
\label{sec:dust-star}
We explore the dust-to-stellar mass ratio, $\Sigma_{\rm dust}$/$\Sigma_{\rm star}$, also called the specific dust mass.
The mean value of $\Sigma_{\rm dust}$/$\Sigma_{\rm star}$ for our resolved sample is log($\Sigma_{\rm dust}$/$\Sigma_{\rm star}$)~$= -3.12 \pm 0.39$,
broadly consistent with the dust-to-stellar mass ratio of \textit{Herschel} Reference Survey \citep[HRS,][]{boselli10} galaxies at stellar mass M$_{\rm star} \gtrsim 10^9$~M$_{\odot}$  \citep[][]{devis17a}\footnote{Our range of $\Sigma_{\rm dust}$/$\Sigma_{\rm star}$ values is consistent with that of \citet{devis17a}.}.
Comparisons with the literature require to pay attention to derivation method of the involved quantities and this is particularly important for the dust mass whose values are strongly dependent on adopted SED fitting and dust model \citep[see, e.g.,][]{relano22}.  
However, \citet{devis19} found that dust masses used in \citet{devis17a} derived with MAGPHYS \citep[][]{dacunha08} and the DustPedia ones (see Sect.~\ref{sec:dataset}) are very similar. 
 
Dust is formed by condensation of metals in the atmospheres of evolved stars (seed grains formation in supernova (SN) ejecta and AGB stars)
and can grow its mass in the dense ISM (where the accretion of metals on grains is balanced by SN shock-wave destruction). 
Because of this, relations are expected between the dust content, the stellar mass 
and SF activity, and the gas-phase metallicity.
Figure \ref{fig:mstar-ssfr-dustonstar} shows the trend of $\Sigma_{\rm dust}$/$\Sigma_{\rm star}$ as a function of $\Sigma_{\rm star}$ (left panel) and  
$\Sigma_{\rm SSFR}$ (right panel).
As expected, there is a strong 
anticorrelation between $\Sigma_{\rm dust}$/$\Sigma_{\rm star}$ and $\Sigma_{\rm star}$,
which is already known in the literature both at subkpc scales for, for example, M~31 \citep[][]{viaene14} and at global scales for
samples of galaxies \citep[e.g.,][]{cortese12,clemens13,clark15,calura17,devis17a,orellana17,casasola20,delooze20}, 
and it has been reproduced by both chemical evolutionary models \citep[e.g.,][]{cortese12,calura17,devis17a} 
and cosmological hydro simulations \citep[e.g.,][]{camps16}.
The anticorrelation between $\Sigma_{\rm dust}$/$\Sigma_{\rm star}$ and $\Sigma_{\rm star}$ 
indicates that galaxy regions with a lower stellar mass have higher dust-to-stellar mass ratios with respect to stellar mass than their more massive counterparts.
In C20, where we have defined the global M$_{\rm star}$--M$_{\rm dust}$/M$_{\rm star}$ SR for 432 DustPedia late-type galaxies,
we have proposed the following interpretation:
although both stars and dust are products of SF, 
stellar mass grows with time as the galaxy evolves whereas the amount of dust can decrease as due to astration and destruction \citep[see, e.g.,][]{calura08}.
This explanation can be also applied to individual galaxy regions.
This trend could be seen as a local version of the downsizing of galaxies \citep[e.g.,][]{cowie96,fontanot09}, where massive galaxies have already consumed most of their gas
converting it into stars and most of dust mass was produced during these SF episodes.   
The correlation between $\Sigma_{\rm dust}$/$\Sigma_{\rm star}$ and $\Sigma_{\rm SSFR}$ is moderate: 
$\Sigma_{\rm dust}$/$\Sigma_{\rm star}$ increases with increasing $\Sigma_{\rm SSFR}$, according to global analyses \citep[e.g.,][]{dacunha10,delooze20}.
This relation provides indications on the role of recent SF activity in the determination of dust content.

Figure~\ref{fig:oh-dustonstar} shows $\Sigma_{\rm dust}$/$\Sigma_{\rm star}$ vs 12~+~log (O/H): we find a decrease by a factor of $\sim$3, from $\approx$0.001 at 12~+~log(O/H) $\sim$8.3 to $\approx$0.0004 at 12~+~log(O/H)~$\sim$~8.8.
A similar trend is also found by \citet{vilchez19}, although characterized by a decreasing by an order of magnitude.
The anticorrelation of $\Sigma_{\rm dust}$/$\Sigma_{\rm star}$ with metallicity might simply reflect that with the stellar mass, since both quantities are well known to correlate: the dust mass is consumed in more evolved, high-metallicity and high-mass, galaxies.

\subsection{Dust-to-total gas mass ratio}
\label{sec:dgr}
We study the dust-to-total gas mass ratio (DGR = $\Sigma_{\rm dust}/\Sigma_{\rm tot\,gas}$).
The mean values of DGR for our resolved sample are log(DGR)~$= -2.20 \pm 0.04$ and $-2.23 \pm 0.21$ for constant and 
metallicity-dependent $X_{\rm CO}$, respectively.
They are consistent within errors with the global values found for DustPedia late-type galaxies (C20)
and other samples \citep[e.g., KINGFISH,][]{aniano20}. 

\begin{figure*}
\includegraphics[width=0.5\textwidth]{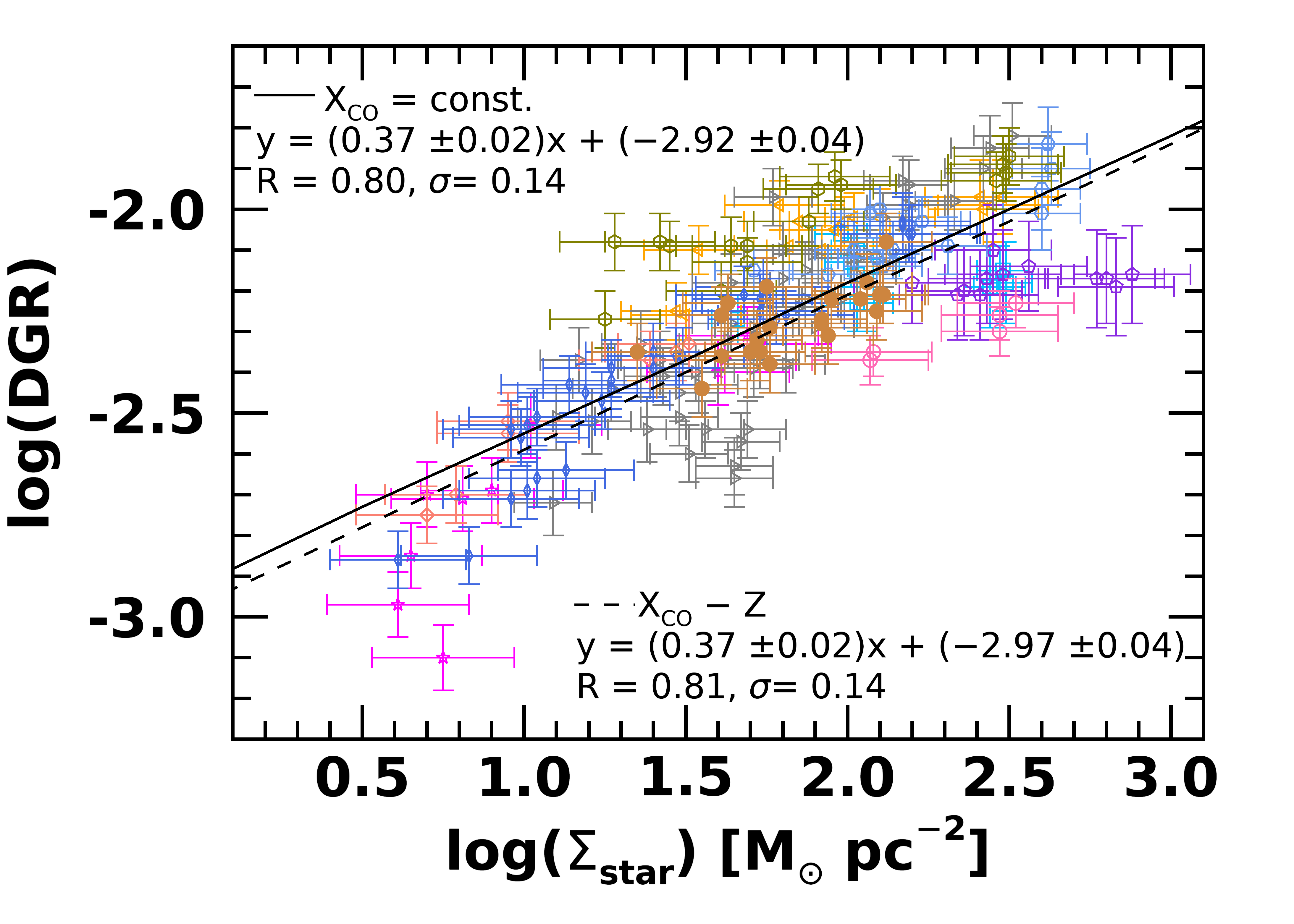}  
\includegraphics[width=0.5\textwidth]{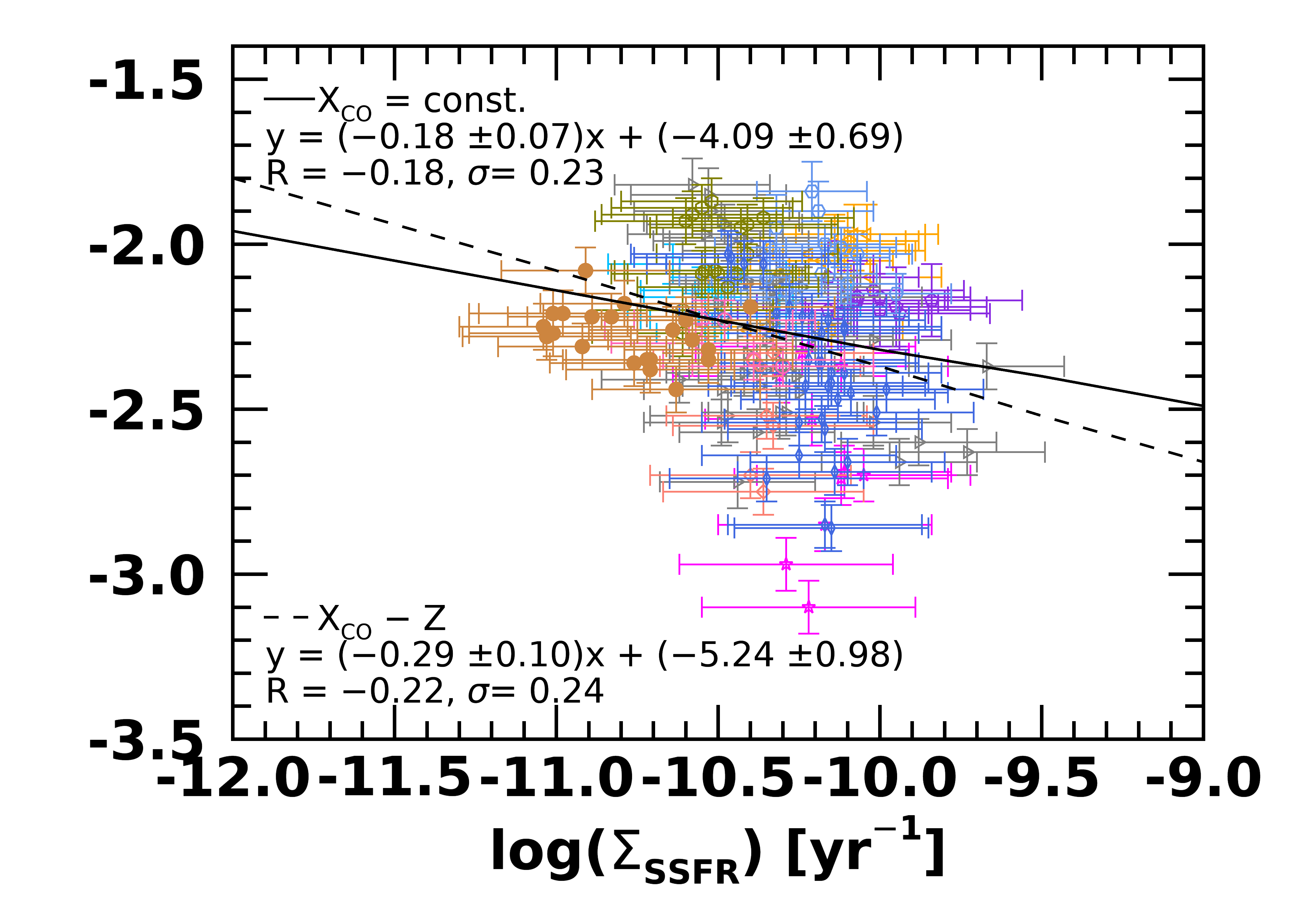}  
\caption{Logarithm of dust-to-gas mass ratio as a function of logarithm of $\Sigma_{\rm star}$ (left panel) and $\Sigma_{\rm SSFR}$ (right panel).
Data points refer to the assumption of the constant $X_{\rm CO}$, fit lines to both prescriptions on $X_{\rm CO}$ (see legend).
Symbols are same as Fig.~\ref{fig:all}.}
\label{fig:dgr-mstar-ssfr}
\end{figure*}

\begin{figure*}
\includegraphics[width=0.5\textwidth]{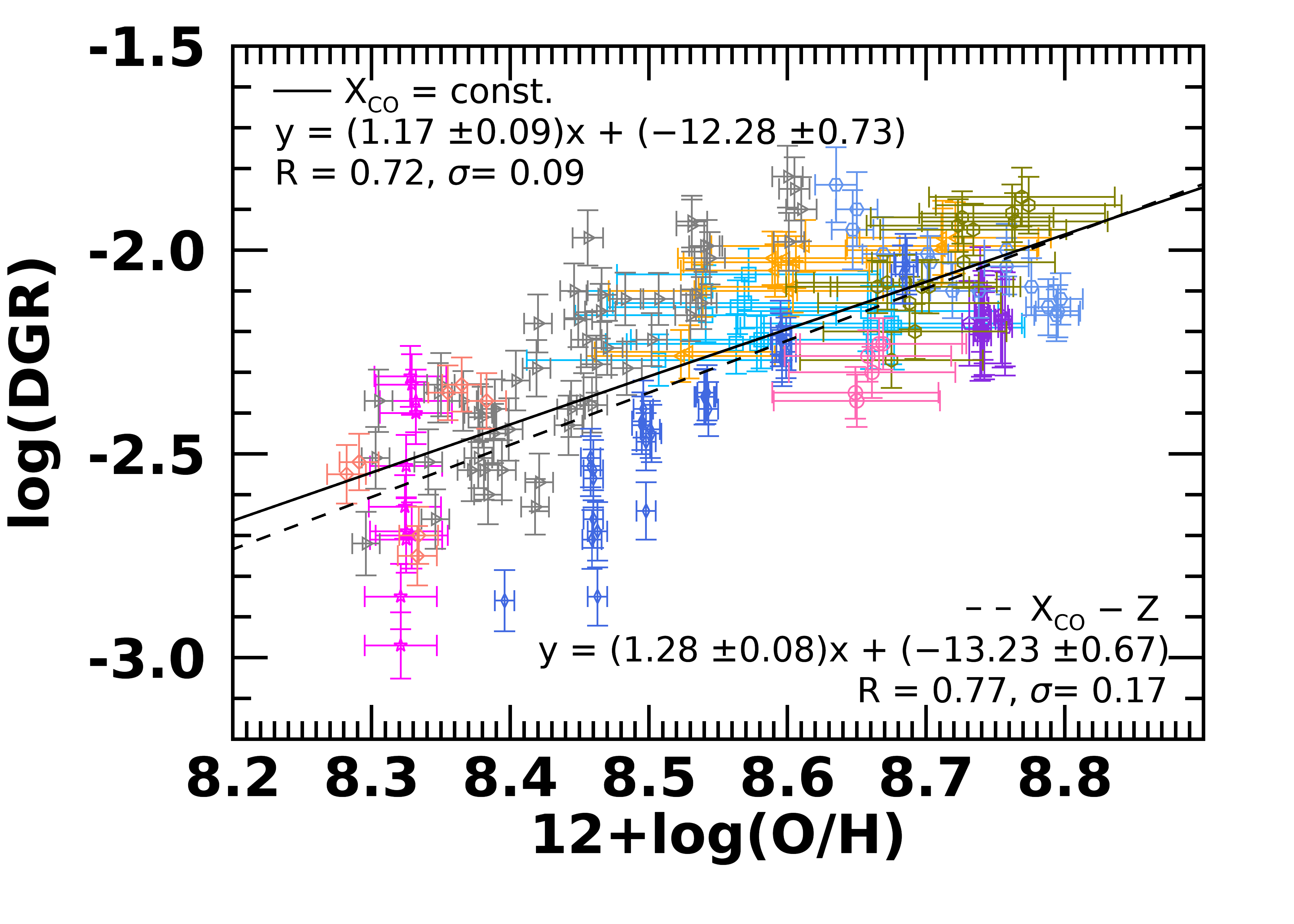}  
\includegraphics[width=0.5\textwidth]{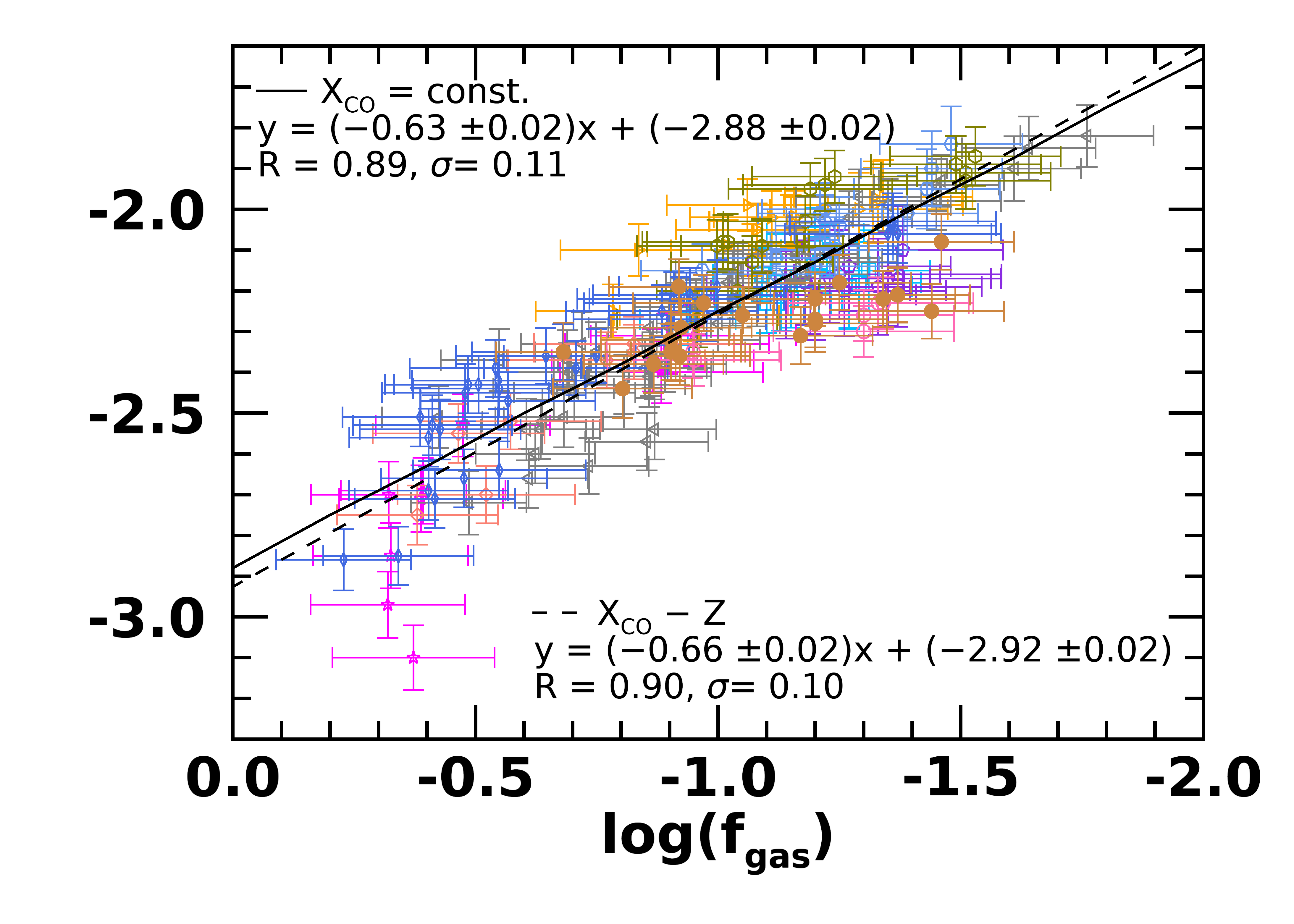}  
\caption{Same as Fig.~\ref{fig:dgr-mstar-ssfr}, but as a function of 12~+~log~(O/H) (left panel) and logarithm of $f_{\rm gas}$ (right panel).}
\label{fig:dgr-z}
\end{figure*}

Figure \ref{fig:dgr-mstar-ssfr} shows DGR as a function of $\Sigma_{\rm star}$ (left panel) and $\Sigma_{\rm SSFR}$ (right panel).
DGR shows a strong correlation with $\Sigma_{\rm star}$, while it is not correlated (or is very weakly correlated)
with $\Sigma_{\rm SSFR}$.
Strong relations of the DGR with both stellar mass and SSFR are usually found for global values;
they are generally interpreted by saying
that less massive galaxies are vigorously 
forming stars, while most of their metals have not been locked up in dust grains \citep[see, e.g.,][]{delooze20}.
In our study, where DGR is computed also using information on the molecular gas, we still find a strong correlation with the stellar mass. Instead, we do not see any correlation with the resolved SSFR.
Thus, the anticorrelation between DGR and $\Sigma_{\rm SSFR}$ 
is valid only if averaged over considerable areas ($\gtrsim$~kpc or global scales).
This agrees with the recent resolved analysis performed by \citet{abdurrouf22} finding  
no significant anticorrelation between dust and SSFR for ten nearby galaxies 
(some in common with our sample).

Figure~\ref{fig:dgr-z} shows the trend of DGR vs 12~+~log~(O/H) (left panel) and $f_{\rm gas}$ (right panel). 
We find an increasing DGR with increasing metallicity.
The DGR depends on metallicity as 
DGR~$\propto$~(O/H)$^{1.2}$ and DGR~$\propto$~(O/H)$^{1.3}$,
assuming constant and metallicity-dependent $X_{\rm CO}$, respectively.
Our results are consistent with several observational works finding that the DGR is well represented by a power law with a slope of about 1
or higher at high metallicity \citep[12 + log (O/H)~$\gtrsim$~8.3, e.g.,][]{james02,draine07,leroy11,galliano08,sandstrom13,giannetti17}, and 
with theoretical expectations including dust grain growth \citep[e.g.,][]{asano13,zhukovska14,feldmann15,aoyama17,devis17b,mcKinnon18}. 
We stress that the single power law with superlinear slope provides the best description of the
DGR-metallicity relation for the full DustPedia late-type galaxy sample, testing different metallicity calibrations \citep[][]{devis19}.
A superlinear slope indicates that the stellar dust alone cannot explain the resulting DGR-metallicity relation.  
Some spatially resolved studies instead found a DGR-metallicity relation showing a two slopes behavior, 
with a break at a critical metallicity \citep[e.g.,][]{relano18,vilchez19}, consistently with the results of \citet{remy-ruyer14} 
based on the global analysis of nearby galaxies covering a 2 dex metallicity range (from 12+log(O/H) = 7.1 to 9.1).
Below this critical metallicity, 
the DGR-metallicity relation becomes steeper with respect to higher metallicities. 
The existence of a critical metallicity able to explain the change of slope in the DGR-metallicity relation has been 
theoretically predicted by \citeauthor{asano13} (\citeyear{asano13}, see also \citeauthor{popping17} \citeyear{popping17}).
It defines the two regimes dominated by 
stellar dust production and grain growth in the ISM, and its existence is invoked when stardust production equals grain grown.    
In this regard, very recently \citet{galliano22} defined a critical metallicity regime that is in the range from 12~+~log(O/H)~$\sim$8 to $\sim$8.3 for nearby galaxies \citep[see also][]{galliano18}. 
We recall that our sample has metallicity values ranging from 12~+~log(O/H)~=~8.3 to 8.8 ($\sim$20$\%$ have metallicities lower than 8.4).
DGR increases with decreasing $f_{\rm gas}$ (right panel of Fig.~\ref{fig:dgr-z}) following a strong relation.
This anticorrelation says that galaxy regions with lower $f_{\rm gas}$, which implies that they are more evolved and more quiescent
having converted most of their gas reservoir into stars, tend to have higher DGR.    
    
\begin{figure*}
\includegraphics[width=0.5\textwidth]{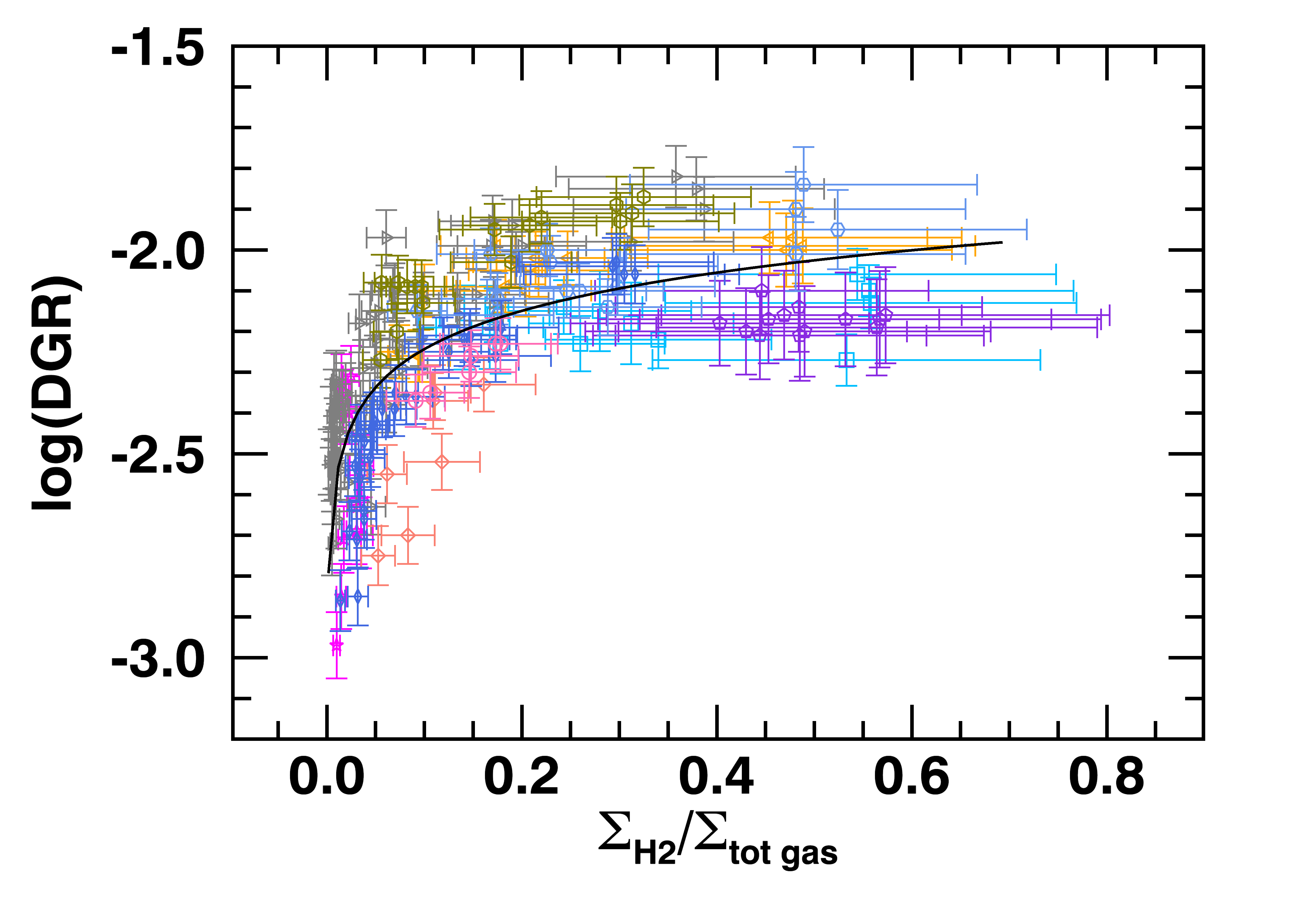}  
\includegraphics[width=0.5\textwidth]{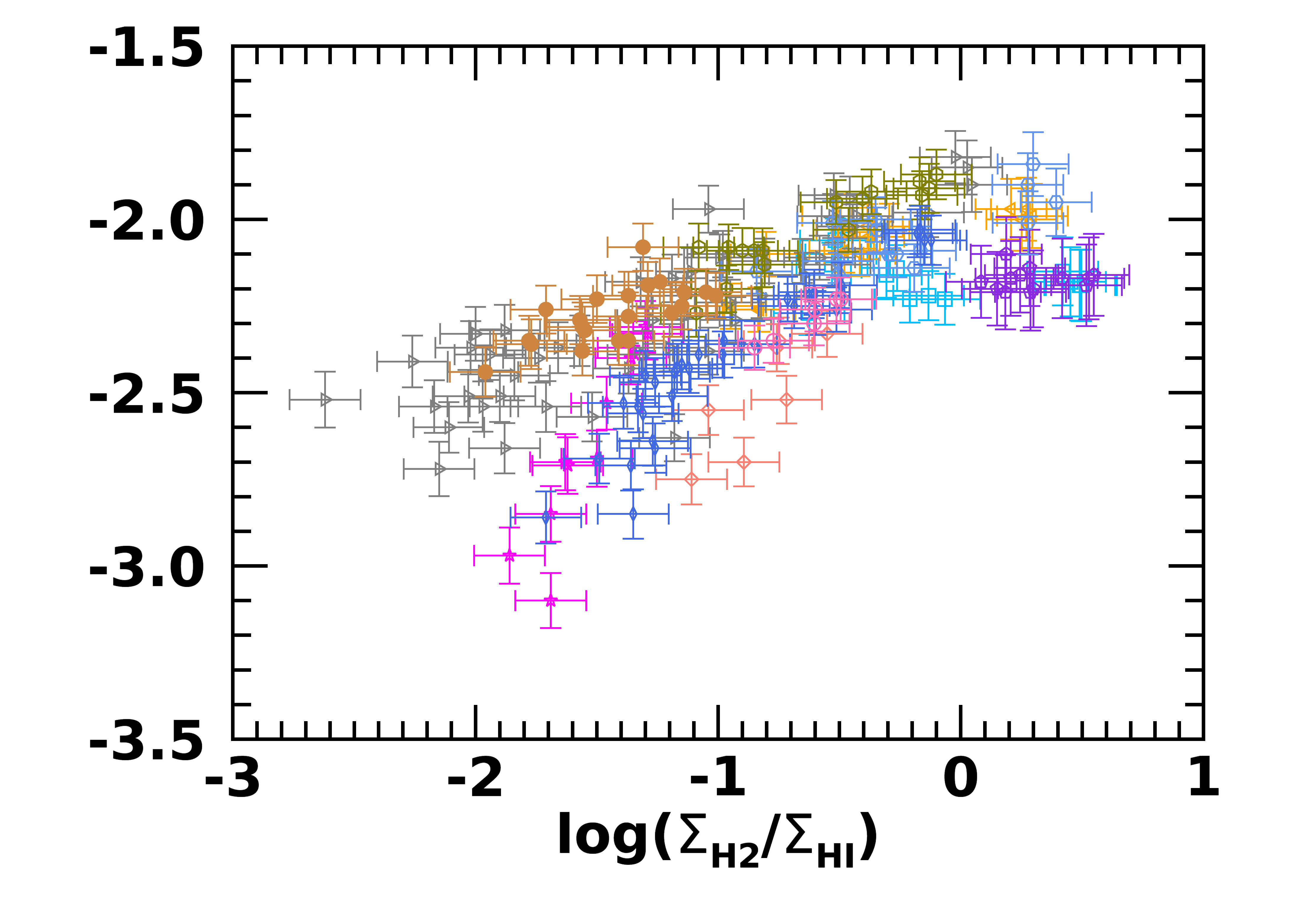}  
\caption{
Same as Fig.~\ref{fig:dgr-mstar-ssfr}, but as a function of $\Sigma_{\rm H2}$/$\Sigma_{\rm tot\,gas}$ (left panel) and logarithm of 
$\Sigma_{\rm H2}/\Sigma_{\rm HI}$ (right panel).
In the left panel, the drawn black line is the fit to the data and it only helps to follow the trend (the corresponding equation is not provided).}
\label{fig:dgr-h2ontgas}
\end{figure*}

We also explore DGR as a function of $\Sigma_{\rm H2}$/$\Sigma_{\rm tot\,gas}$ and $\Sigma_{\rm H2}/\Sigma_{\rm HI}$
following \citet{vilchez19} and C20, respectively. 
\citet{asano13} predicted that the accretion time for grain growth is proportional to the inverse of the fraction of the cold 
component of the ISM, and $\Sigma_{\rm H2}$/$\Sigma_{\rm tot\,gas}$ is a proxy of this  fraction. 
The left panel of Fig.~\ref{fig:dgr-h2ontgas} shows DGR vs $\Sigma_{\rm H2}$/$\Sigma_{\rm tot\,gas}$ across sample galaxies (with constant $X_{\rm CO}$).
The DGR tends to increase with increasing $\Sigma_{\rm H2}$/$\Sigma_{\rm tot\,gas}$ for $\Sigma_{\rm H2}$/$\Sigma_{\rm tot\,gas} \gtrsim$~0.1, 
while it vertically drops down for lower molecular gas fractions.
This suggests a nondependence of low DGR (log(DGR)~$\lesssim -2.3$) on the grain growth.
It is consistent with the results of \citet{vilchez19} and supports the idea that the trend of DGR as a function of $\Sigma_{\rm H2}$/$\Sigma_{\rm tot\,gas}$
is associated to the grain growth that should occur when the DGR is sufficiently high \citep[see also, e.g., ][]{chiang18}.
The right panel of Fig.~\ref{fig:dgr-h2ontgas} displays the trend of DGR as a function of $\Sigma_{\rm H2}/\Sigma_{\rm HI}$:
in general, DGR tends to increase with increasing $\Sigma_{\rm H2}/\Sigma_{\rm HI}$. 
We also note a sort of plateau or the beginning of a decreasing of DGR at $\Sigma_{\rm H2}/\Sigma_{\rm HI} \approx 1$.  
With our dataset, we are not able to draw this trend up to $\Sigma_{\rm H2}/\Sigma_{\rm HI} \sim 10$ and higher, but the behavior of DGR 
at $\Sigma_{\rm H2}/\Sigma_{\rm HI} \approx 1$ seems to be consistent with the hill-like shape published in C20.
The increasing part of the trend shown in the right panel of Fig.~\ref{fig:dgr-h2ontgas} corresponds to galaxy regions 
where H{\sc i} dominates and dust is yet in small quantities with respect to the total gas, and it is predicted 
from theoretical models \citep[e.g.,][]{krumholz09,magrini12,wong13}.
An increasing $\Sigma_{\rm H2}/\Sigma_{\rm HI}$ indeed implies that the atomic gas is transformed into molecular gas and then into stars,
and these latter in turn produce an increase of the hydrostatic pressure in the disk, triggering the formation of dense molecular clouds.
The central and right-hand parts of the plot are instead unexpected, and no simulations or theoretical models, to our knowledge, make predictions on them.
In C20 we proposed some explanations for this decreasing trend invoking dust destroyed by events (e.g., shocks, collisional sputtering) 
and/or reduced due to differential consumption of dust and gas, different emission efficiency of dust grains producing an underestimation
of dust mass, or the use of the same $X_{\rm CO}$  for different H$_2$-dominated regions of different galaxies. 

\begin{figure*}
\includegraphics[width=0.5\textwidth]{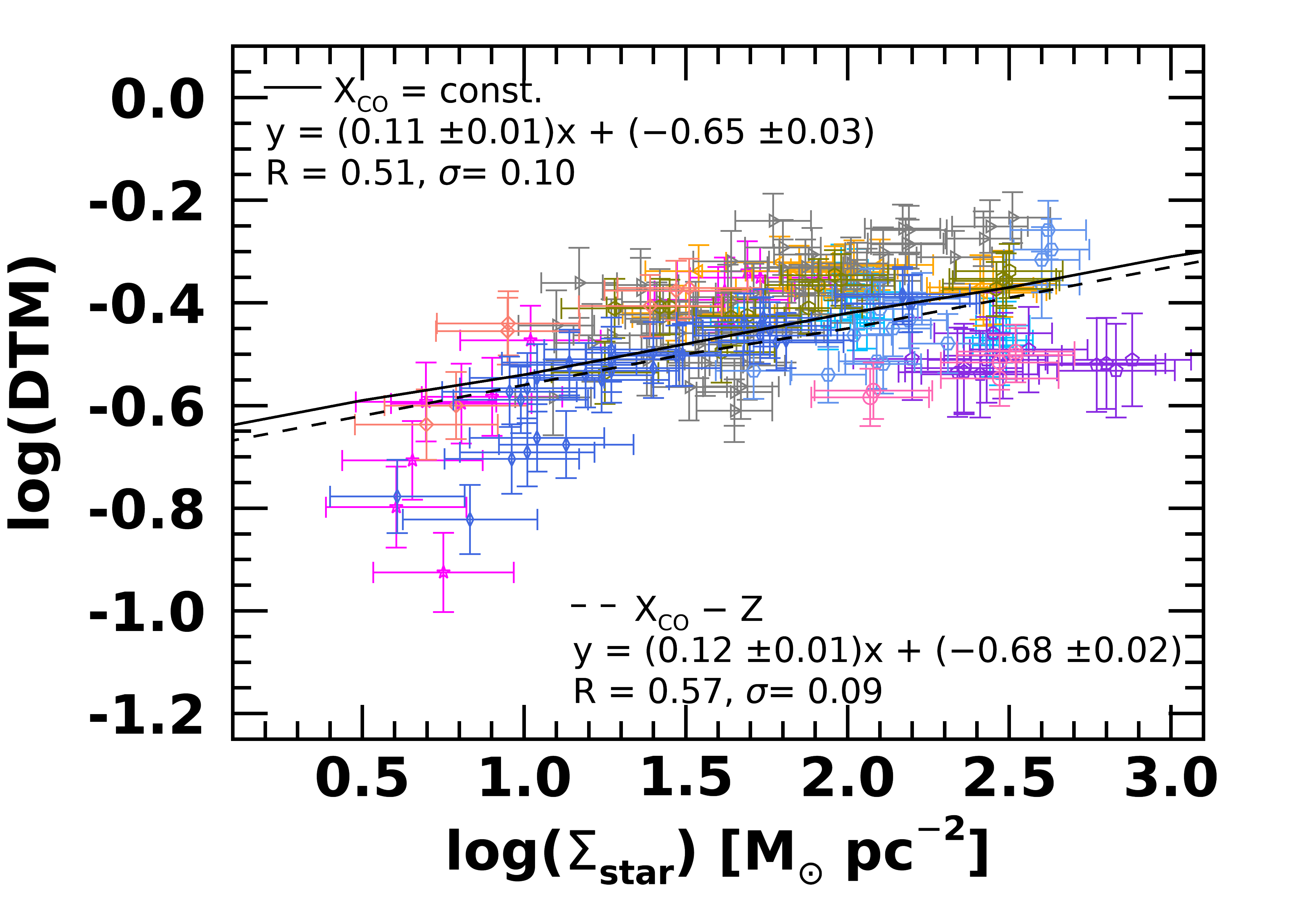}  
\includegraphics[width=0.5\textwidth]{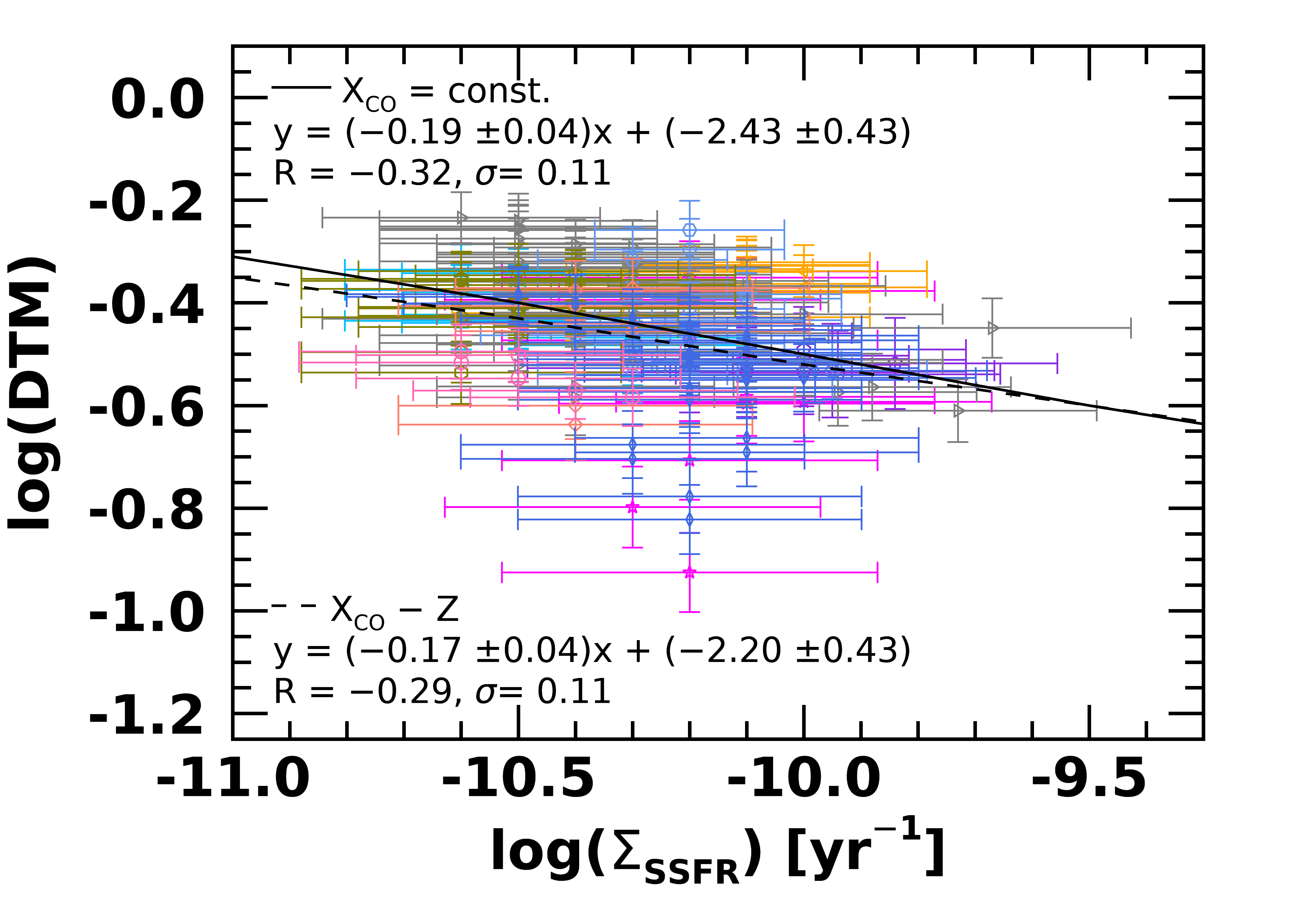}  
\caption{Logarithm of dust-to-metal ratio as a function of logarithm of $\Sigma_{\rm star}$ (left panel) and of $\Sigma_{\rm SSFR}$ (right panel).
Data points refer to constant $X_{\rm CO}$, fit lines to both assumptions on $X_{\rm CO}$ (see legend in Fig.~\ref{fig:dgr-mstar-ssfr} and Table~\ref{tab:ratios}).
Symbols are same as Fig.~\ref{fig:all}.}
\label{fig:dtm-mstar-ssfr}
\end{figure*}

\begin{figure*}
\includegraphics[width=0.5\textwidth]{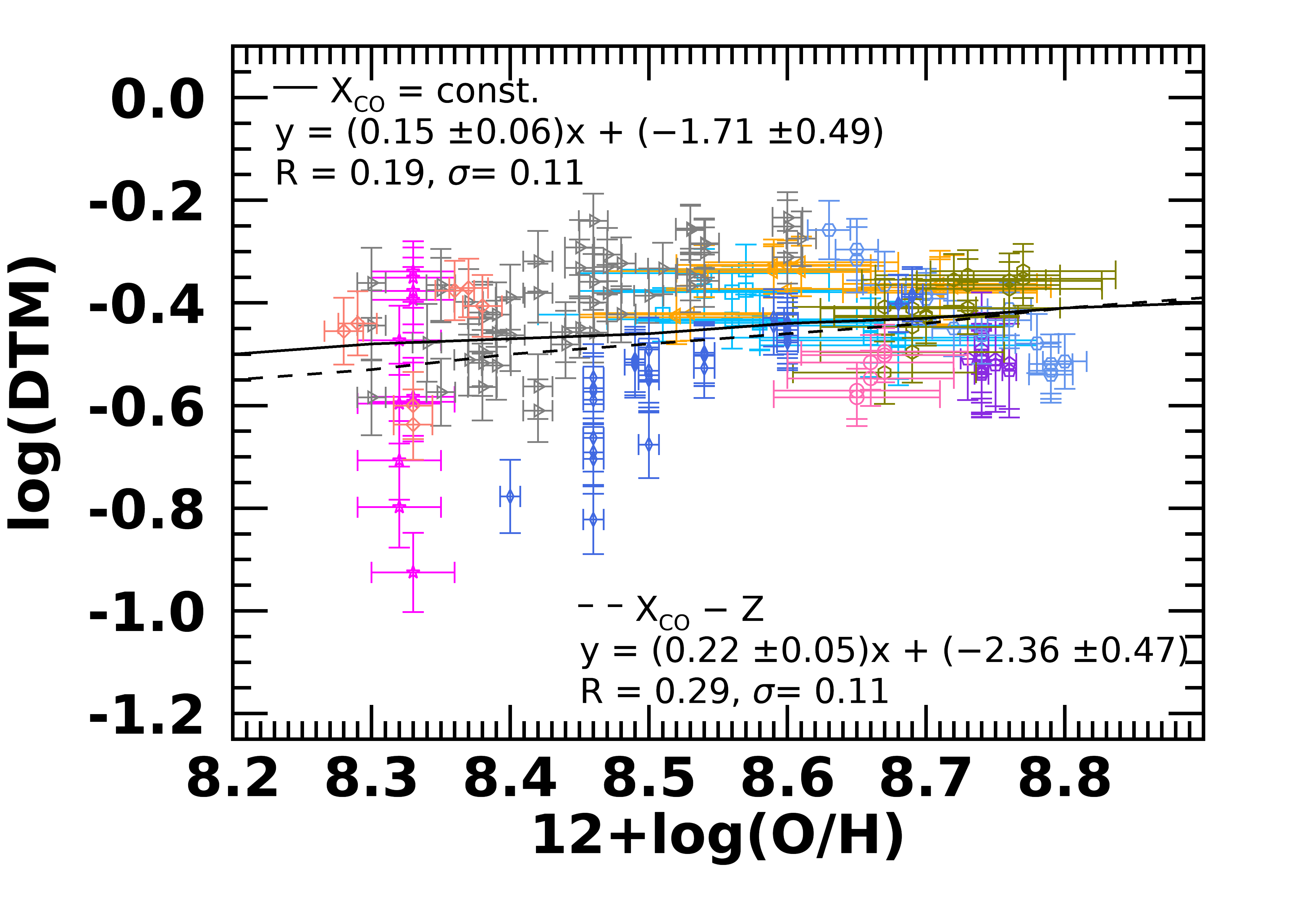} 
\includegraphics[width=0.5\textwidth]{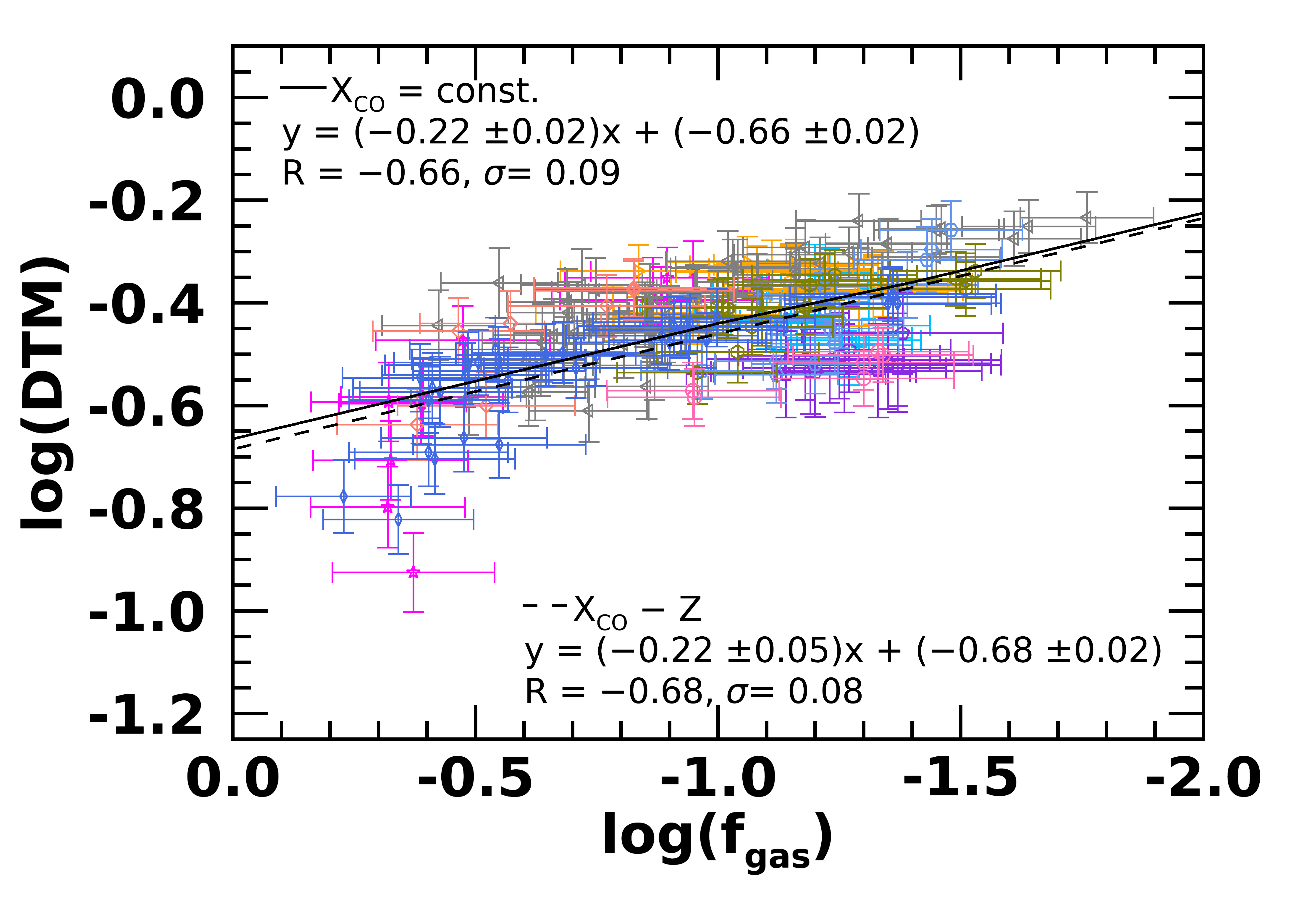}  
\caption{Same as Fig.~\ref{fig:dtm-mstar-ssfr}, but as a function of 12~+~log~(O/H) (left panel) and logarithmic of $f_{\rm gas}$ (right panel).}
\label{fig:dtm-z}
\end{figure*}

\subsection{Dust-to-metal ratio}
\label{sec:dtm}
We explore the ratio between the dust mass and the total amount of metals,
the dust-to-metal ratio (DTM).
Following \citet{devis19}, the DTM is defined as DTM = $\Sigma_{\rm dust}$/$\Sigma_{Z \rm(gas + dust)}$, where 
$\Sigma_{Z({\rm gas + dust})} = f_{Z} \times \Sigma_{\rm gas} + \Sigma_{\rm dust}$ with $f_{Z}$ fraction of metals by mass calculated using 
$f_{Z} = 27.36 \times 10^{(12+\rm{log(O/H}) - 12)}$. 
The DTM is a measure of the fraction of metals locked up in the interstellar dust grains, and its behavior provides information 
on the efficiency of dust production and destruction mechanisms.
By simplifying the general theoretical framework \citep[e.g.,][]{asano13}, the DTM is expected to remain constant 
if the dust is mainly produced by stellar sources, to increase as galaxies evolve enriching the ISM with metals, or to decrease if the dust 
is destroyed via SN shocks.    
The mean values of DTM ratios for our resolved sample are log(DTM)~$= -0.43 \pm 0.10$ and $-0.46 \pm 0.10$ for constant and 
metallicity-dependent $X_{\rm CO}$, respectively.
These values are consistent within errors with the high end of DTM ratios of nearby 
galaxies\footnote{For nearby galaxies typically $-0.90 \lesssim$ log(DTM) $\lesssim -0.40$ \citep[e.g.,][]{delooze20}.} 
corresponding, for instance, to non-H{\sc i}-deficient galaxies
of the HRS sample \citep[see][for the definition of non-H{\sc i}-deficient galaxies]{devis17a}.

Figures~\ref{fig:dtm-mstar-ssfr} and  \ref{fig:dtm-z} show DTM against $\Sigma_{\rm star}$, $\Sigma_{\rm SSFR}$,
12~+~log~(O/H), and $f_{\rm gas}$.
The trends as a function of $\Sigma_{\rm star}$, $\Sigma_{\rm SSFR}$, and $f_{\rm gas}$ are broadly similar to 
those found for DGR but with flatter slopes and lower correlation coefficients (see Table~\ref{tab:ratios}).
This suggests that DGR and DTM are quite interchangeable quantities with respect to $\Sigma_{\rm star}$, $\Sigma_{\rm SSFR}$, 
and $f_{\rm gas}$, and we can reasonable assume that the underlying physics is approximatively the same (see Sect.~\ref{sec:dgr}).
DTM against 12~+~log~(O/H) (left panel of Fig.~\ref{fig:dtm-z}) instead reveals no clear trend, with DTM remaining approximately constant
across the metallicity range.   
This might indicate that the critical metallicity threshold above which grain growth becomes 
efficient in the dust production is not important in the explored range of metallicity (see Sect.~\ref{sec:dgr}).  
The absence of a clear trend of DTM vs 12~+~log~(O/H) also characterizes global analyses of galaxy samples 
with or without few low-metallicity systems \citep[e.g.,][]{devis19,delooze20}.     
   
\begin{table}
\caption{\label{tab:ratio-values} 
Mean values of dust-to-stellar, dust-to-gas, and dust-to-metal ratios for the galaxy sample at 3.4~kpc.}
\centering
\begin{tabular}{lc}
\hline\hline
Ratio & Value  \\
\hline
log($\Sigma_{\rm dust}$)/log($\Sigma_{\rm star}$)	& $-3.12 \pm 0.39$ \\
log(DGR), Const. $X_{\rm CO}$ 				& $-2.20 \pm 0.04$ \\
log(DGR), $Z$--dep. $X_{\rm CO}$ 				& $-2.23 \pm 0.21$ \\
log(DTM), Const. $X_{\rm CO}$ 				& $-0.43 \pm 0.10$ \\
log(DTM), $Z$--dep. $X_{\rm CO}$  				& $-0.46 \pm 0.10$ \\
\hline							
\hline
\end{tabular}
\end{table}

\begin{table*}
\caption{\label{tab:ratios} 
Main properties of the linear fits to data shown in Figs.~\ref{fig:mstar-ssfr-dustonstar}--\ref{fig:dtm-z}.}
\centering
\begin{tabular}{lcccc}
\hline\hline
Scaling relation $(x - y)$$^{(1)}$ & $\textit{m}$$^{(2)}$ & $\textit{q}$$^{(2)}$	& $R$ ($\sigma$, n. pts)$^{(2)}$	& $X_{\rm CO}$$^{(3)}$\\
\hline
log($\Sigma_{\rm star}$) -- log($\Sigma_{\rm dust}$)/log($\Sigma_{\rm star}$)	& $-0.56\pm0.02$	& $-2.19\pm0.04$	& $-0.82$ (0.22, 303) 	 & --\\
log($\Sigma_{\rm SSFR}$) -- log($\Sigma_{\rm dust}$)/log($\Sigma_{\rm star}$)	& $0.63\pm0.06$	& $3.30\pm0.62$	& 0.52 (0.32, 299) & --\\
log(12~+~log(O/H) -- log($\Sigma_{\rm dust}$)/log($\Sigma_{\rm star}$)		& $-0.80\pm0.09$	& $3.64\pm0.75$	& $-0.56$ (0.17, 181) 	 & --\\
log($\Sigma_{\rm star}$) -- log(DGR)	& $0.37\pm0.02$	& $-2.92\pm0.04$	& $0.80$ (0.14, 202) 	 & Const.\\
log($\Sigma_{\rm star}$) -- log(DGR)	& $0.37\pm0.02$	& $-2.97\pm0.04$	& $0.81$ (0.14, 181) 	& $Z$--dep.\\
log($\Sigma_{\rm SSFR}$) -- log(DGR)	& $-0.18\pm0.07$	& $-4.09\pm0.69$	& $-0.18$ (0.23, 202) 	 & const.\\
log($\Sigma_{\rm SSFR}$) -- log(DGR)	& $-0.29\pm0.10$	& $-5.24\pm0.98$	& $-0.22$ (0.24, 181) 	& $Z$--dep.\\
12~+~log(O/H) -- log(DGR)	& $1.17\pm0.09$	& $-12.28\pm0.73$	& 0.72 (0.09, 181) 	 & Const.\\
12~+~log(O/H) -- log(DGR)	& $1.28\pm0.08$	& $-13.23\pm0.67$	& 0.77 (0.17, 181) 	 & $Z$--dep.\\
$f_{\rm gas}$ -- log(DGR)		& $-0.63\pm0.02$	& $-2.88\pm0.02$	& 0.89 (0.11, 202) 	 & Const.\\
$f_{\rm gas}$ -- log(DGR)		& $-0.66\pm0.02$	& $-2.92\pm0.02$	& 0.90 (0.10, 181) 	 & $Z$--dep.\\
log($\Sigma_{\rm star}$) -- log(DTM)	& $0.11\pm0.01$	& $-0.65\pm0.03$	& $0.51$ (0.10, 181) 	 & Const.\\
log($\Sigma_{\rm star}$) -- log(DTM)	& $0.12\pm0.01$	& $-0.68\pm0.02$	& $0.57$ (0.09, 181) 	& $Z$--dep.\\
log($\Sigma_{\rm SSFR}$) -- log(DTM)	& $-0.19\pm0.04$	& $-2.43\pm0.43$	& $-0.32$ (0.11, 181) 	 & Const.\\
log($\Sigma_{\rm SSFR}$) -- log(DTM)	& $-0.17\pm0.04$	& $-2.20\pm0.43$	& $-0.29$ (0.11, 181) 	& $Z$--dep.\\
12~+~log(O/H) -- log(DTM)	& $0.15\pm0.06$	& $-1.71\pm0.49$	& 0.19 (0.11, 181) 	 & Const.\\
12~+~log(O/H) -- log(DTM)	& $0.22\pm0.05$	& $-2.36\pm0.47$	& 0.29 (0.11, 181) 	 & $Z$--dep.\\
$f_{\rm gas}$ -- log(DTM)	& 	$-0.22\pm0.02$	& $-0.66\pm0.02$	& $-0.66$ (0.09, 181) 	 & Const.\\
$f_{\rm gas}$ -- log(DTM)	& 	$-0.22\pm0.05$	& $-0.68\pm0.02$	& $-0.68$ (0.08, 181) 	 & $Z$--dep.\\
\hline
\hline
\end{tabular}
\tablefoot{
$^{(1)}$ Explored scaling relation.
$^{(2)}$ Slope $m$, intercept $q$, Pearson correlation coefficient $R$, dispersion $\sigma$, and number of pixels.
$^{(3)}$ The two assumptions on $X_{\rm CO}$: ``Const.'' corresponds to constant $X_{\rm CO}$, ``$Z$--dep.'' to metallicity-dependent $X_{\rm CO}$ (see Sect.~\ref{sec:gas}).
}
\end{table*}

\section{Discussion}
\label{sec:discussion} 

\subsection{The nonuniversality of resolved scaling relations}
\label{sec:nonuniversality} 
All the studied SRs show moderate or strong correlations except the $\Sigma_{\rm dust}$--$\Sigma_{\rm HI}$ SR that 
does not exist for most galaxies.
Since mass radial profiles of many quantities (e.g., dust, stars, gas) in spiral galaxies have been drawn  
and generally characterized with an exponential decline of their surface density with radius
\citep[e.g.,][]{alton98,xilouris99,bianchi07,leroy08,leroy09,munoz09,schruba11,degeyter14,hunt15,smith16,casasola17}, 
a positive correlation between these properties is expected within individual galaxies.
We find that each galaxy is characterized by distinct SRs on the explored scales.
The studied SRs are therefore, on average, strong correlations though not universal.
These variations found for individual galaxies, not being driven by the sampled scale, seem to be affected by local processes and 
galaxy peculiarities (e.g., an overabundance of \hi).
The nonuniversality of the SRs at subkpc/kpc scales has been already found by other authors
referred to a restricted set of SRs, the SF relations \citep[e.g.,][]{vulcani19,ellison21,thorp22}. 
We extend the nonuniversality to other SRs involving dust, \hi, and total baryonic content. 

\citet{leroy13} cautioned that the KS relation tends to show a uniform behavior when the studied sample 
is dominated by star-forming disk galaxies.
However, our sample shows a wide range of behaviors in terms of SRs although it consists of galaxies belonging 
to the same subclass of objects (that is, large, face-on, spiral galaxies) and all the derived quantities have been treated 
with the same methodology.
The nonuniversal KS relation we find is consistent with a KS relation 
varying significantly from cloud to cloud in the Milky Way found by some authors
\citep[e.g.,][]{gutermuth11,lada13,willis15,pokhrel20,pokhrel21}.
However, \citet{pokhrel21} found a tight, linear correlation between $\Sigma_{\rm {gas}}$ and $\Sigma_{\rm {SFR}}$ normalized 
by the gas free-fall time ($t_{ff}$) for 12 Galactic nearby ($<$1.5~kpc) molecular clouds.
This relation is the same in all studied molecular clouds spanning a huge range of properties (e.g., clouds without massive stars or massive stellar feedback, clouds being sites of ongoing massive SF, large complexes comparable to those observed in nearby galaxies).
Most theoretical models predicting the existence of a KS relation for single clouds also predict a dependence on the gas $t_{ff}$
\citep[e.g.,][]{krumholz05,federrath12,padoan12,krumholz19}, which in turn depends on the volume density.  
The finding of a universal intercloud KS relation, which link SFR to the volume density, supports models in which SF is regulated 
by local processes (e.g., turbulence, stellar feedback, protostellar outflows), while disfavors models in which SF is driven only by galaxy properties 
or phenomena on galactic scales (e.g., SN feedback).
The dependence on local processes is in line with our results showing different SRs (not only the KS one) found for individual galaxies
at subkpc/kpc scales.

All SRs involving only H$_2$ gas mass are potentially affected by the assumptions on $X_{\rm CO}$.
On the contrary, SRs involving total (H$_2$ + \hi) gas are, on average, less dependent 
on these assumptions, suggesting that adding \hi\ tends to cancel out or mitigate 
the variations observed in the SRs as a function of $X_{\rm CO}$.

It is well known that the H$_{2}$ gas typically dominates in the central regions of the optical disk of galaxies while \hi\ outside of it, 
extending, on average, far away beyond $r_{25}$, by a factor between 2 and 4 
\citep[$r_{\rm HI} = 2-4r_{25}$, e.g.,][]{wong02,bigiel08,casasola17}.
This is the most frequently observed trend in galaxies and can explain the very weak or absent correlation
between dust (present within the disk of galaxies) and \hi\ gas at subkpc/kpc scales.
However, we have to mention that a variety of ISM morphologies has been found  
\citep[][]{dib21}, including a central gaseous hole, both in CO and \hi, as, for instance, in the Milky Way 
\citep[][]{misiriotis06} and NGC~3031 \citep[][present in our sample]{casasola07}.
The idea that recently emerged is that the role of H{\sc i} in regulating SF may have been down-played so far \citep[see, e.g., the discussion in][]{ellison21}.
\citeauthor{bacchini19a}~(\citeyear{bacchini19a}, \citeyear{bacchini19b})
defined a tight correlation between volume densities of H{\sc i} and SFR in disk galaxies, including the Milky Way, and 
both \citet{saintonge16} and \citet{morselli20} found that the MS can be explained by their cold gas reservoir as observed in the H{\sc i} line,
both globally and with spatially resolved measurements. 
More recently,  \citet{morselli21} estimated the redshift evolution of the H$_{2}$/H{\sc i} mass ratio within galaxies finding that H{\sc i} 
should not be neglected at high redshift, as commonly done, but rather we should reevaluate its role \citep[see also][]{roychowdhury15}. 

\subsection{The resolved vs global scaling relations}
\label{sec:global}
The SRs studied in this paper hold starting from the scale of 0.3~kpc, with some exceptions (see Appendix~\ref{sec:peculiar}).
If a breaking down scale for these SRs exists, it is below 0.3~kpc.
When all galaxies are evaluated together at the common scale of 3.4~kpc, differences due to peculiarities of individual galaxies,
such as an abundant \hi\ content (e.g., NGC~7793) and/or an anomalous dust mass/distribution (e.g., NGC~300), are cancelled out.
The resulting SRs are consistent with those typically characterizing whole galaxies.
The scale of 3.4~kpc therefore provides a view of the behavior of SRs at global scales.

The good match between subgalactic and global SRs should be in line with a 
scenario where main physical processes regulating the properties and evolution of galaxies occur at subkpc/kpc scales and drive the behavior of galaxies as a whole. 
In particular, the agreement between the SF relationships studied at different scales has been recently 
discussed in the hypothesis of self-regulation of the SF process \citep[e.g.,][]{zaragoza19,zaragoza20,barrera21,sanchez21}.
This scenario could be applied to all SRs we explore.
The hypothesis of self-regulation of the SF process assumes that since stars impact their environment, the rate at which gas can collapse 
to make new stars is also affected by the previous generation of SF.
The molecular clouds, once collapsed, follow, both individually and locally, the KS relation.  
The different slopes of this relation present in the literature would be explained by the fact that 
this relationship is ``smoothed'' at subkpc/kpc scales including different galaxy regions, that is those actively forming stars, 
those not forming stars, and those with stars formed by the previous generations of clouds. 
Timescales also play a role in the scenario of self-regulation of the SF process. 
Many works found that the KS relation holds when it is averaged on timescales of $\sim$30--100~Myr \citep[e.g.,][]{silk97,ostriker10,kruijssen19},
much larger than those of the SF process \citep[for instance, a typical mean density within GMCs of $n_{\rm H} = 100$~cm$^{-3}$
implies a typical $t_{ff}$~$\sim$$2-4\times10^{6}$~yr, see, e.g.,][]{li19b}. 
\citet{kruijssen19} inferred that the decorrelation between the molecular gas and high-mass SF
on the spatial scale of GMCs in NGC~300 implies rapid evolutionary cycling 
between clouds, SF, and feedback. 
They demonstrated that SF is regulated by efficient stellar feedback arising from radiation and stellar winds, before SN 
explosions can occur.

If small-scale processes drive the SF and a large set of SRs at different scales, this does not mean that also 
global processes and global galaxy properties do not play a role \citep[see, e.g.,][]{abdurrouf22}. 
Galaxies with strong bars, interactions between galaxies and with the environment, mergers, 
gas inflows/outflows, and/or the presence of an AGN can affect the gas fraction and therefore the SF of the entire galaxy 
\citep[e.g.,][]{casasola04,poggianti17,sanchez18,sanchez20,mancillas19}.
However, as pointed out by \citet{sanchez21}, they influence the whole galaxy in an indirect way, via the physical processes 
that affect the SF.

\subsection{Ratios and their evolution}
\label{sec:ratios-discussion}
The mean values and evolution of specific dust mass, DGR, and DTM for nearby galaxies at subgalactic scales are consistent with those derived 
for whole galaxies.
In particular, their evolution is in agreement with what is globally observed both for nearby and high-redshift galaxies 
\citep[e.g.,][]{remy-ruyer14,calura17,casasola20,popping22}.
\citet{popping22} especially found little to no evolution in the observed trends between DGR and DTM and gas-phase metallicity from $z = 0$ to 5.
They concluded that this is indicative of a balance between the formation and destruction of dust already at $z = 5$, when the Universe was 1.2-Gyr old.
These observational results are consistent with the predictions of the model of \citet{li19a} finding a little evolution in the DGR-metallicity relationship in the redshift range $z = 0 - 6$. 

At subgalactic scales the specific dust mass is primarily correlated with $\Sigma_{\rm star}$, 
while DGR and DTM are better correlated with $f_{\rm gas}$ though correlations with 12~+~log(O/H) 
and $\Sigma_{\rm star}$ are nonnegligible and very weak or absent correlations are found for DGR and DTM with $\Sigma_{\rm SSFR}$. 
Global analyses of galaxies at different redshifts show instead that DGR and DTM depend most sensitively on the gas-phase metallicity, 
and there are important secondary relationships between these ratios and $\Sigma_{\rm star}$ and $f_{\rm gas}$ and moderate correlations
with $\Sigma_{\rm SSFR}$ \citep[e.g.,][]{remy-ruyer14,li19a}.
The explored ratios, all involving the dust content, therefore 
would seem to be set and governed by subgalactic evolutionary processes that then would drive the galaxy evolution as a whole.

\section{Conclusions}
\label{sec:conclusions}  
From the analysis of an unprecedented set of ten pixel-by-pixel SRs between dust, H$_2$, \hi, total gas, stars, total baryonic content, SFR, 
and some ratios of these quantities for 18 dust-resolved DustPedia galaxies at scales between 0.3 and 3.4~kpc, studied within the optical disk and exploring
both constant and metallicity-dependent $X_{\rm CO}$, we draw the following conclusions.
\begin{itemize}
\item All the explored SRs are moderate or strong correlations except the $\Sigma_{\rm dust}$--$\Sigma_{\rm HI}$ SR 
that is negligible or very weak for most galaxies.
We do not define universal SRs at scales below 3.4~kpc, in the range 0.3--3.4~kpc, because each galaxy is characterized by distinct SRs, affected by local processes and galaxy peculiarities. 
For a given galaxy, the slopes and correlation coefficients of the SRs are approximately independent on scale, in the range 0.3--3.4~kpc. 
The SRs hold, on average, starting from 0.3~kpc, and, if a breaking down scale exists, it is below 0.3~kpc.
\item 
By studying all galaxies together at the scale of 3.4~kpc, differences due to peculiarities of individual galaxies are cancelled out and the corresponding SRs are compatible with those of whole galaxies.
The consistency between subgalactic and global SRs may be compatible with a picture where the main physical processes regulating the properties and evolution of galaxies locally occur.
In this perspective, our results are in line with the scenario of self-regulation of the SF.
\item 
The most striking result emerges from the ISM SRs. 
The $\Sigma_{\rm dust}$--$\Sigma_{\rm tot\,gas}$ SR is a good correlation at all scales,
while the $\Sigma_{\rm dust}$--$\Sigma_{\rm H2}$ and $\Sigma_{\rm dust}$--$\Sigma_{\rm HI}$ SRs have different behaviors based on scale.
The atomic gas is globally a very good tracer of dust within the disk of spiral galaxies, while it is almost always 
not correlated with dust at subkpc/kpc scales.
The H$_2$ gas is globally a good tracer of dust within the disk of spiral galaxies, and the dust--H$_2$ correlation improves further at subkpc/kpc scales.
\item 
All SRs involving only H$_2$ mass are potentially affected by the assumption on $X_{\rm CO}$, however the results found under different assumptions are similar, 
in terms of slopes and strength. 
SRs involving (H$_2$~+~\hi) gas are less dependent on the prescription on $X_{\rm CO}$.
\item
The mean values and the evolution of specific dust mass, DGR, and DTM for nearby galaxies at subgalactic scales
are consistent with those derived for whole galaxies from $z = 0$ to 5, suggesting that also these ratios are mainly set by local processes.
We find that DGR depends on metallicity as 
DGR~$\propto$~(O/H)$^{1.2-1.3}$
and this is consistent with theoretical expectations including dust grain growth.
DGR against $\Sigma_{\rm H2}$/$\Sigma_{\rm tot\,gas}$ shows that low DGR values do not depend on the grain growth, 
and DTM against 12~+~log~(O/H) that the critical metallicity threshold above which grain growth becomes 
efficient in the dust production is not crucial in the explored range of metallicity.
\end{itemize}

Our results underline the importance of focusing on resolved local galaxies in the general picture of galaxy evolution.
In this respect, the DustPedia collaboration is planning a set of studies at finer resolution (e.g., Casasola et al., in prep.; Salvestrini et al. in prep.).
The newly rediscovered important of HI in the SF process is particularly pertinent with current and incoming facilities 
devoted also to trace \hi\ emission in galaxies, such as the operative MeerKAT and the future ngVLA and SKA.
In one of these incoming papers, we will present the \hi\ properties in nearby galaxies with new high-resolution and high-sensitivity MeerKAT observations. 
These works provide both new observational constraints for theoretical models on structure, formation, and evolution of galaxies 
and an updated local benchmark for high-redshift studies.

\begin{acknowledgements}
This paper is dedicated to the memory of Jonathan Ivor Davies, mentor and leader of the DustPedia collaboration.
The presented results have been obtained during the COVID-19 pandemic and this work is also dedicated to all the people 
who during these hard times continued working in hospitals, pharmacies, grocery stores, transports, and other essential services, 
allowing us to continue our research work in astronomy.  
We are grateful to the anonymous referee for the rapidity and pertinence of his/her comments and suggestions improved the quality of this manuscript.
DustPedia is a collaborative focused research project supported by the European Union under the Seventh Framework Programme (2007-2013) call (proposal no. 606824). 
The participating institutions are: Cardiff University, UK; National Observatory of Athens, Greece; Ghent University, Belgium; Universit\'{e} Paris Sud, France; 
National Institute for Astrophysics, Italy and CEA (Paris), France. 
We acknowledge funding from the INAF main stream 2018 program ``Gas-DustPedia: A definitive view of the ISM in the Local Universe''.
VC, SB, and EC acknowledge the support from grant PRIN MIUR 2017 - 20173ML3WW$\_$001.
JF acknowledges financial support from the UNAM-DGAPA-PAPIIT IN111620 grant, M\'exico.
This paper makes use of the following ALMA data: ADS/JAO.ALMA${\#}$2017.1.00886.L and ADS/JAO.ALMA${\#}$2017.1.00129.S.
ALMA is a partnership of ESO (representing its member states), NSF (USA) and NINS (Japan), together with NRC (Canada), MOST and ASIAA (Taiwan), 
and KASI (Republic of Korea), in cooperation with the Republic of Chile. 
The Joint ALMA Observatory is operated by ESO, AUI/NRAO and NAOJ. 
In addition, publications from NA authors must include the standard NRAO acknowledgement: 
The National Radio Astronomy Observatory is a facility of the National Science Foundation operated under cooperative agreement by Associated Universities, Inc.
This work made use of HERACLES, ``The HERA CO-Line Extragalactic Survey'' \citep[][]{leroy09}.
This work made use of THINGS, ``The HI Nearby Galaxy Survey'' \citep[][]{walter08}. 
This publication made use of data from COMING, CO Multi-line Imaging of Nearby Galaxies, a legacy project of the Nobeyama 45-m radio telescope.
\end{acknowledgements}

%----------------

\begin{appendix}

\section{Additional figures -- Scaling relations}
\label{sec:add-sample}
In this section, we collect figures showing SRs for the entire galaxy sample 
as done in Fig.~\ref{fig:n6946} for NGC~6946.
Main parameters of the linear fits are collected in Table~\ref{tab:fit-sr}.

\begin{figure*}
\centering
\includegraphics[width=0.33\textwidth]{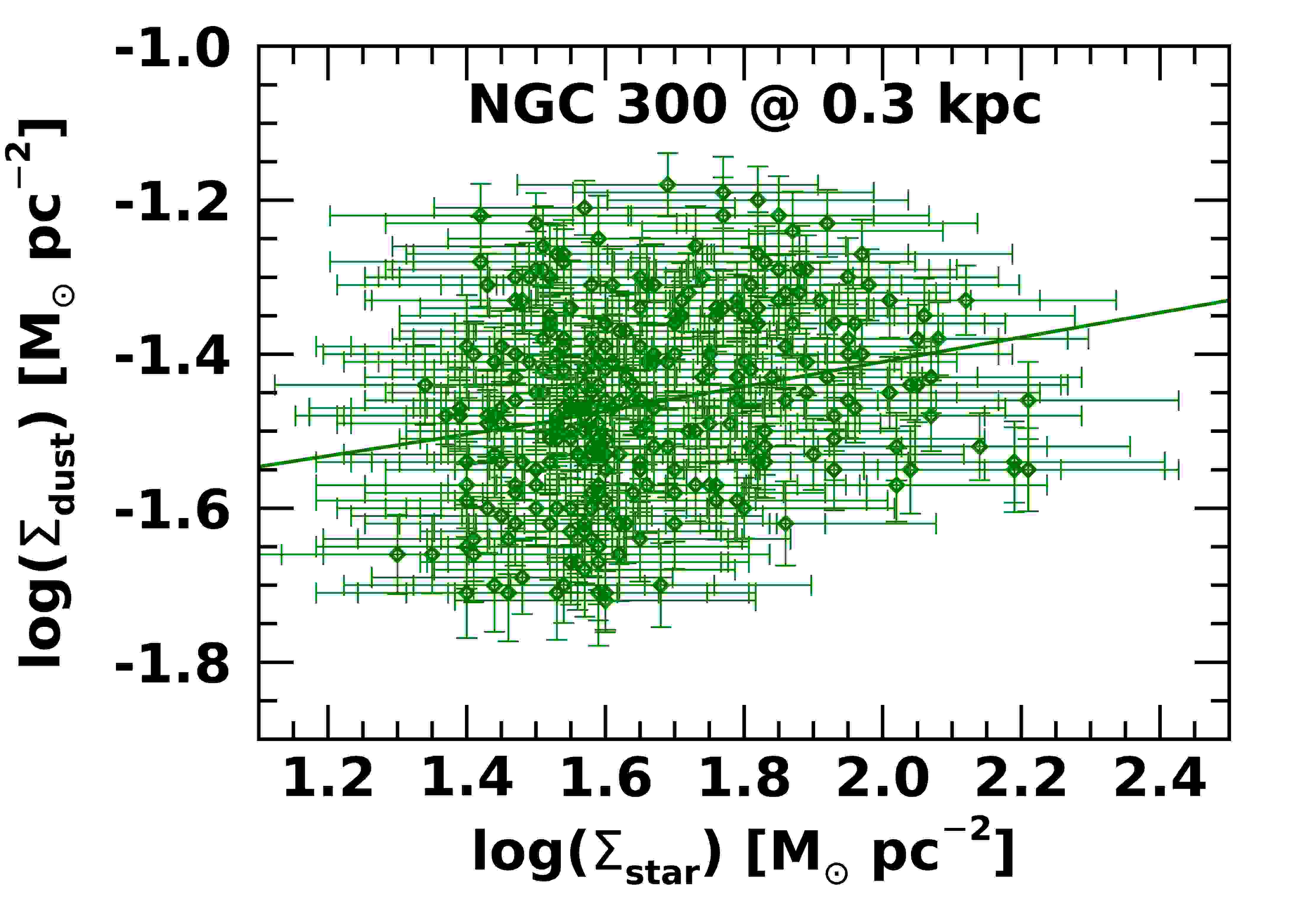}
\includegraphics[width=0.33\textwidth]{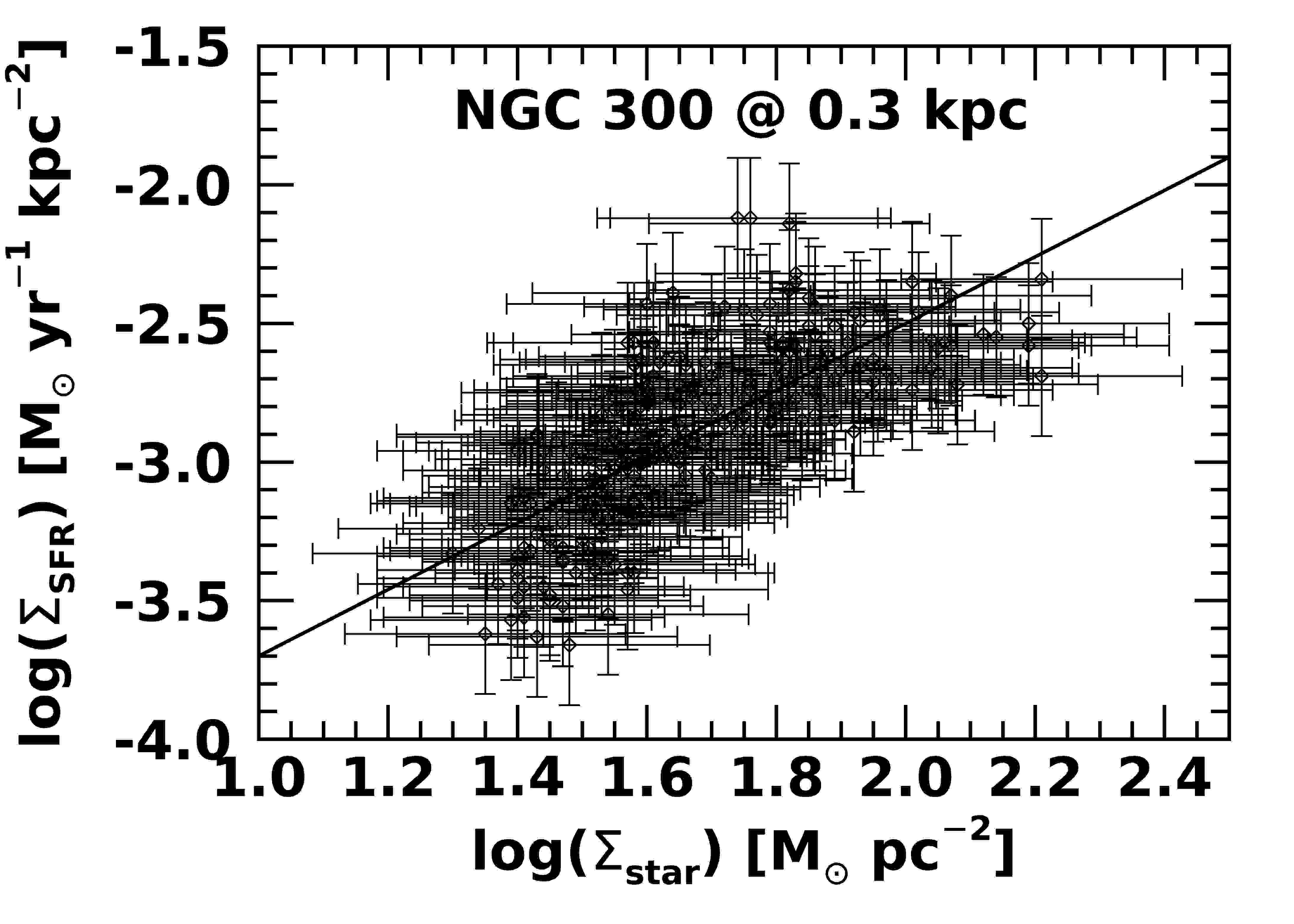}
\includegraphics[width=0.33\textwidth]{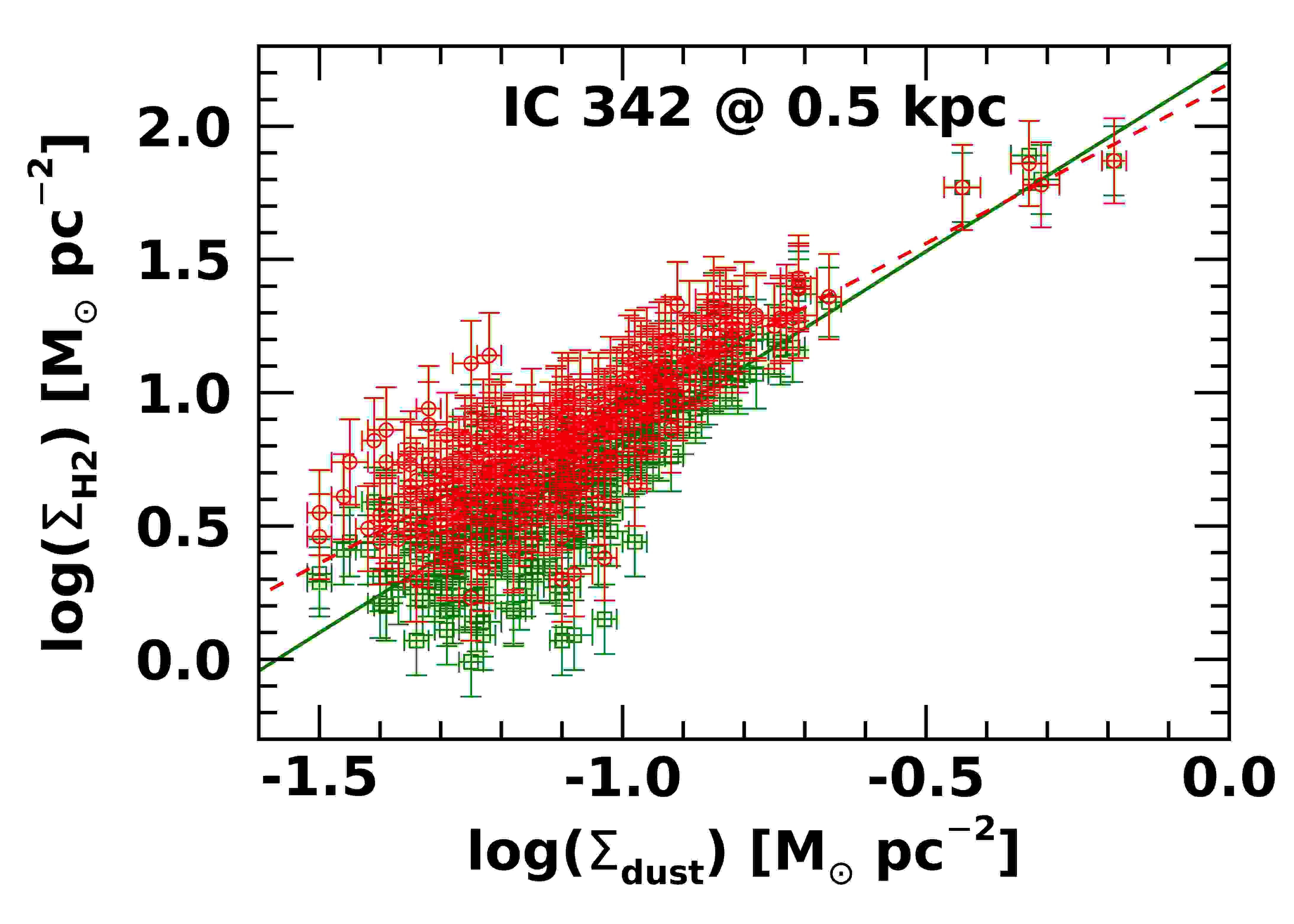}
\includegraphics[width=0.33\textwidth]{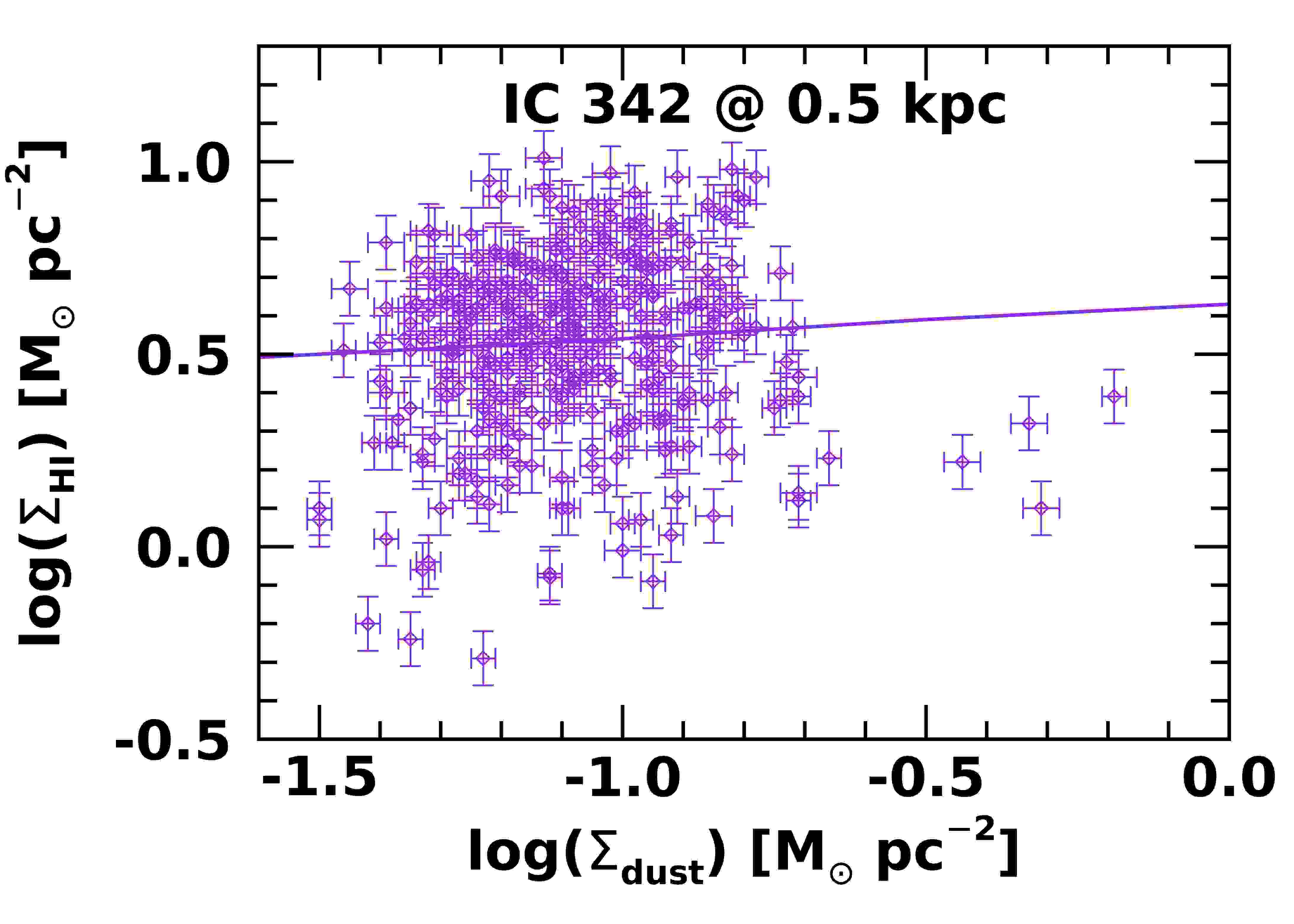}
\includegraphics[width=0.33\textwidth]{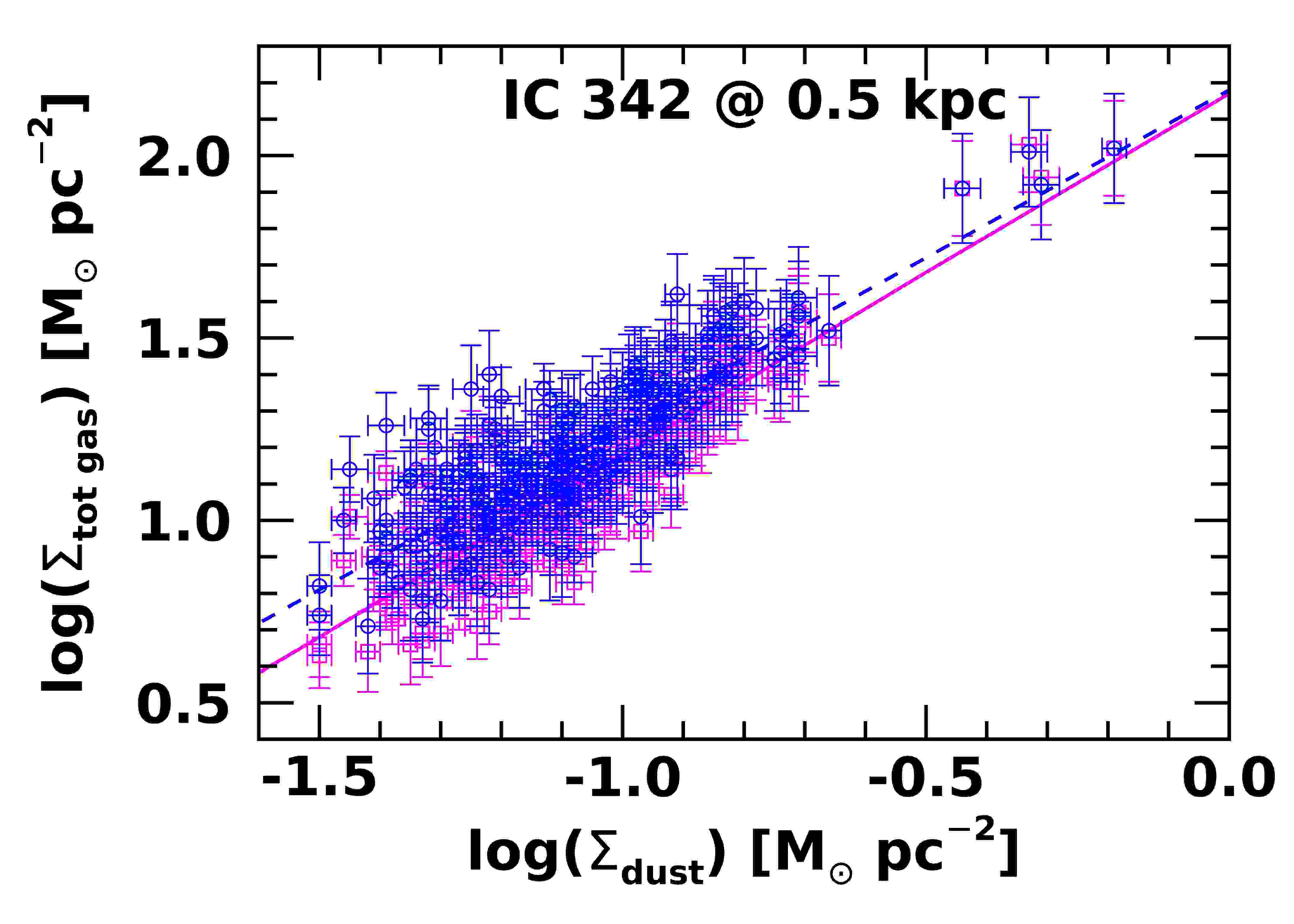}
\includegraphics[width=0.33\textwidth]{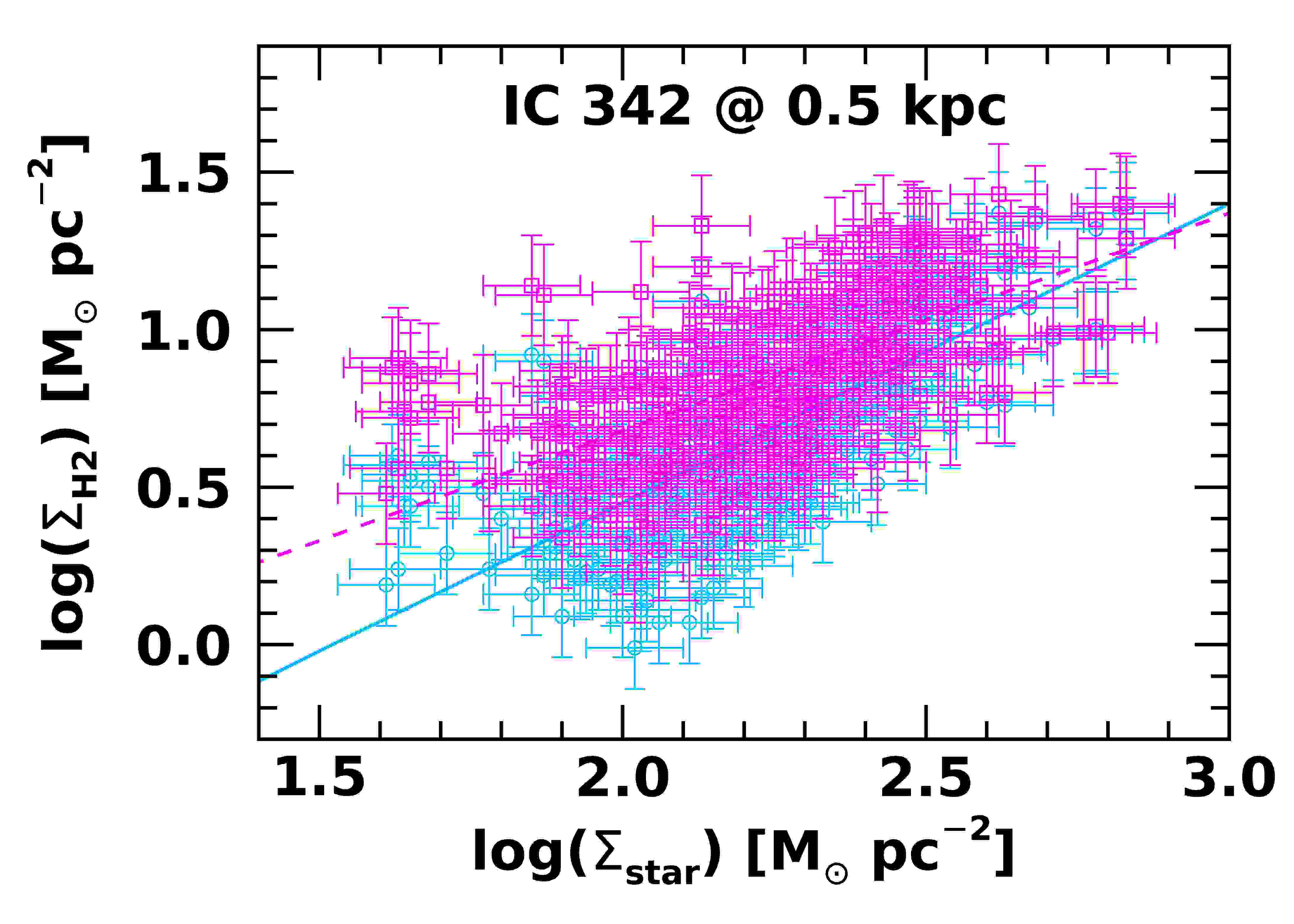}
\includegraphics[width=0.33\textwidth]{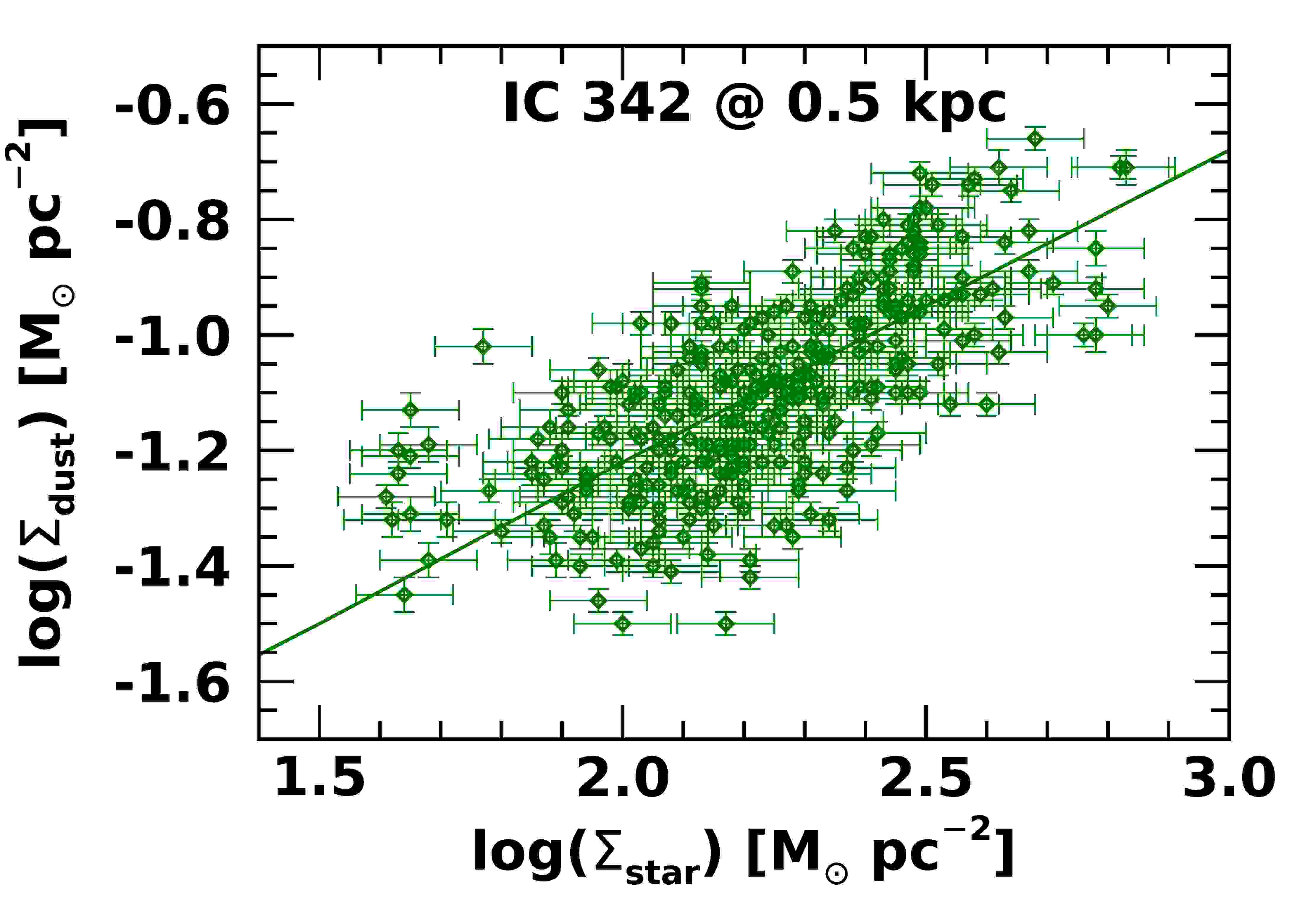}
\includegraphics[width=0.33\textwidth]{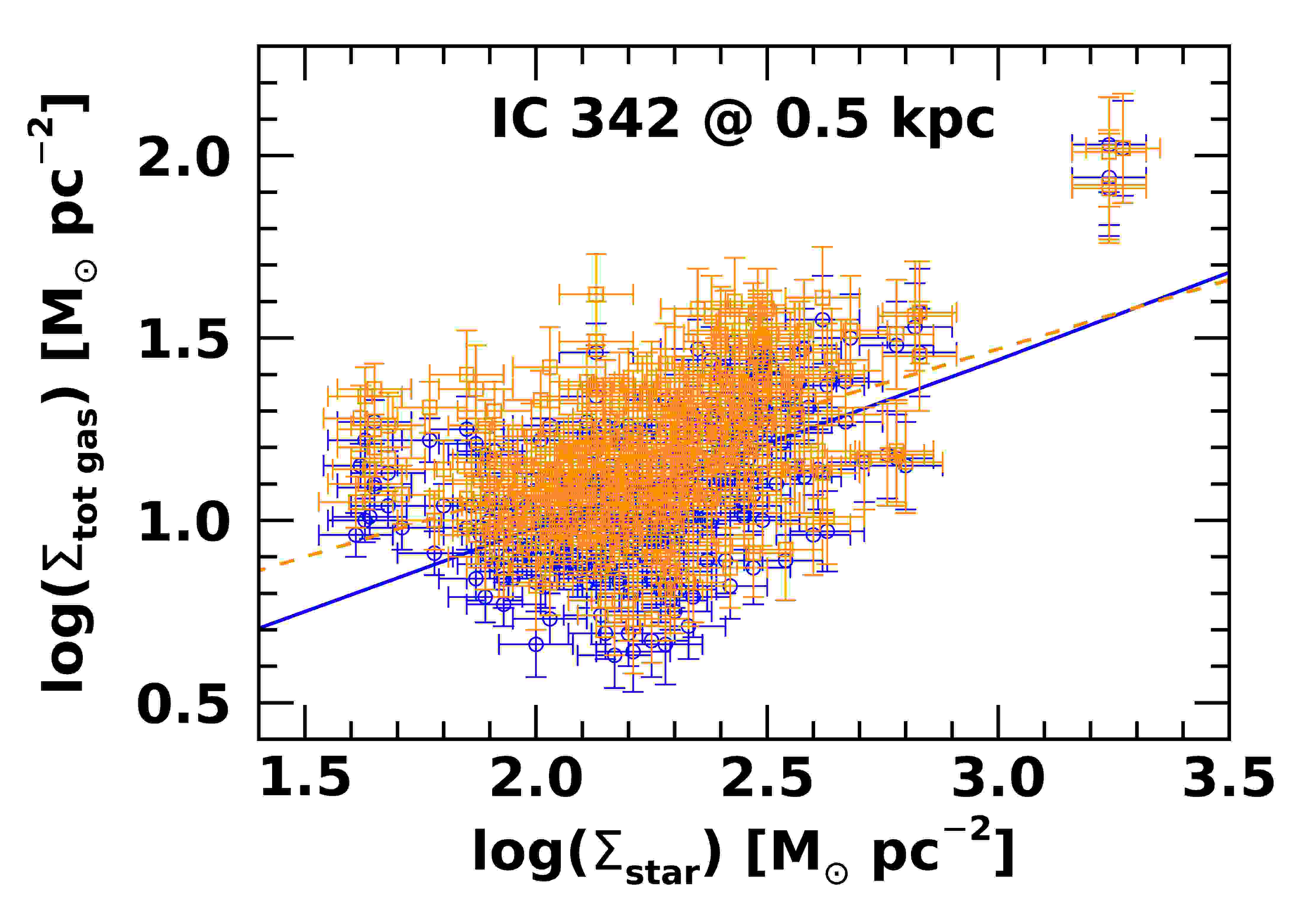}
\includegraphics[width=0.33\textwidth]{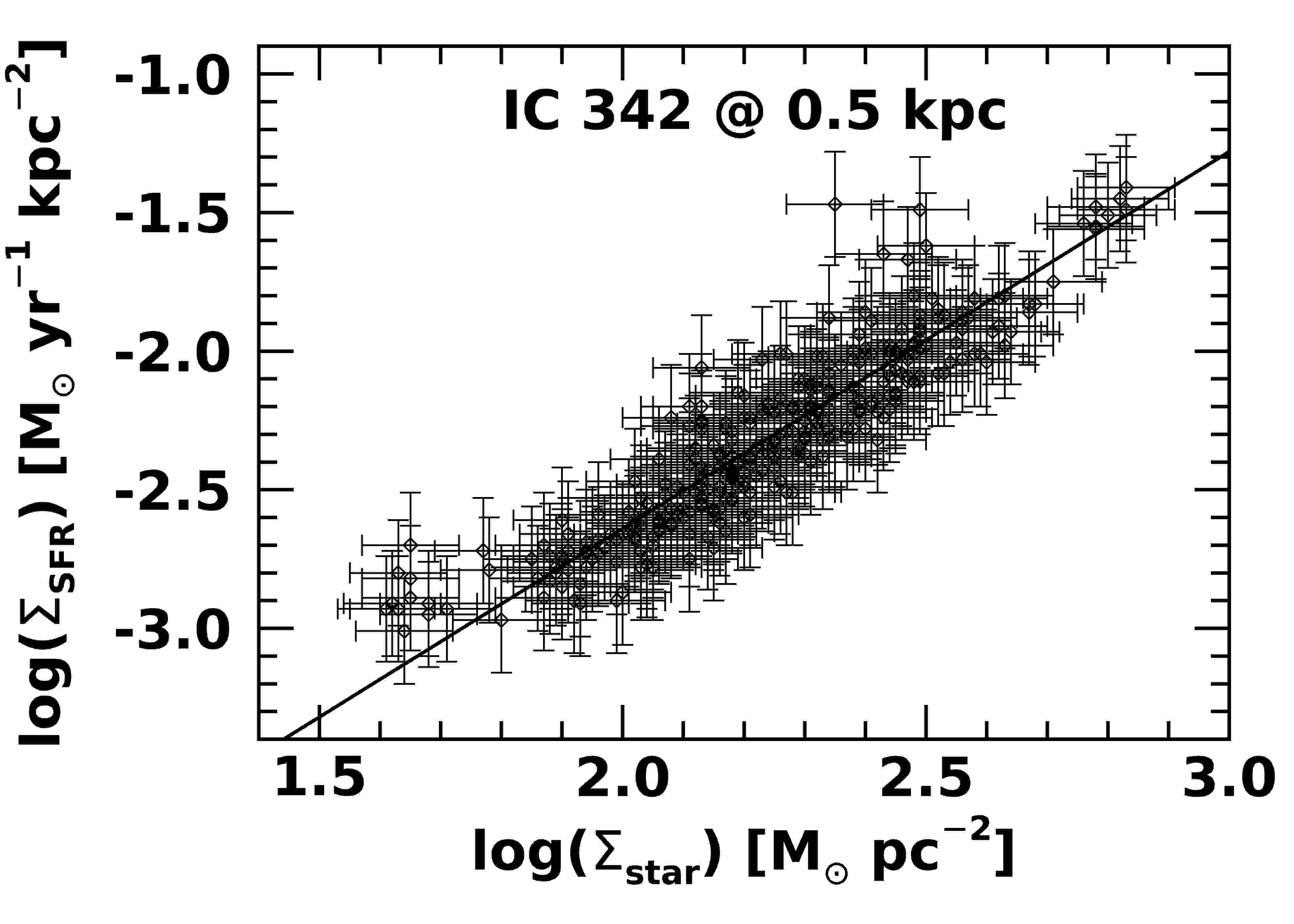}
\includegraphics[width=0.33\textwidth]{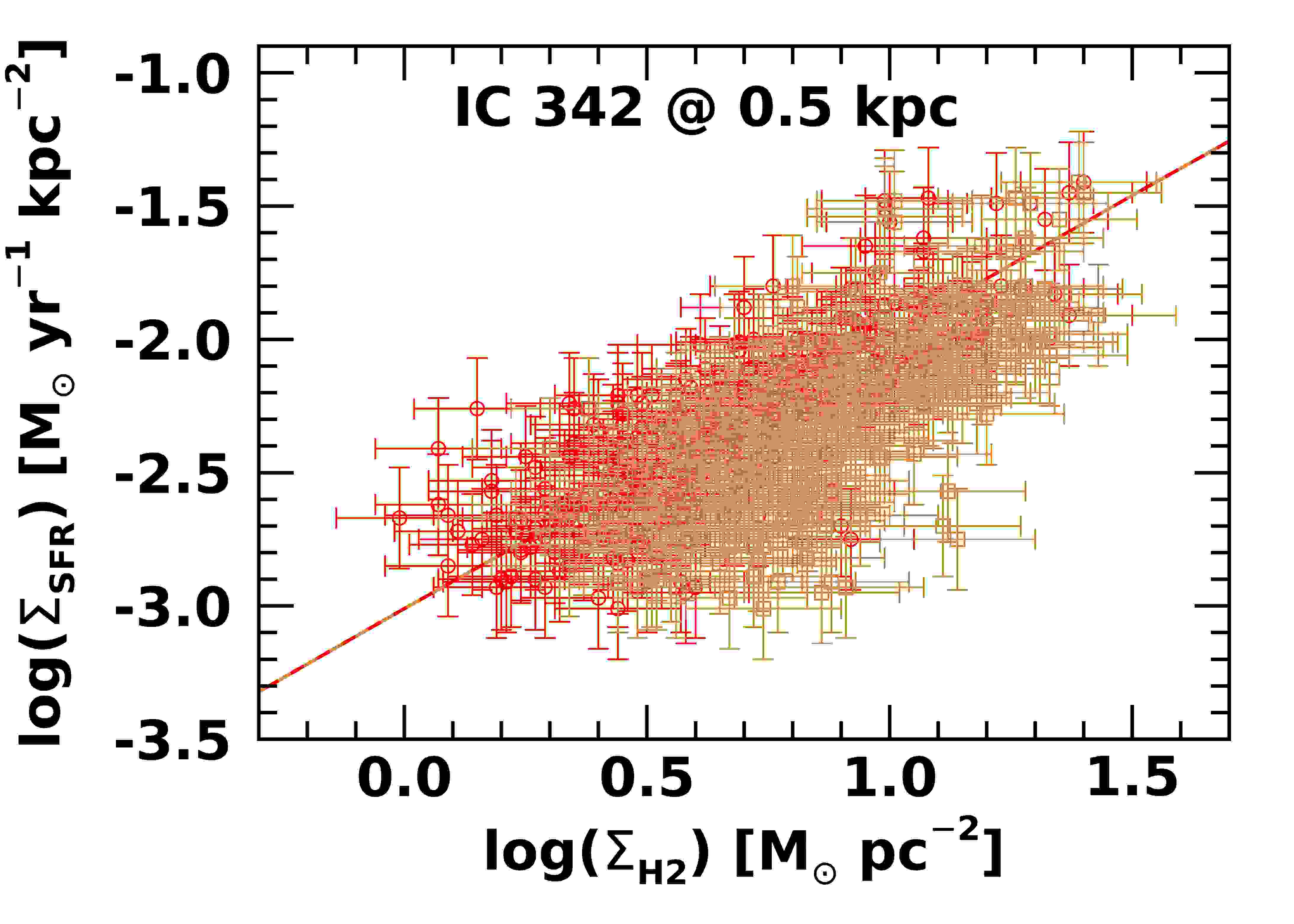}
\includegraphics[width=0.33\textwidth]{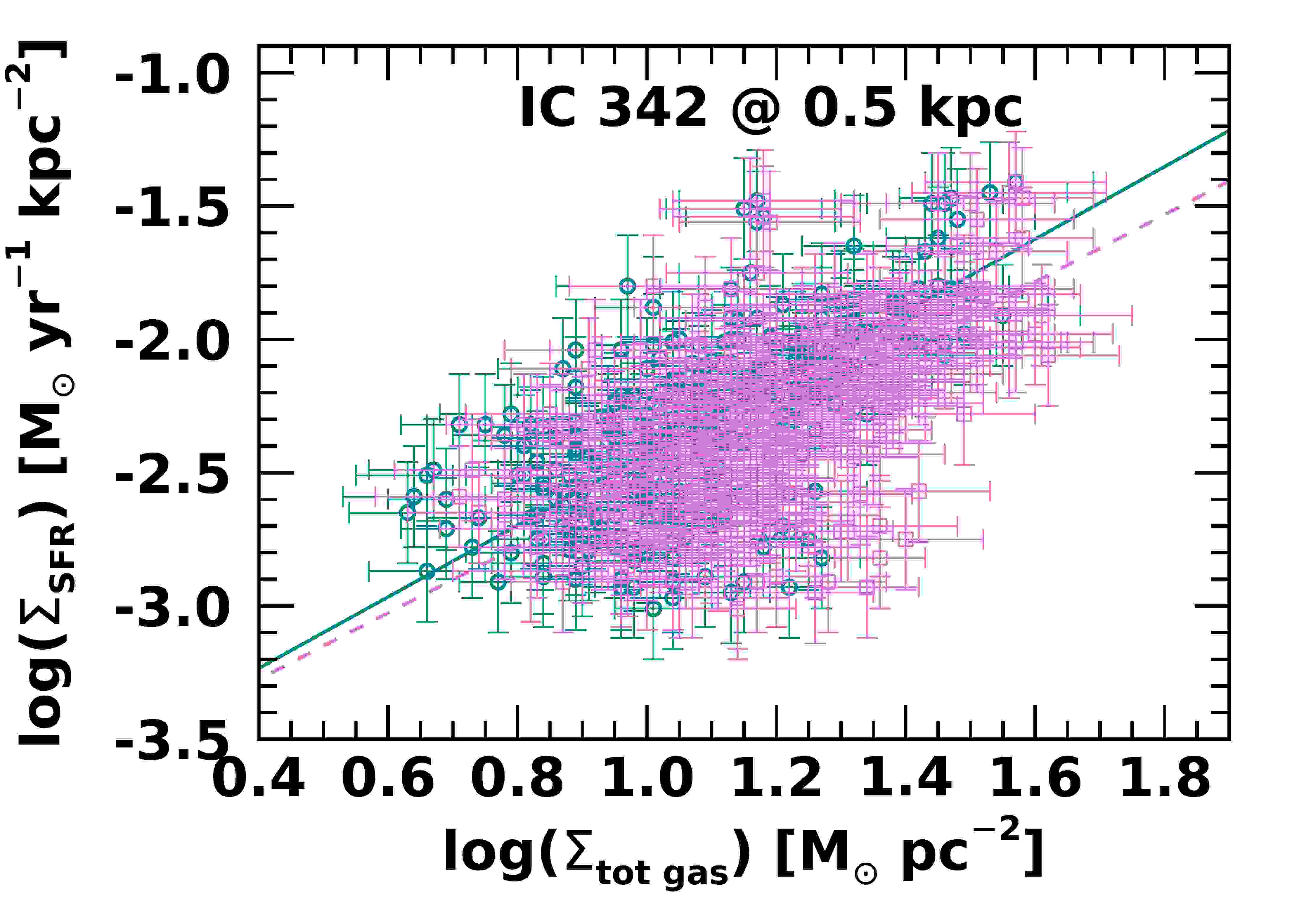}
\includegraphics[width=0.33\textwidth]{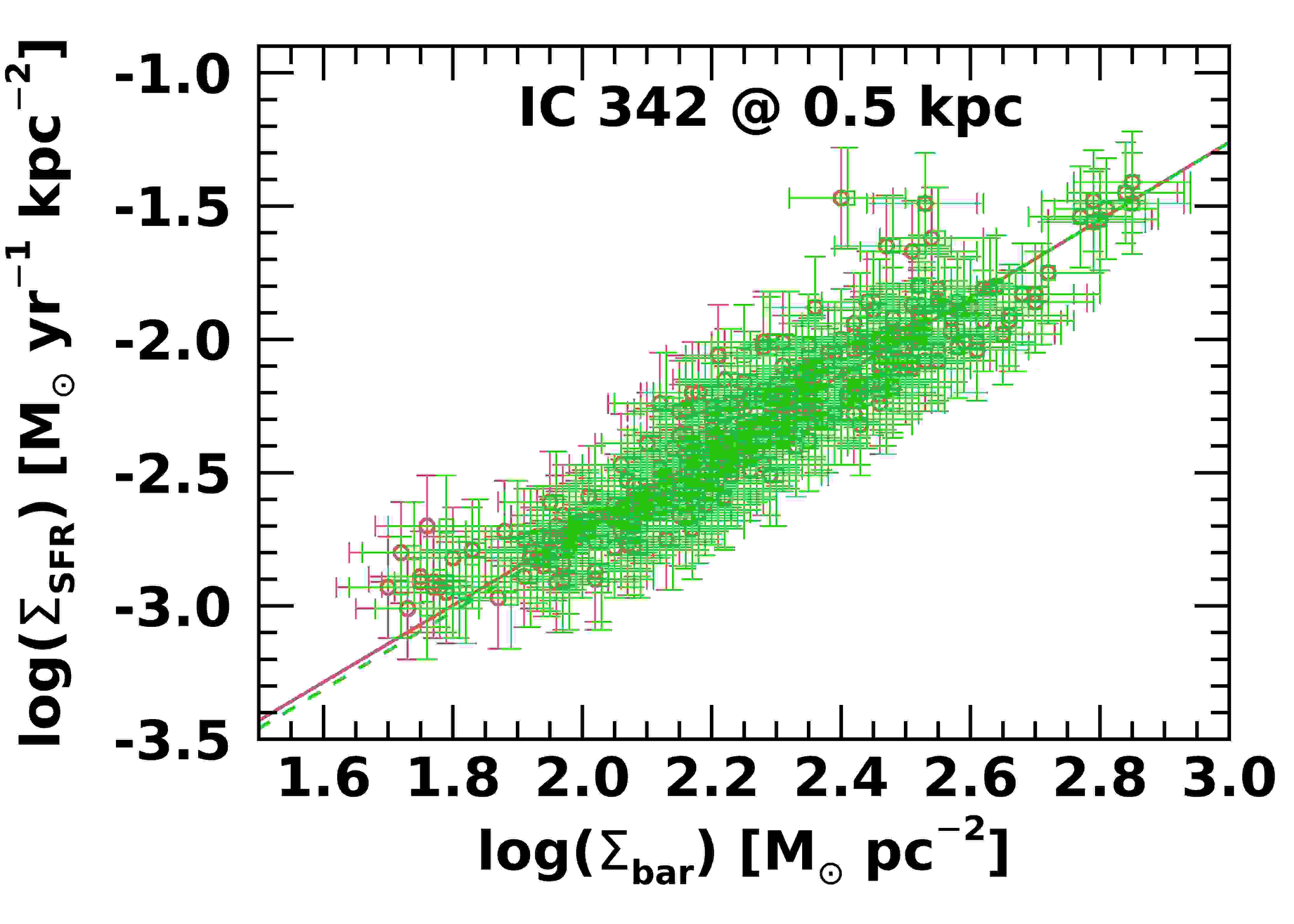}
\includegraphics[width=0.33\textwidth]{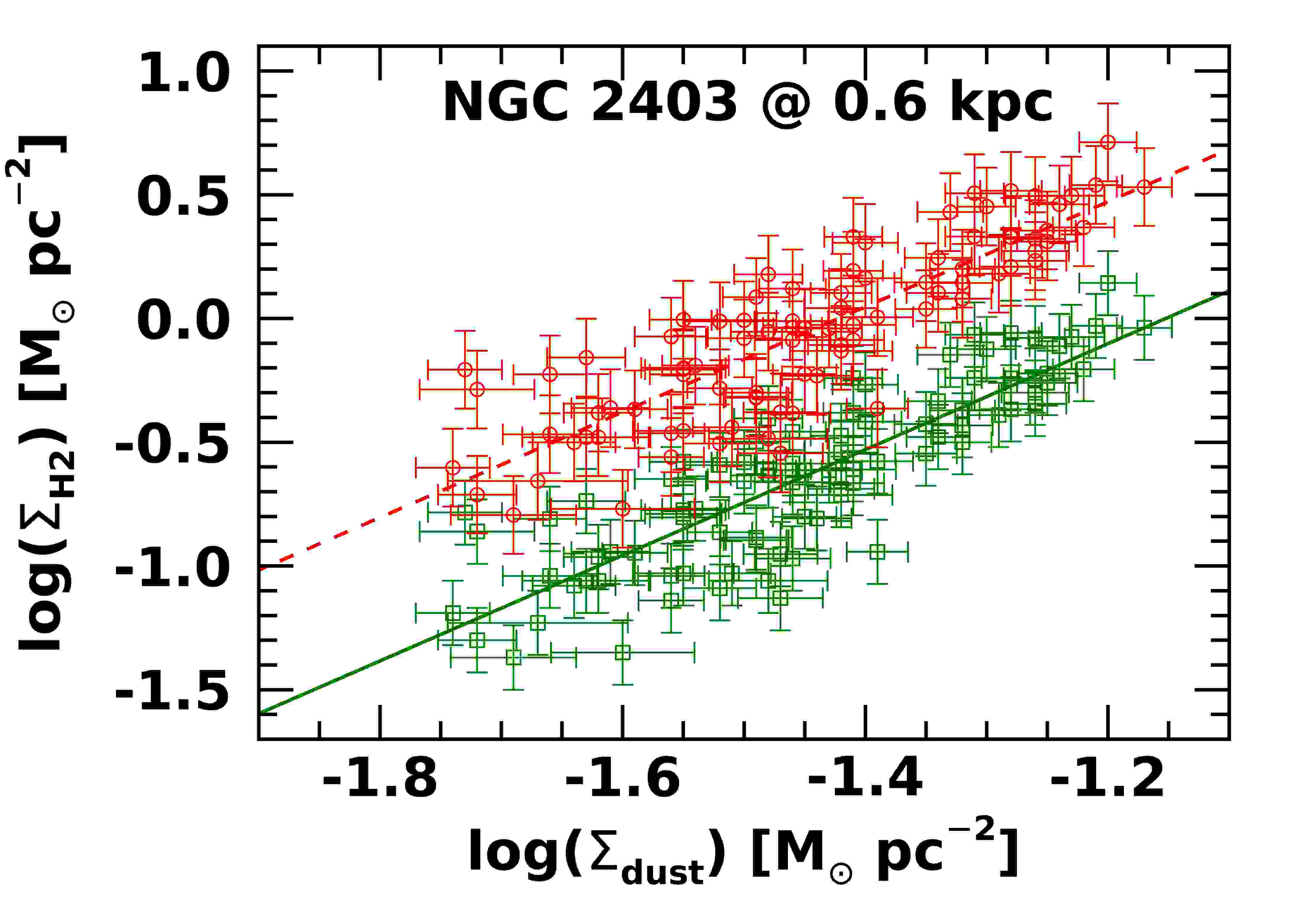}
\includegraphics[width=0.33\textwidth]{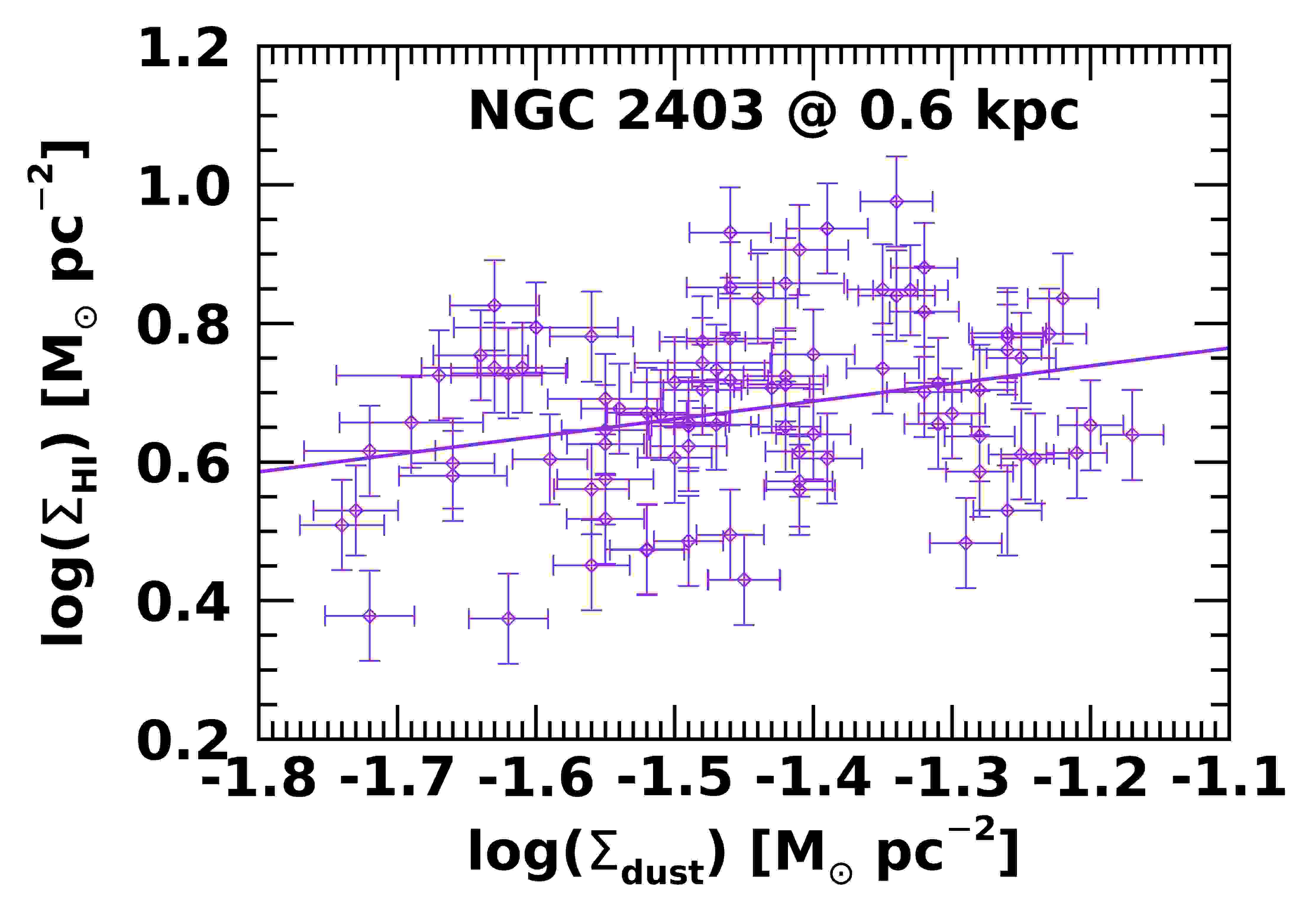}
\includegraphics[width=0.33\textwidth]{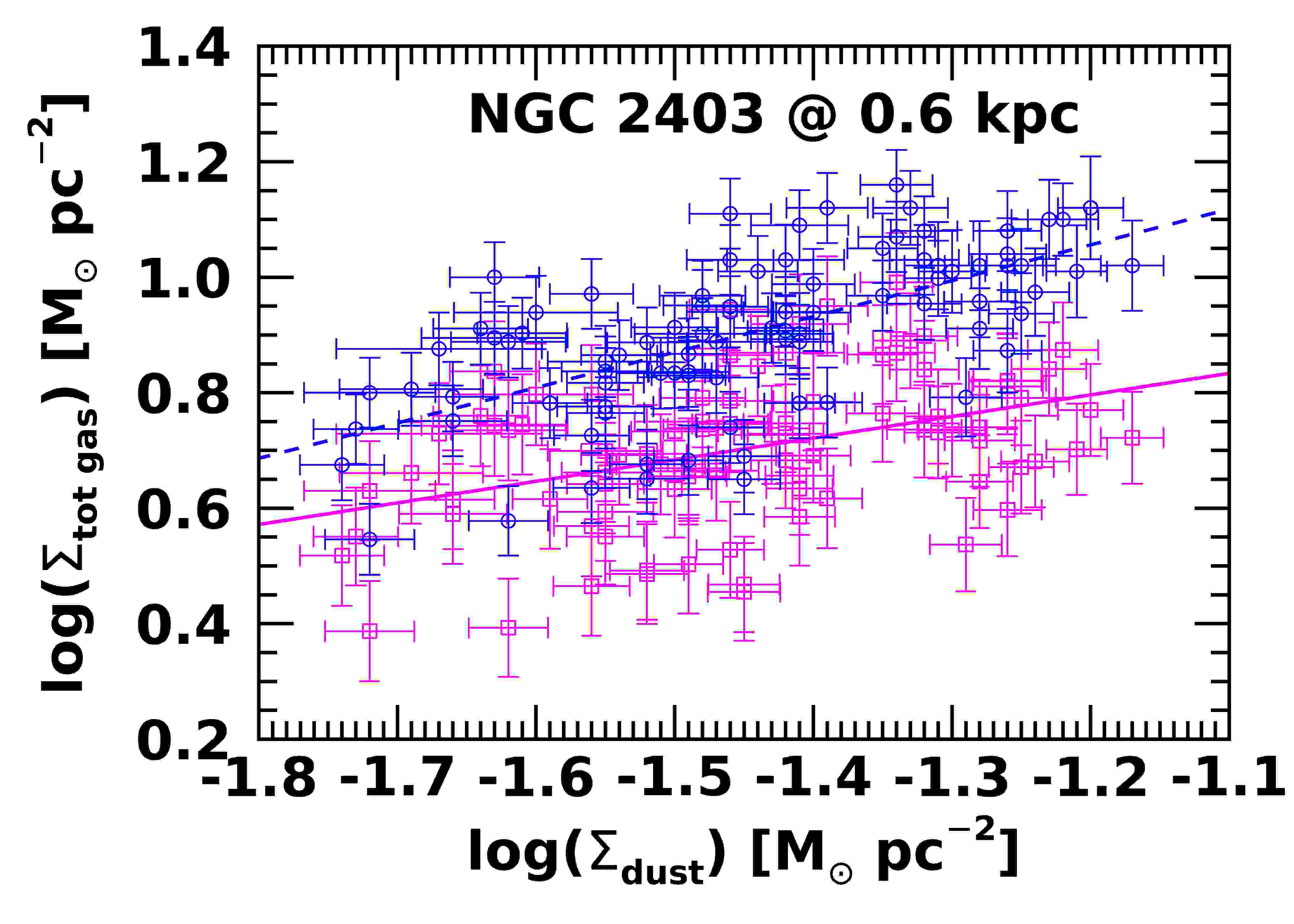}
\caption{
SRs for the entire sample, except for the galaxy NGC~6946 already shown in Fig.~\ref{fig:n6946}.}
\label{fig:add-ism}
\end{figure*}

\begin{figure*}
\centering
\includegraphics[width=0.33\textwidth]{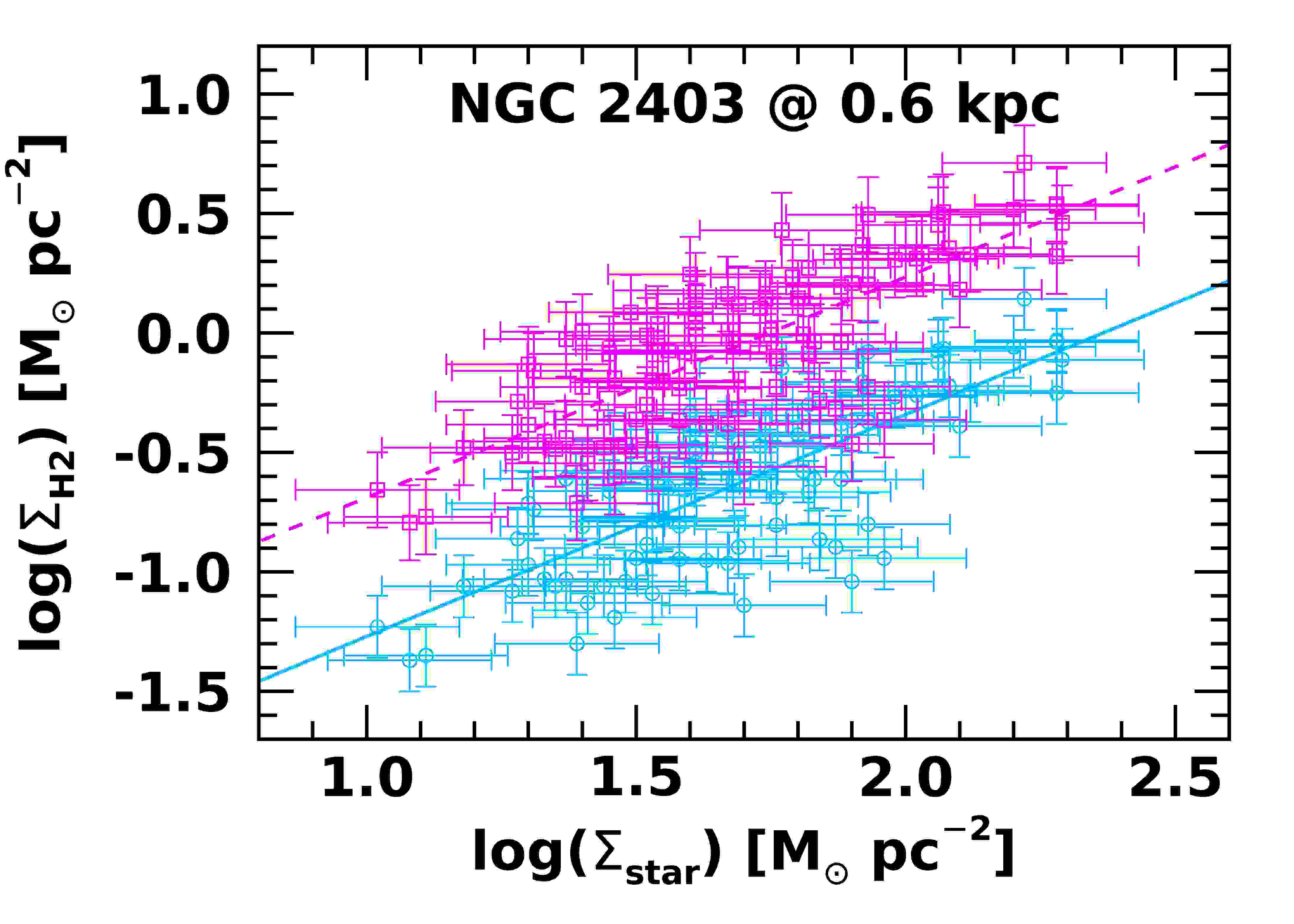}
\includegraphics[width=0.33\textwidth]{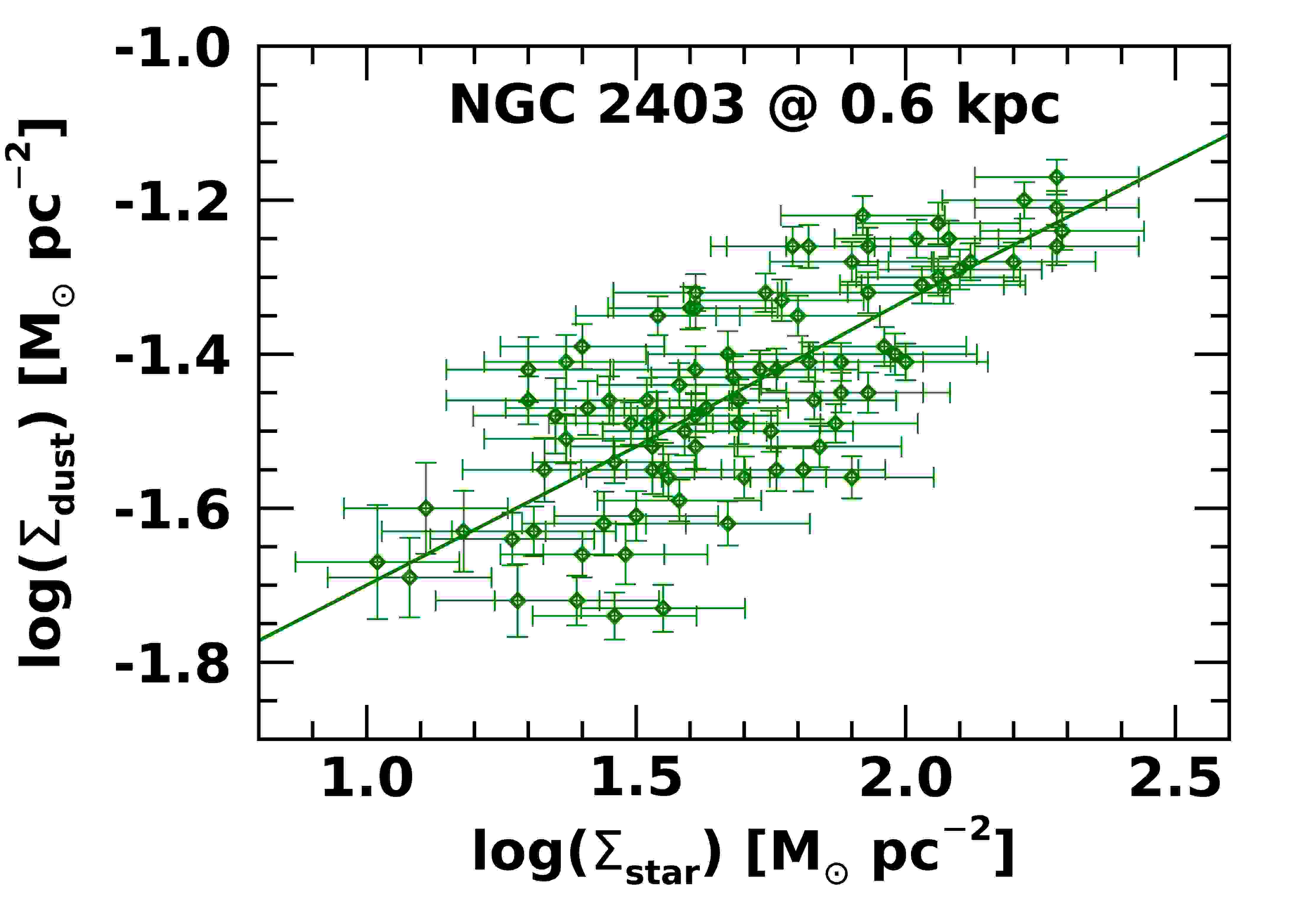}
\includegraphics[width=0.33\textwidth]{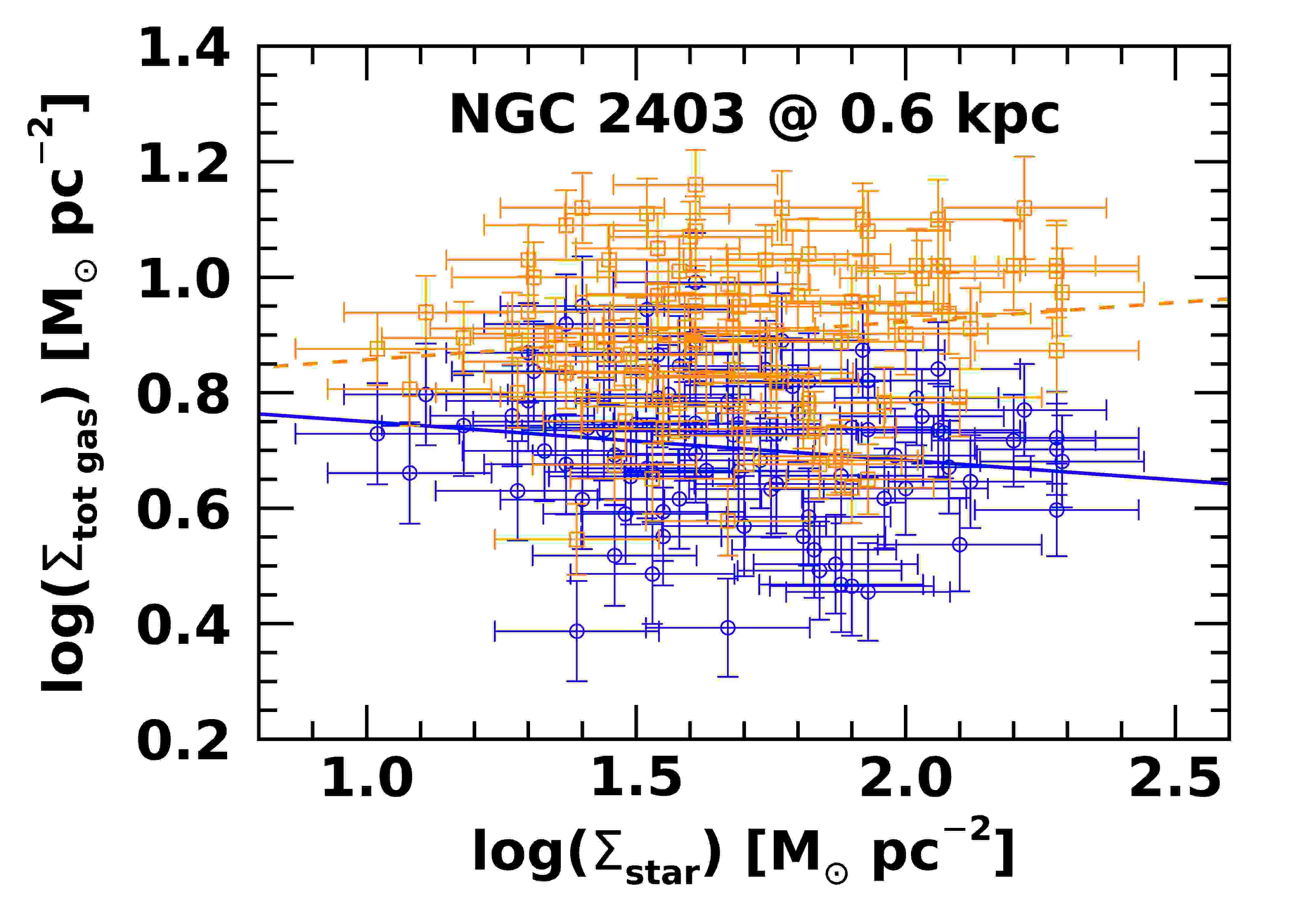}
\includegraphics[width=0.33\textwidth]{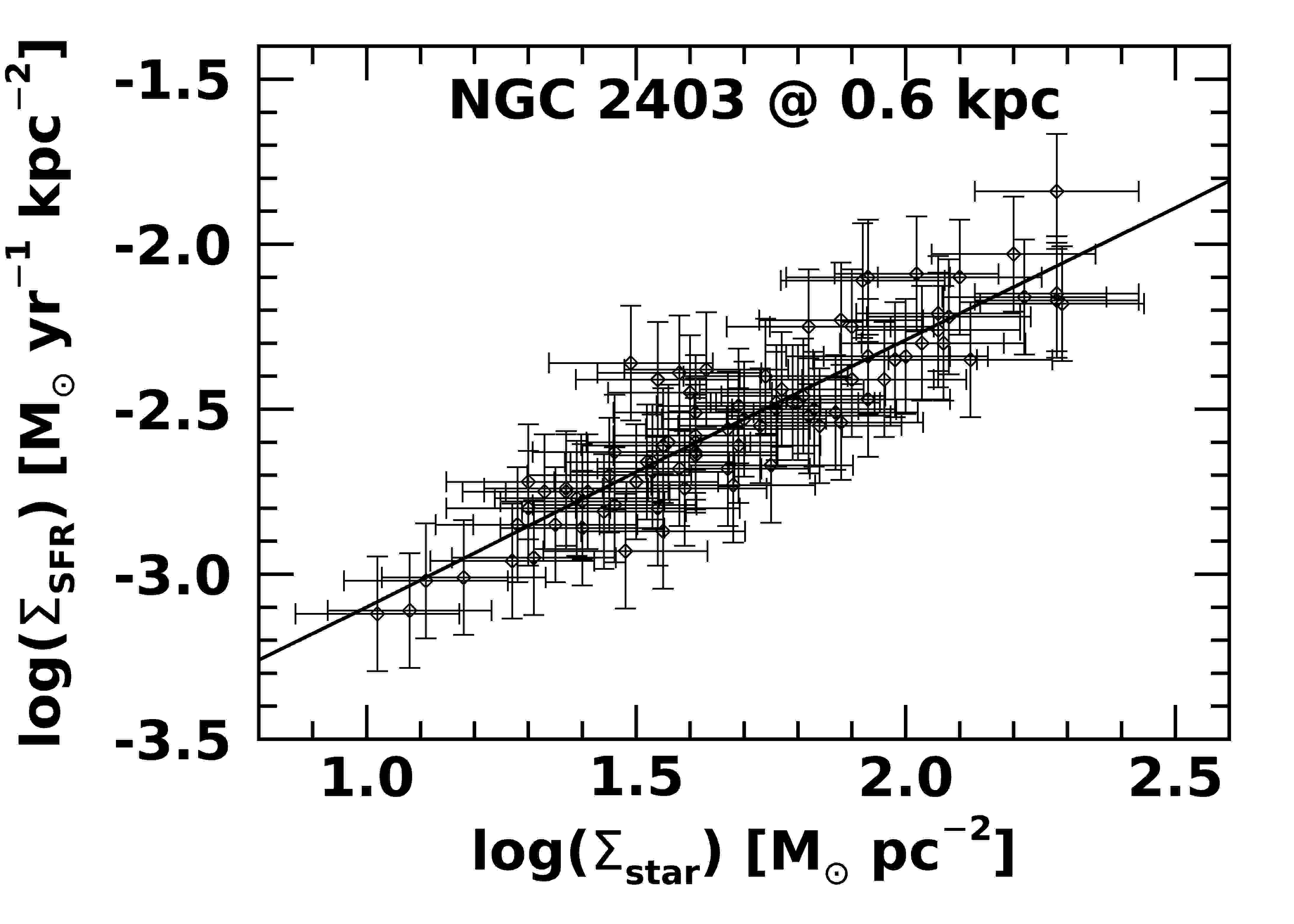}
\includegraphics[width=0.33\textwidth]{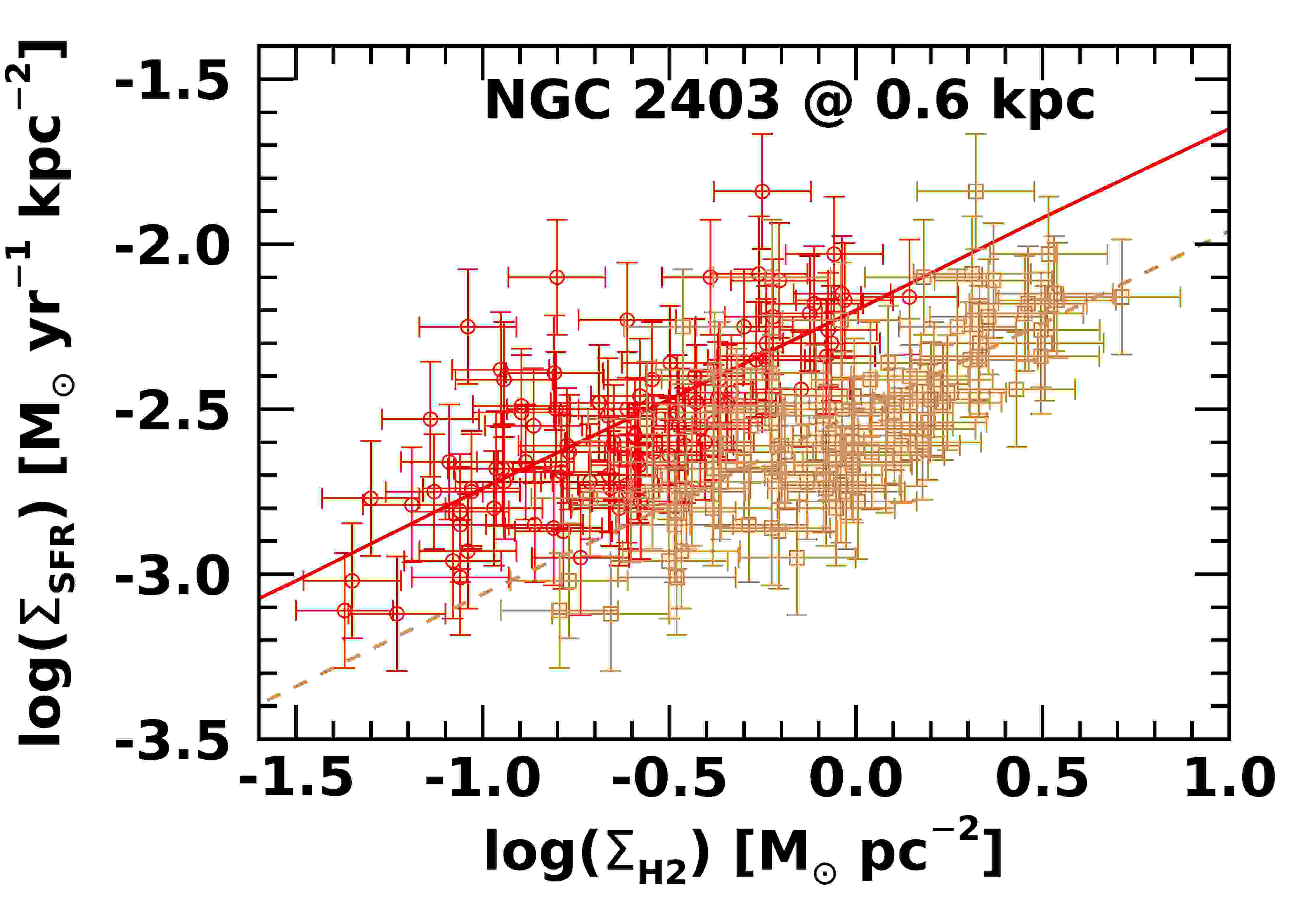}
\includegraphics[width=0.33\textwidth]{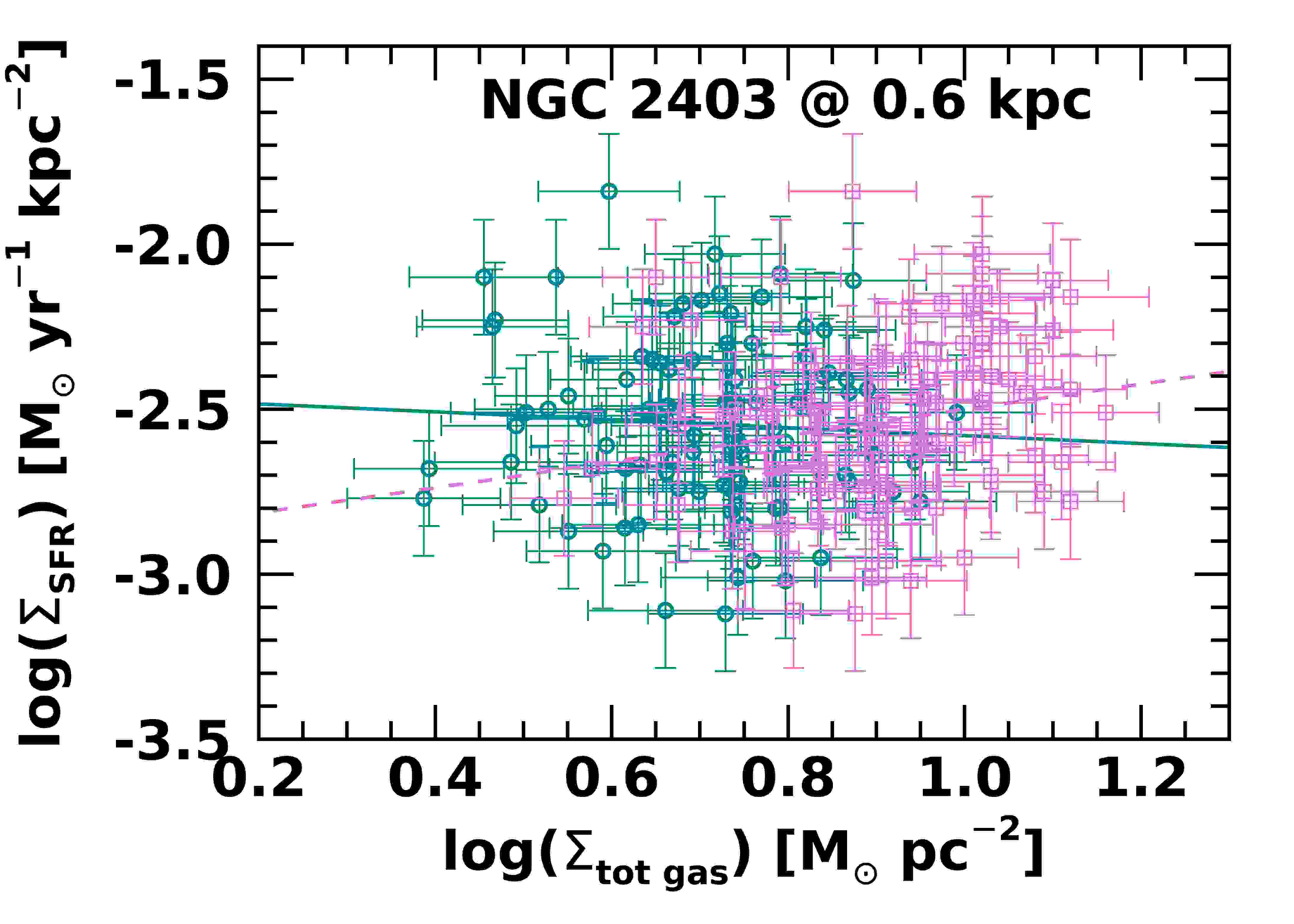}
\includegraphics[width=0.33\textwidth]{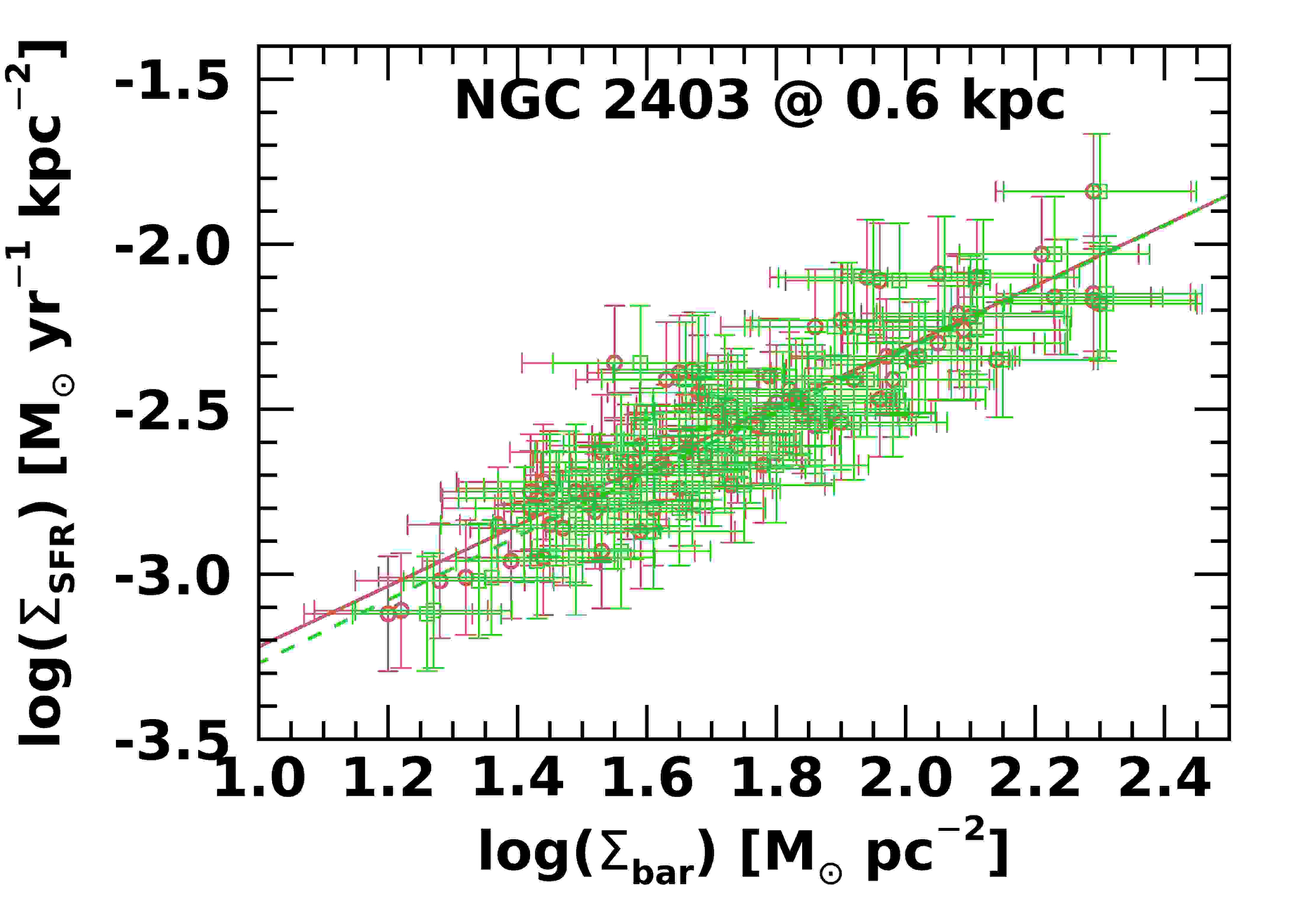}
\includegraphics[width=0.33\textwidth]{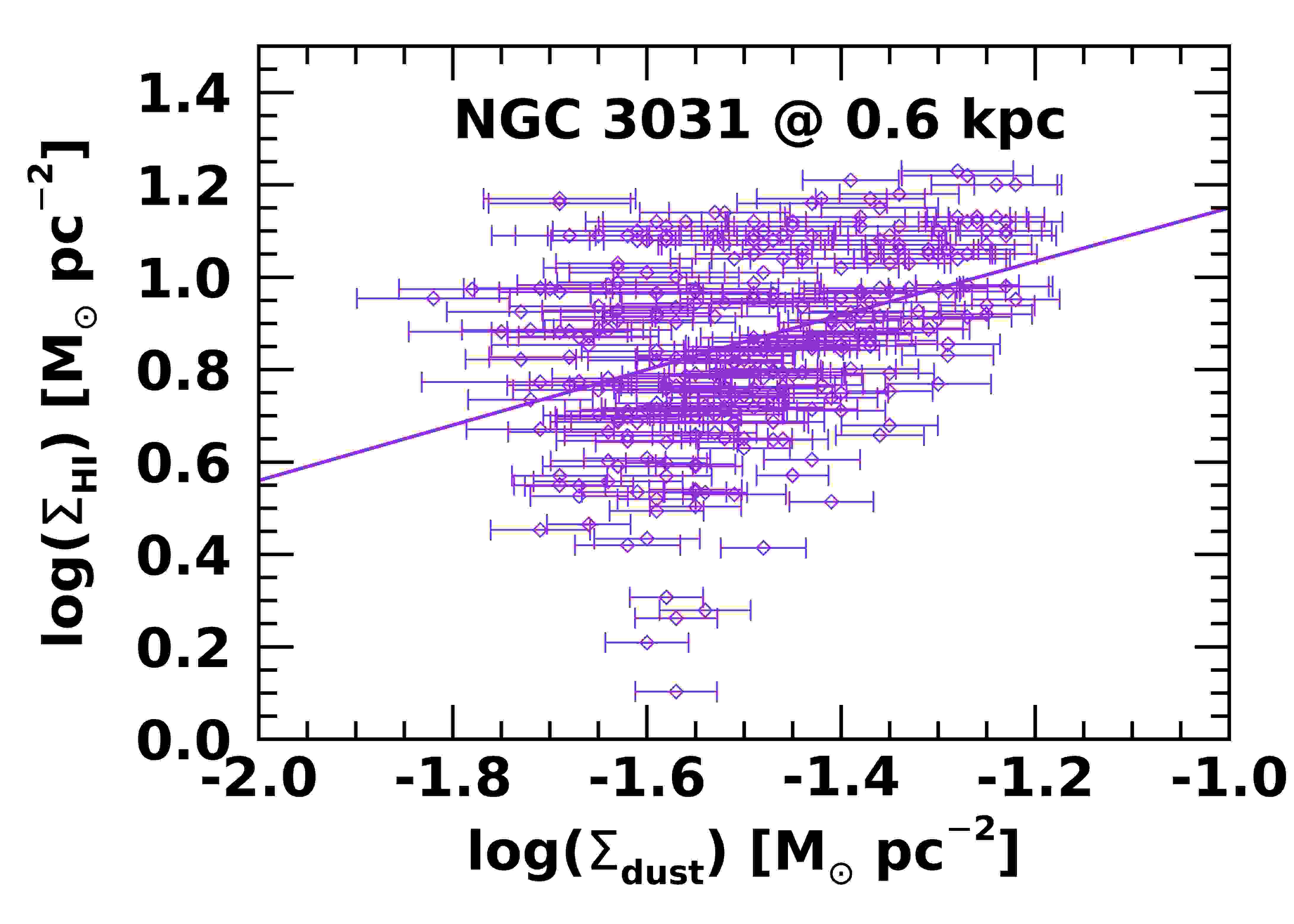}
\includegraphics[width=0.33\textwidth]{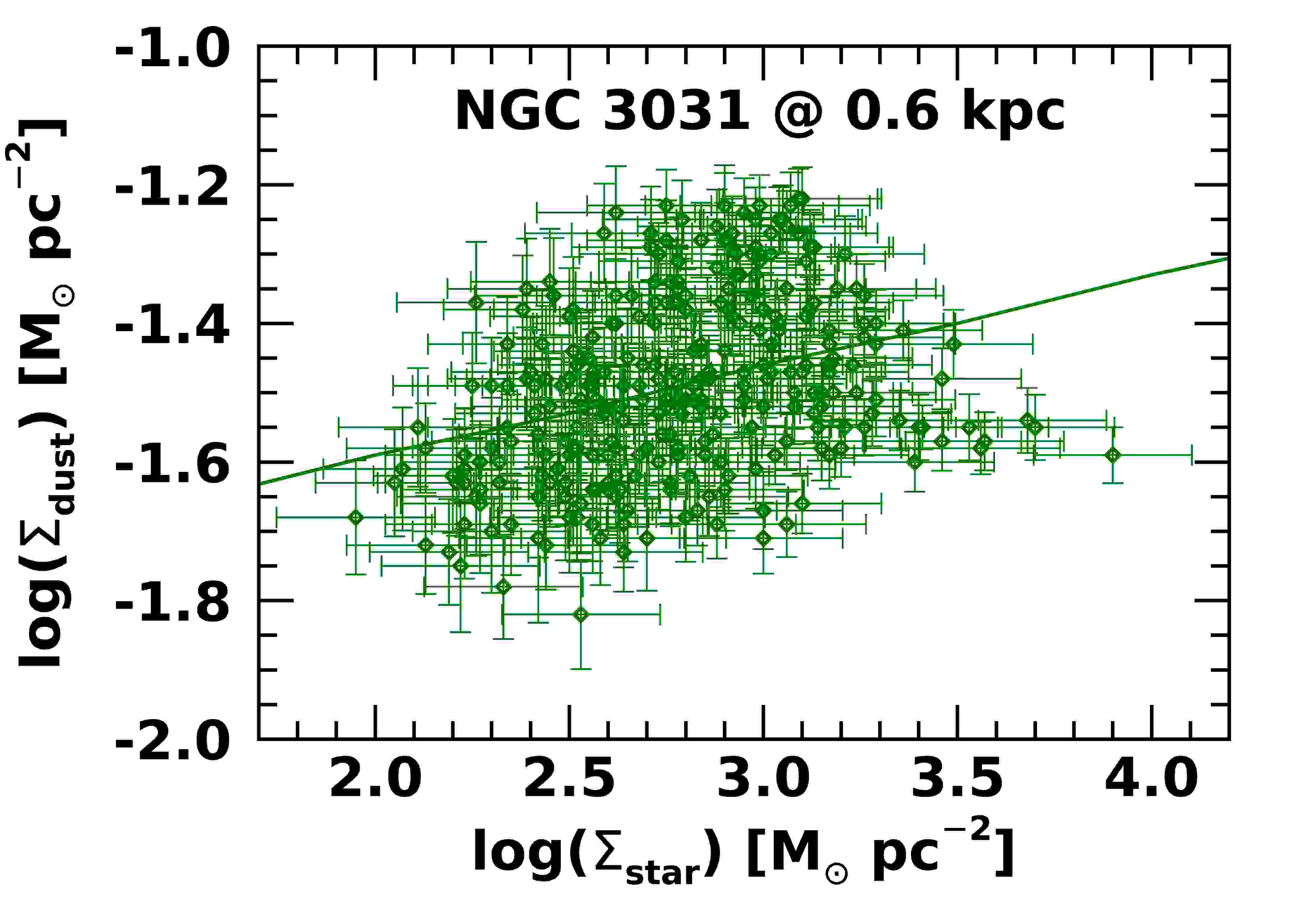}
\includegraphics[width=0.33\textwidth]{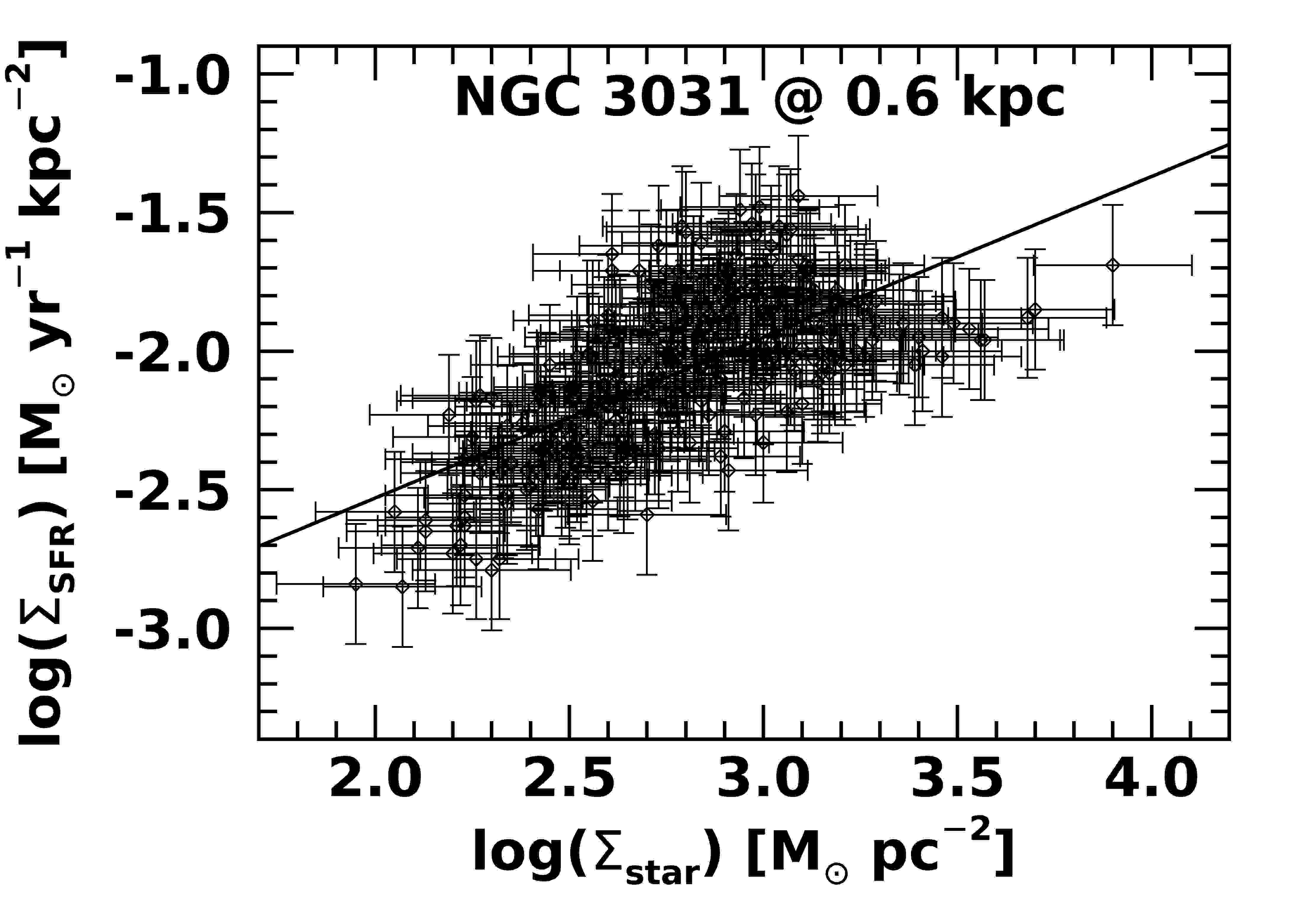}
\includegraphics[width=0.33\textwidth]{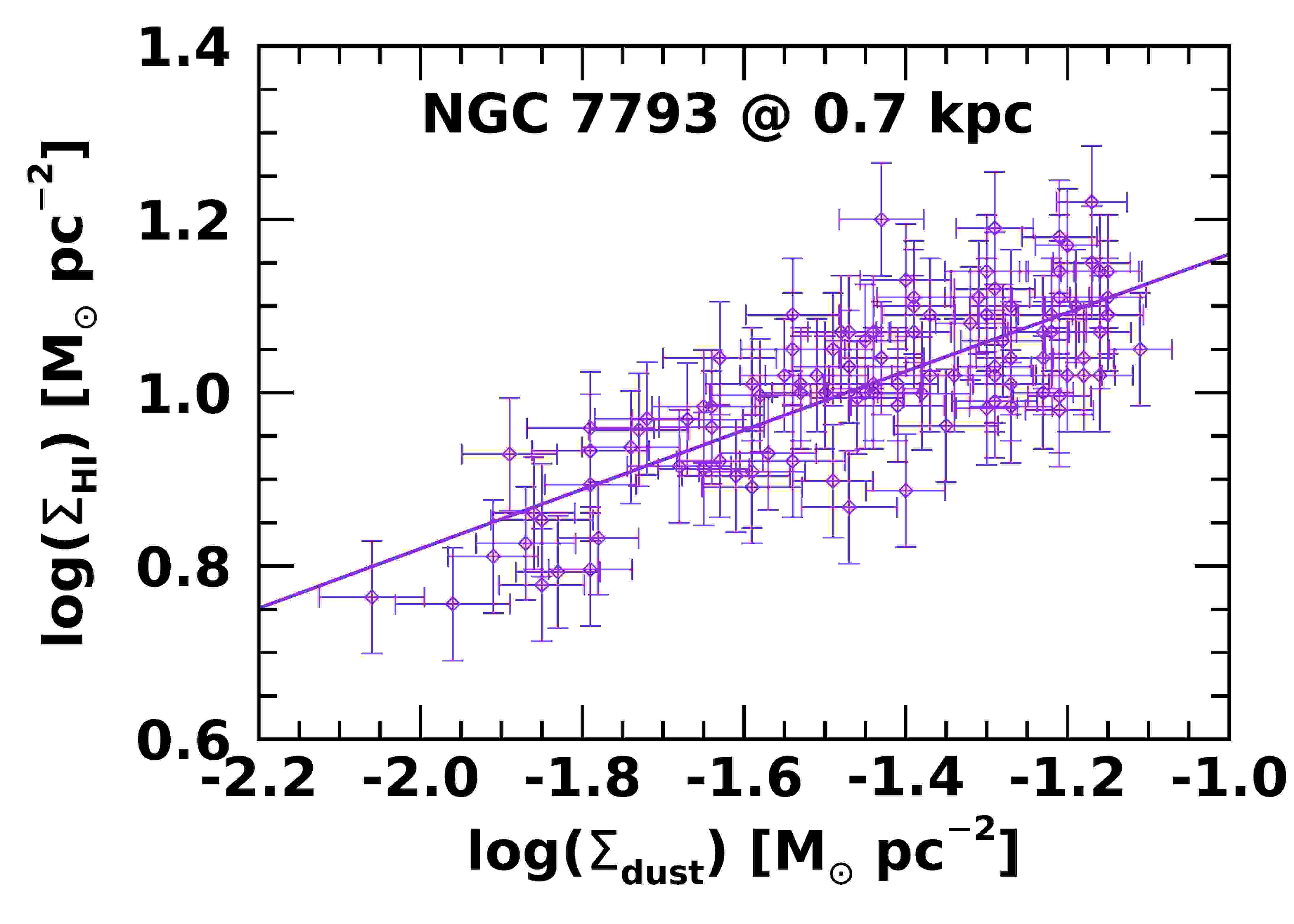}
\includegraphics[width=0.33\textwidth]{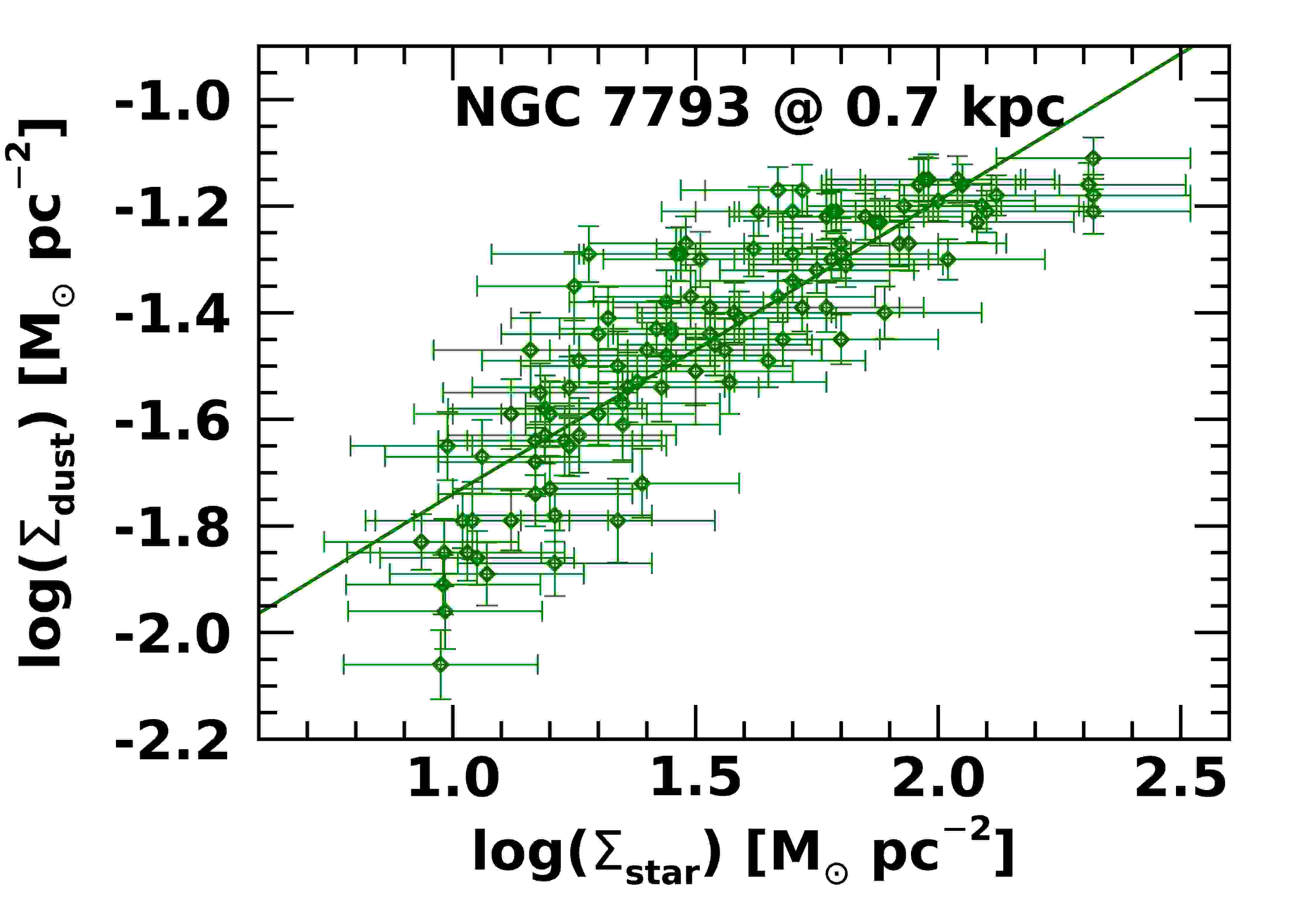}
\includegraphics[width=0.33\textwidth]{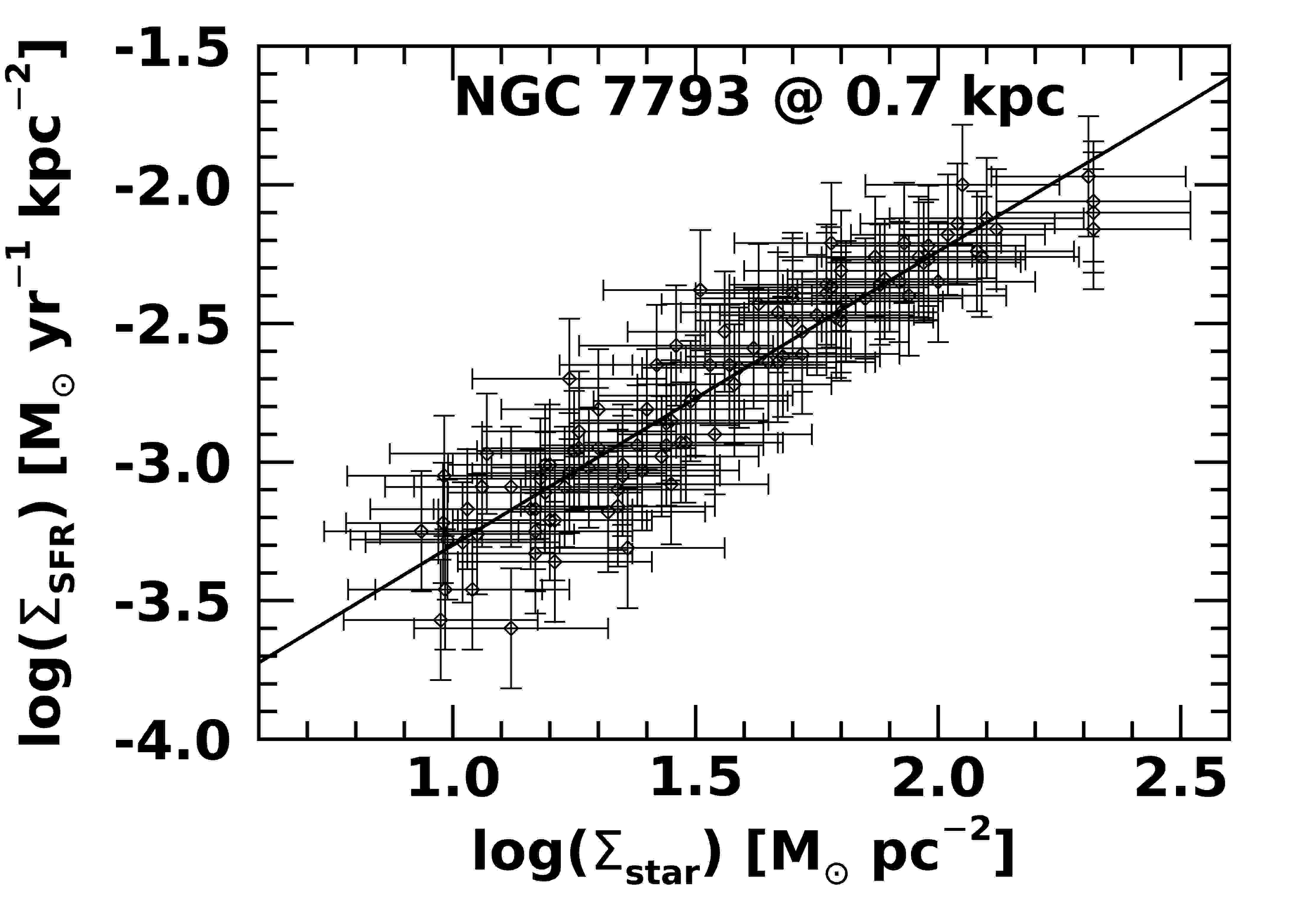}
\includegraphics[width=0.33\textwidth]{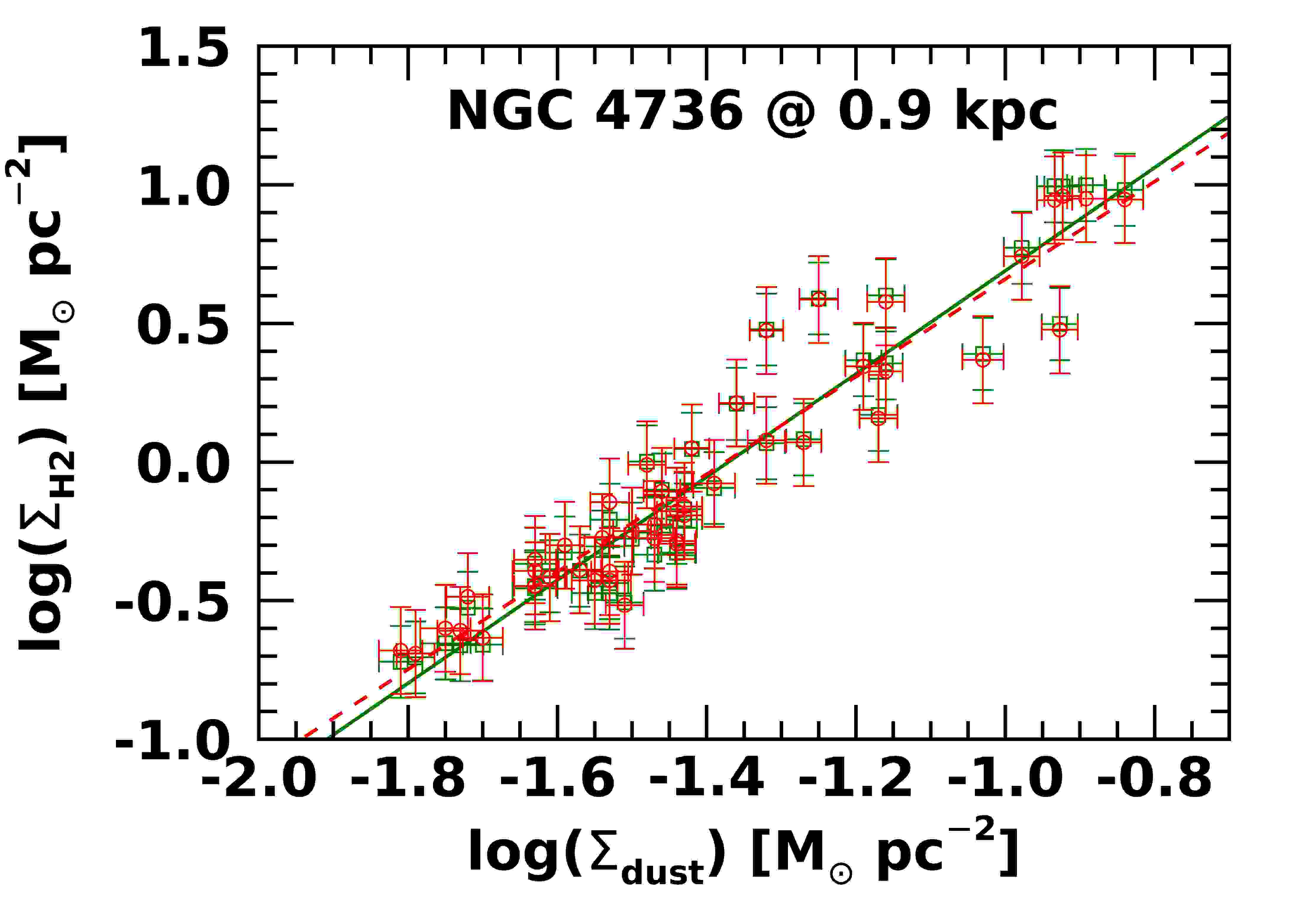}
\includegraphics[width=0.33\textwidth]{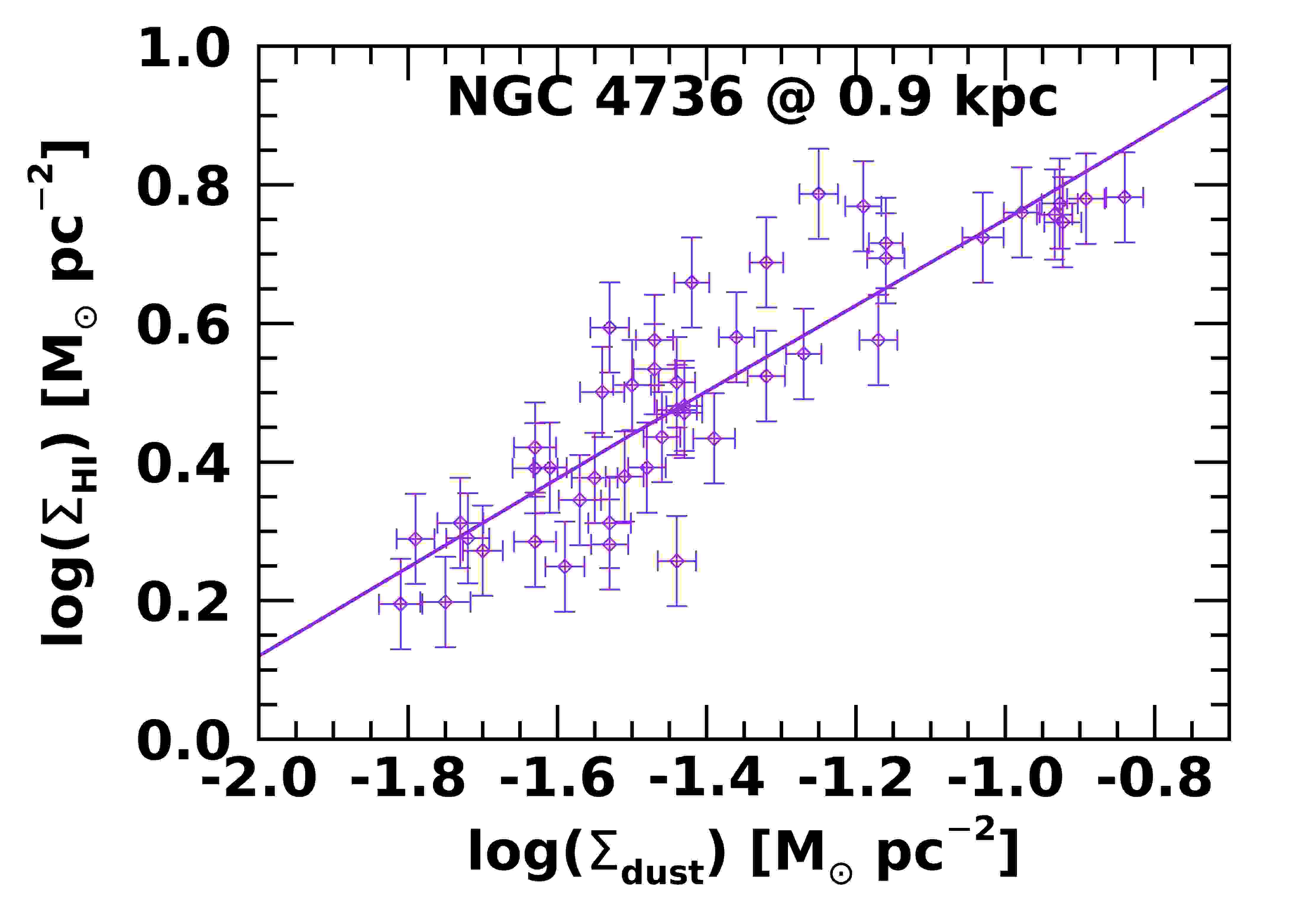}
\caption*{Figure~\ref{fig:add-ism} continued}
\end{figure*}

\begin{figure*}
\centering
\includegraphics[width=0.33\textwidth]{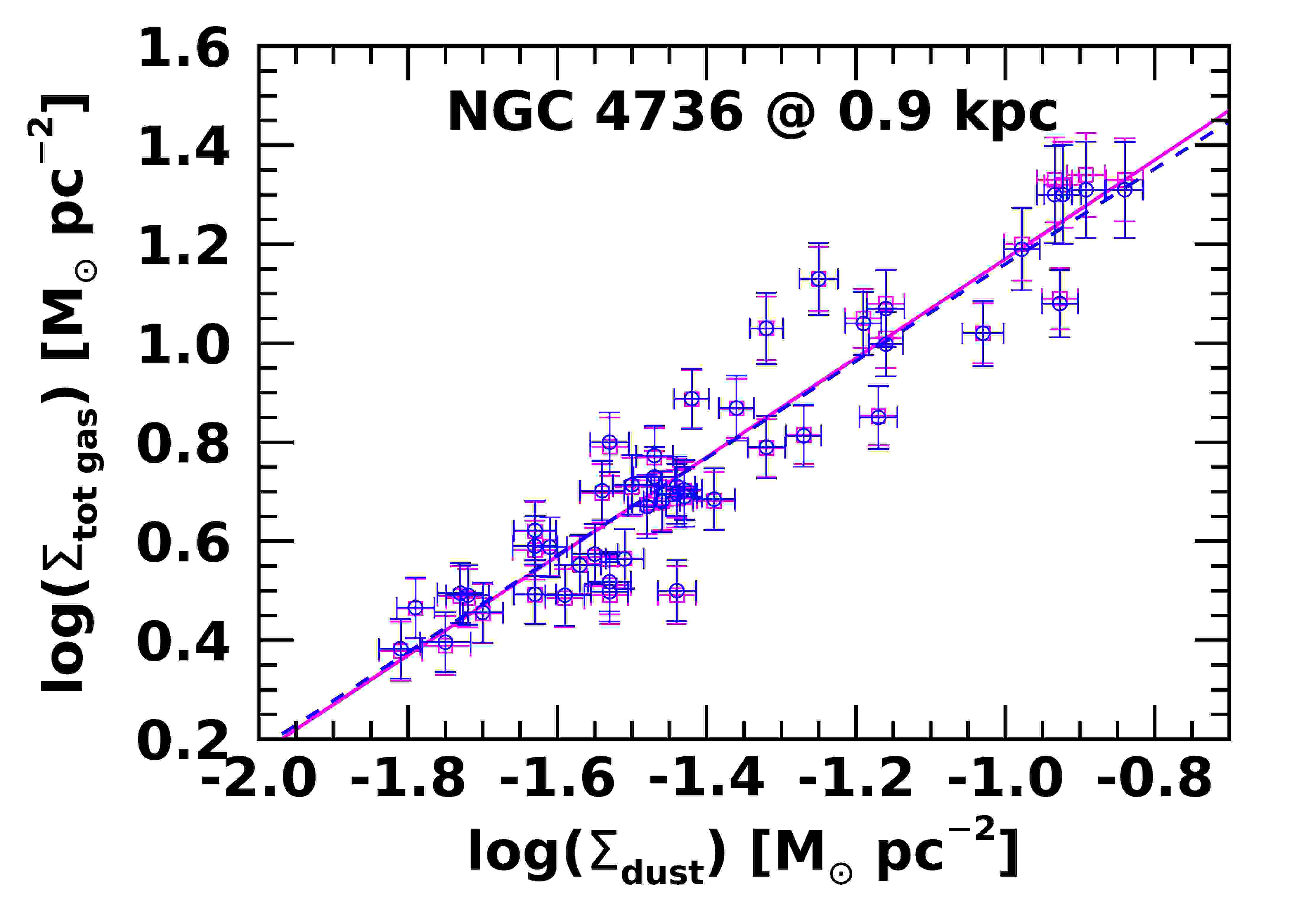}
\includegraphics[width=0.33\textwidth]{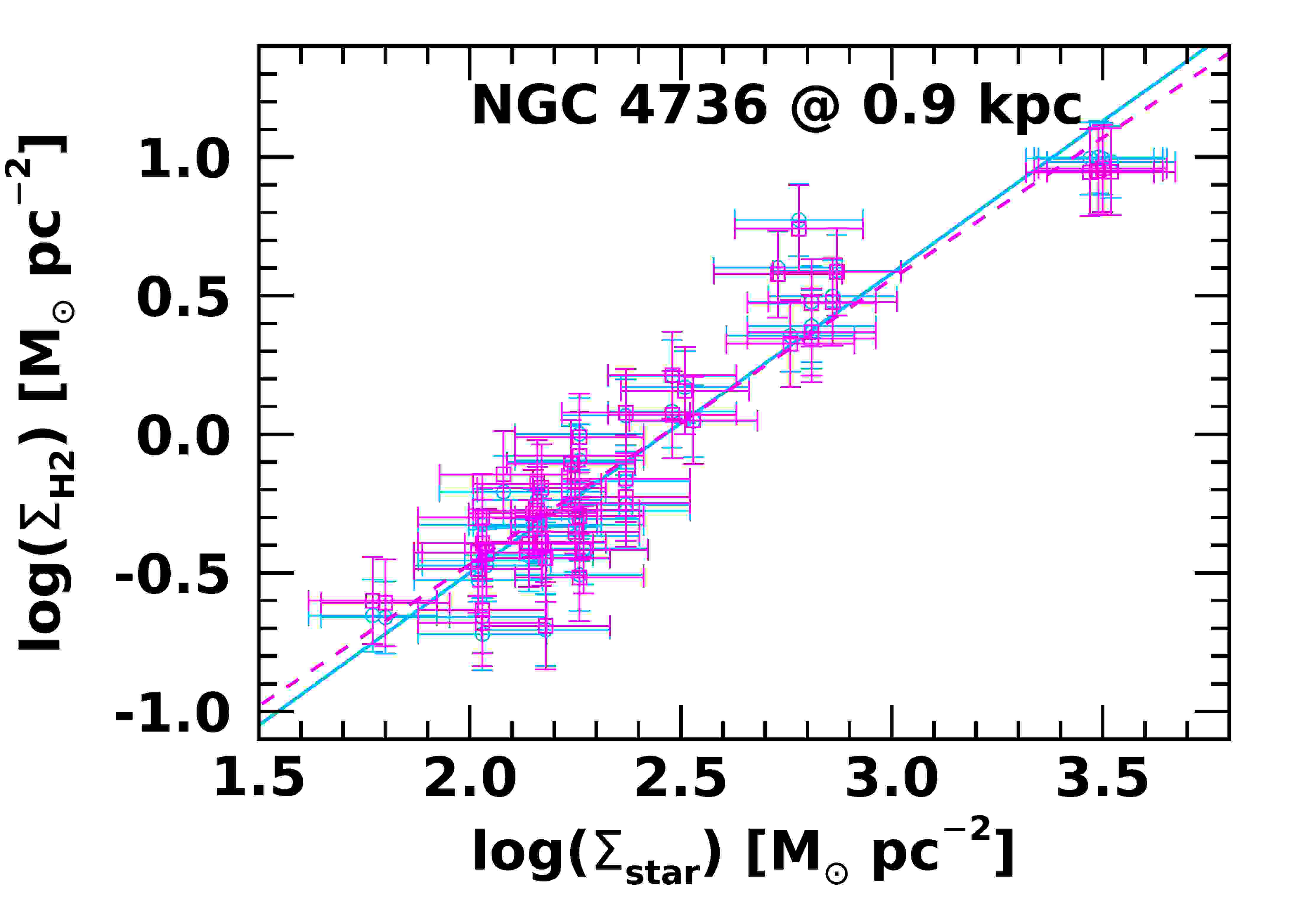}
\includegraphics[width=0.33\textwidth]{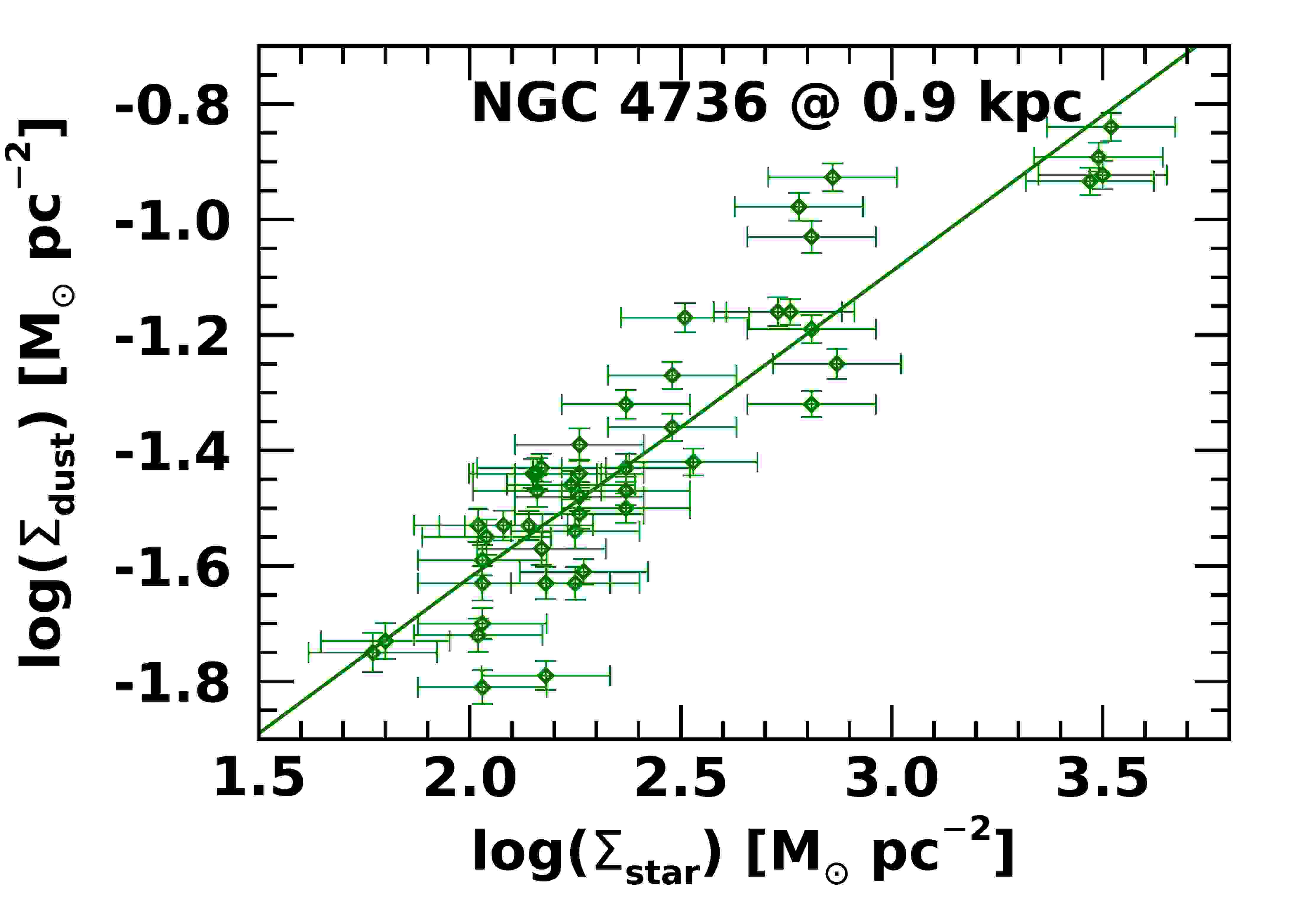}
\includegraphics[width=0.33\textwidth]{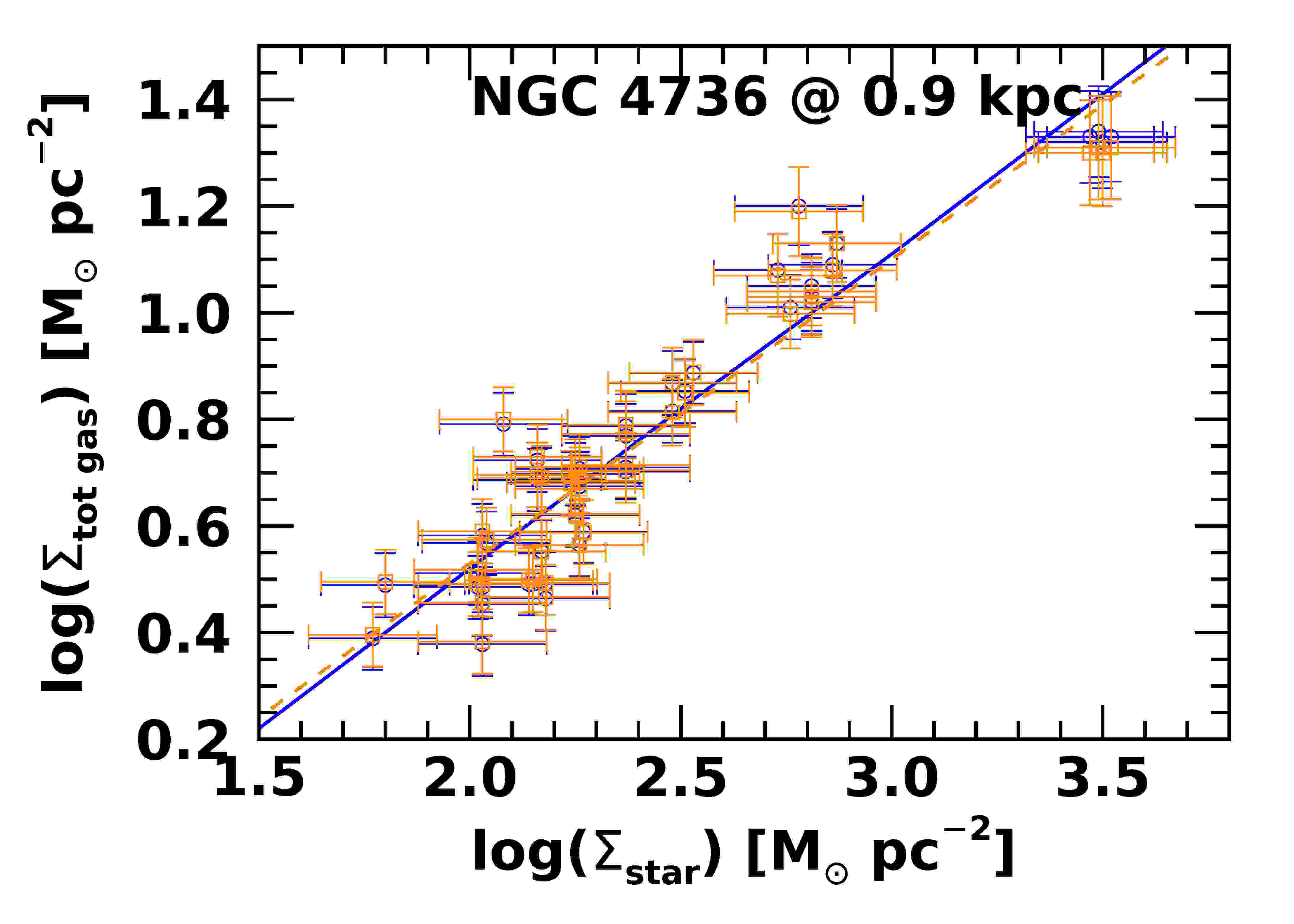}
\includegraphics[width=0.33\textwidth]{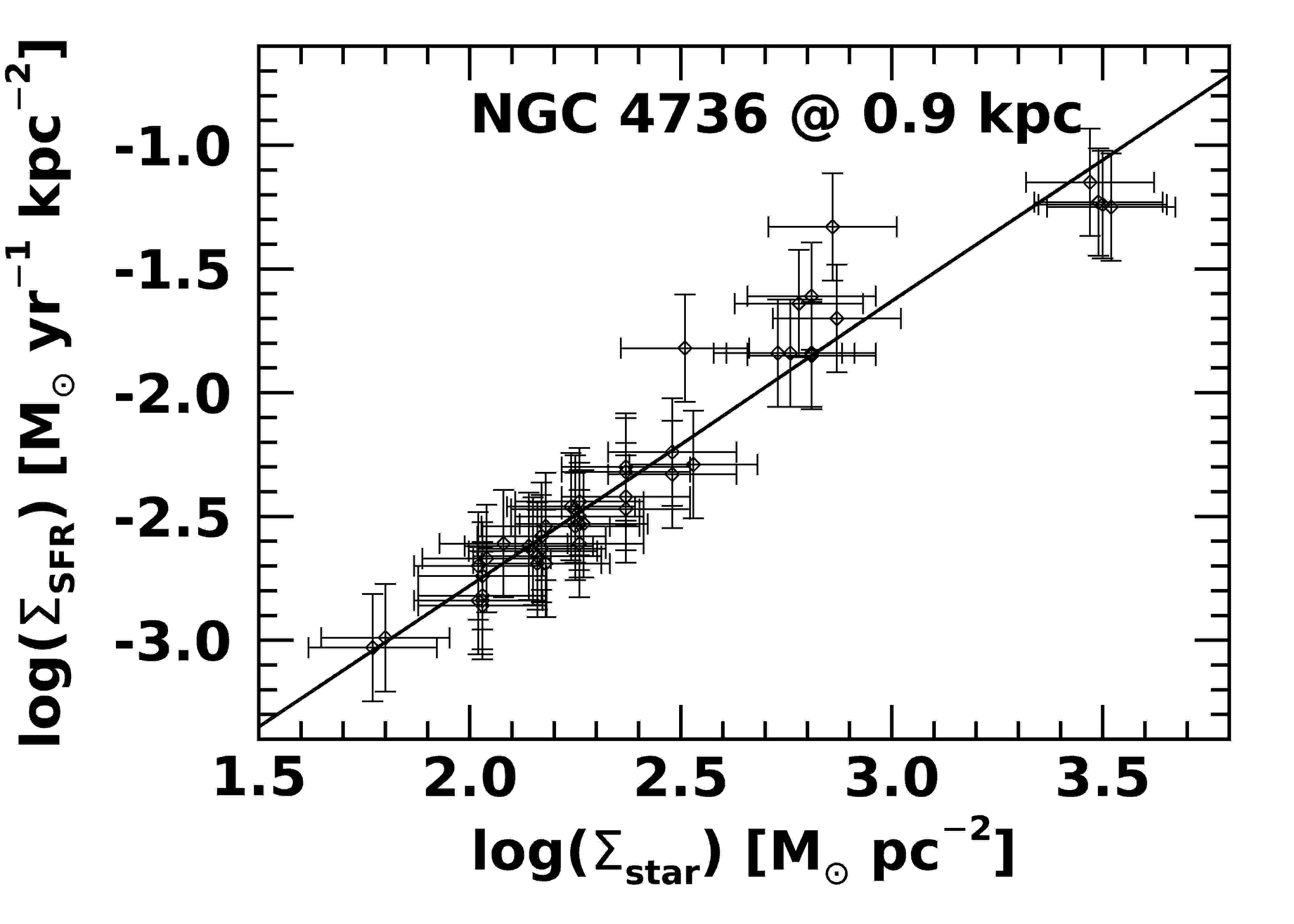}
\includegraphics[width=0.33\textwidth]{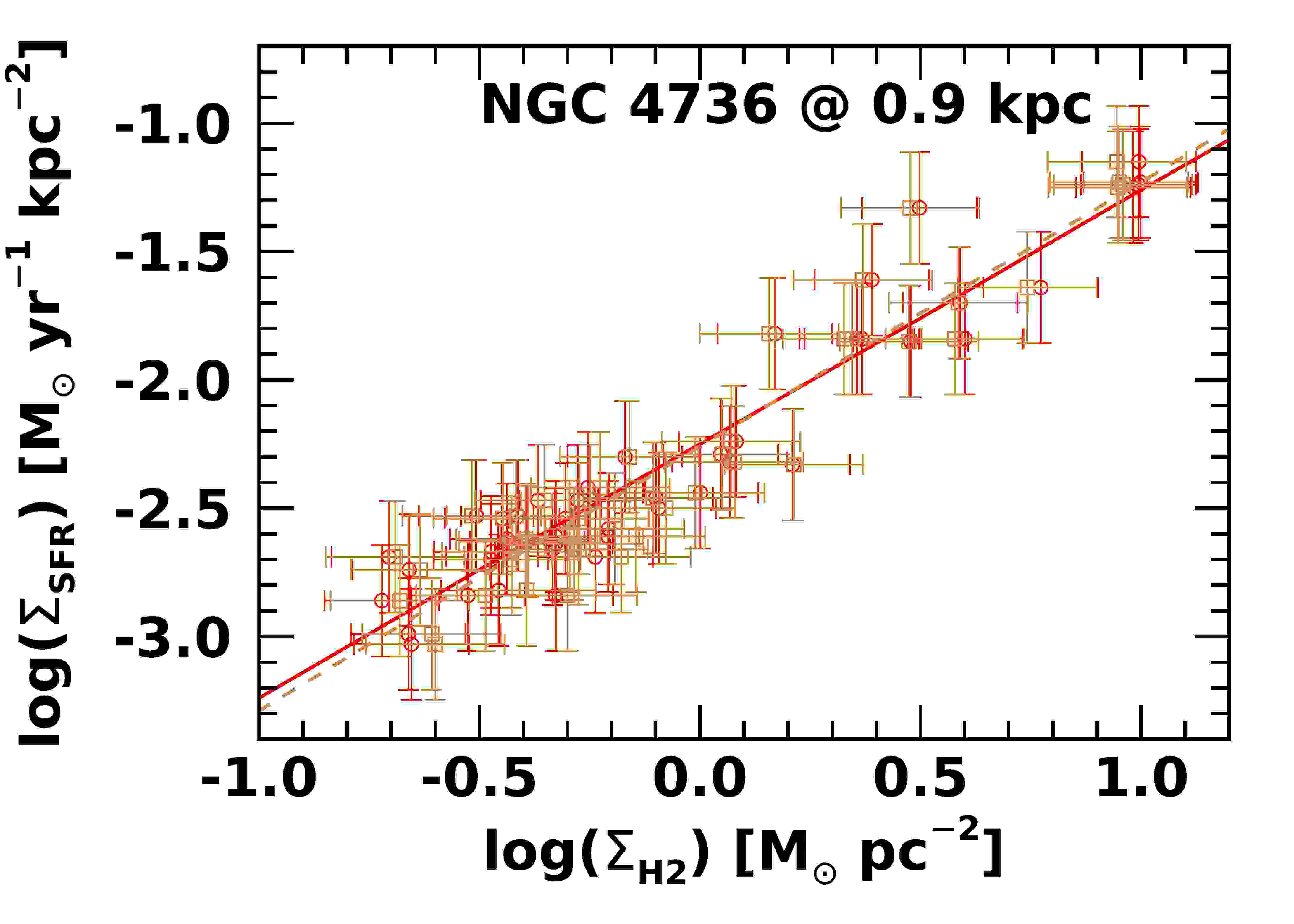}
\includegraphics[width=0.33\textwidth]{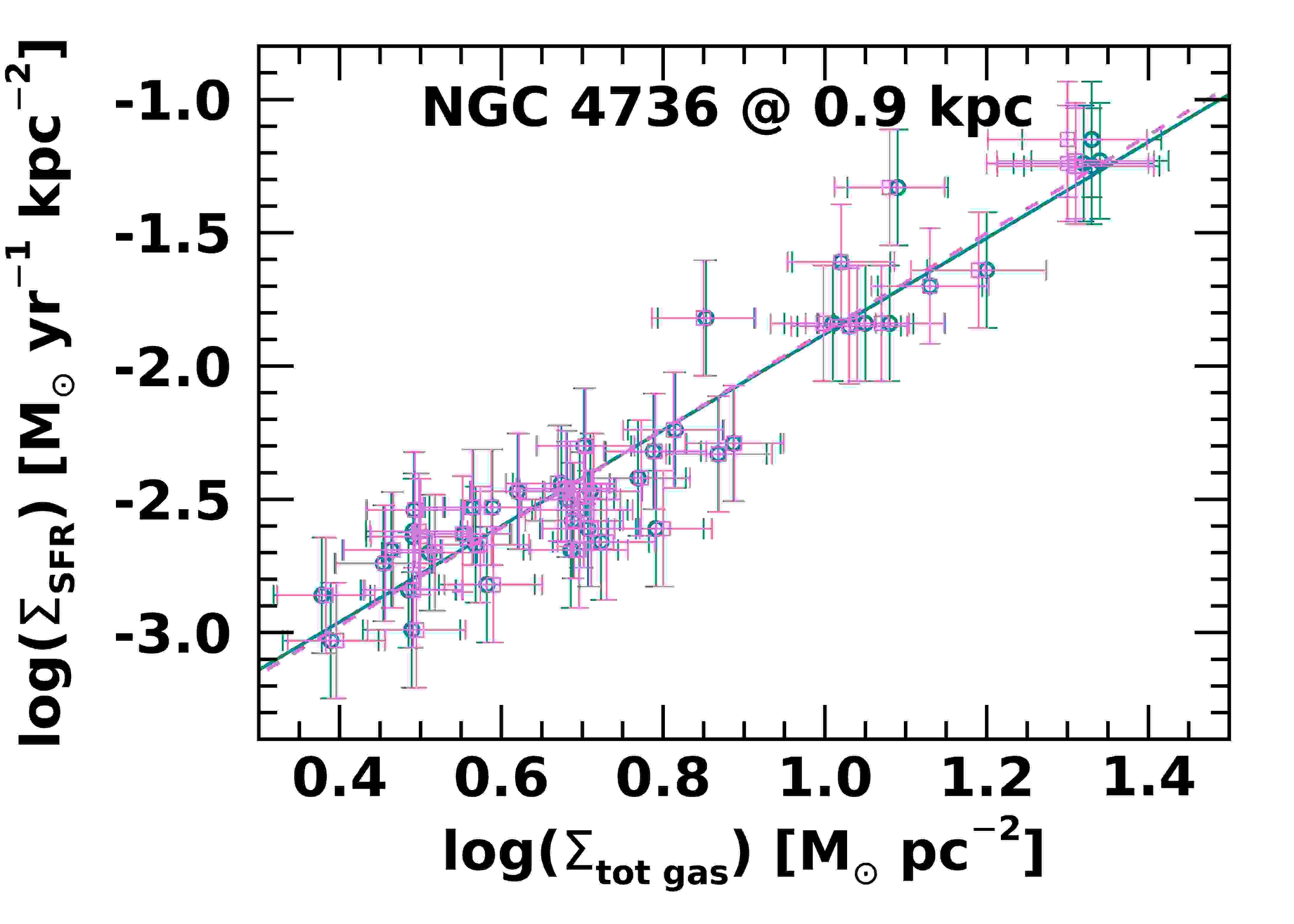}
\includegraphics[width=0.33\textwidth]{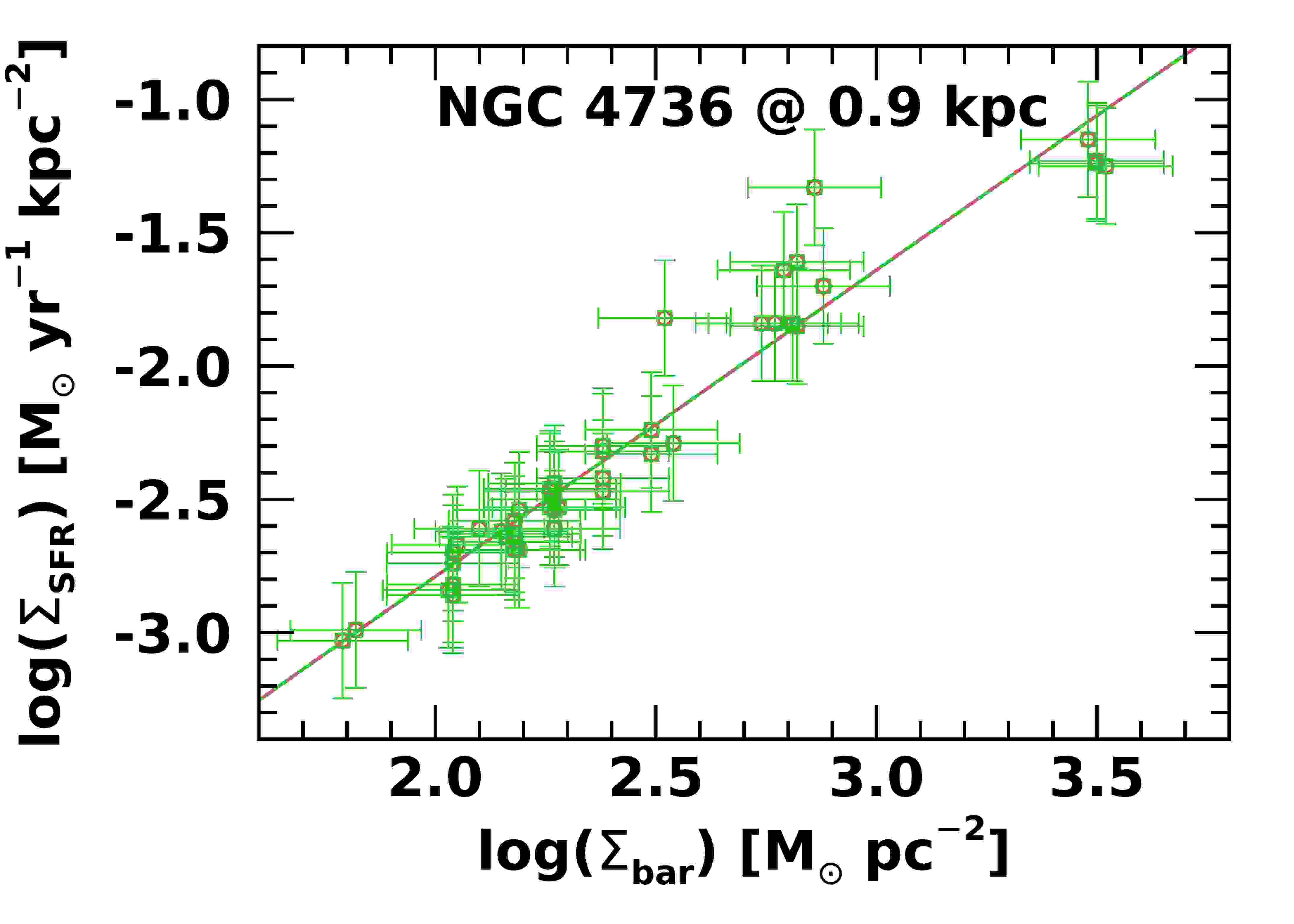}
\includegraphics[width=0.33\textwidth]{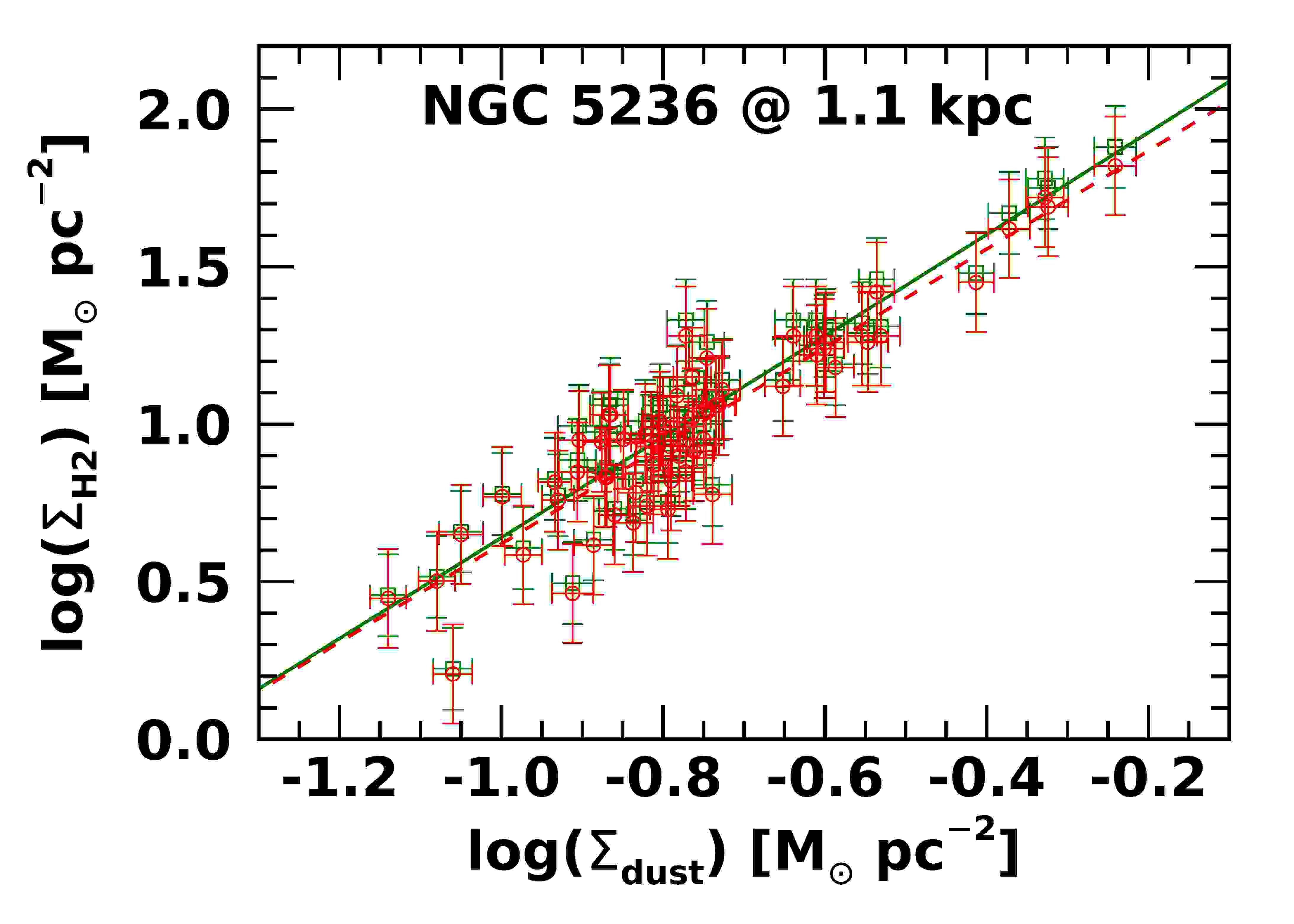}
\includegraphics[width=0.33\textwidth]{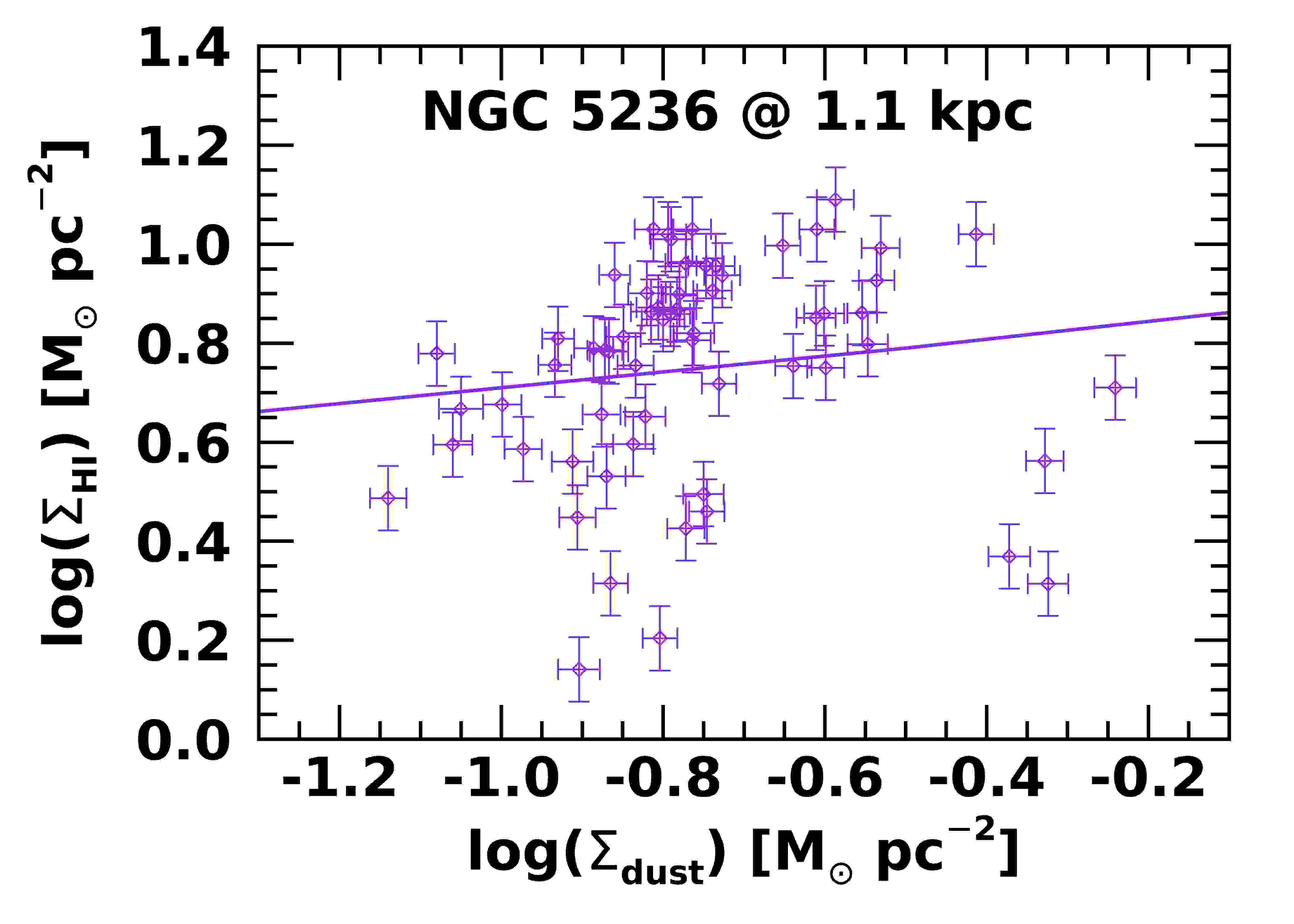}
\includegraphics[width=0.33\textwidth]{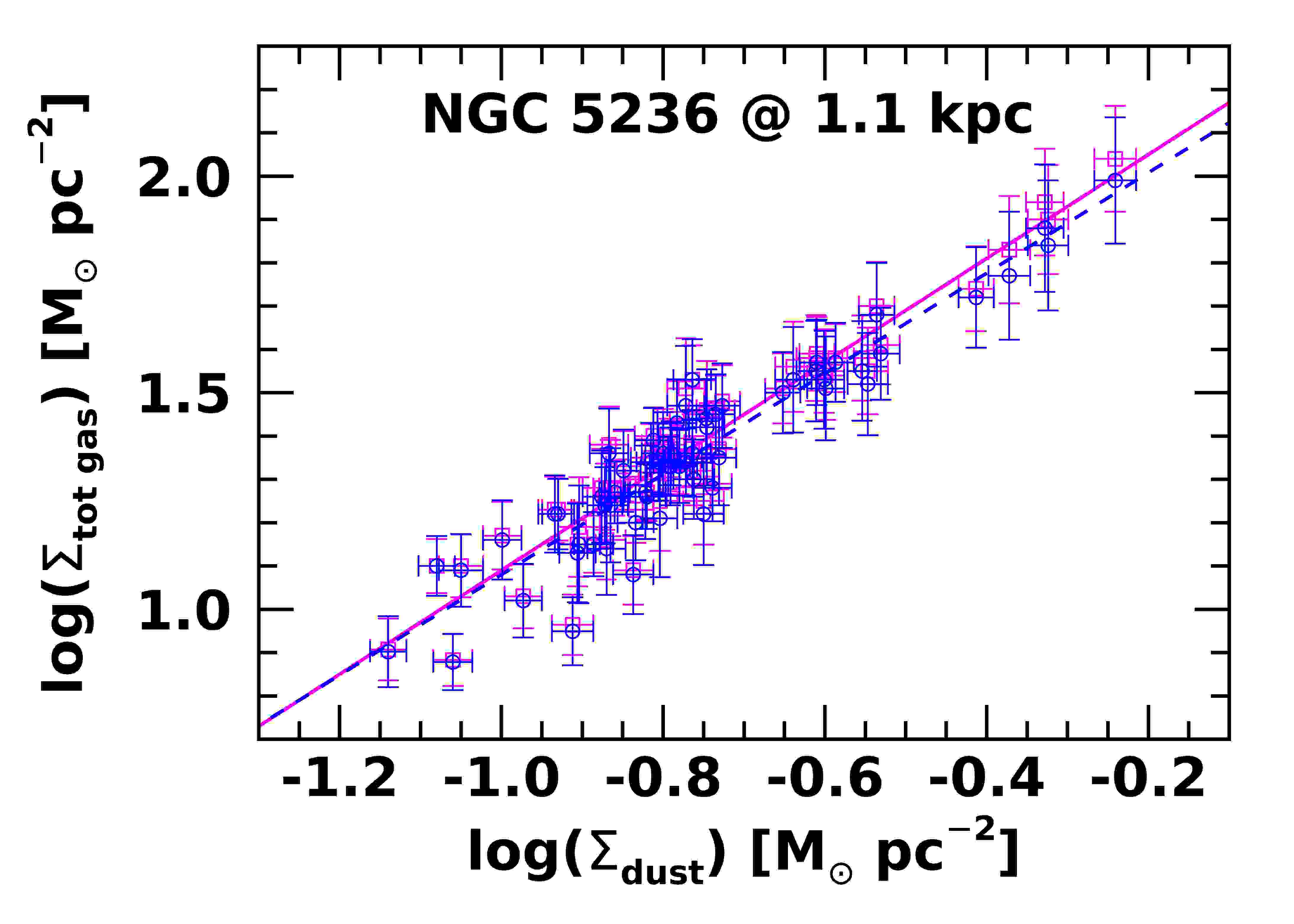}
\includegraphics[width=0.33\textwidth]{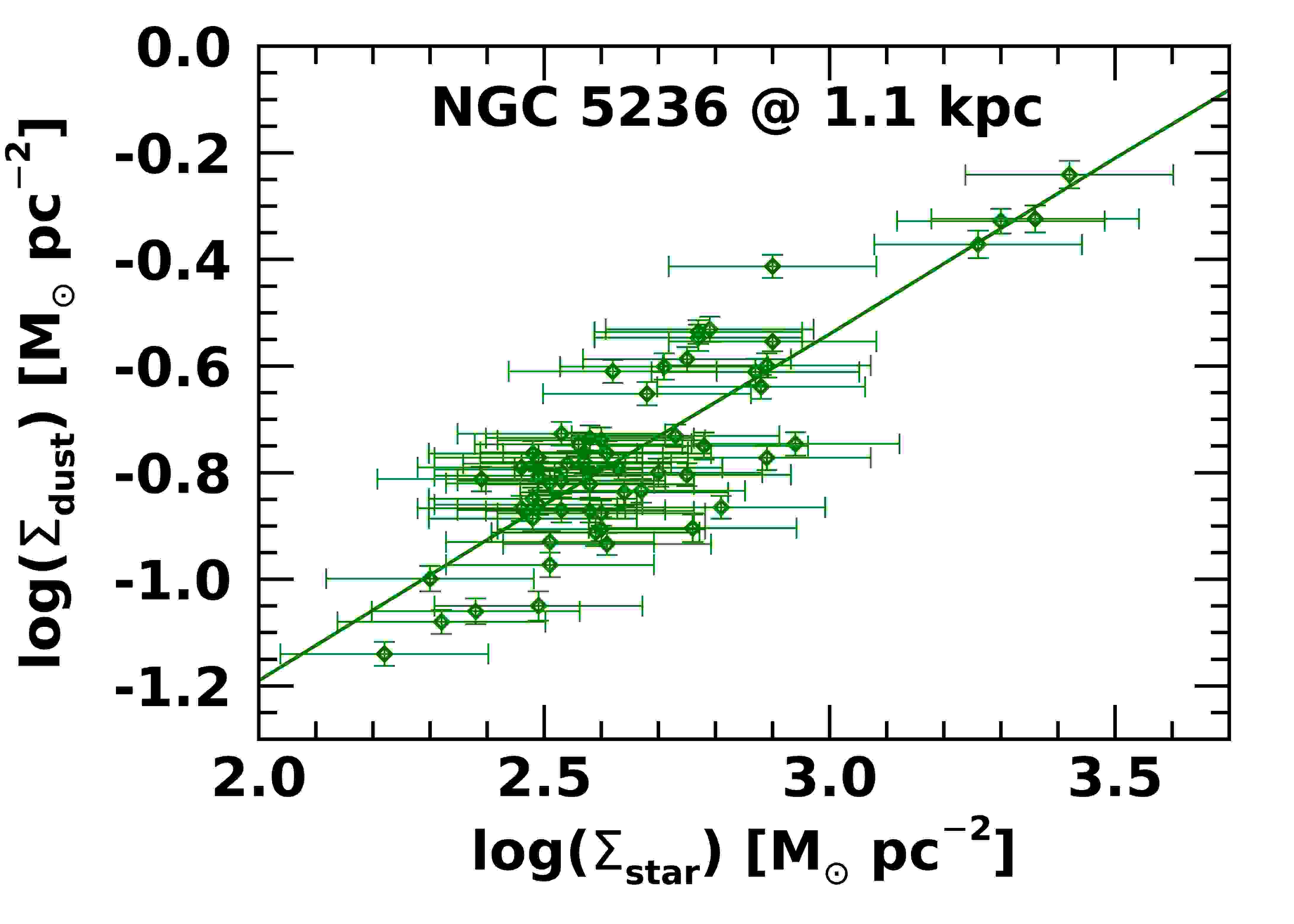}
\includegraphics[width=0.33\textwidth]{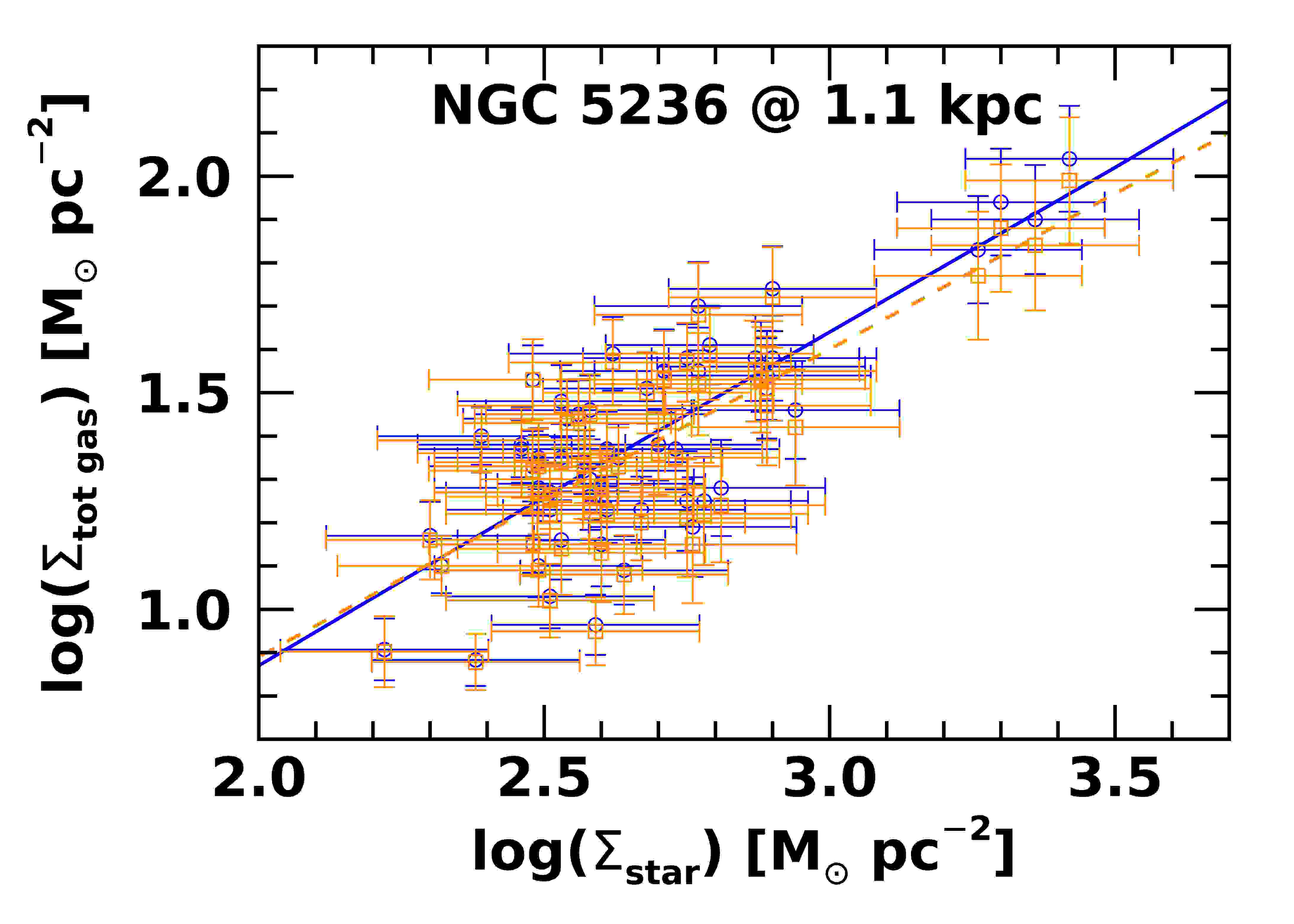}
\includegraphics[width=0.33\textwidth]{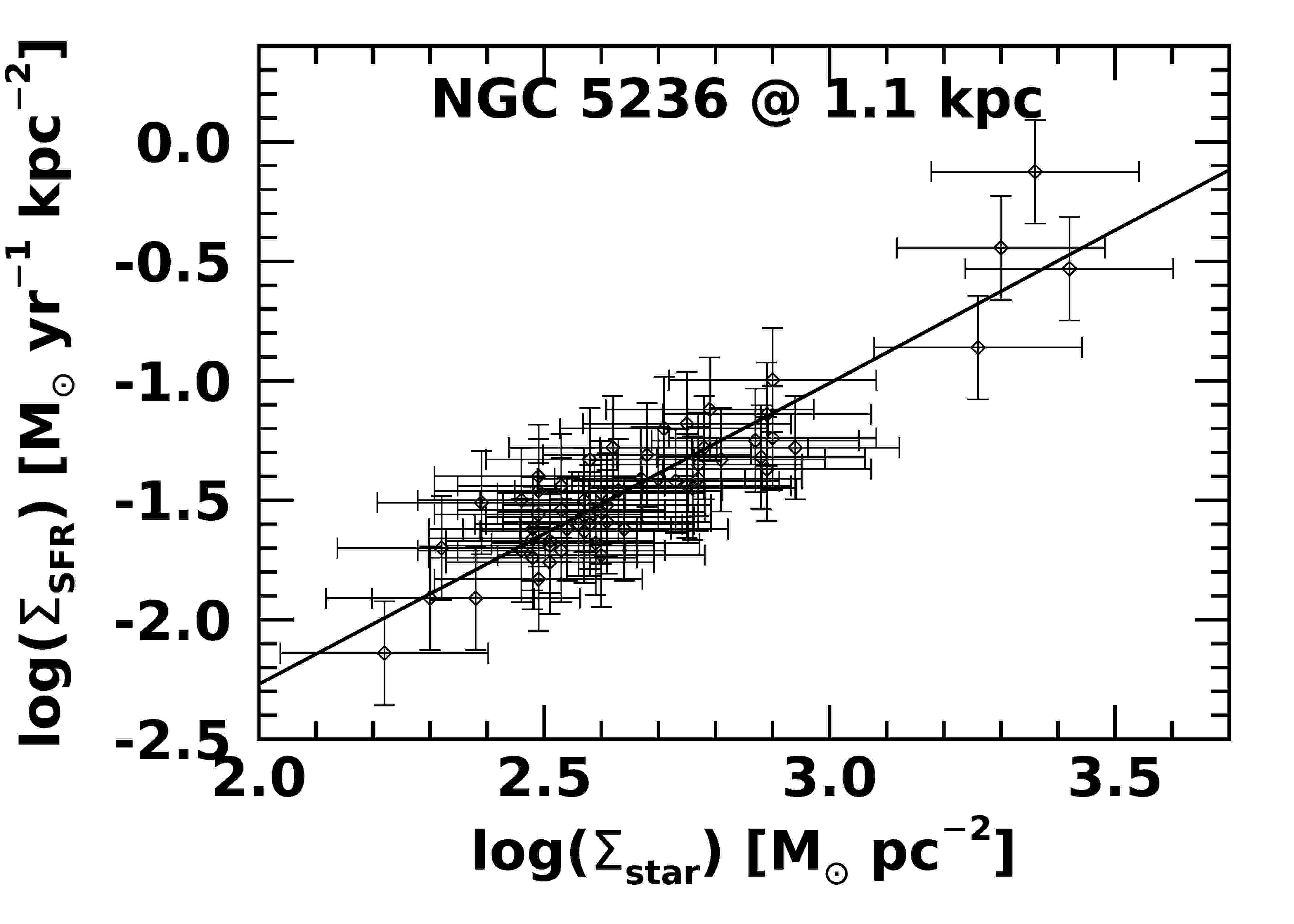}
\includegraphics[width=0.33\textwidth]{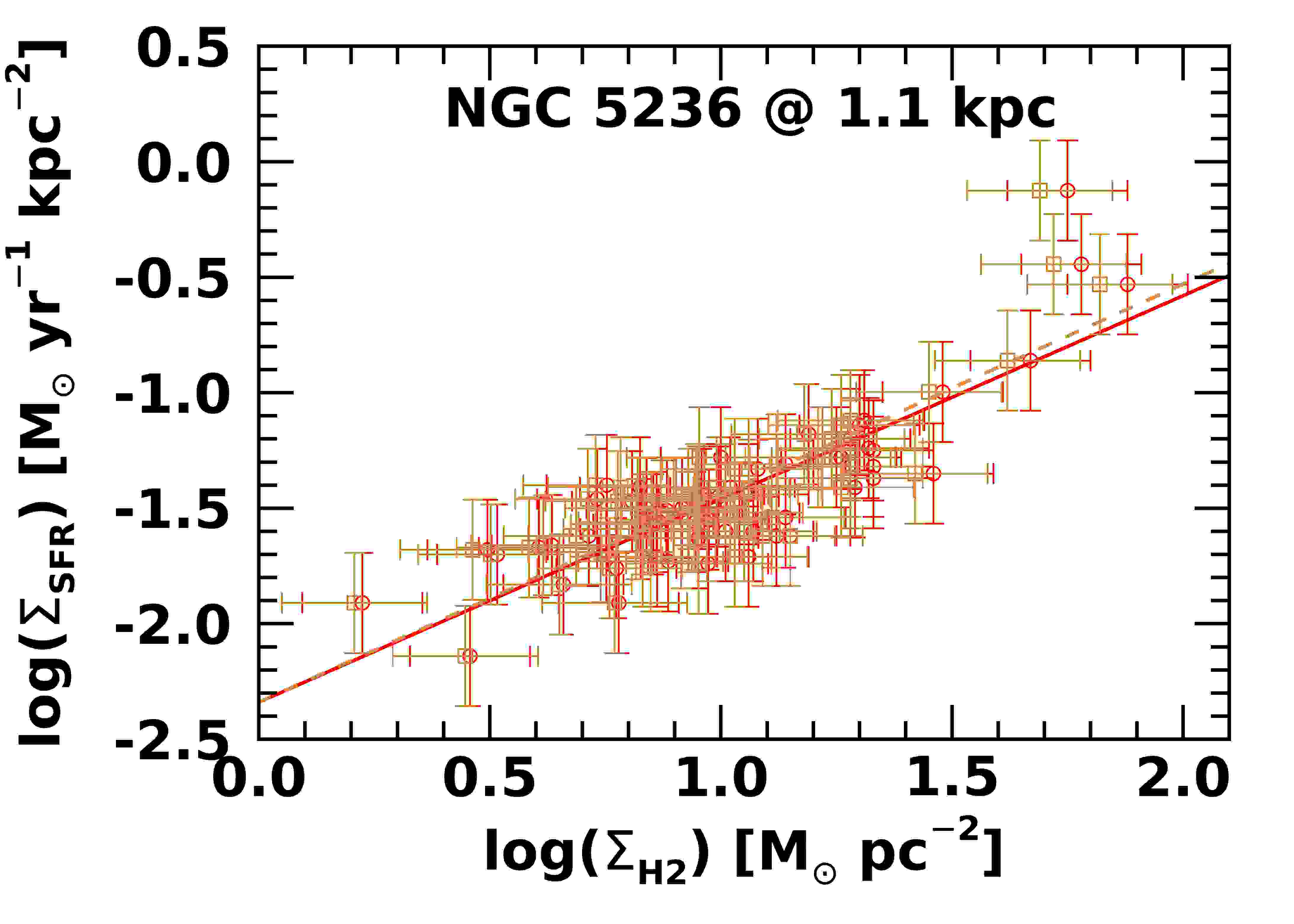}
\caption*{Figure~\ref{fig:add-ism} continued}
\end{figure*}

\begin{figure*}
\centering
\includegraphics[width=0.33\textwidth]{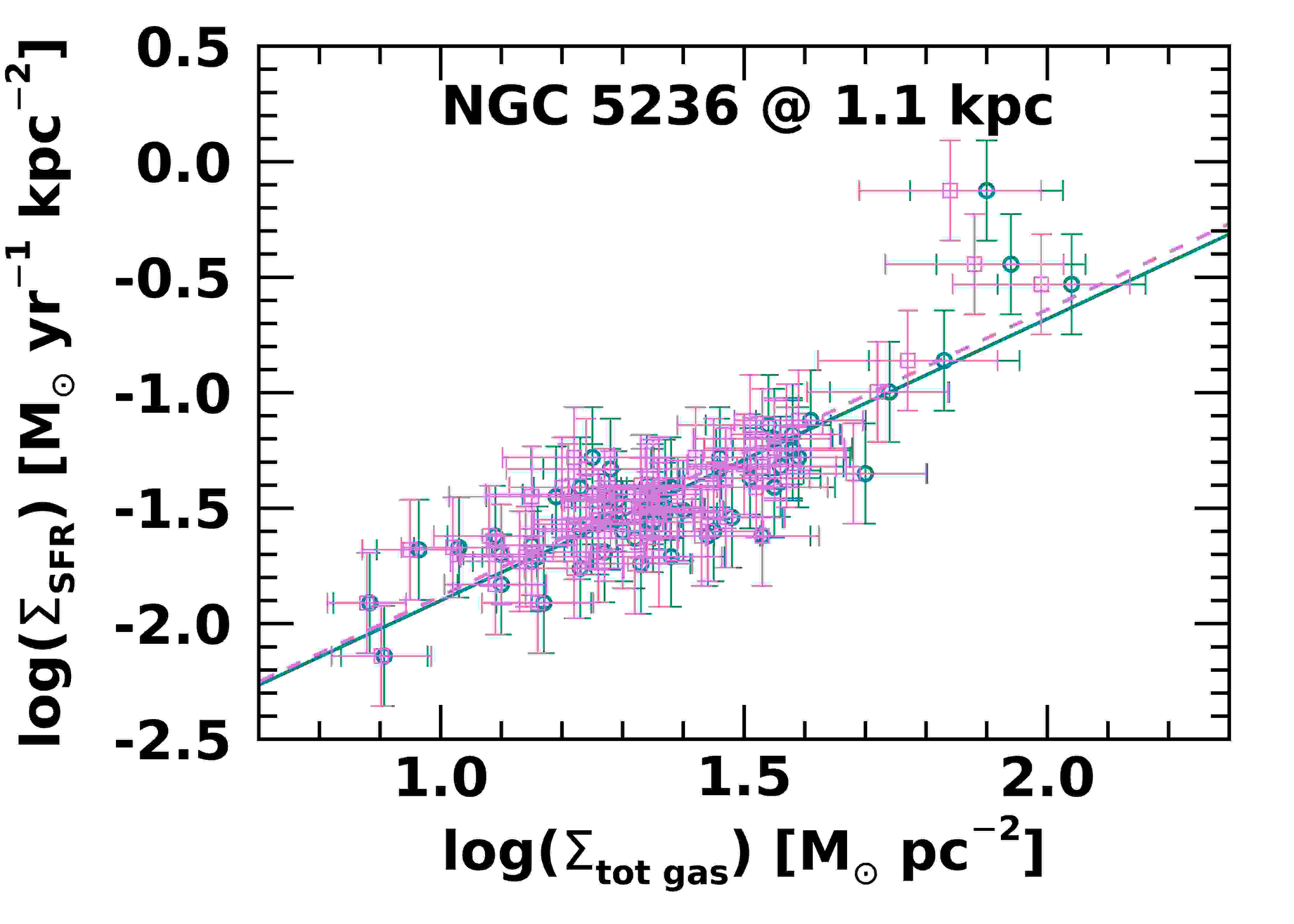}
\includegraphics[width=0.33\textwidth]{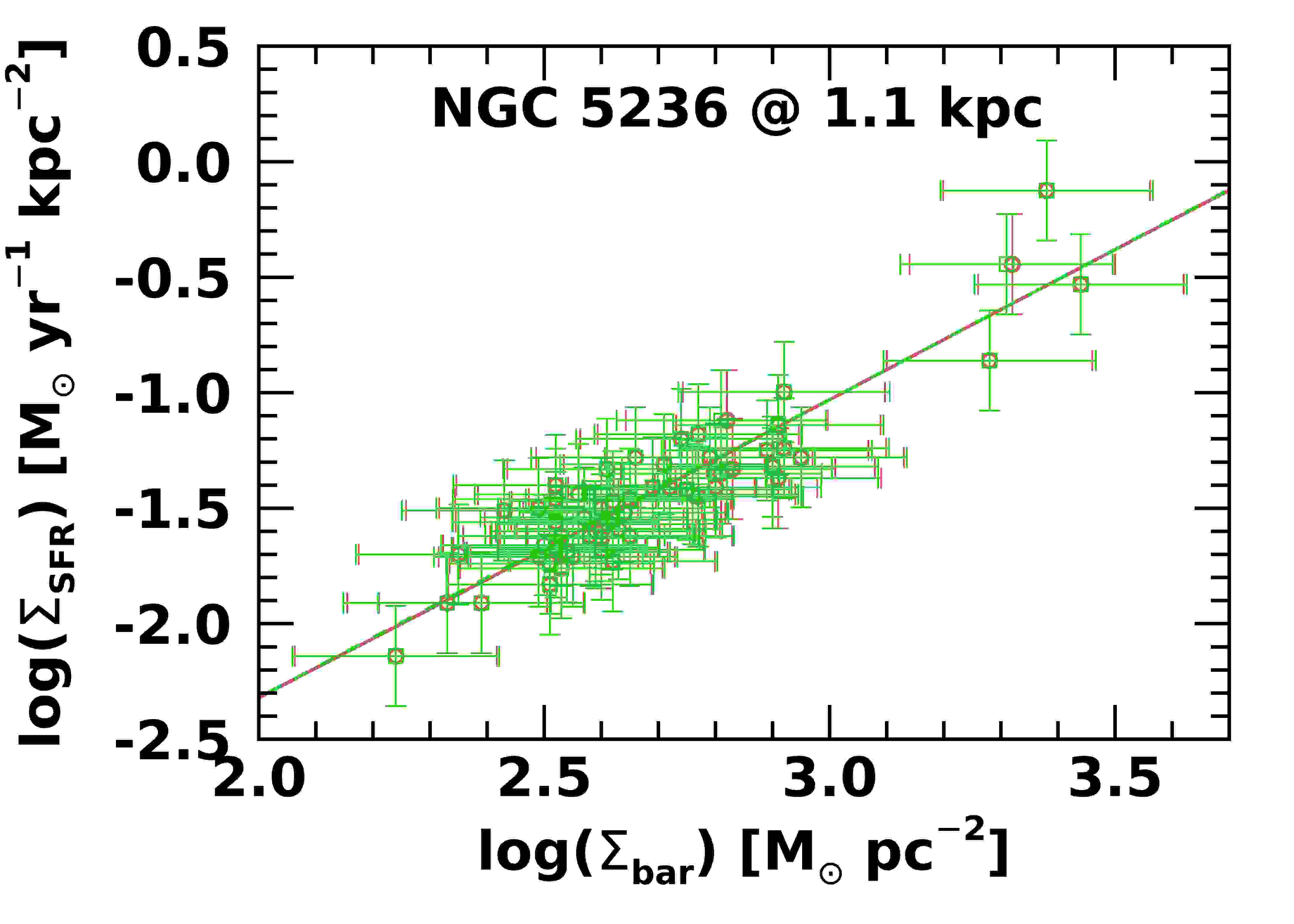}
\includegraphics[width=0.33\textwidth]{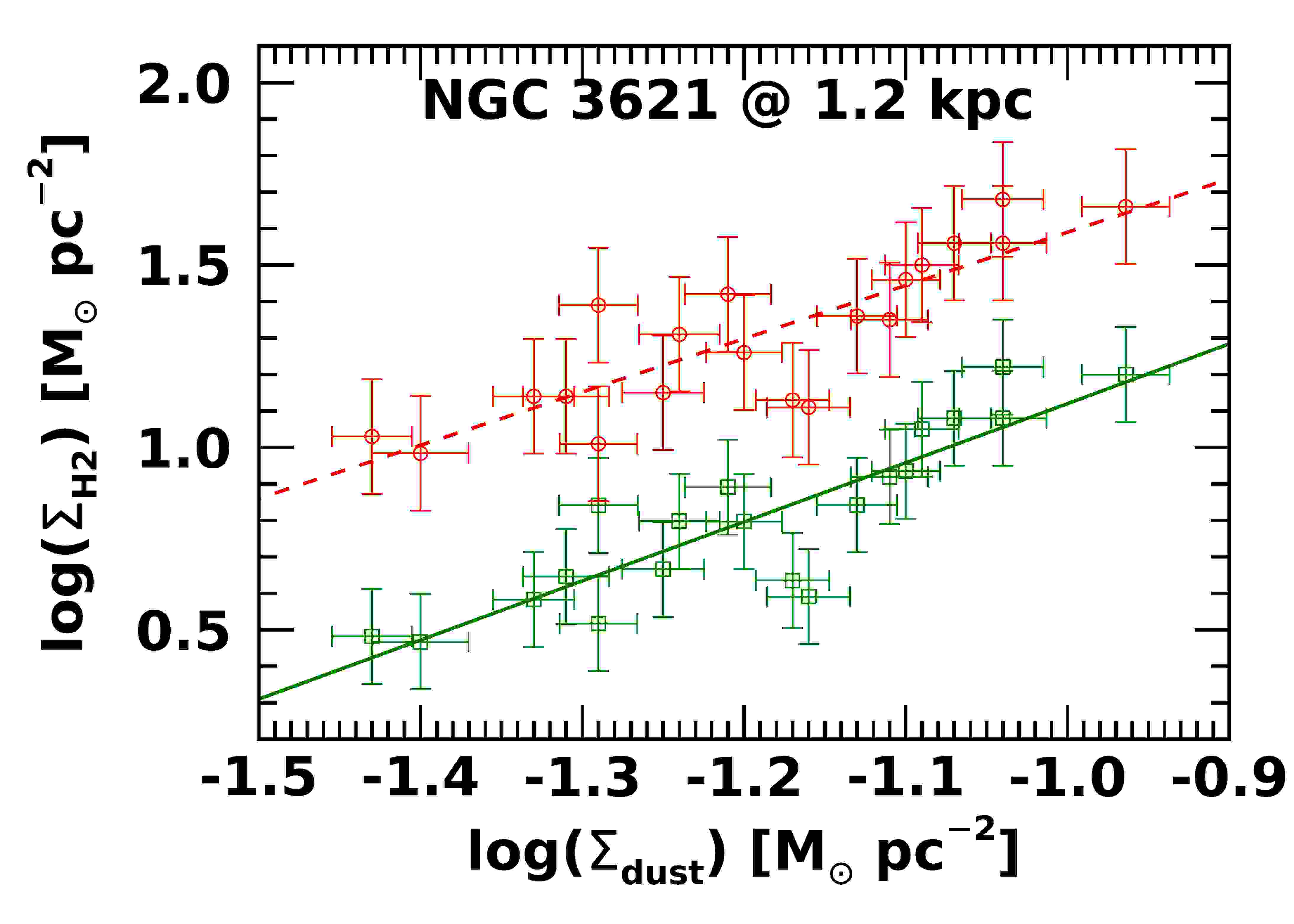}
\includegraphics[width=0.33\textwidth]{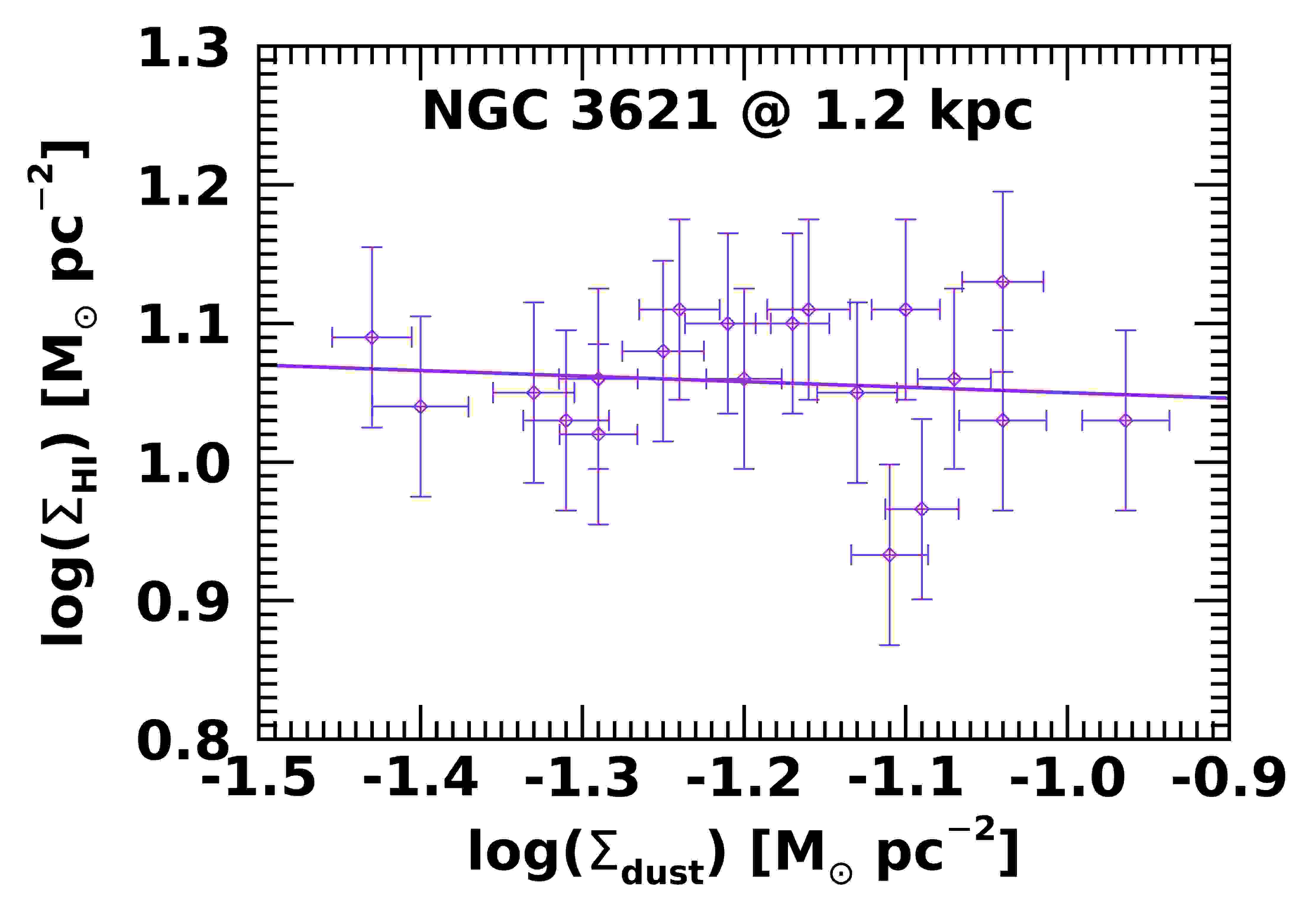}
\includegraphics[width=0.33\textwidth]{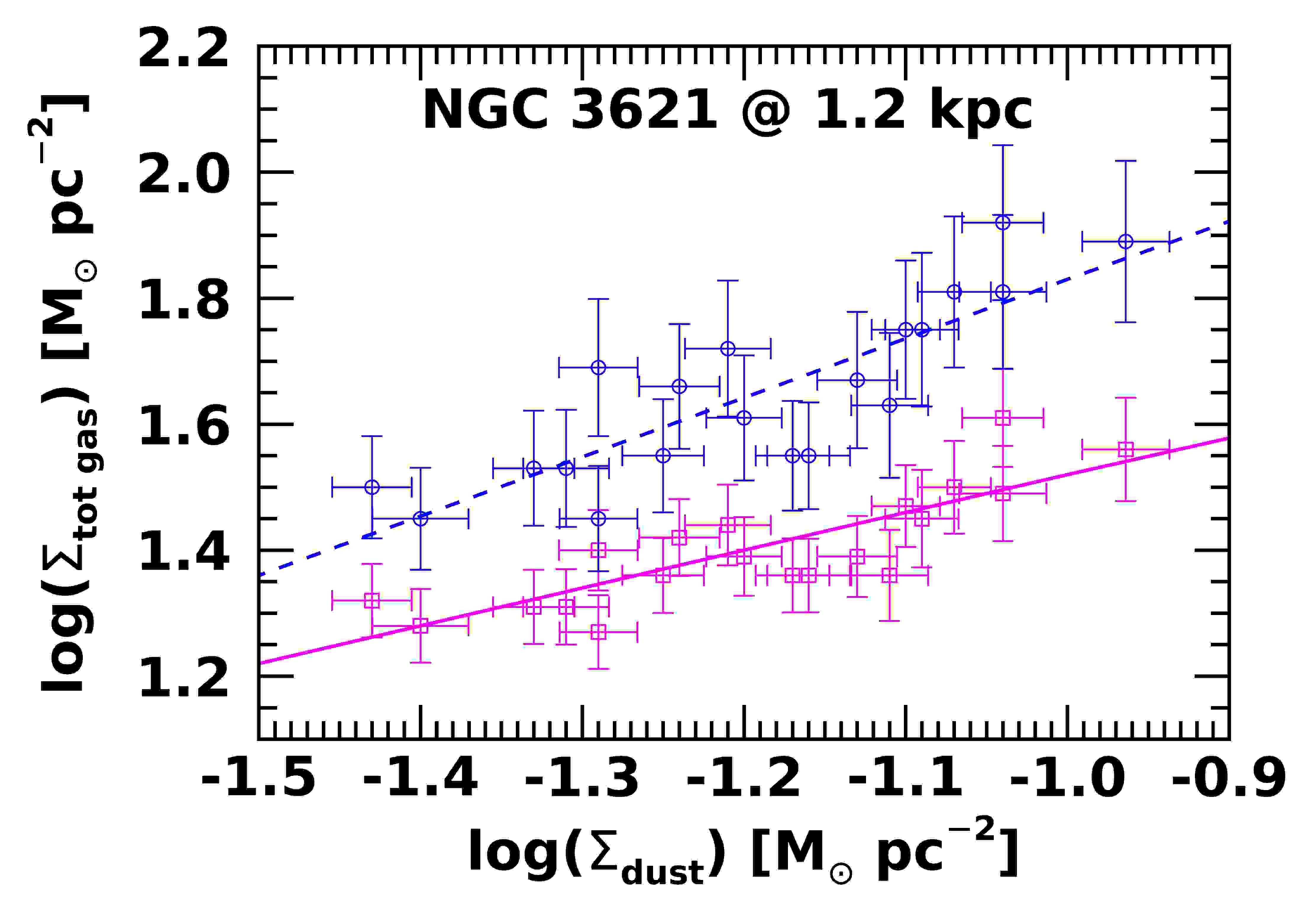}
\includegraphics[width=0.33\textwidth]{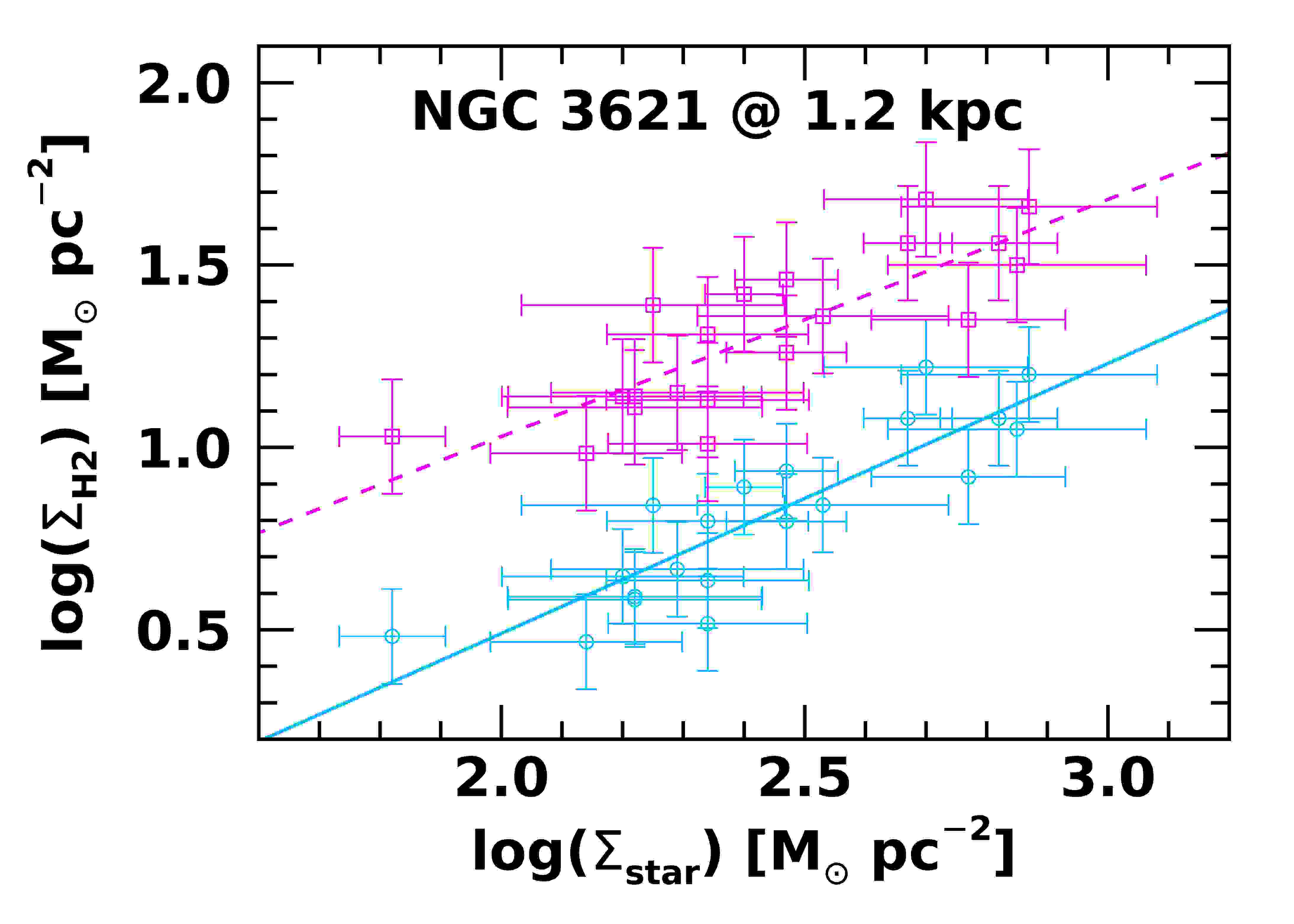}
\includegraphics[width=0.33\textwidth]{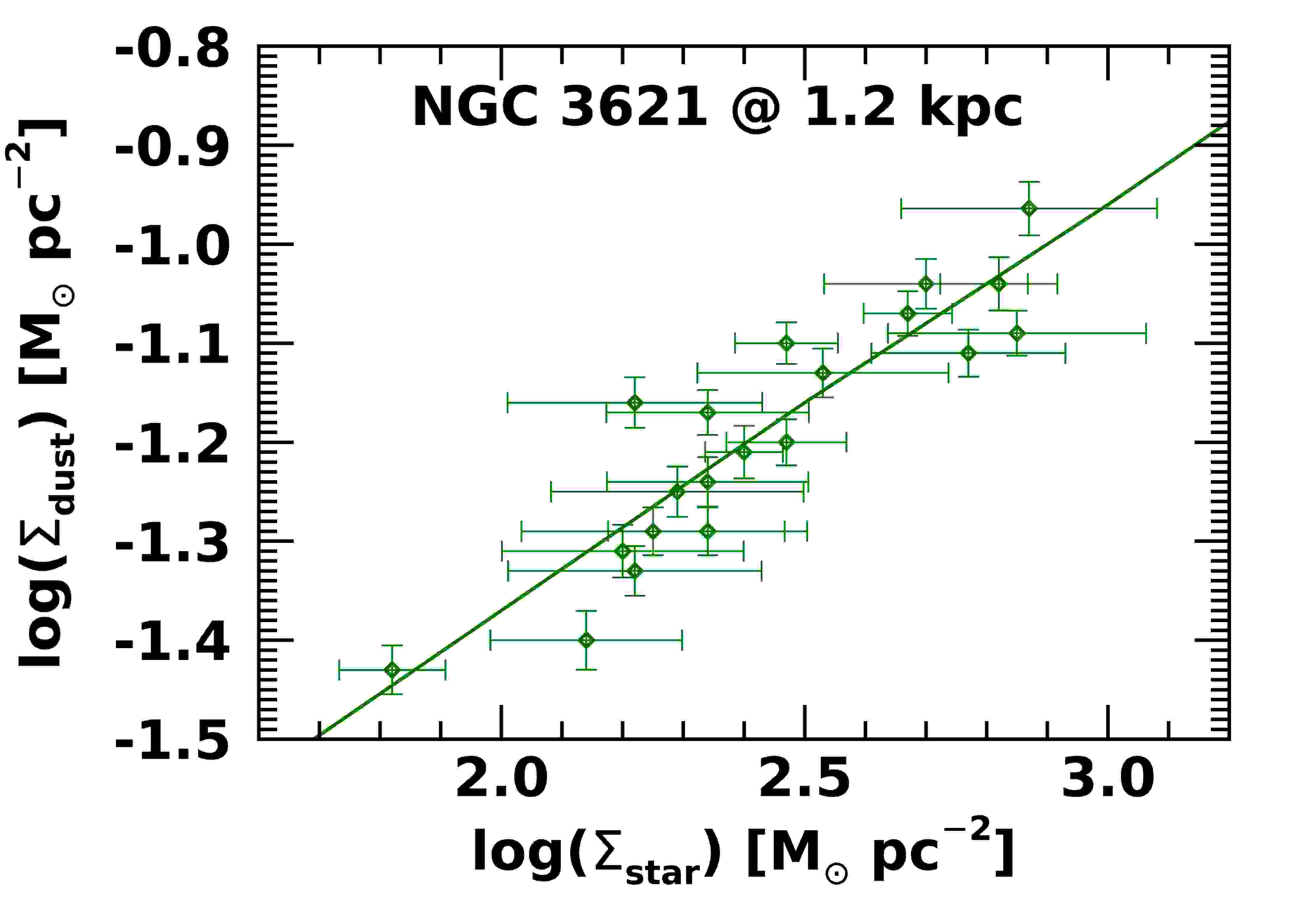}
\includegraphics[width=0.33\textwidth]{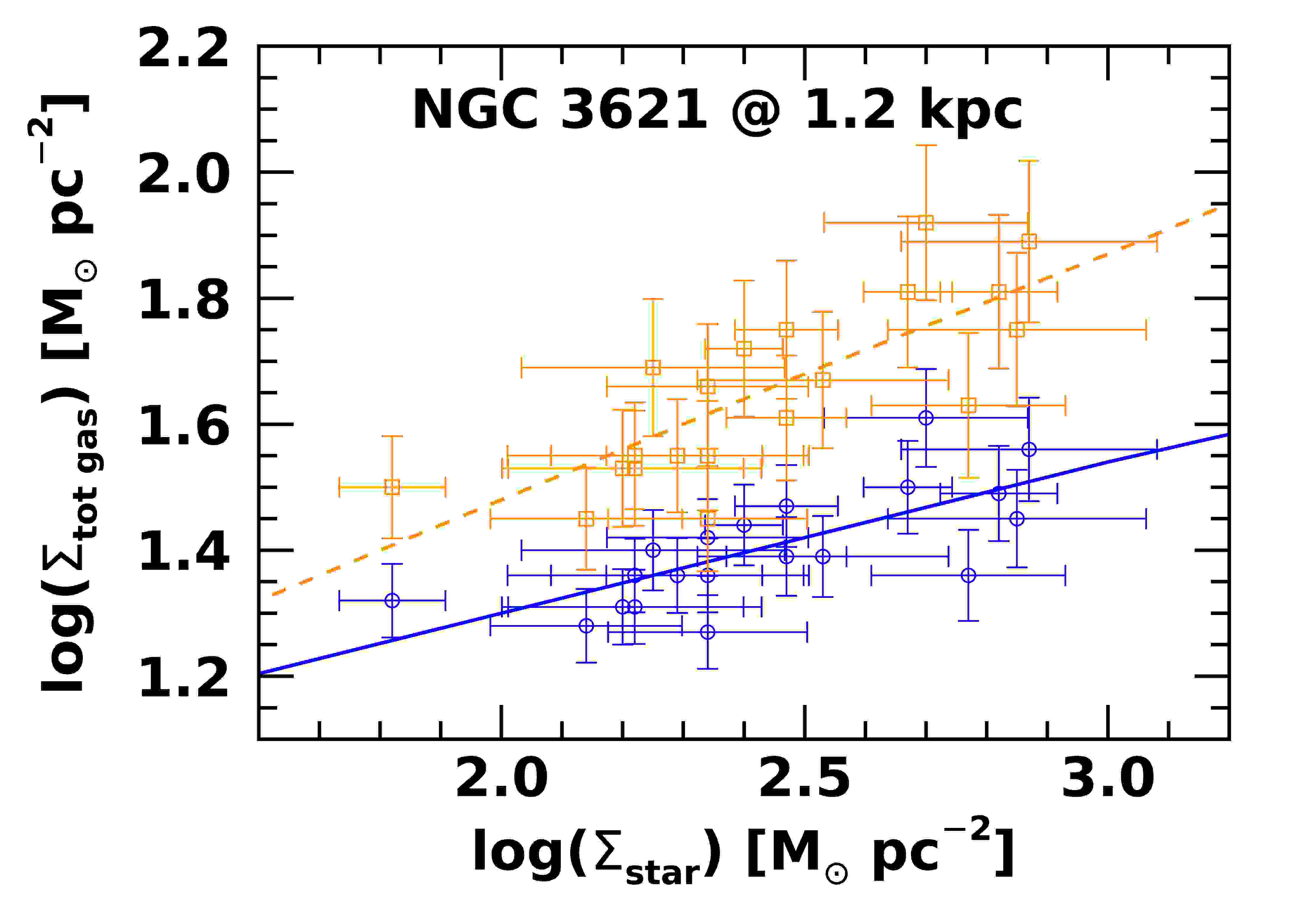}
\includegraphics[width=0.33\textwidth]{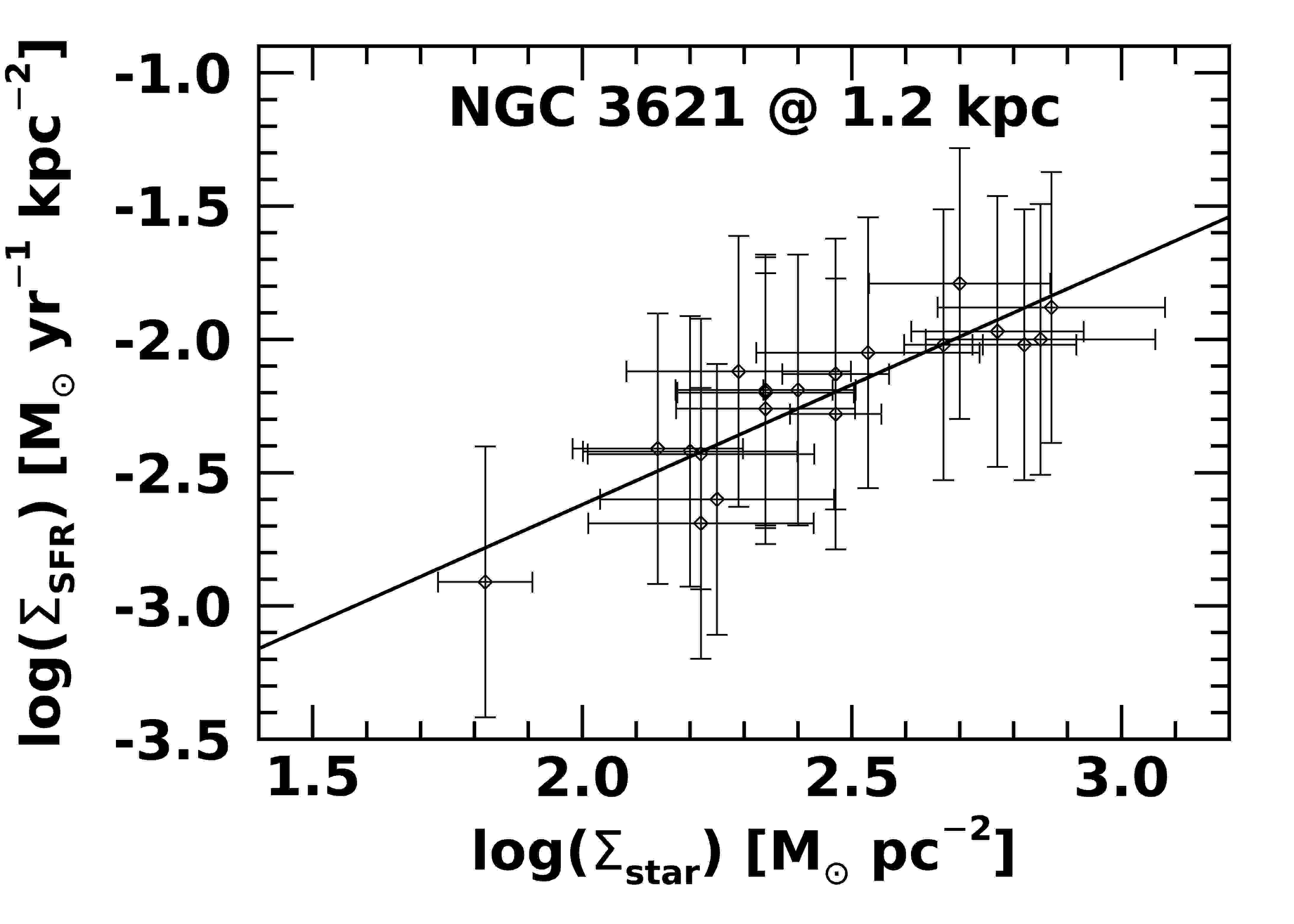}
\includegraphics[width=0.33\textwidth]{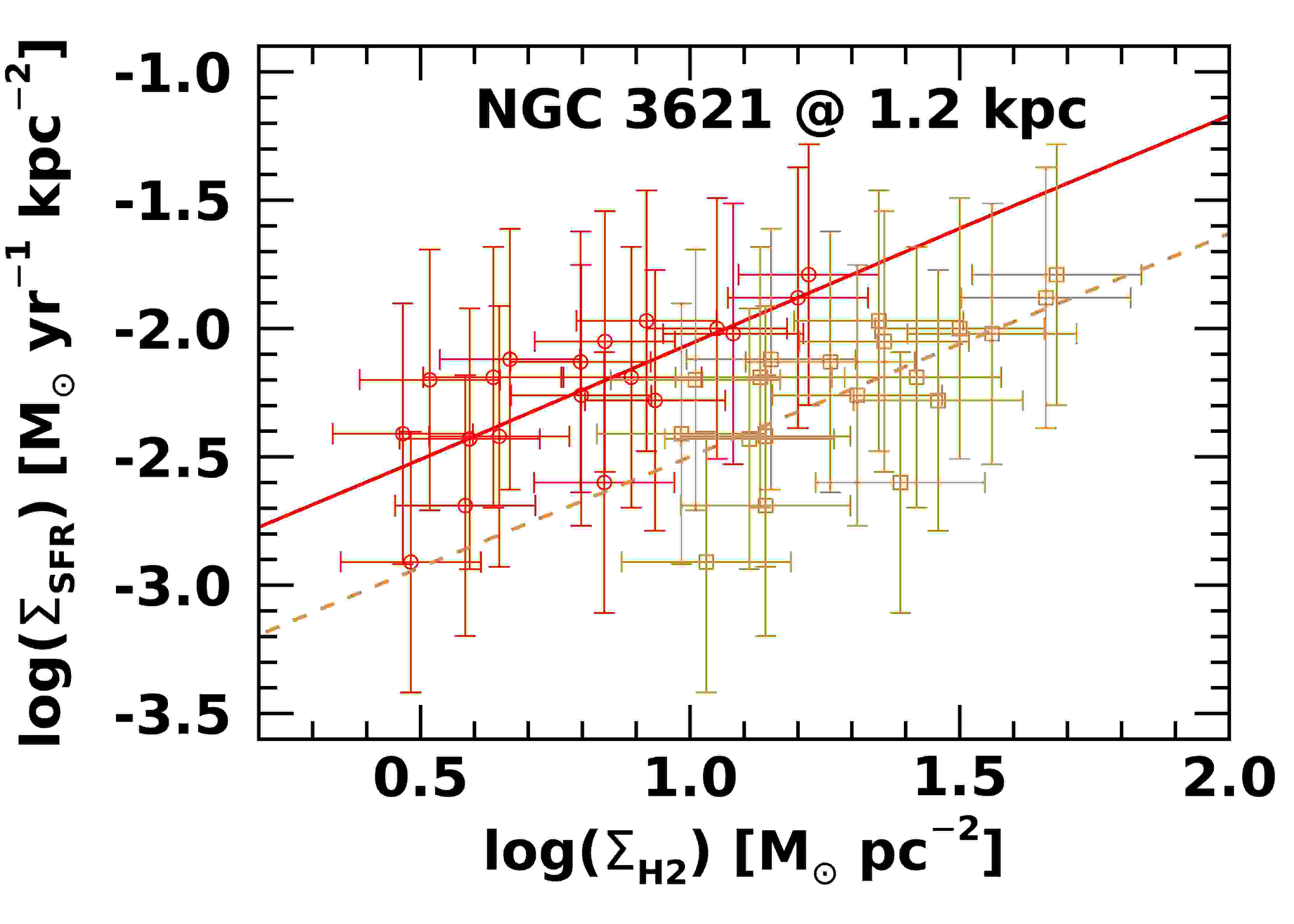}
\includegraphics[width=0.33\textwidth]{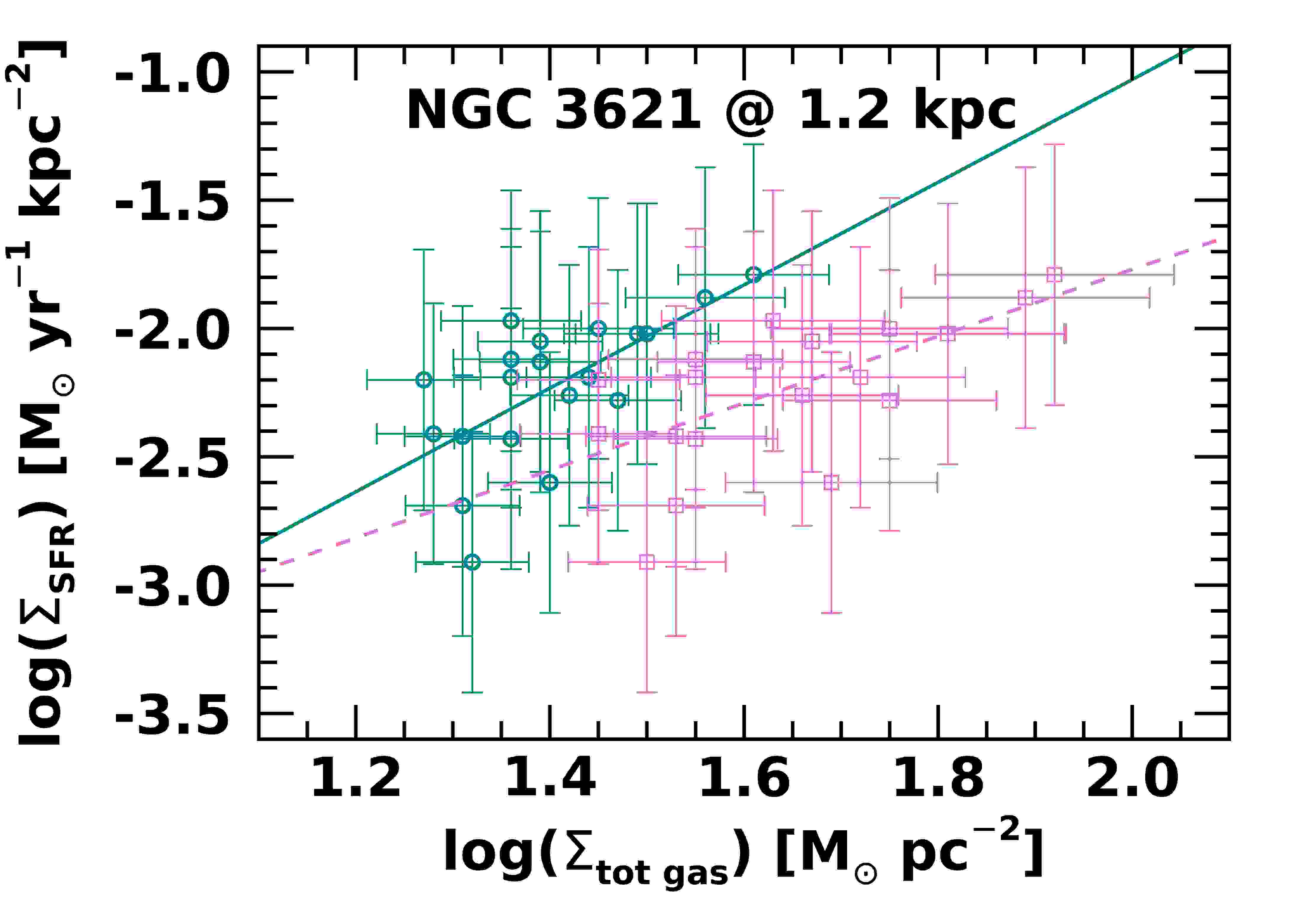}
\includegraphics[width=0.33\textwidth]{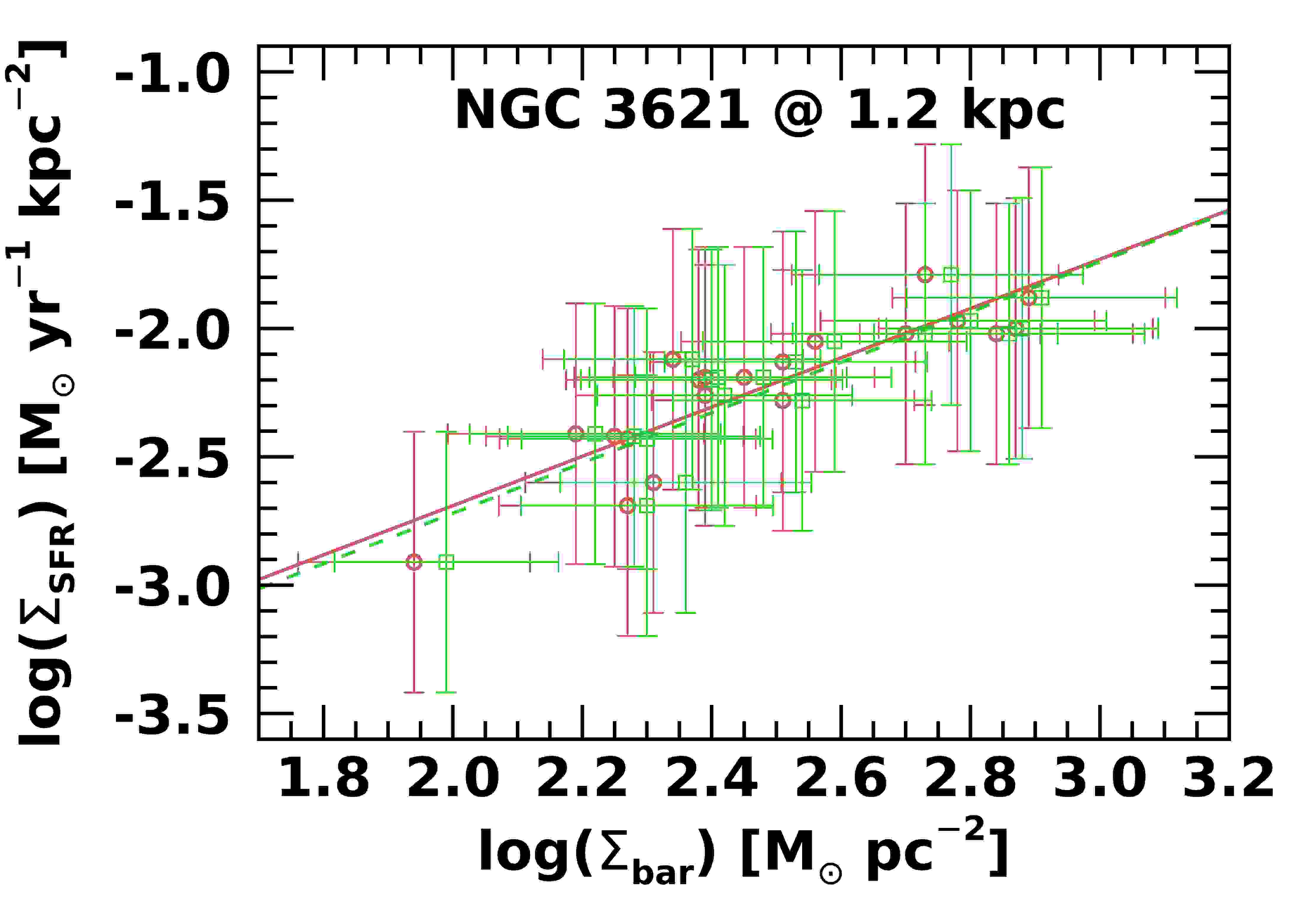}
\includegraphics[width=0.33\textwidth]{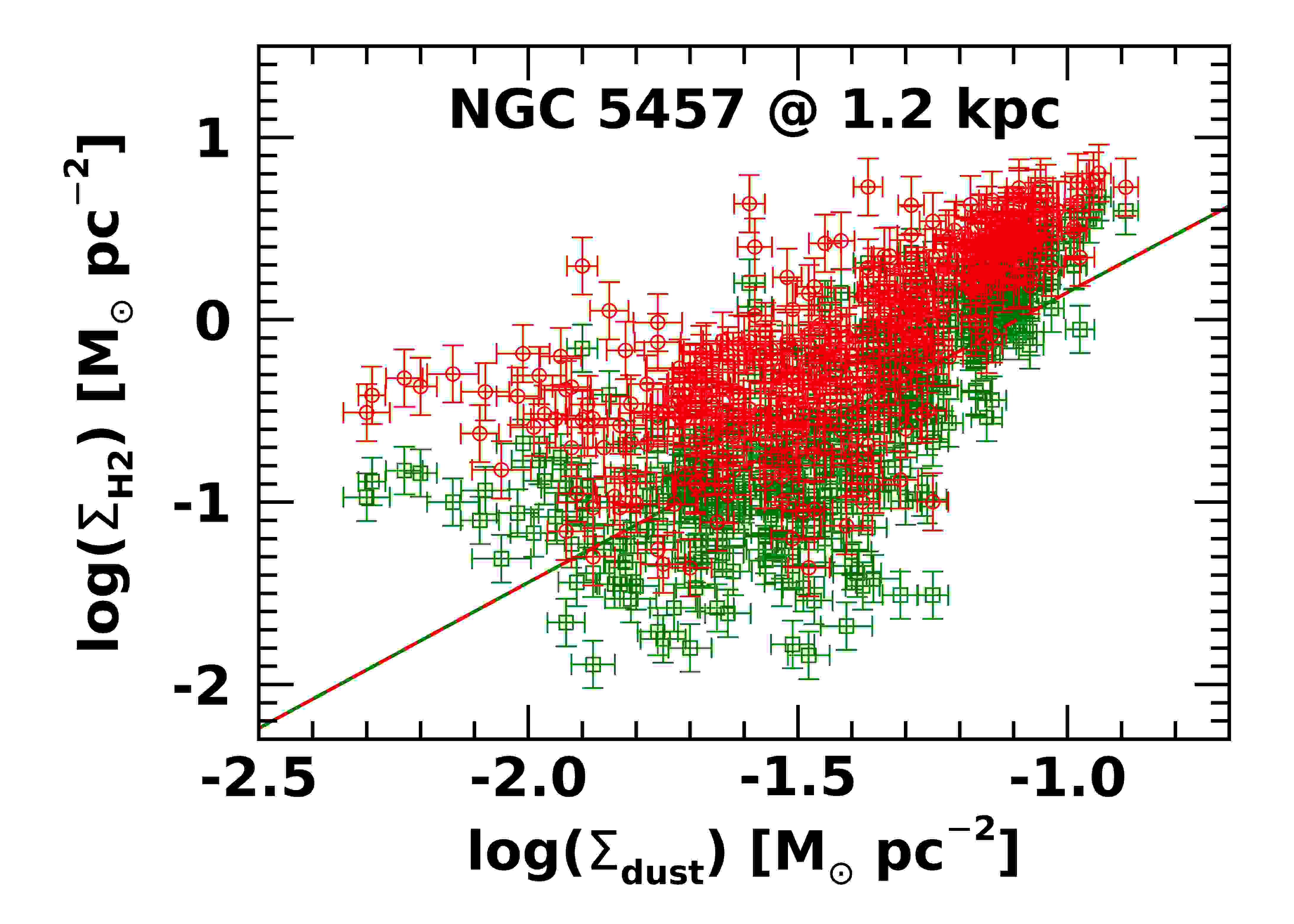}
\includegraphics[width=0.33\textwidth]{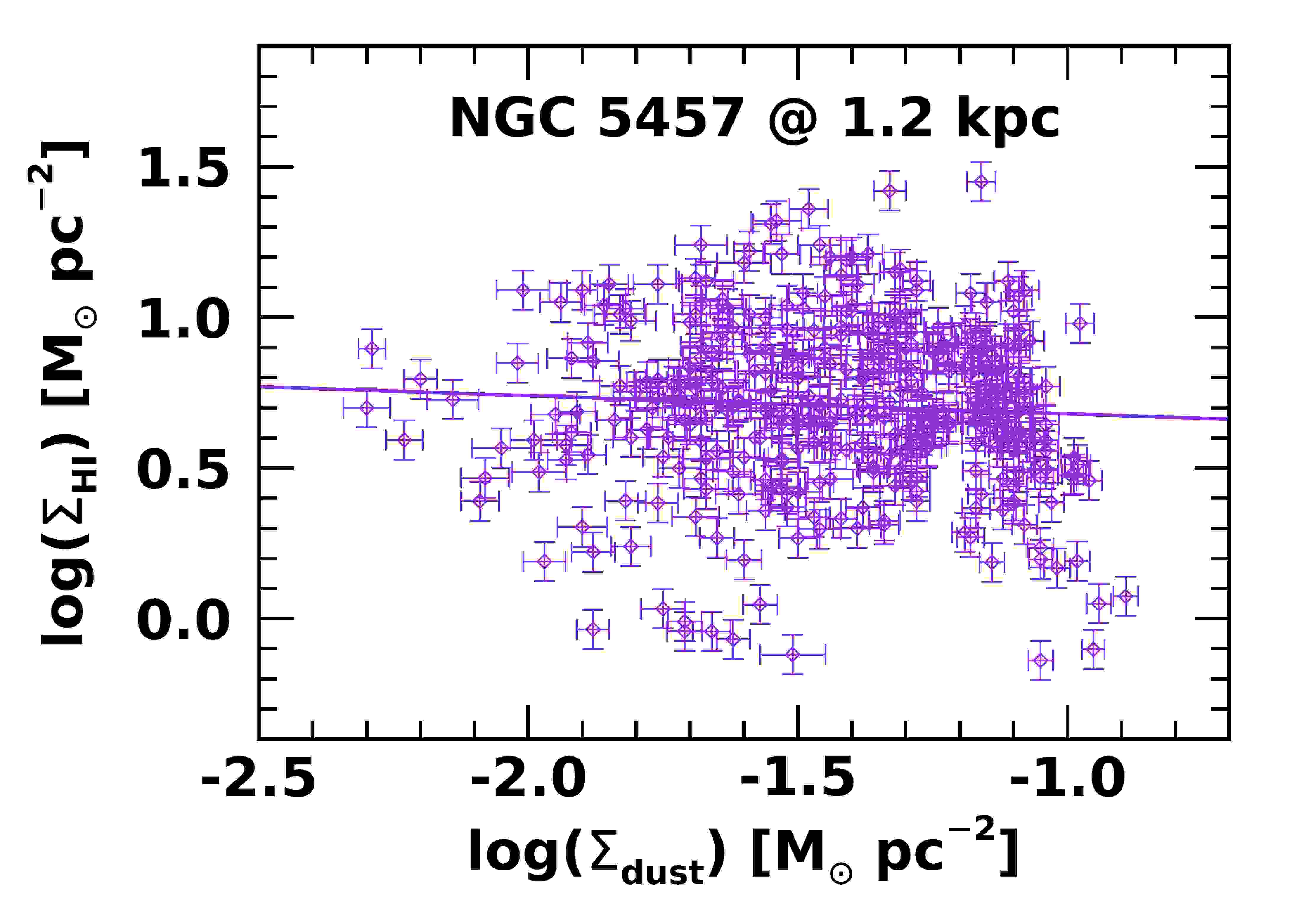}
\includegraphics[width=0.33\textwidth]{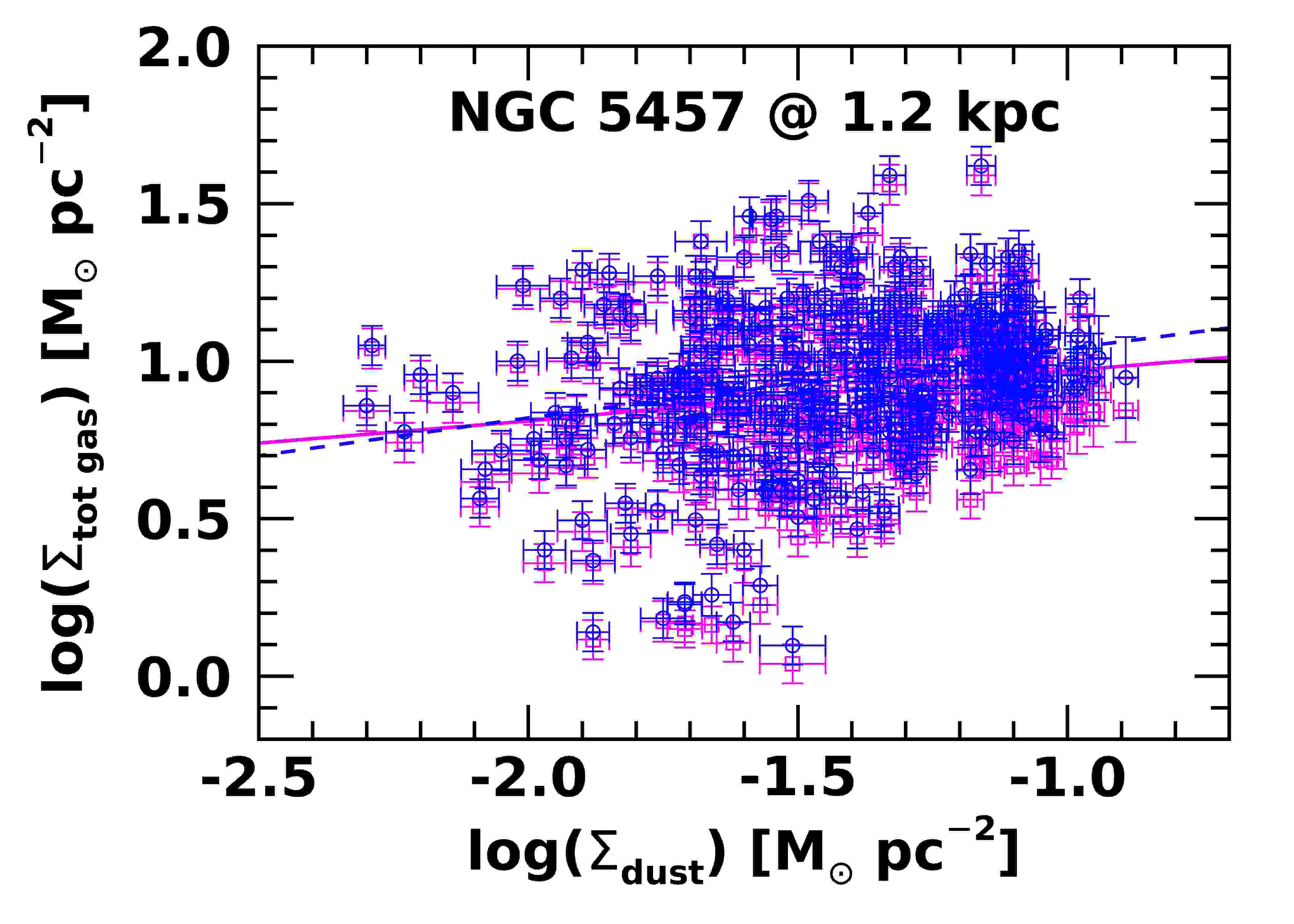}
\caption*{Figure~\ref{fig:add-ism} continued}
\end{figure*}

\begin{figure*}
\centering
\includegraphics[width=0.33\textwidth]{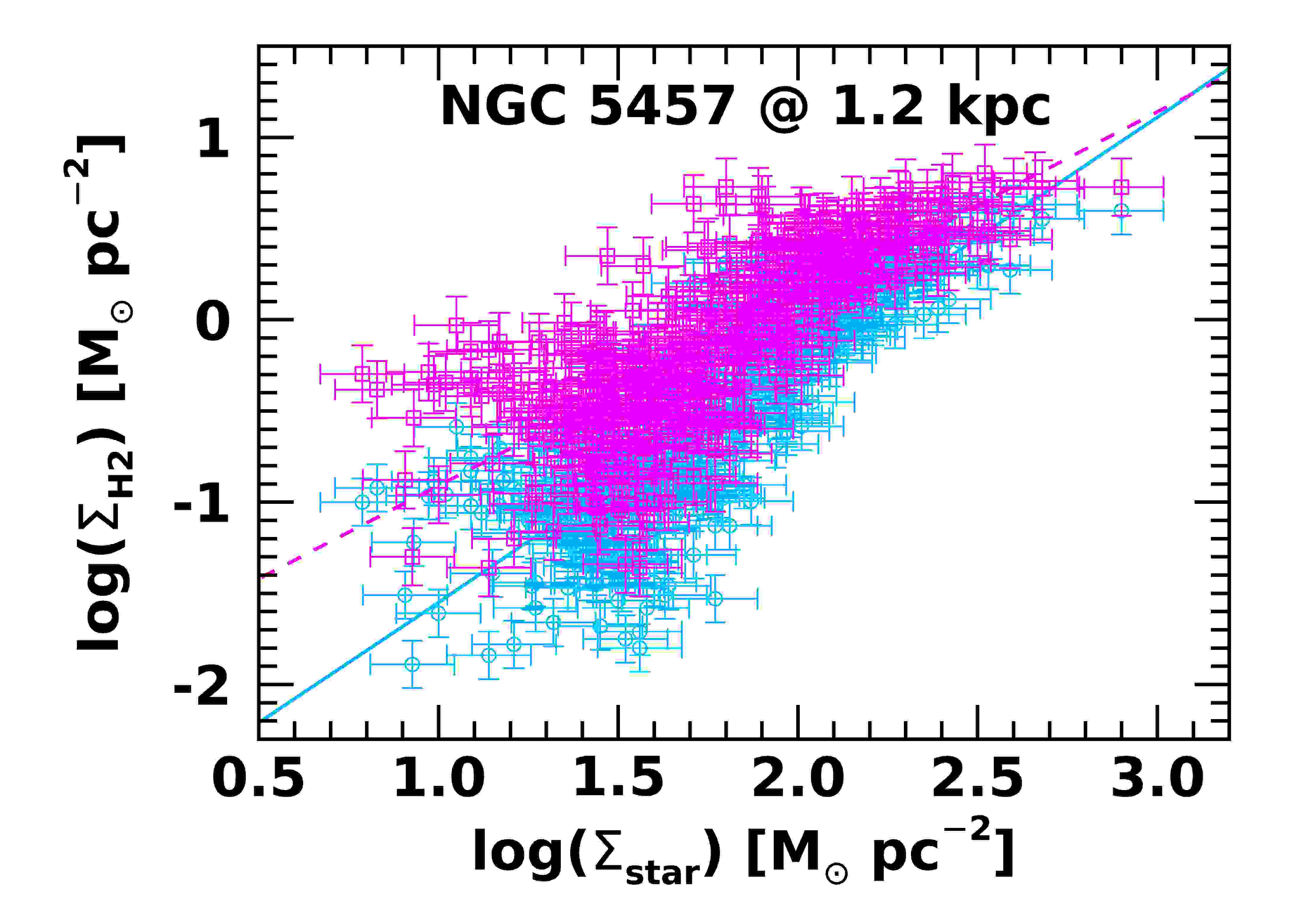}
\includegraphics[width=0.33\textwidth]{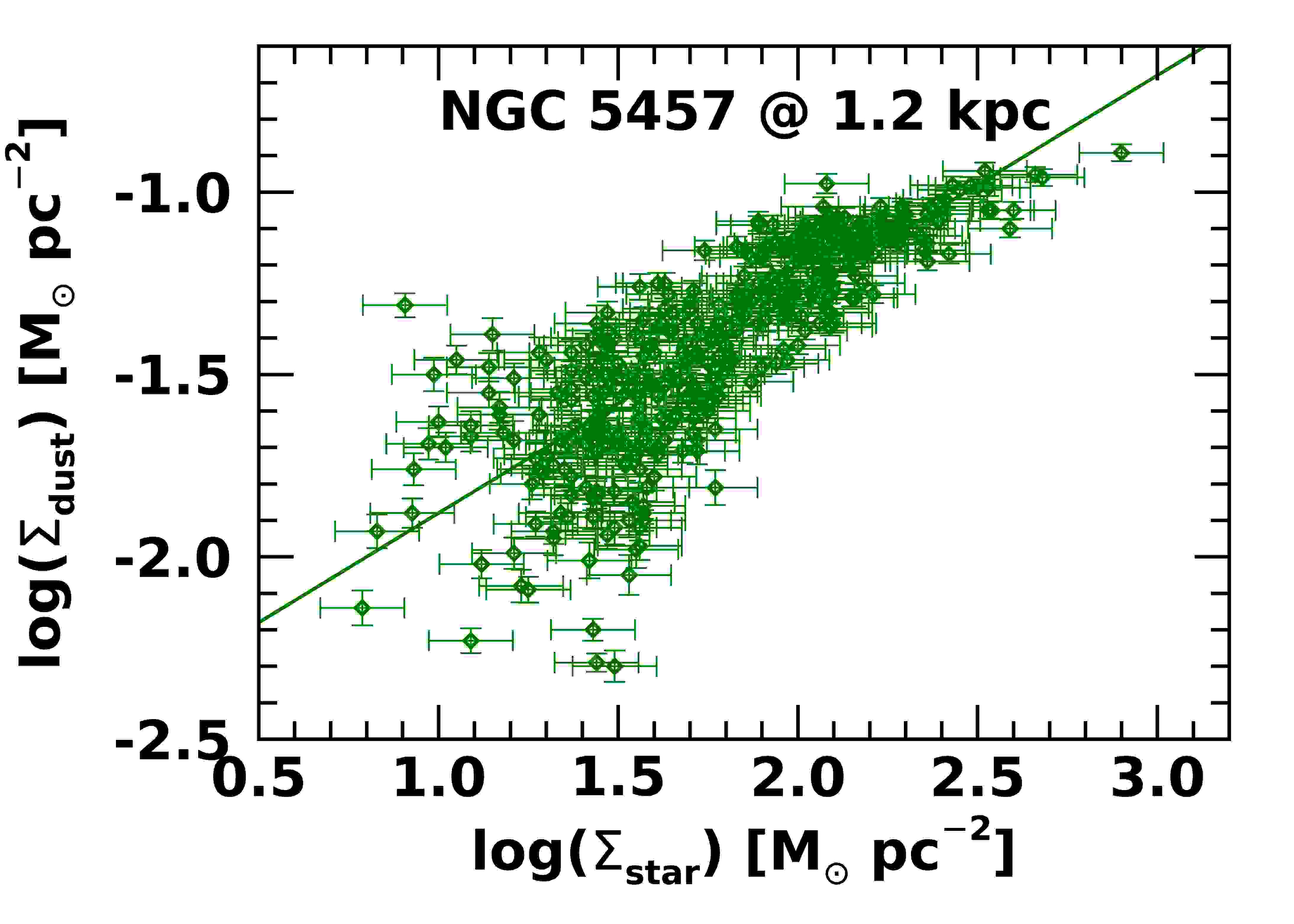}
\includegraphics[width=0.33\textwidth]{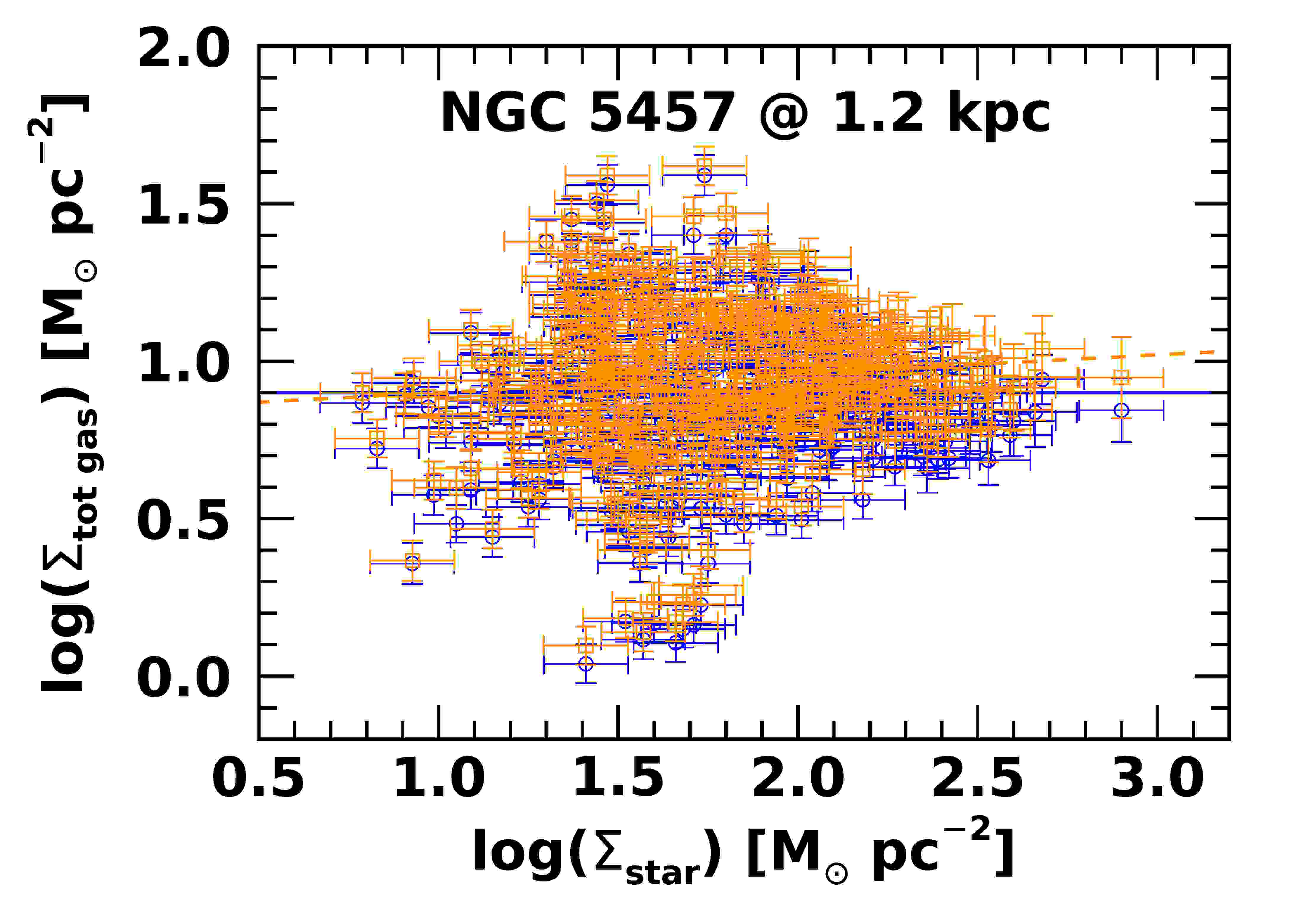}
\includegraphics[width=0.33\textwidth]{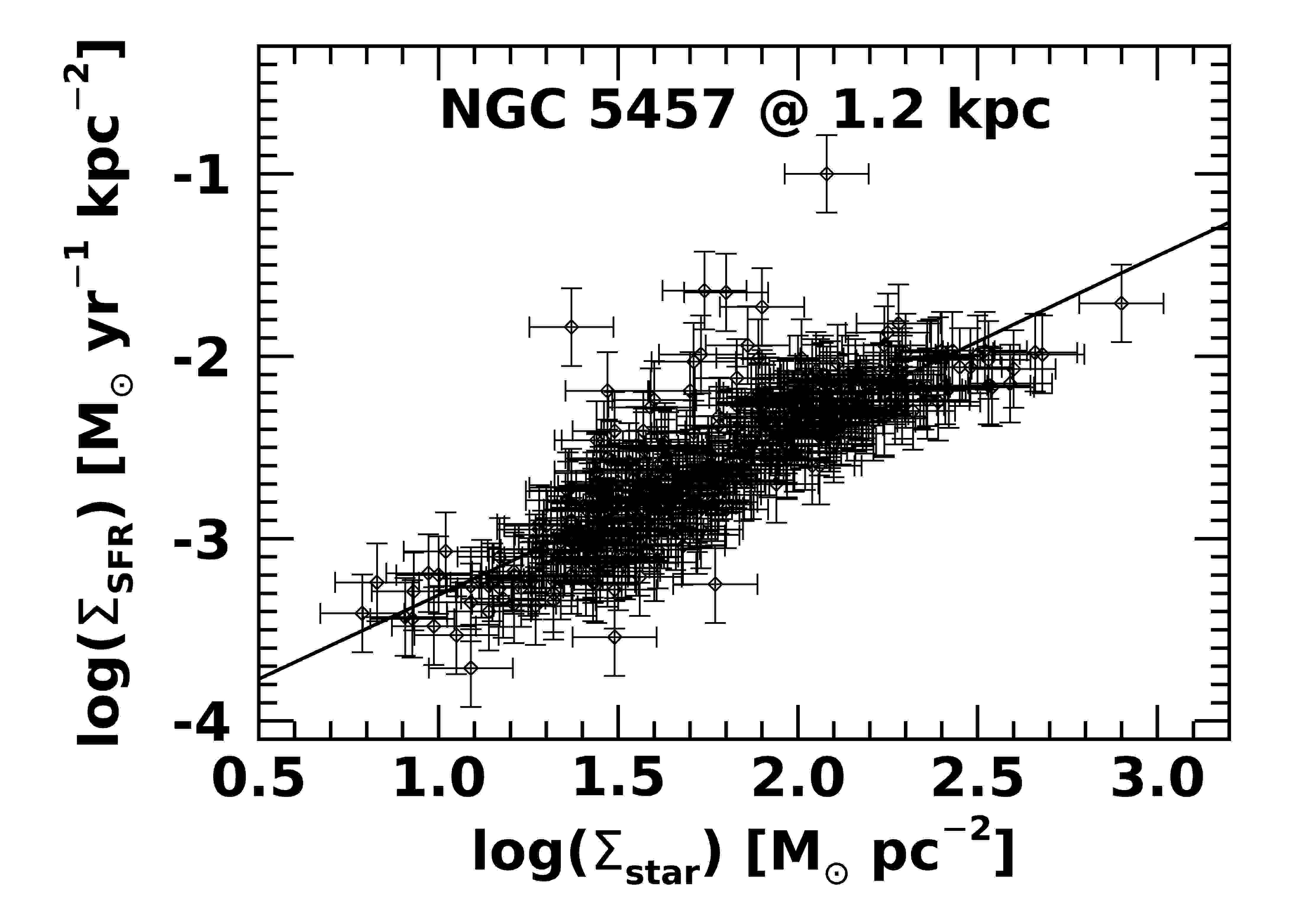}
\includegraphics[width=0.33\textwidth]{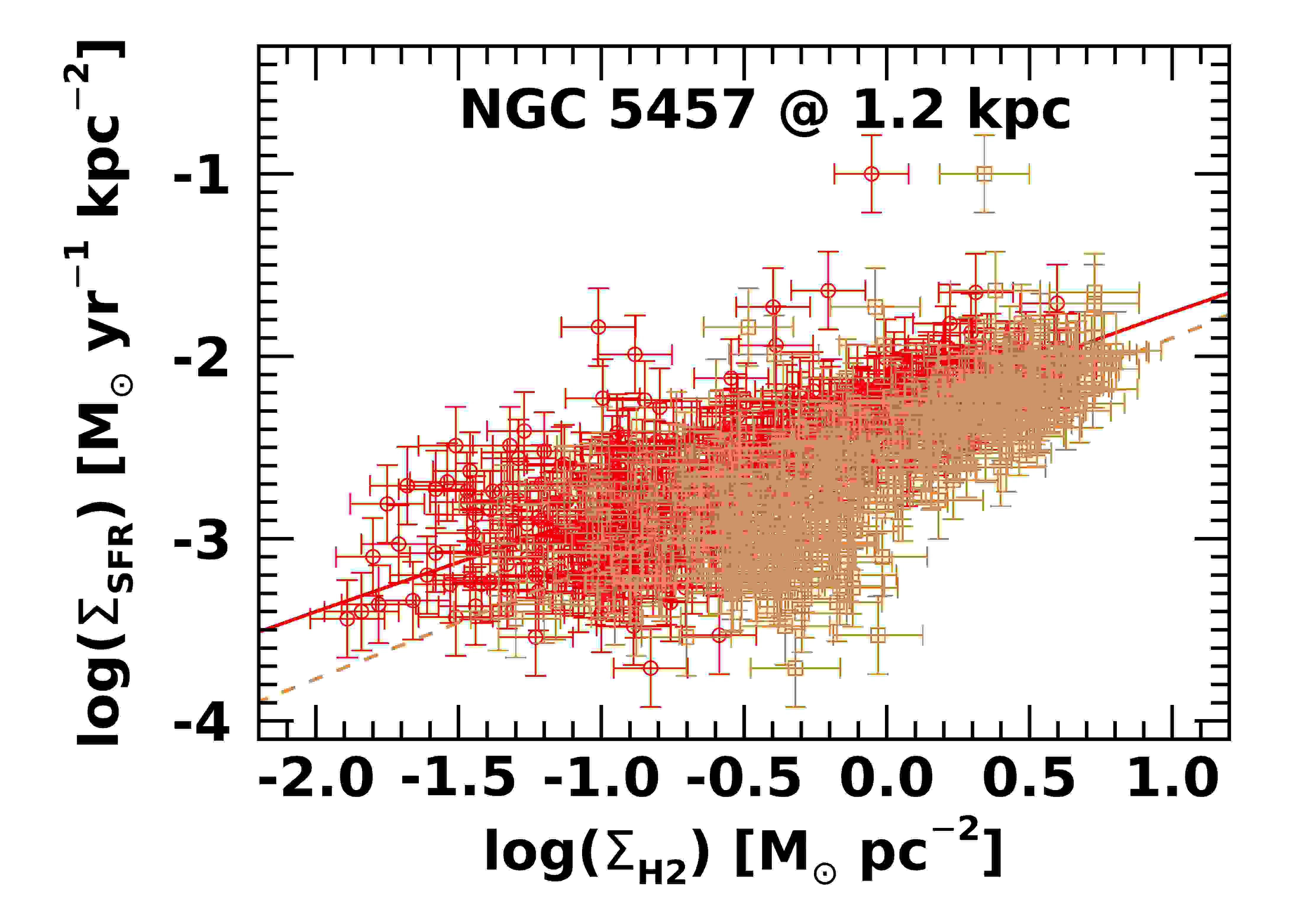}
\includegraphics[width=0.33\textwidth]{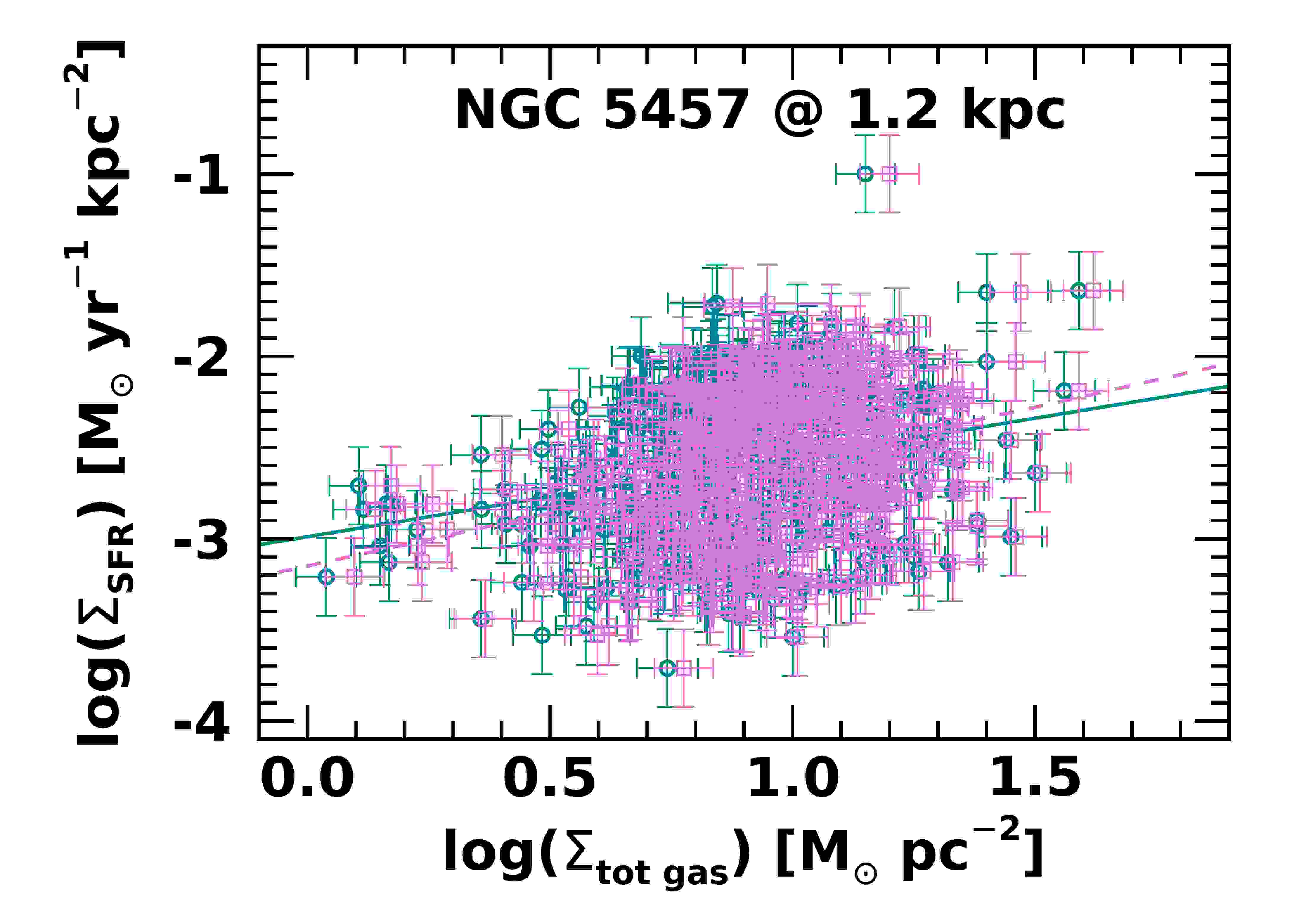}
\includegraphics[width=0.33\textwidth]{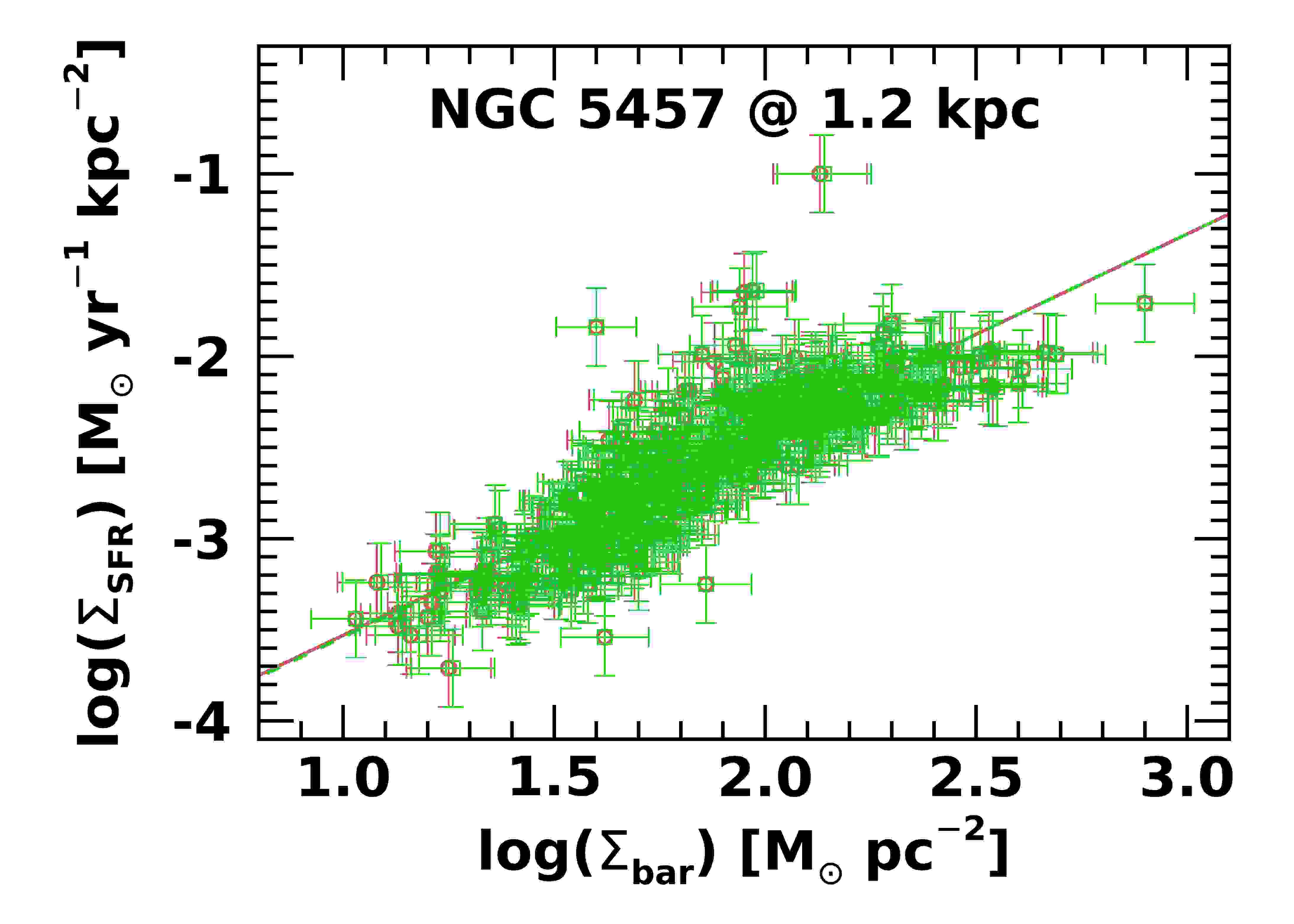}
\includegraphics[width=0.33\textwidth]{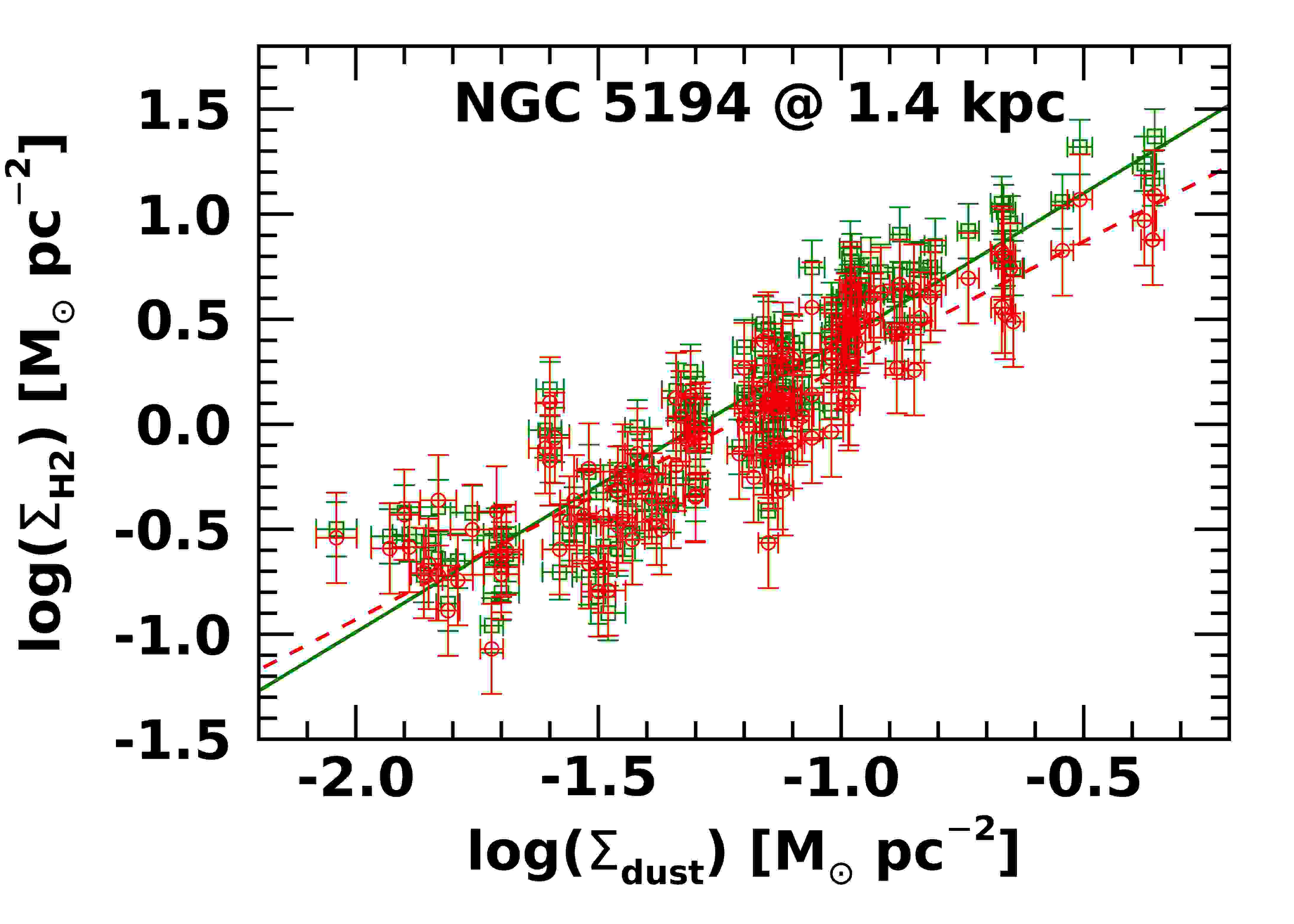}
\includegraphics[width=0.33\textwidth]{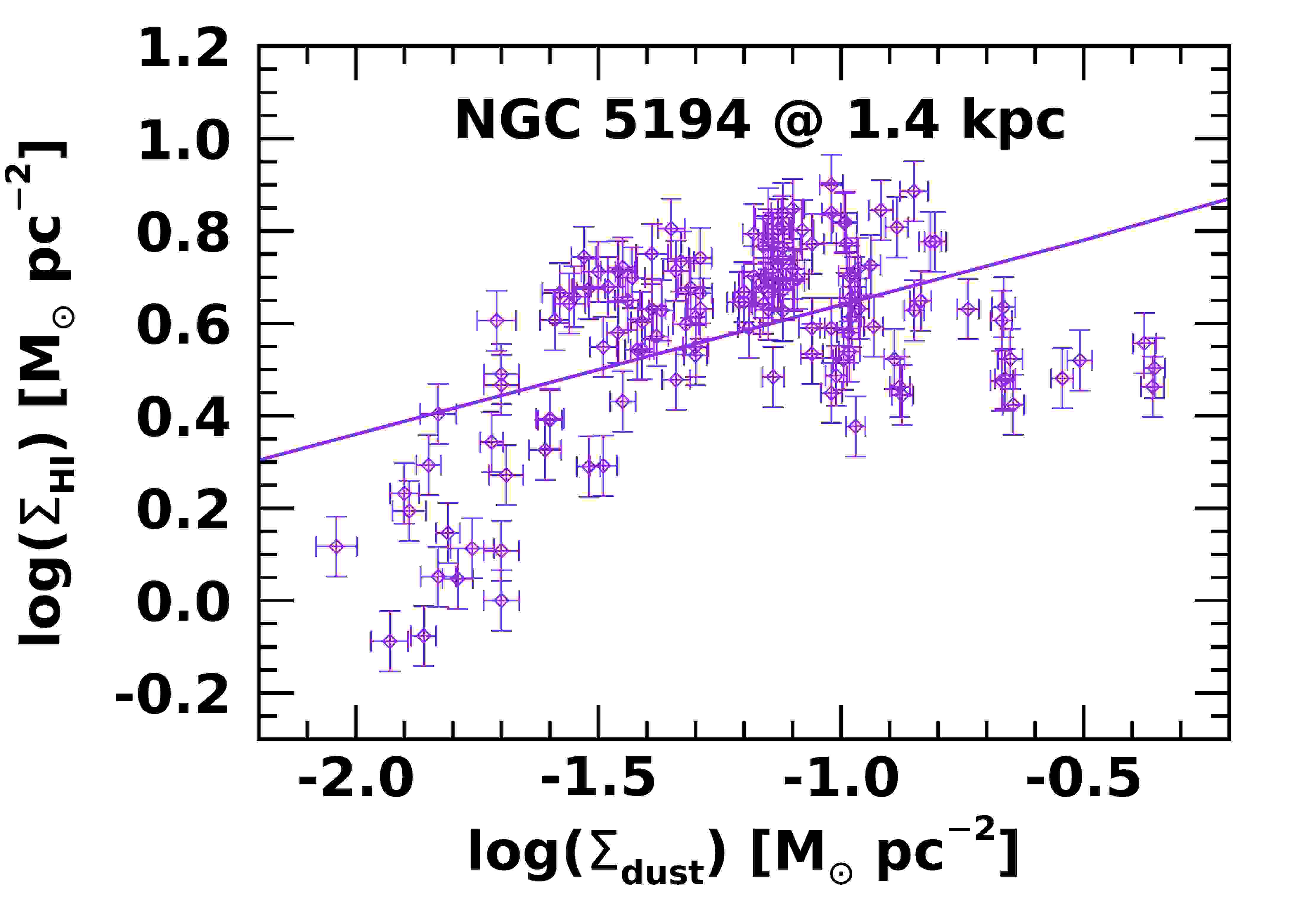}
\includegraphics[width=0.33\textwidth]{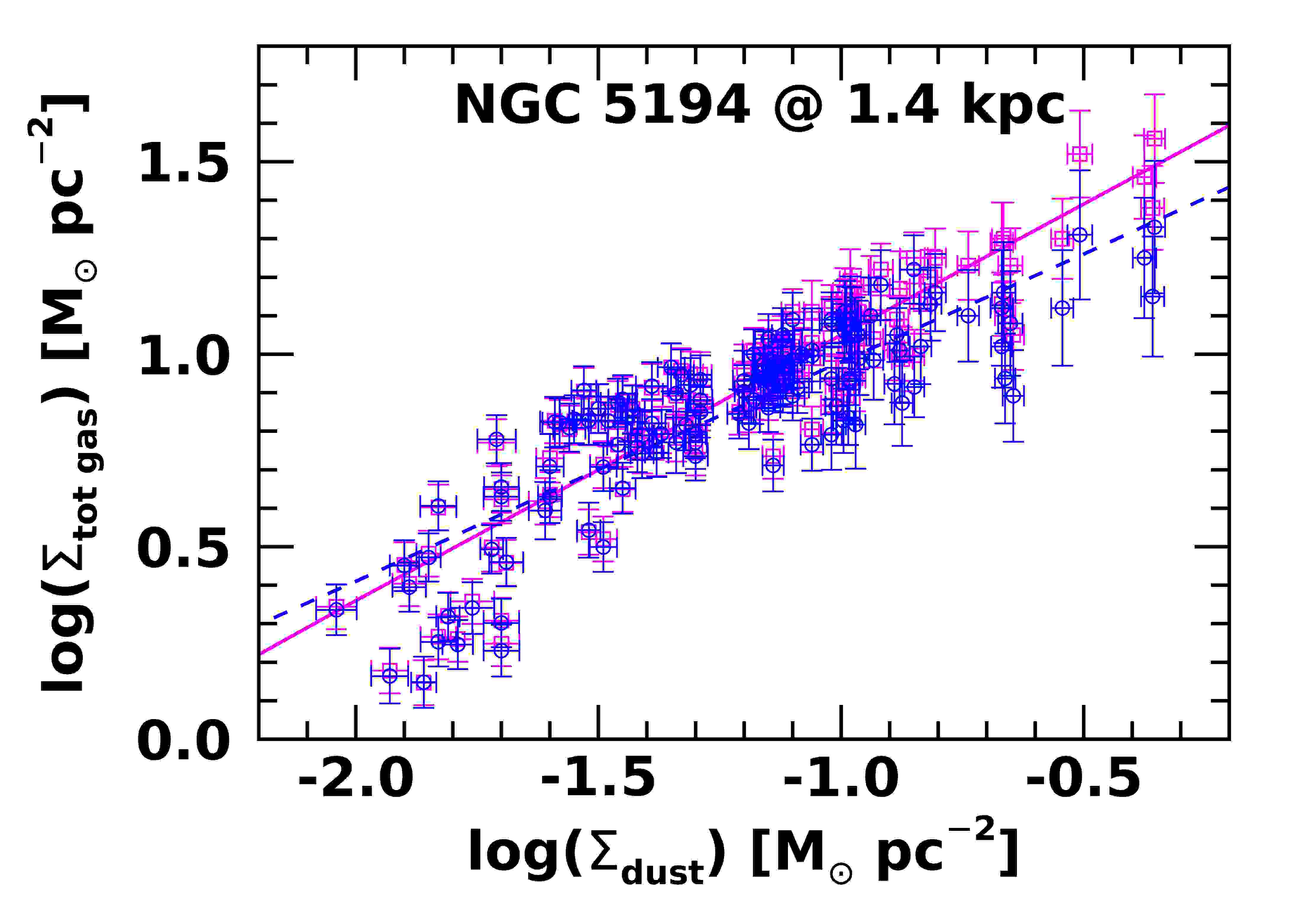}
\includegraphics[width=0.33\textwidth]{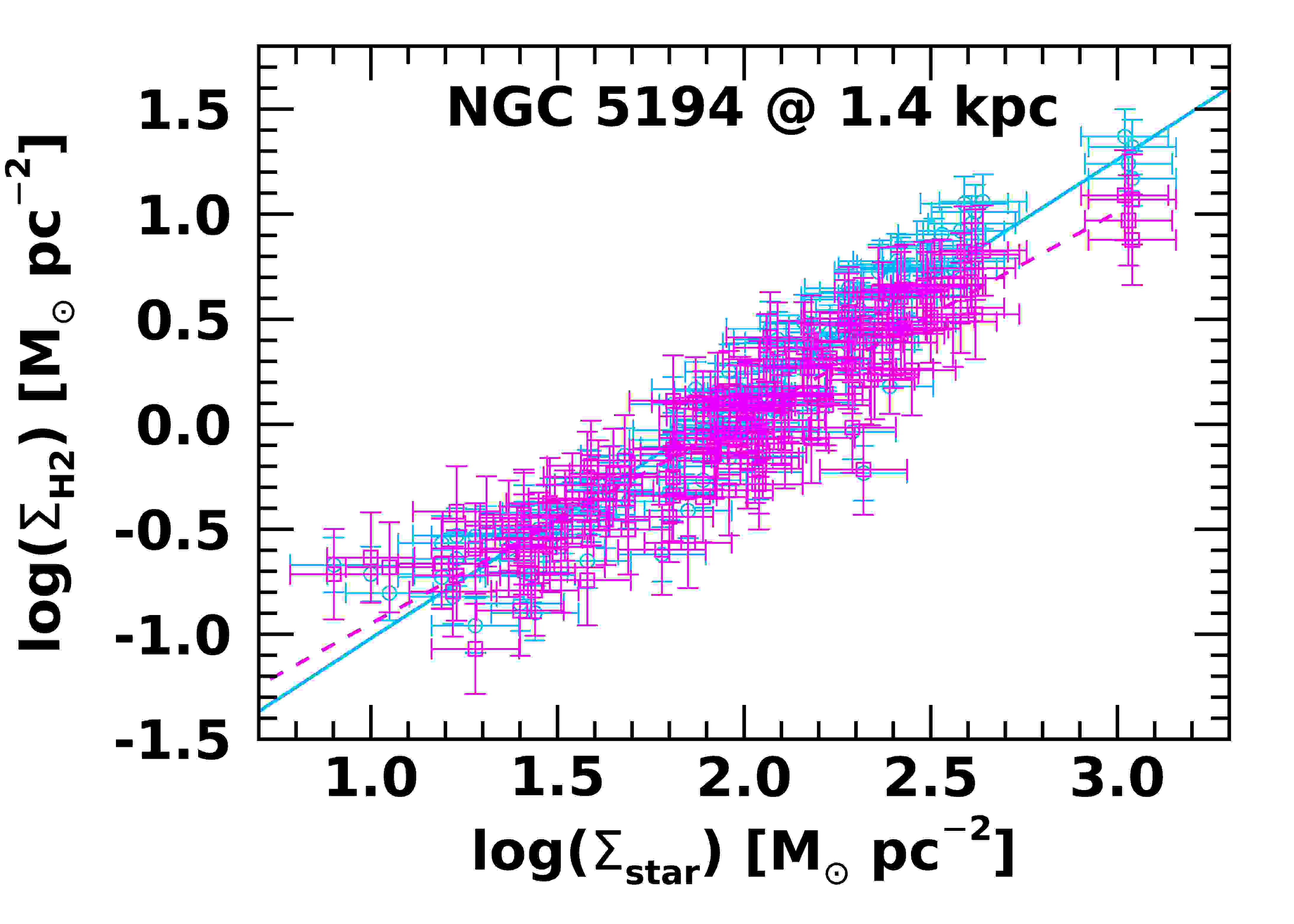}
\includegraphics[width=0.33\textwidth]{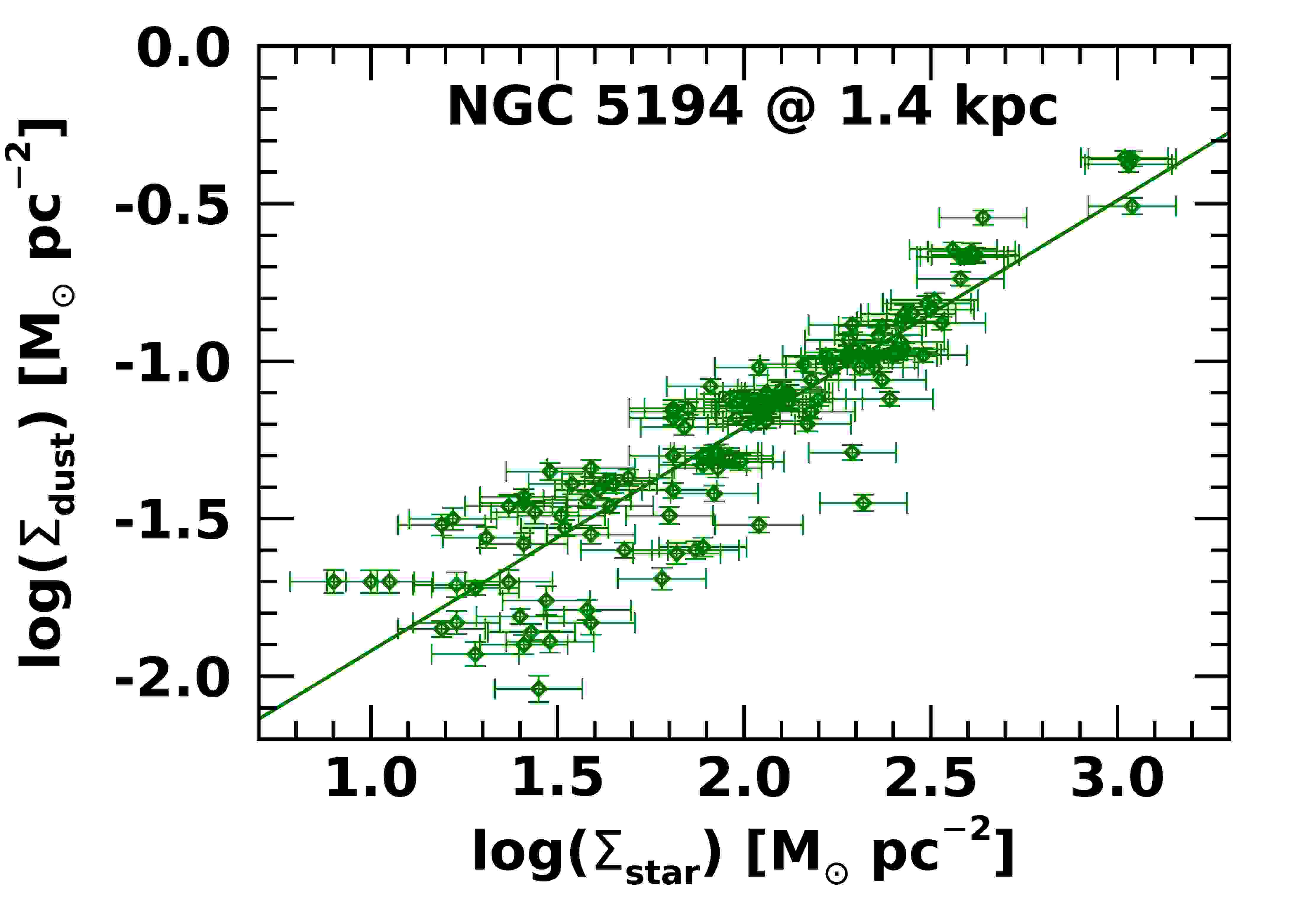}
\includegraphics[width=0.33\textwidth]{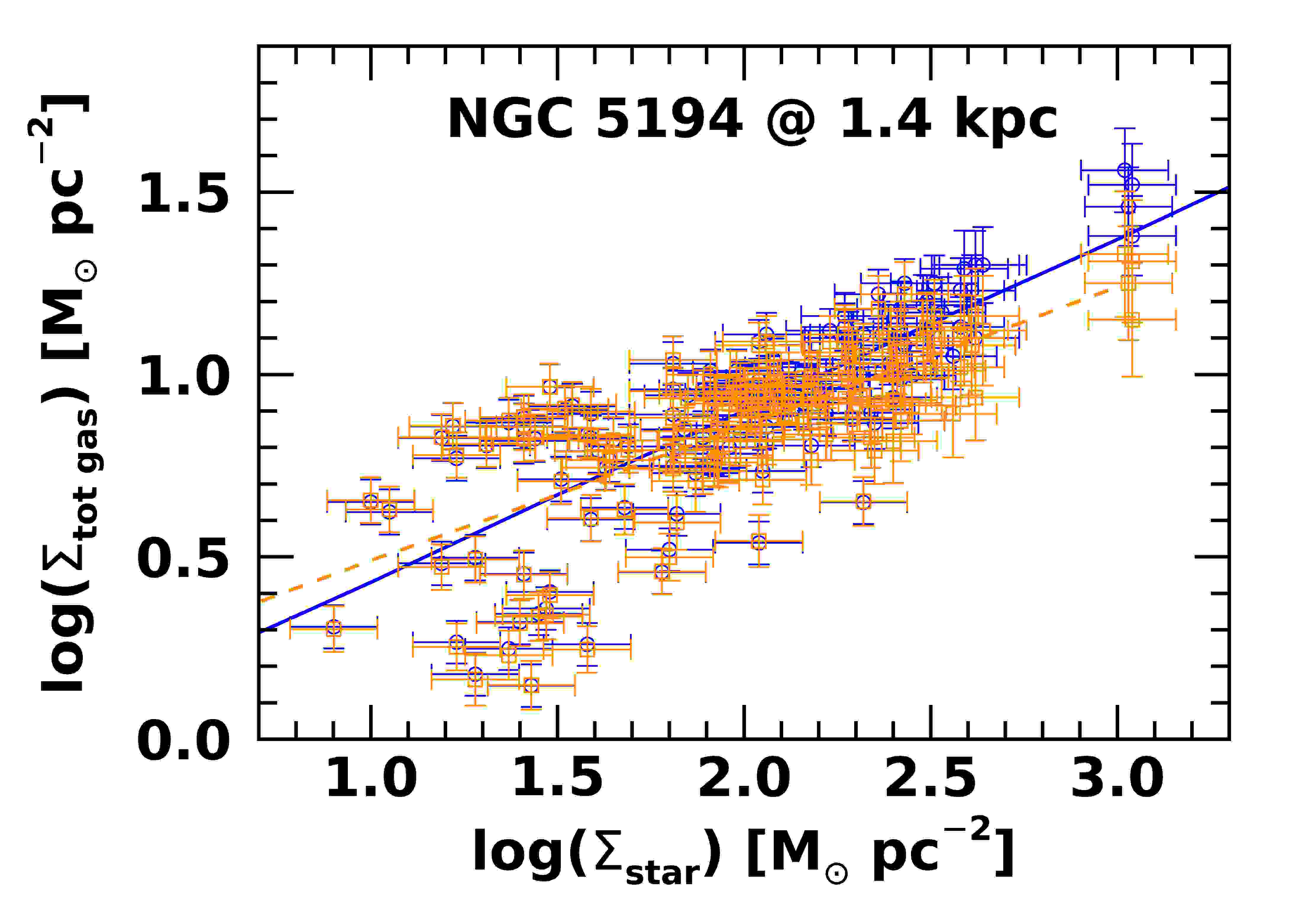}
\includegraphics[width=0.33\textwidth]{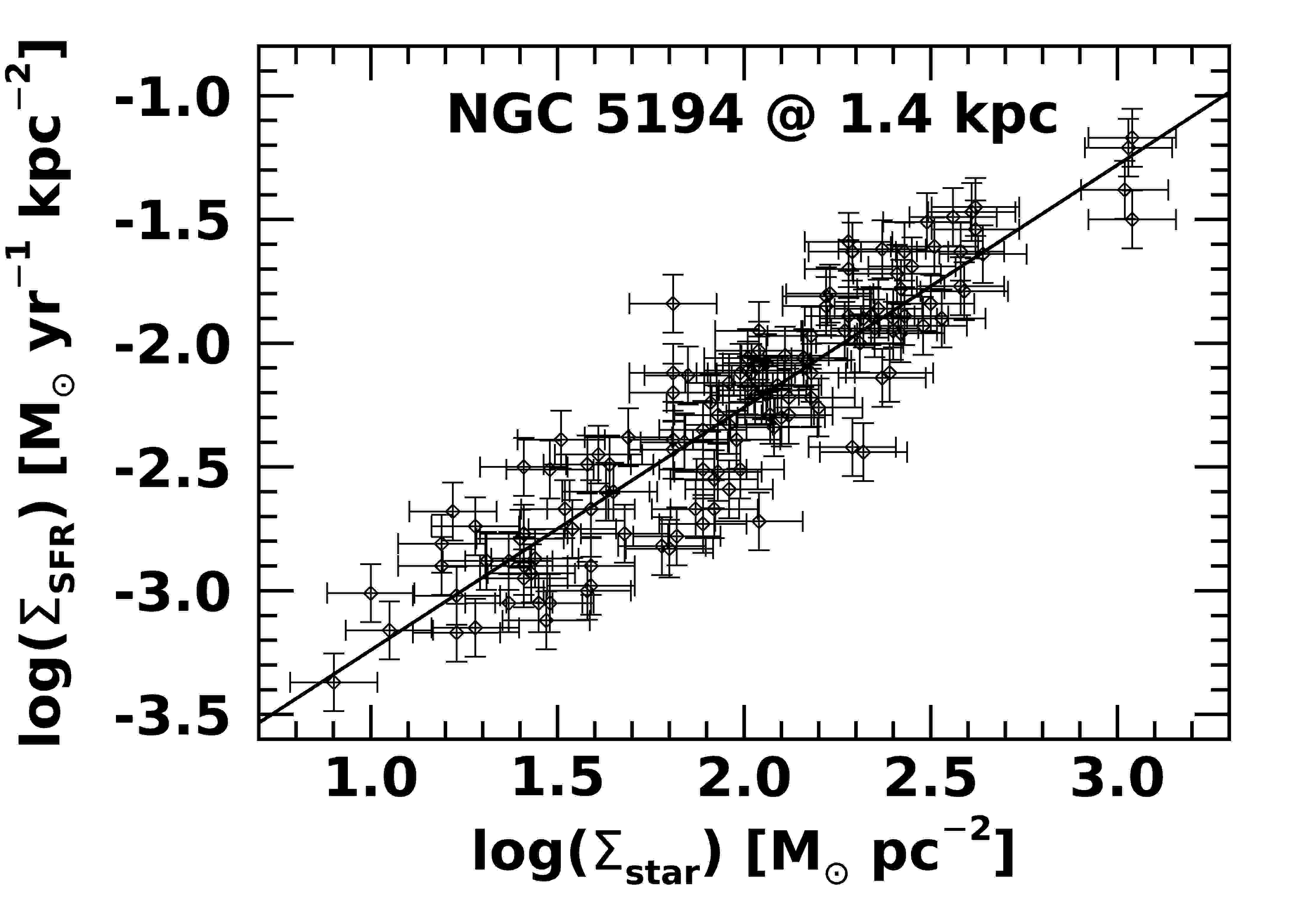}
\includegraphics[width=0.33\textwidth]{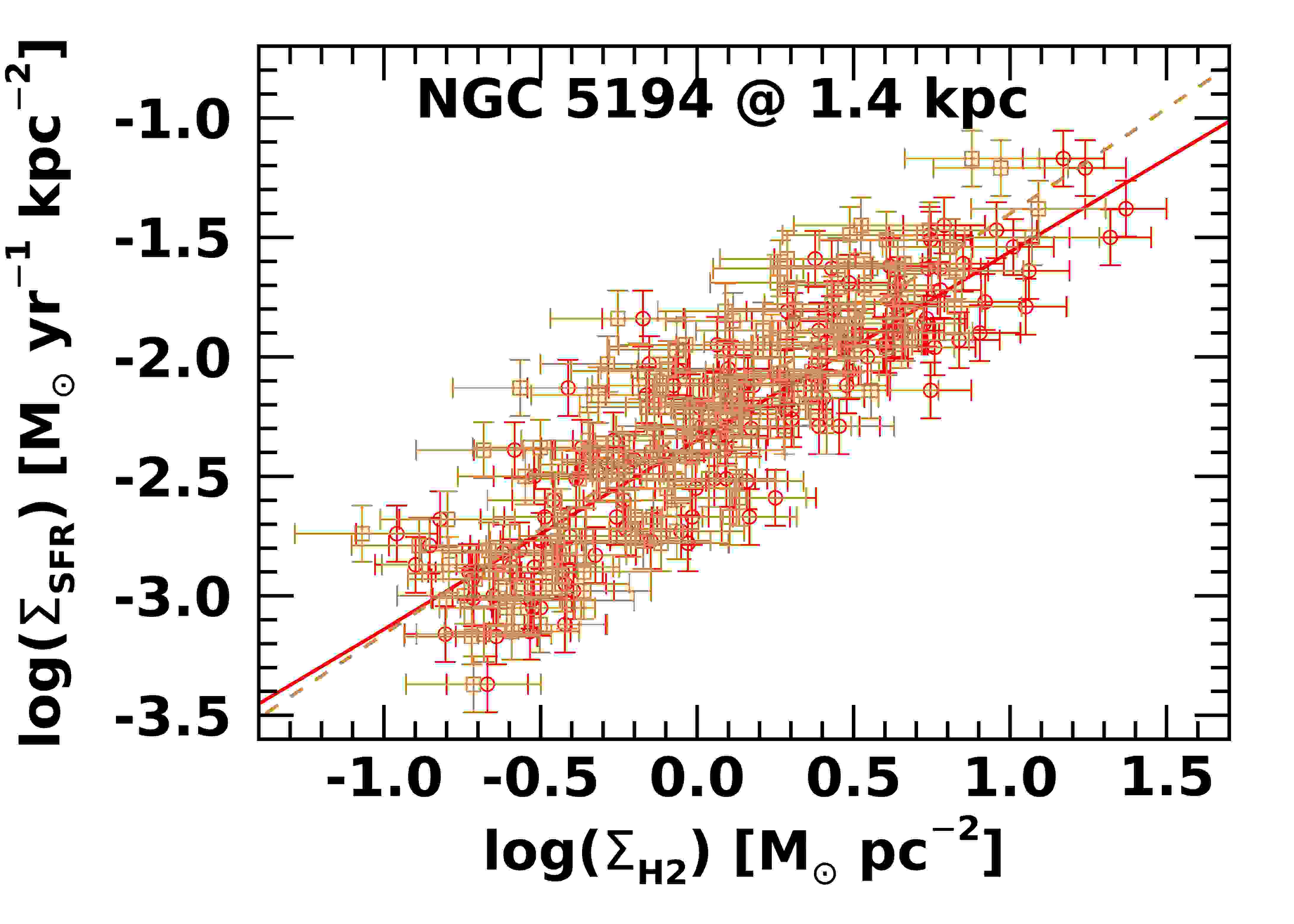}
\caption*{Figure~\ref{fig:add-ism} continued}
\end{figure*}

\begin{figure*}
\centering
\includegraphics[width=0.33\textwidth]{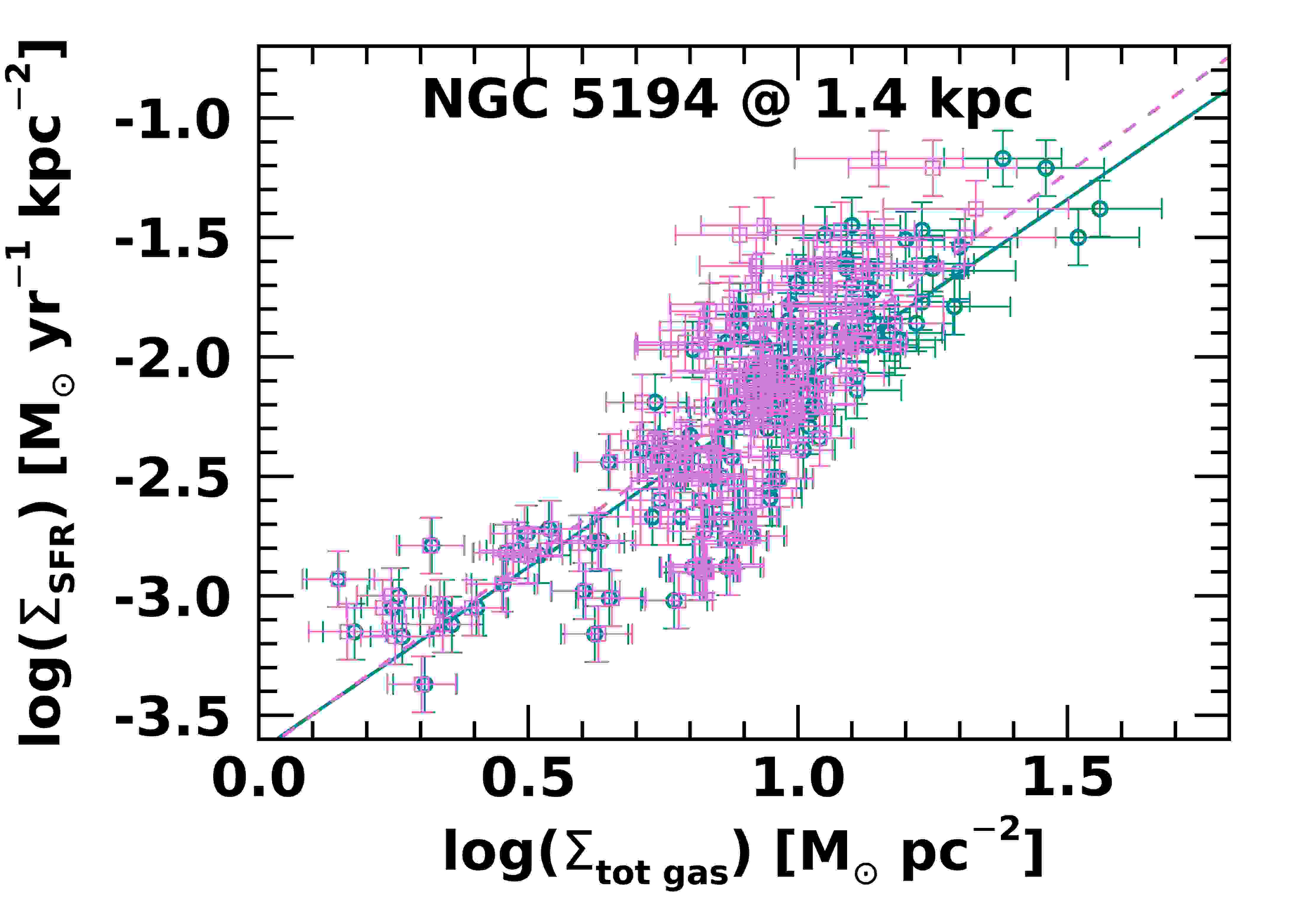}
\includegraphics[width=0.33\textwidth]{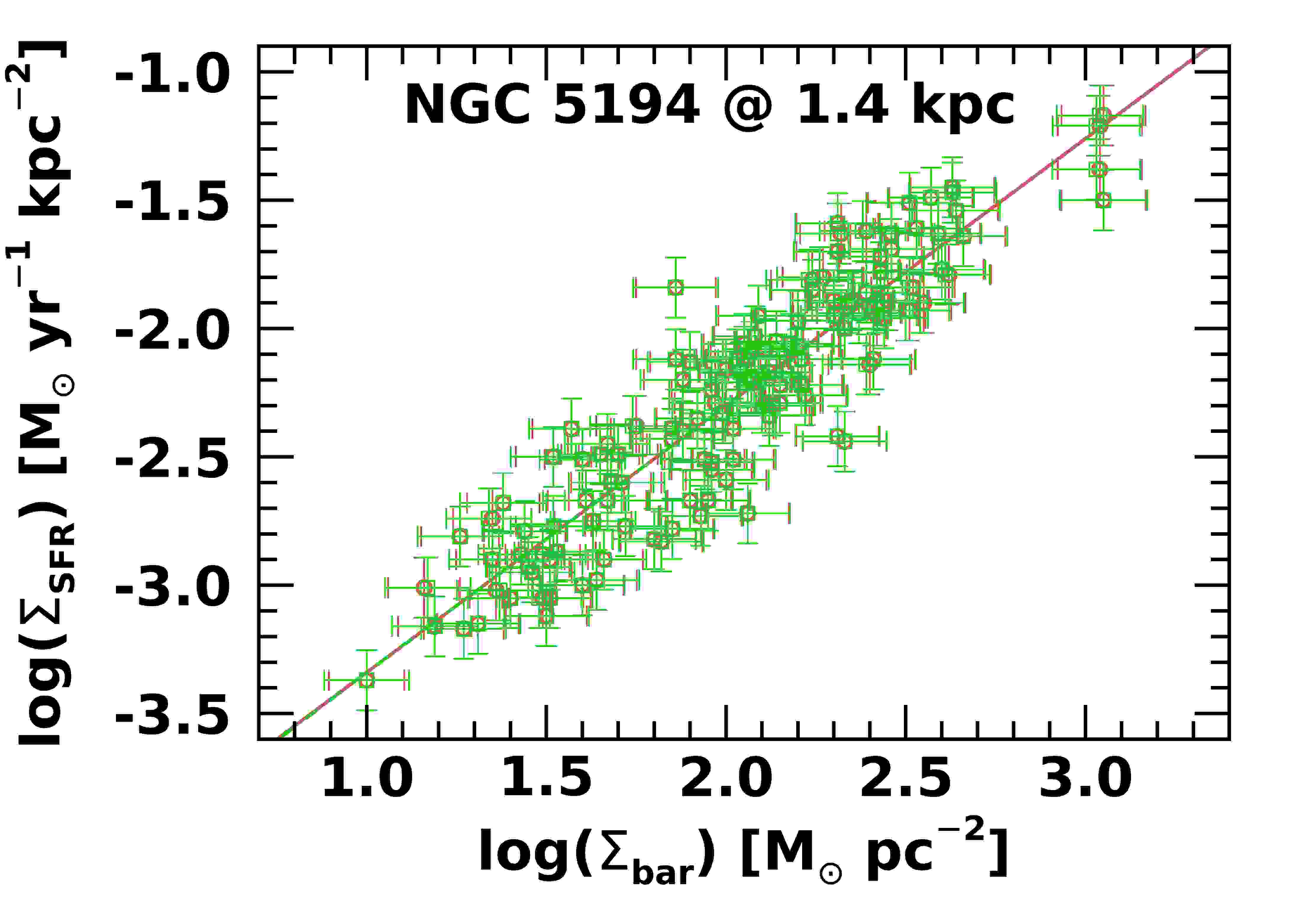}
\includegraphics[width=0.33\textwidth]{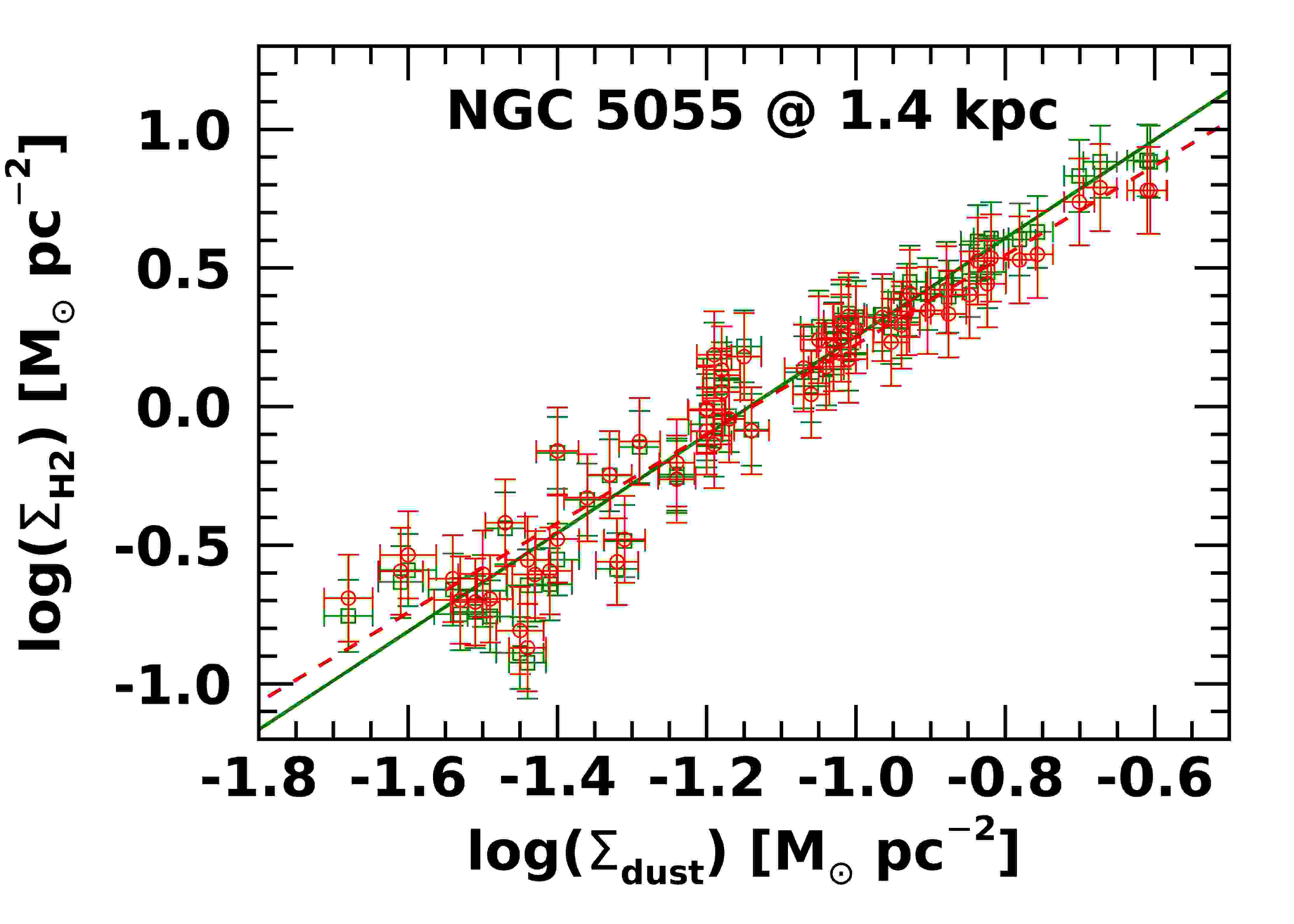}
\includegraphics[width=0.33\textwidth]{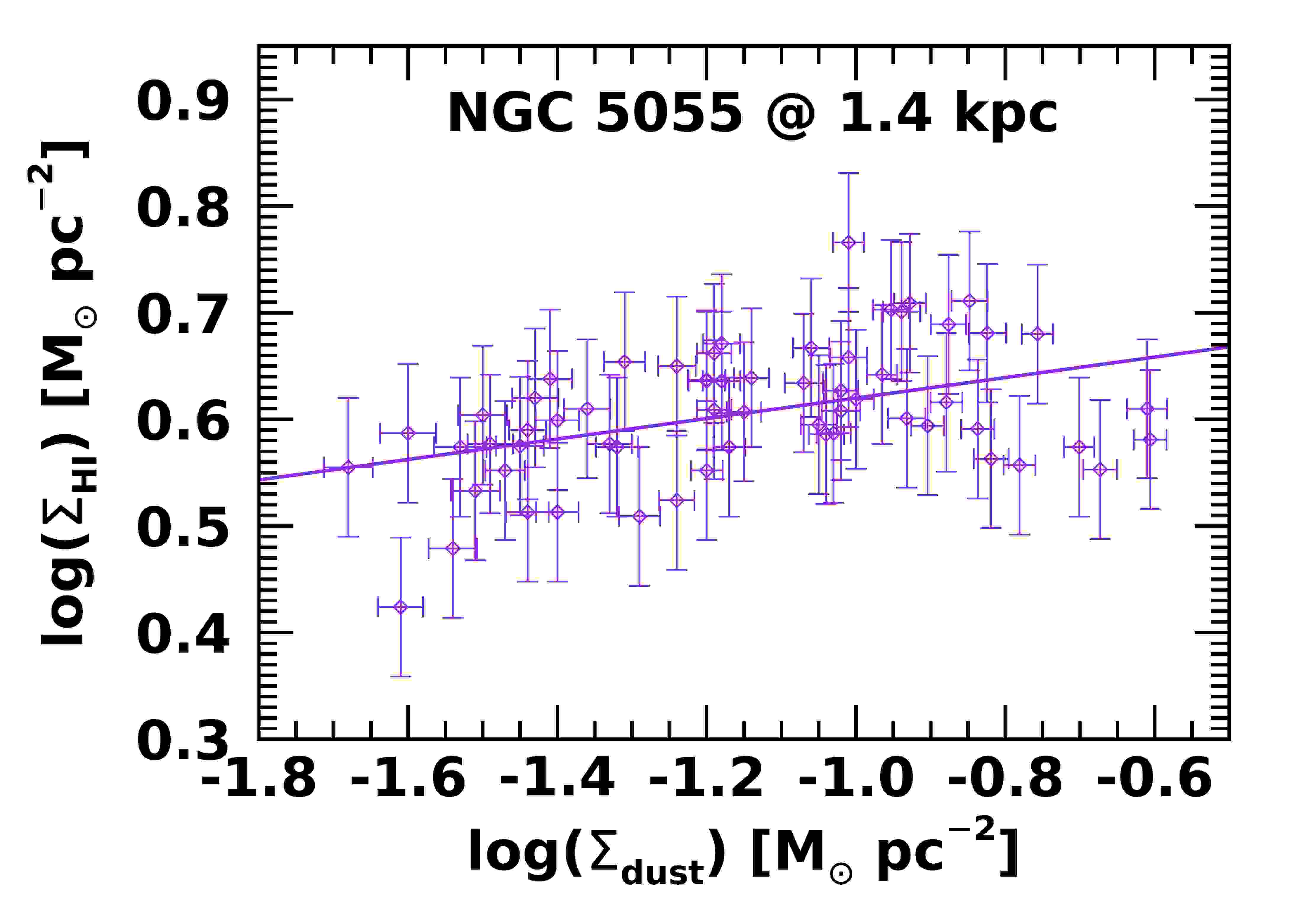}
\includegraphics[width=0.33\textwidth]{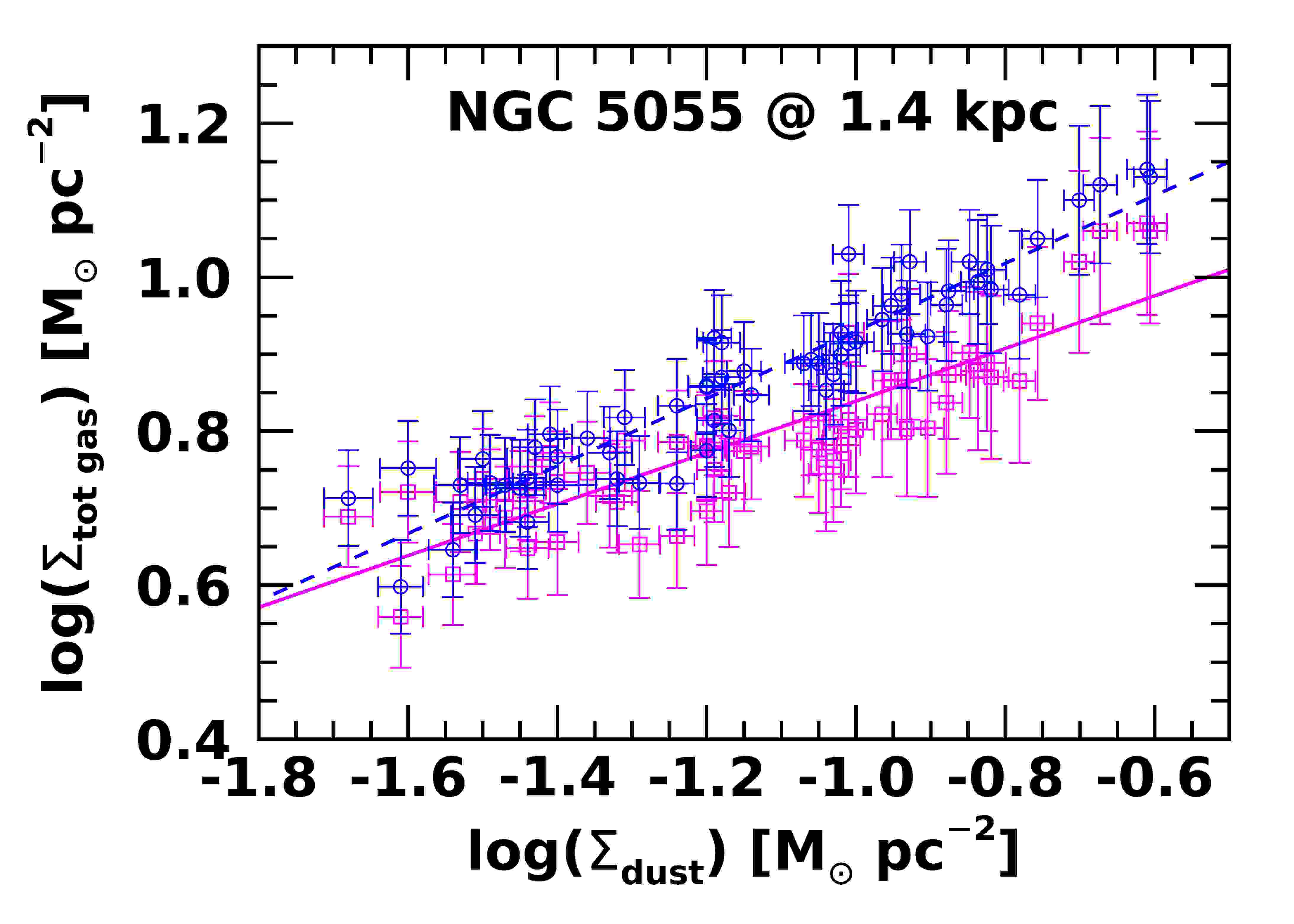}
\includegraphics[width=0.33\textwidth]{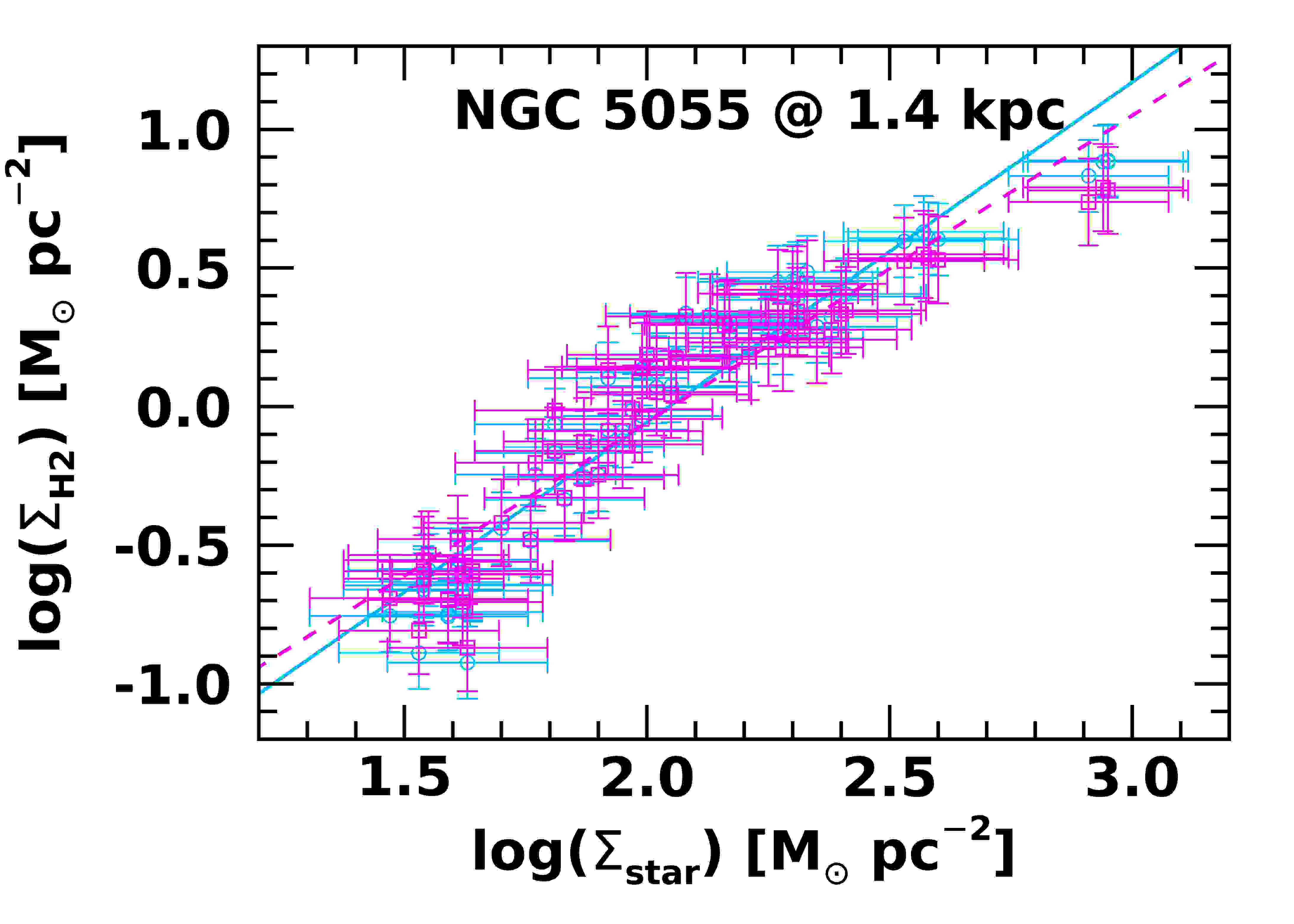}
\includegraphics[width=0.33\textwidth]{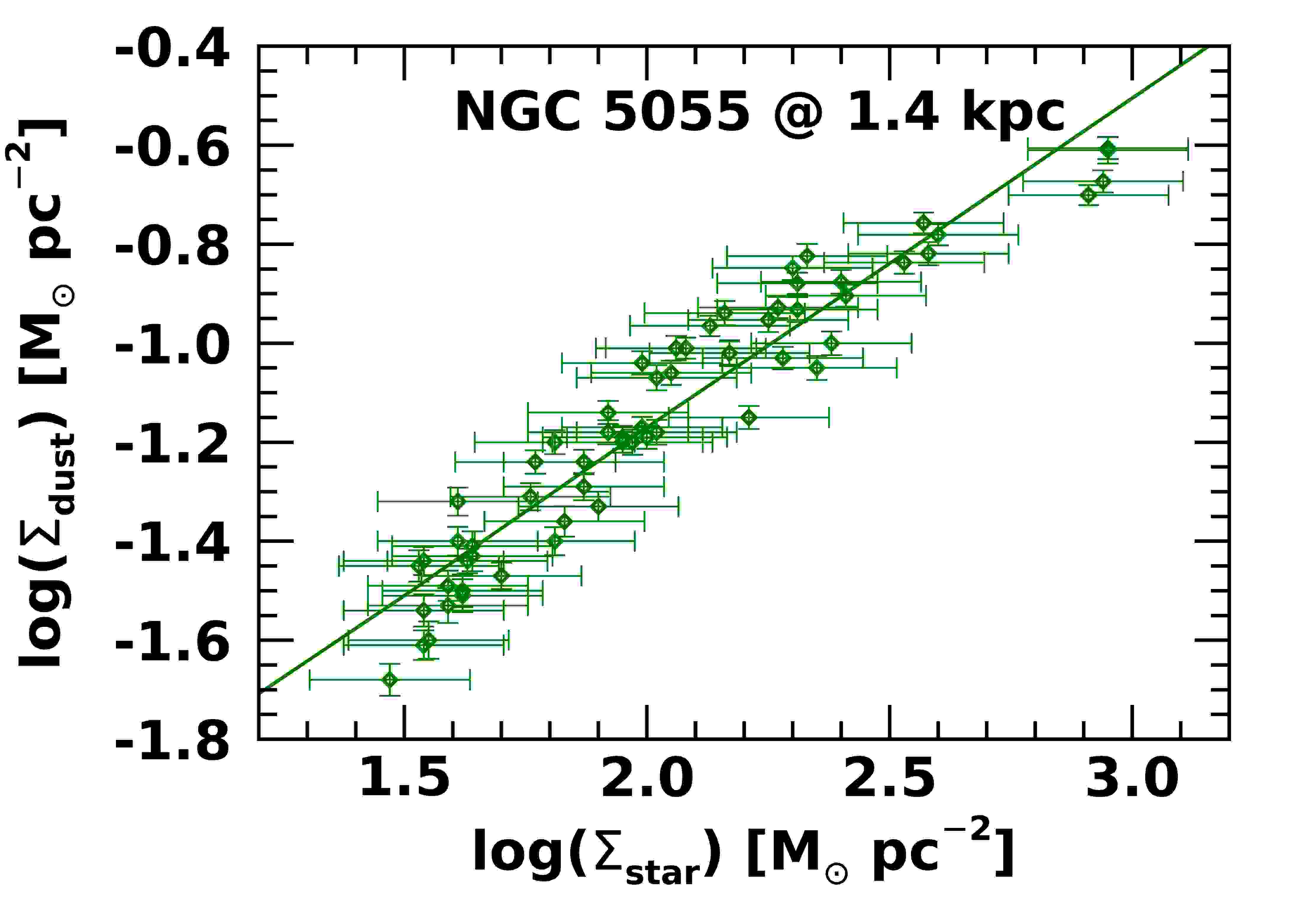}
\includegraphics[width=0.33\textwidth]{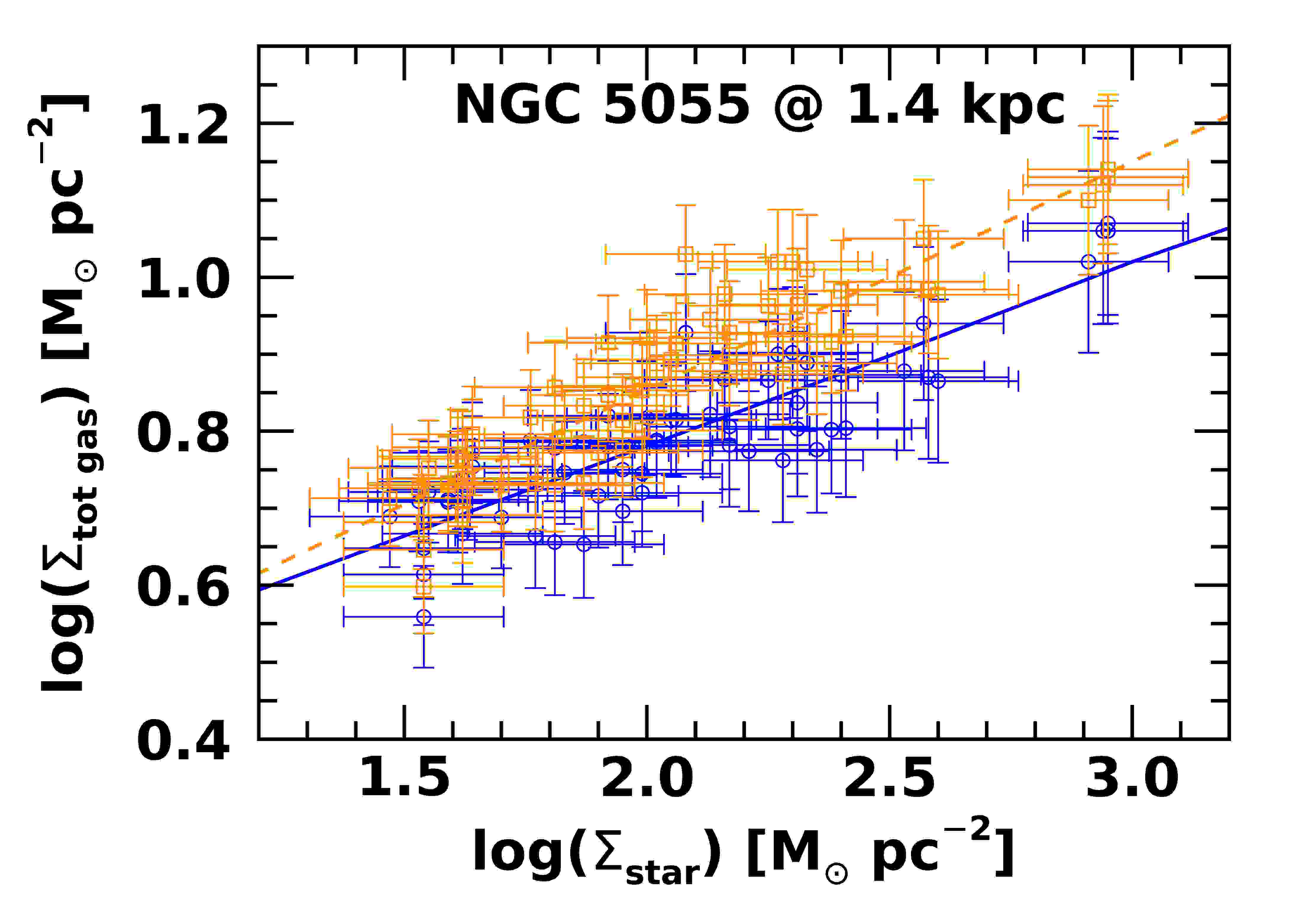}
\includegraphics[width=0.33\textwidth]{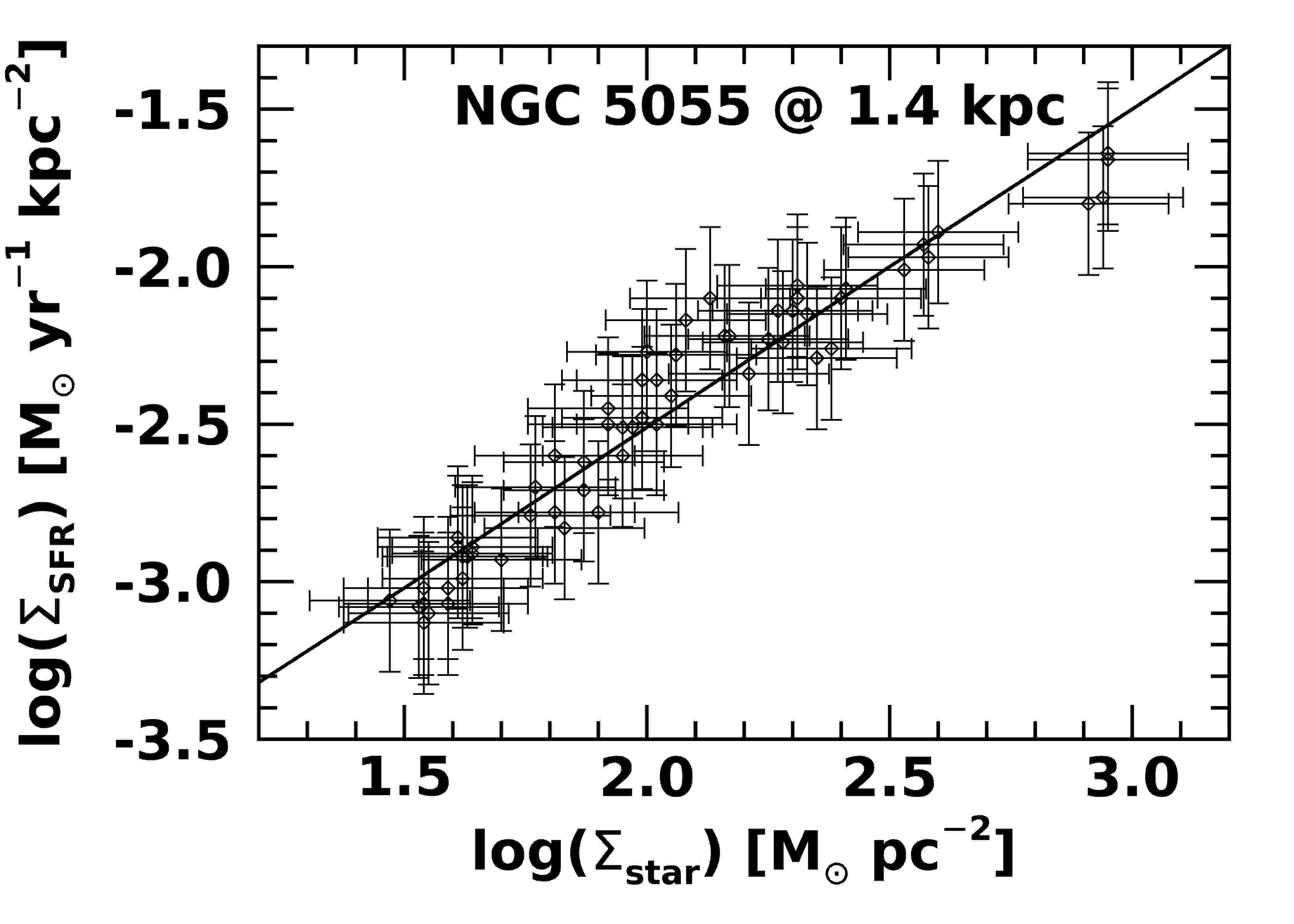}
\includegraphics[width=0.33\textwidth]{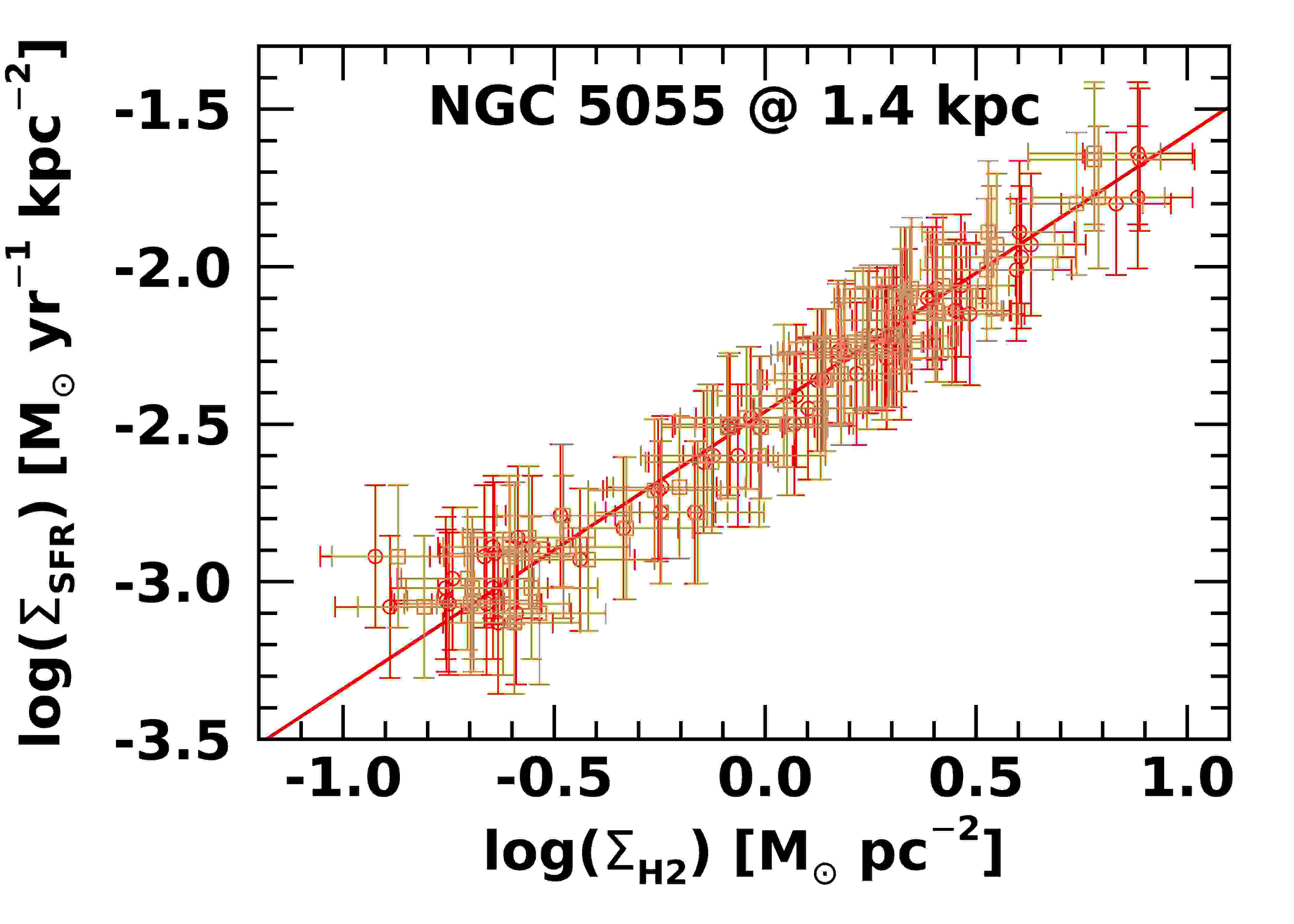}
\includegraphics[width=0.33\textwidth]{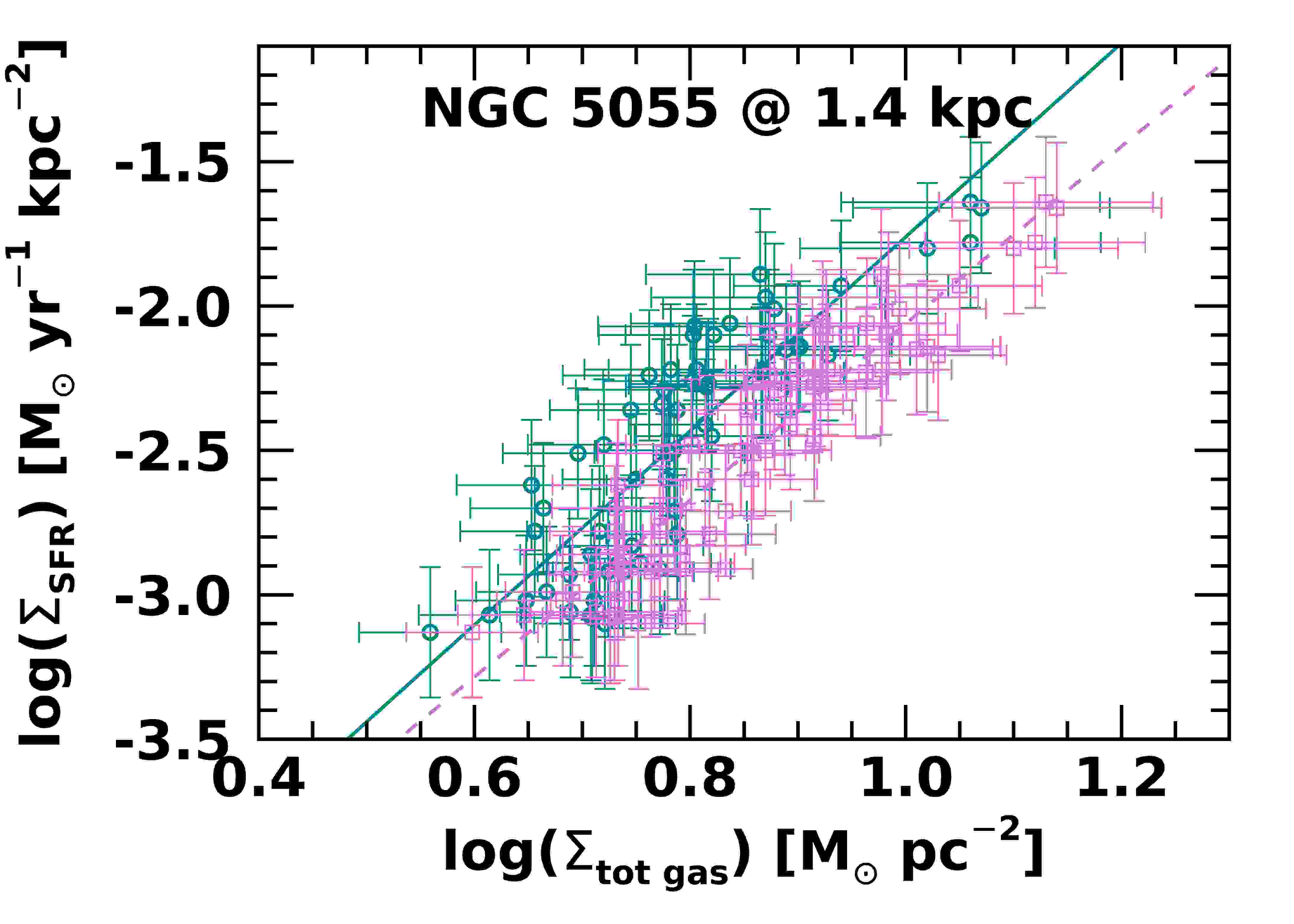}
\includegraphics[width=0.33\textwidth]{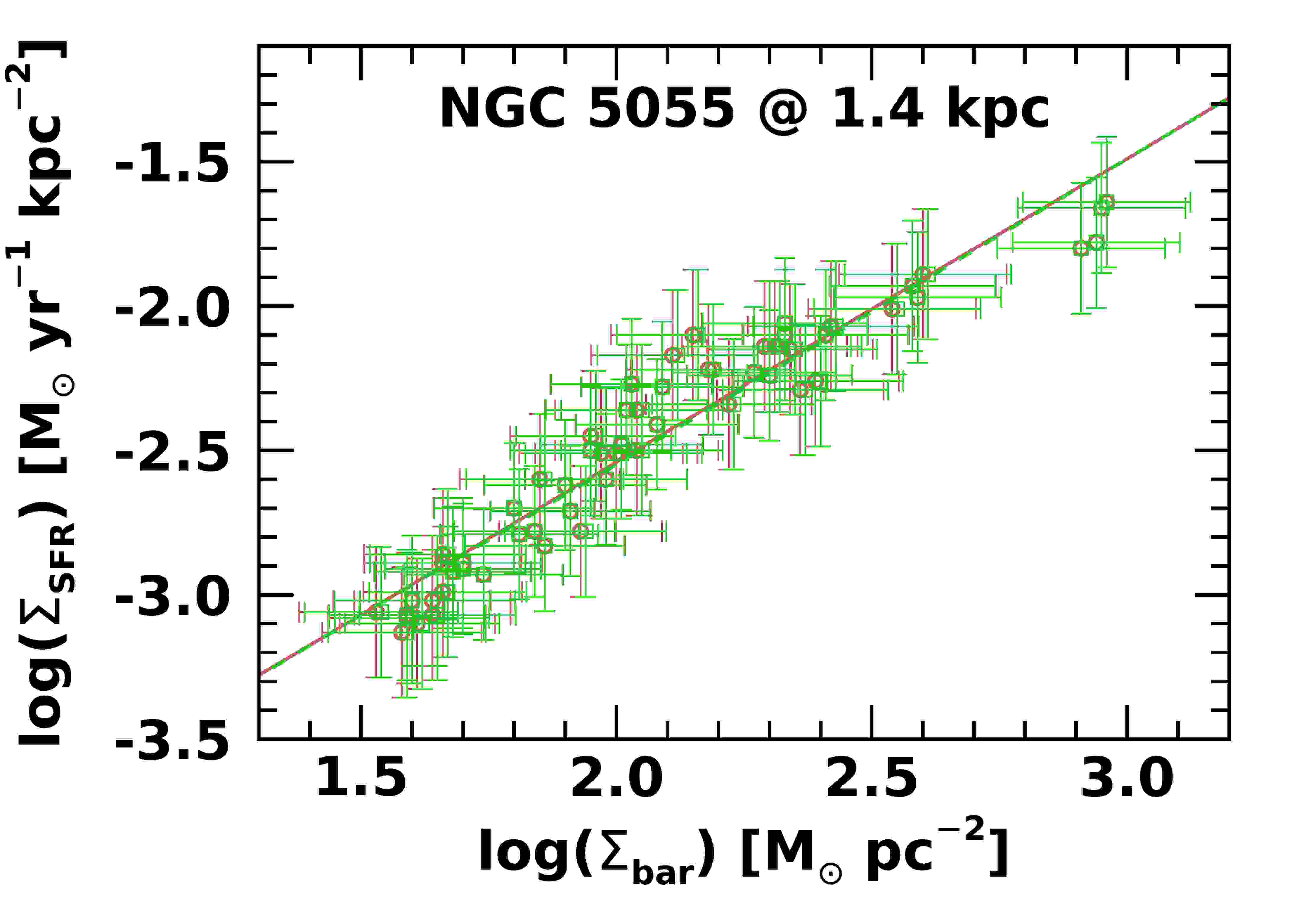}
\includegraphics[width=0.33\textwidth]{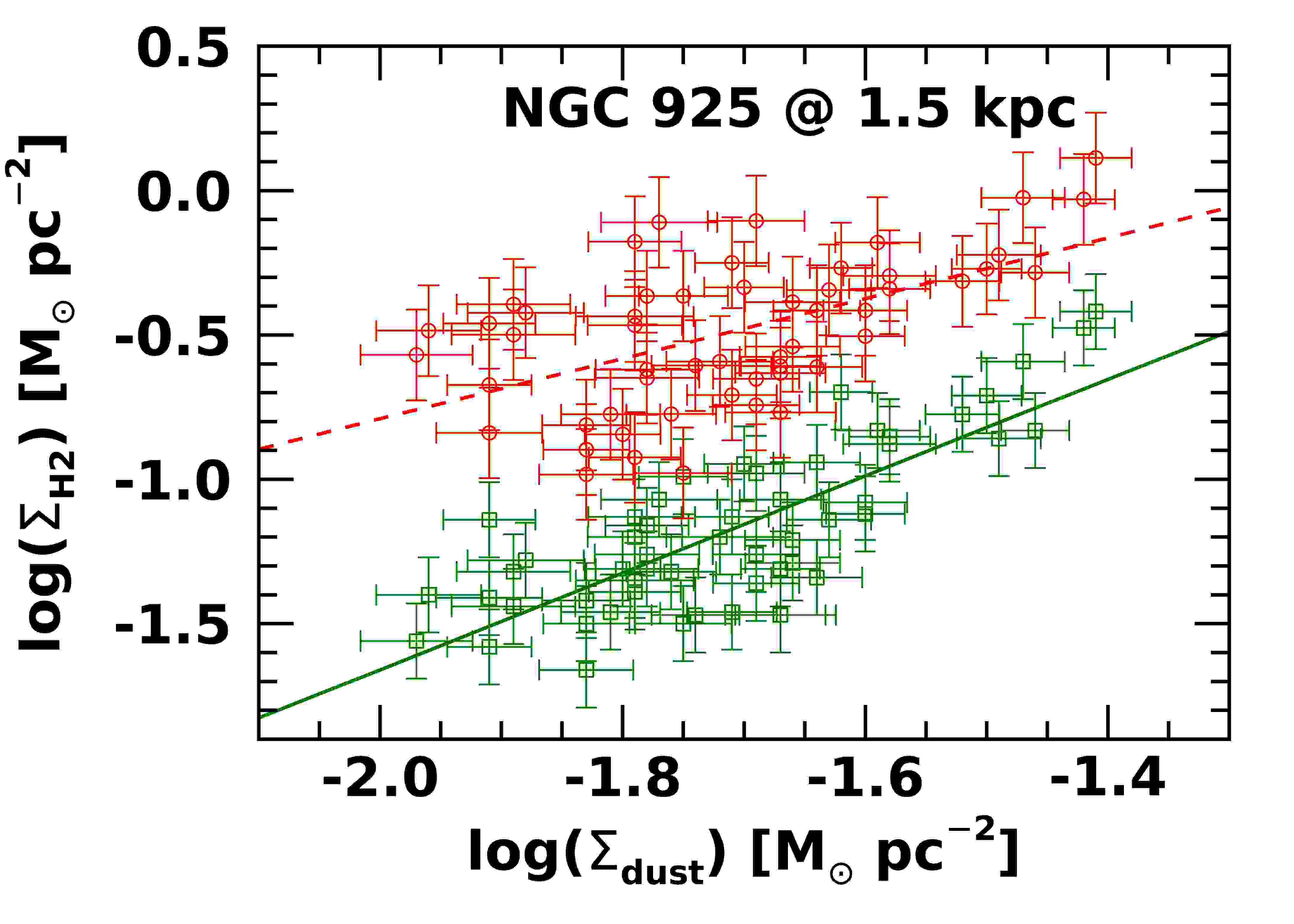}
\includegraphics[width=0.33\textwidth]{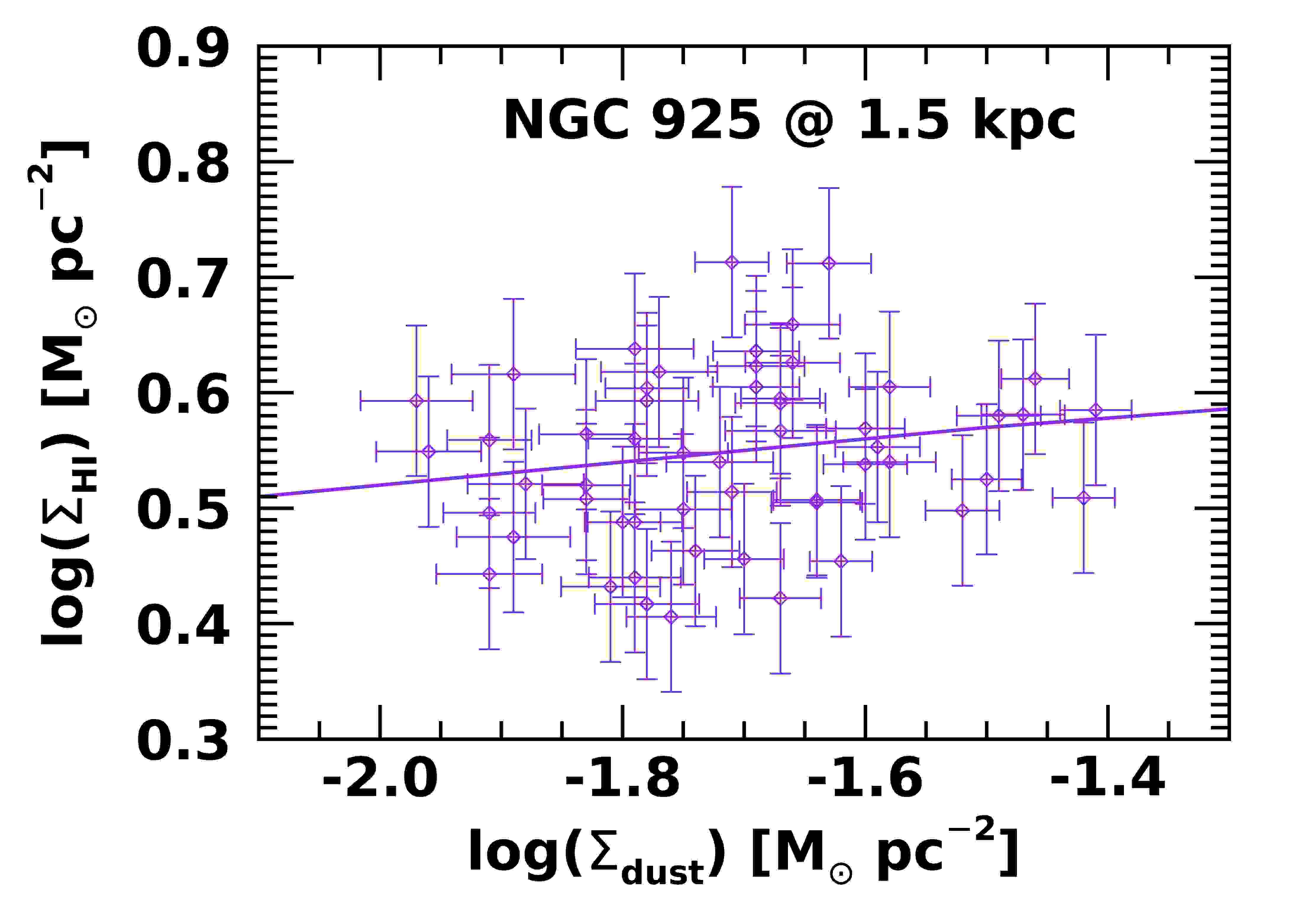}
\includegraphics[width=0.33\textwidth]{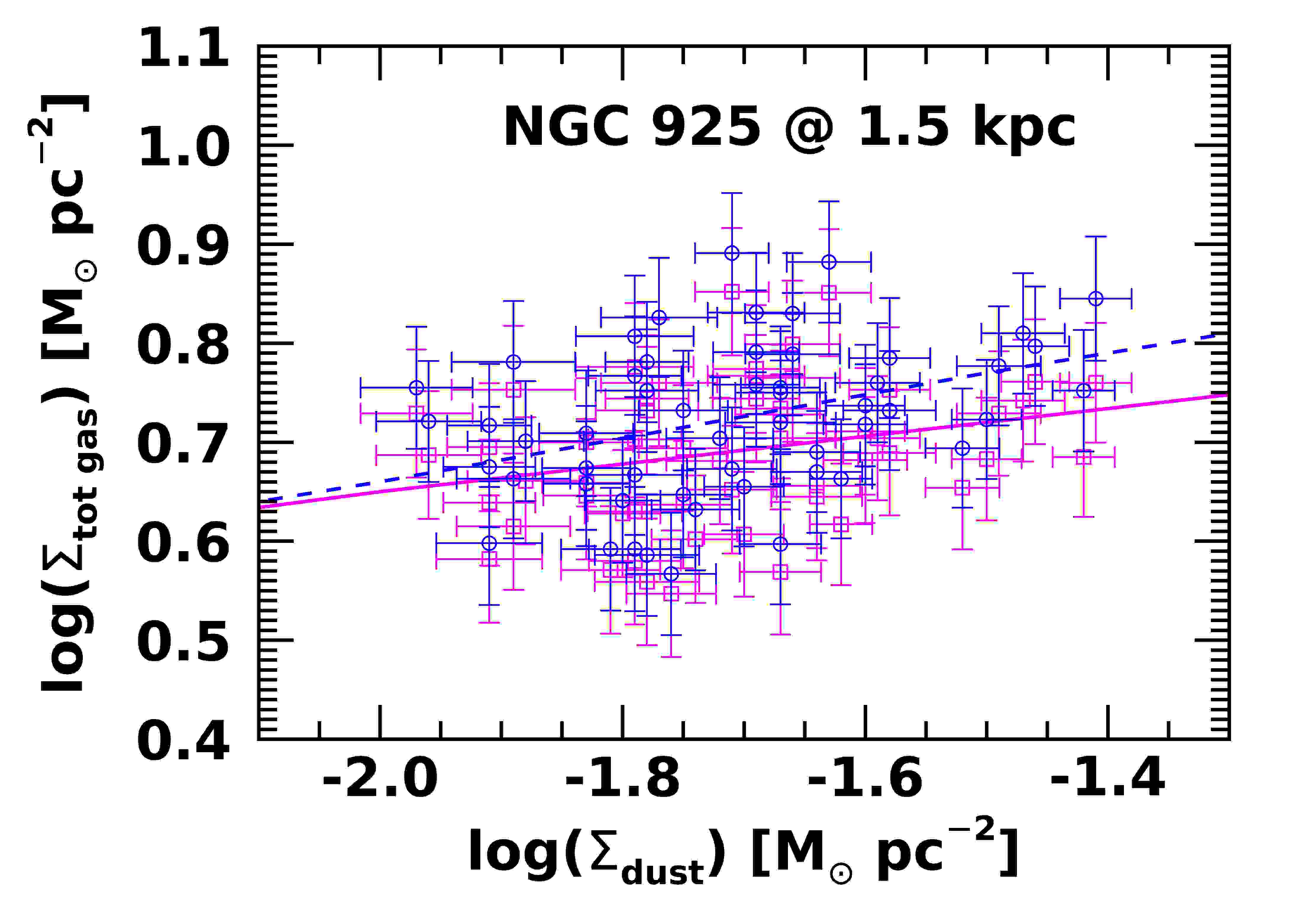}
\caption*{Figure~\ref{fig:add-ism} continued}
\end{figure*}

\begin{figure*}
\centering
\includegraphics[width=0.33\textwidth]{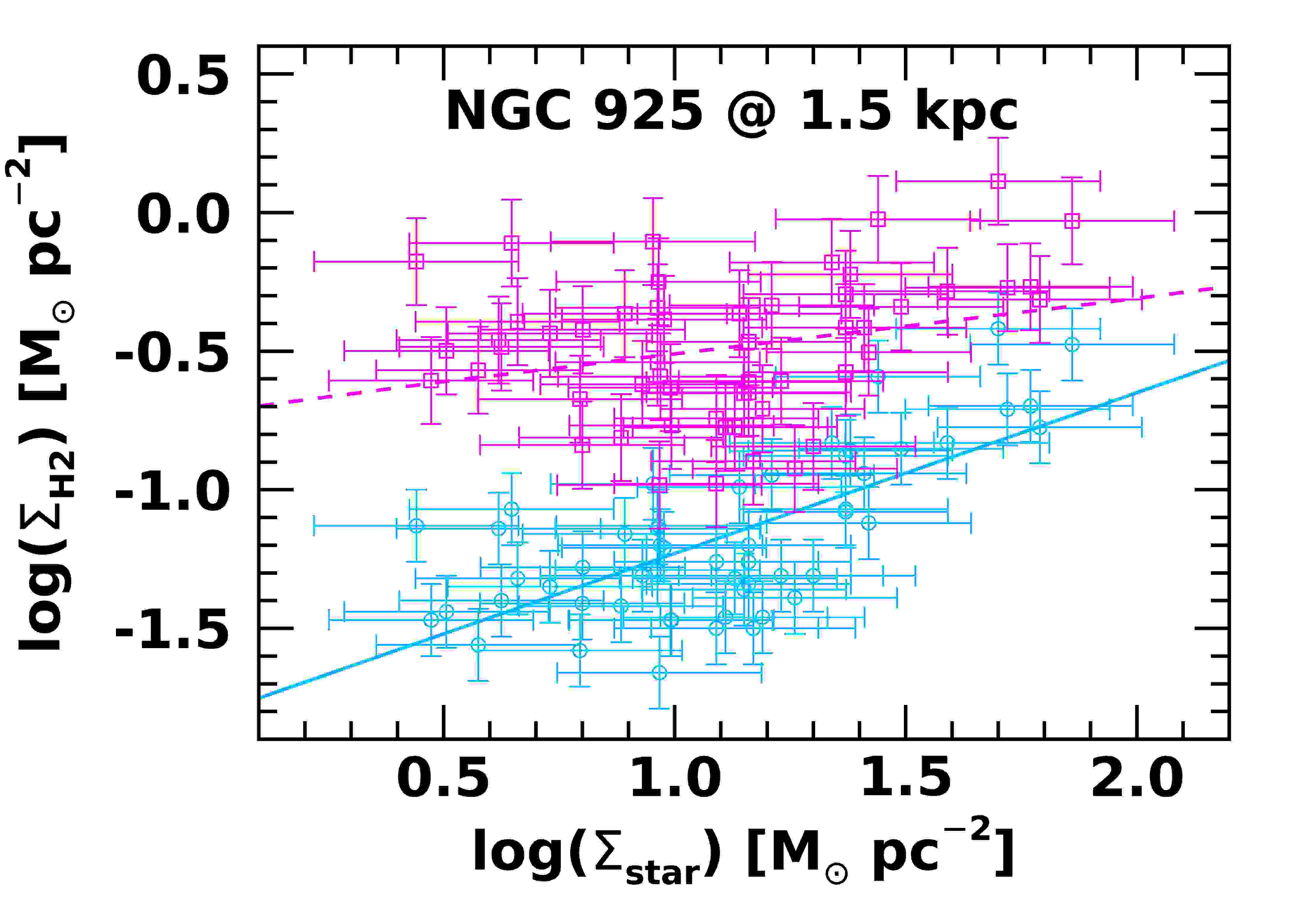}
\includegraphics[width=0.33\textwidth]{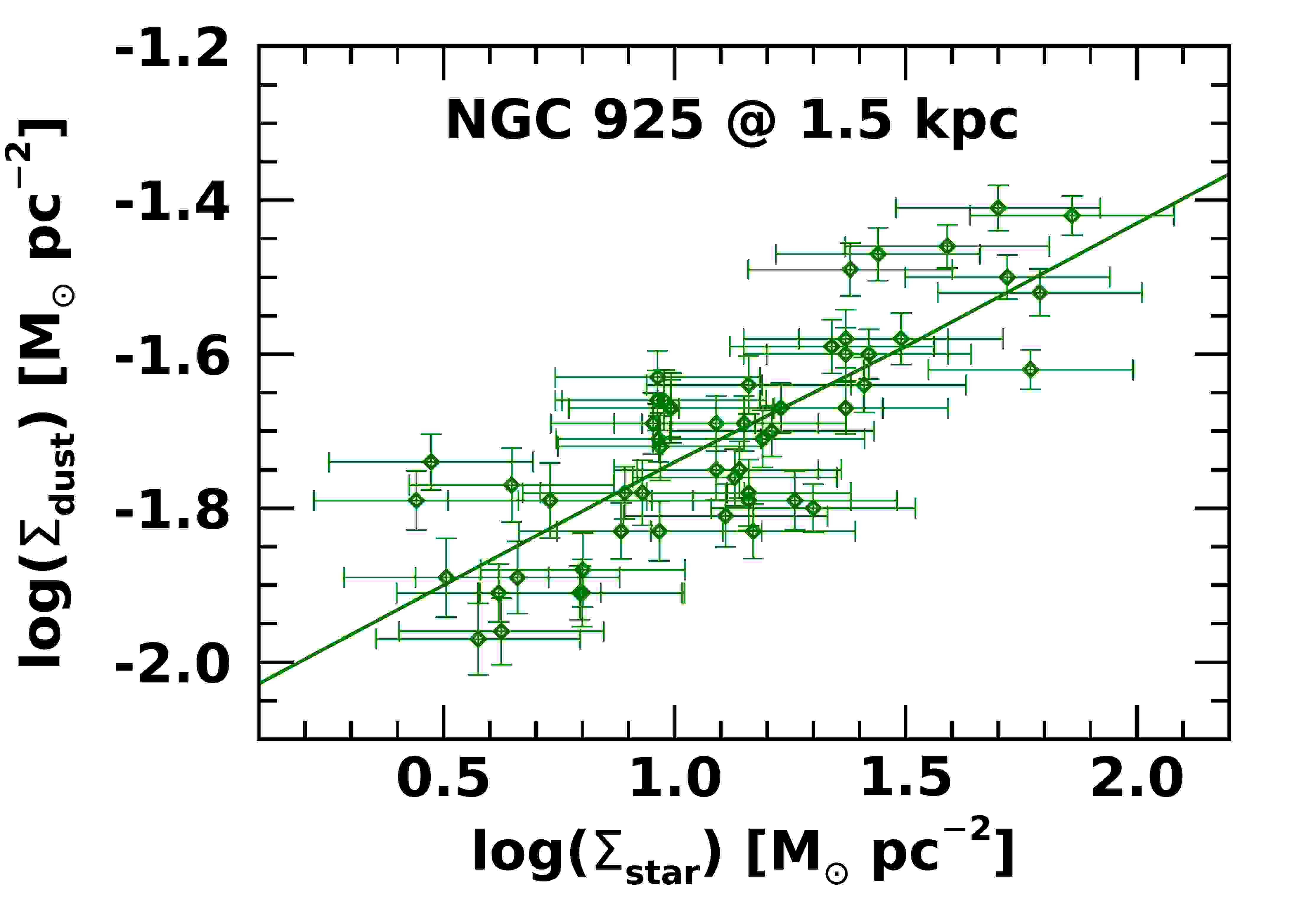}
\includegraphics[width=0.33\textwidth]{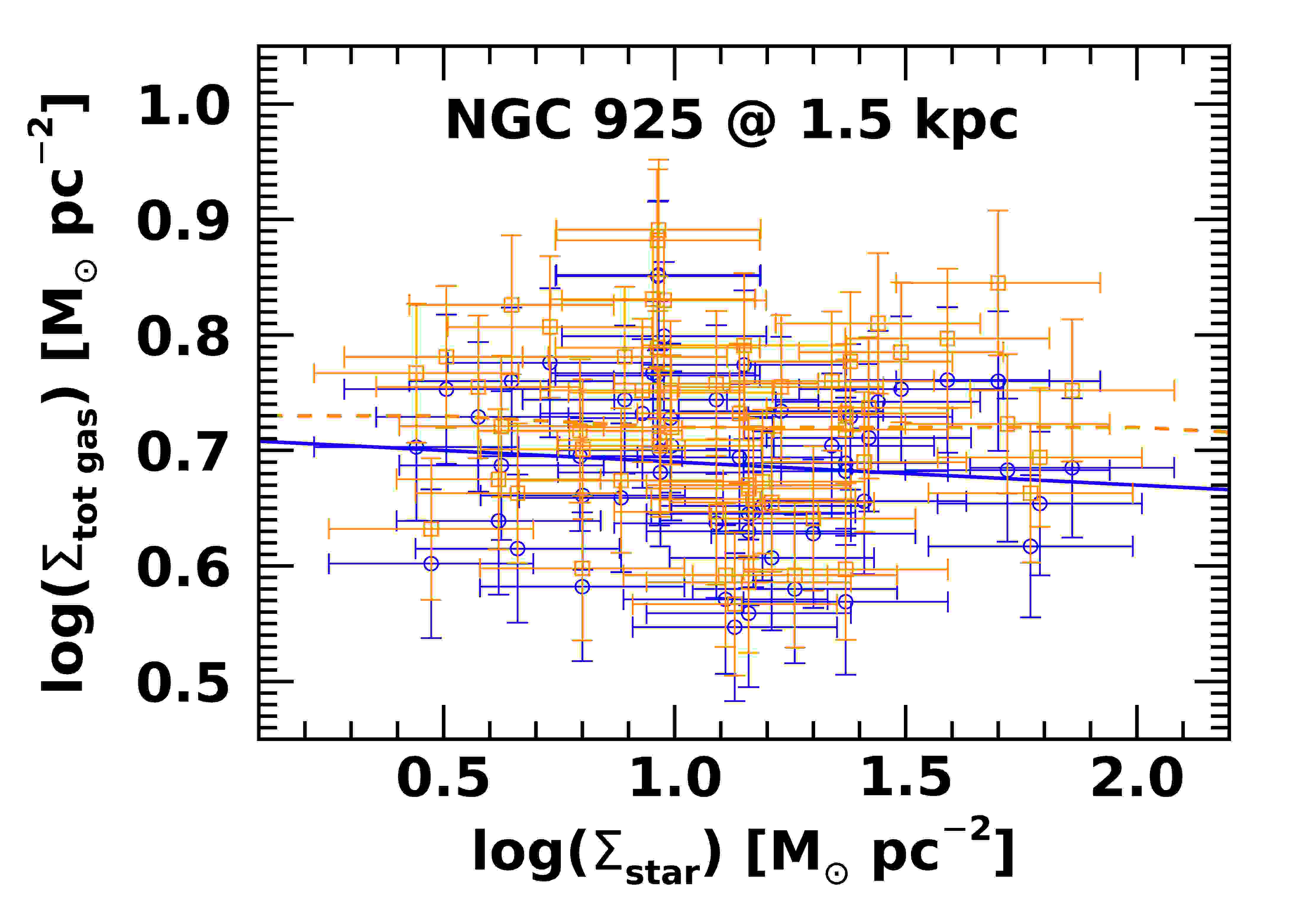}
\includegraphics[width=0.33\textwidth]{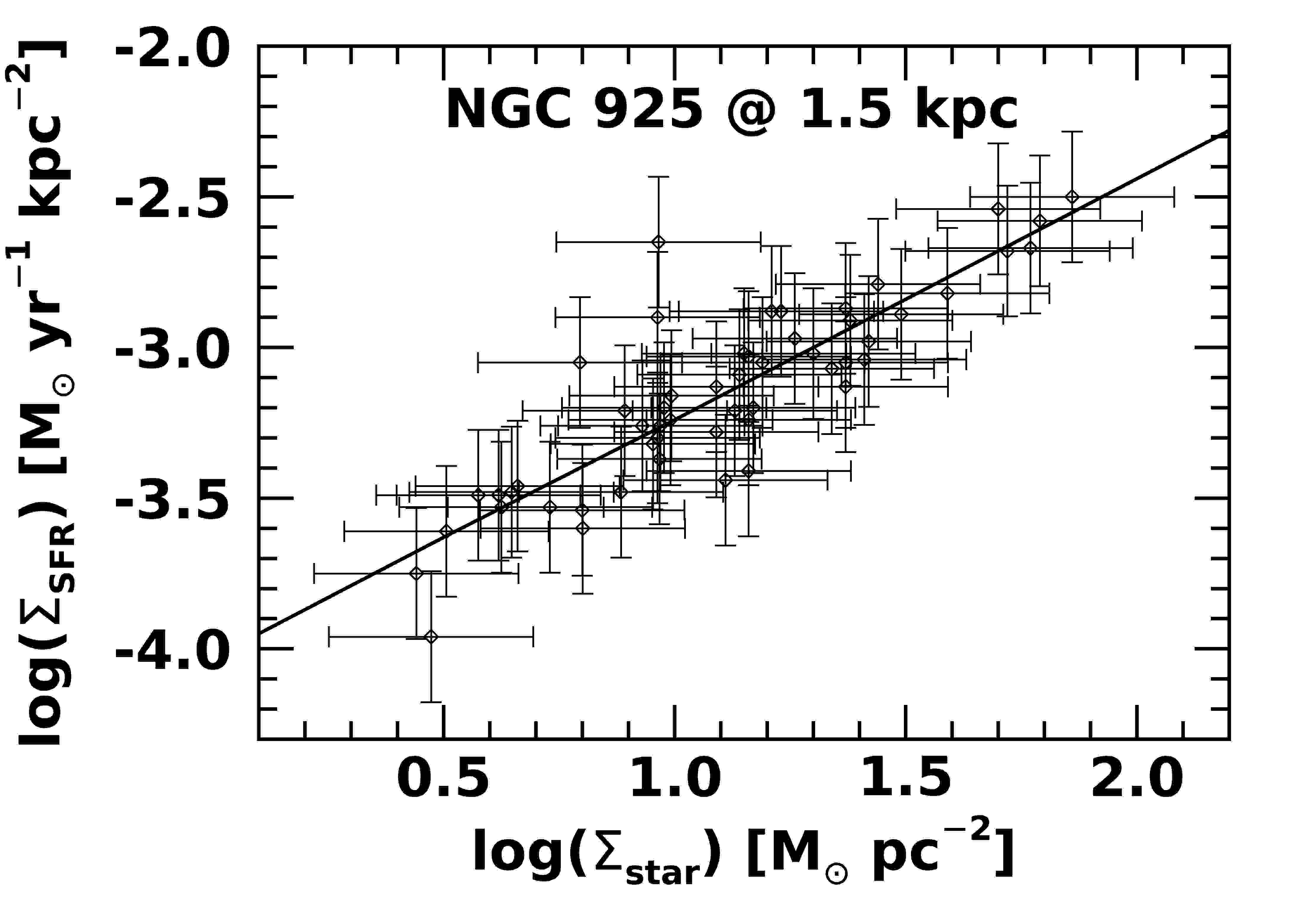}
\includegraphics[width=0.33\textwidth]{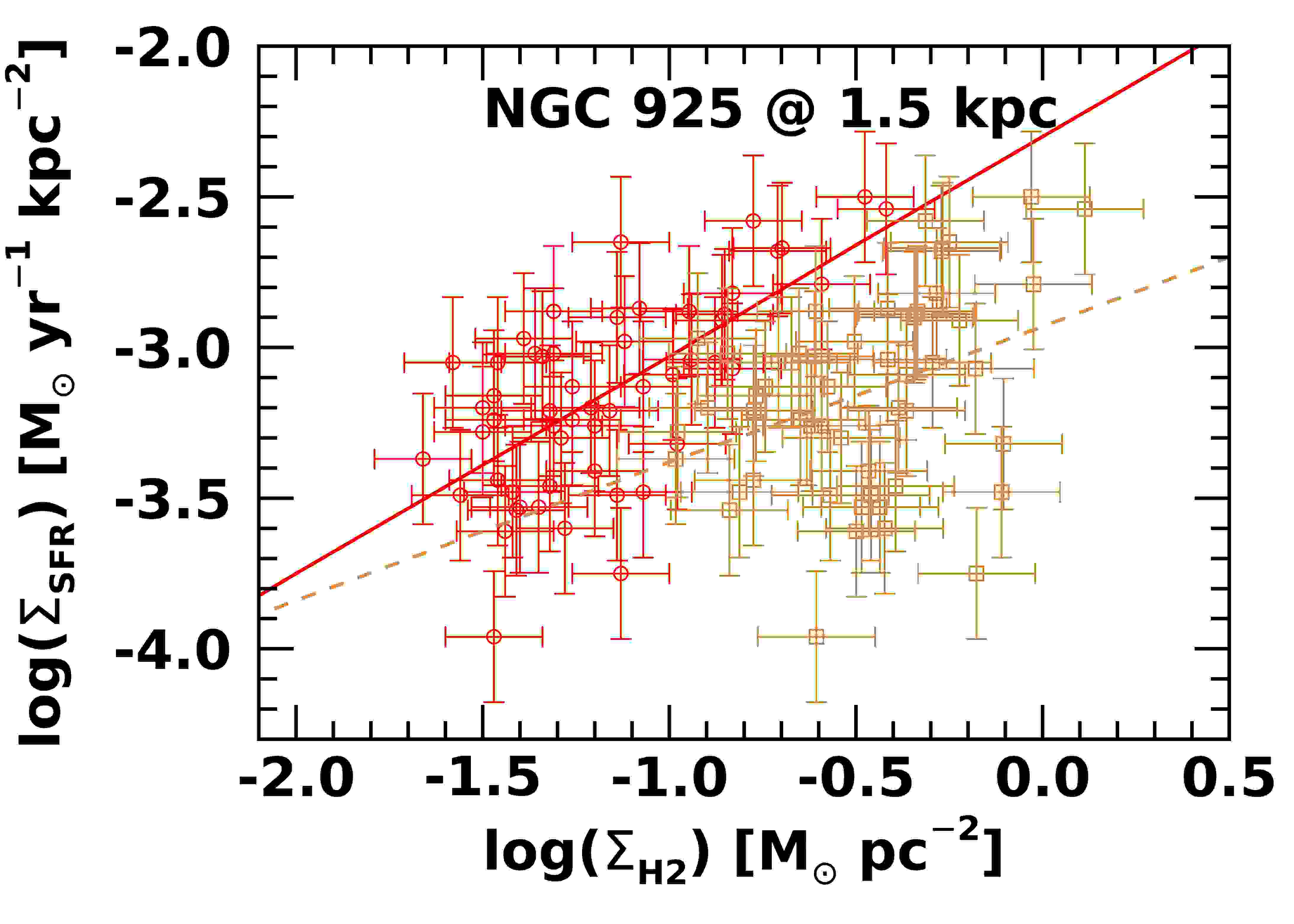}
\includegraphics[width=0.33\textwidth]{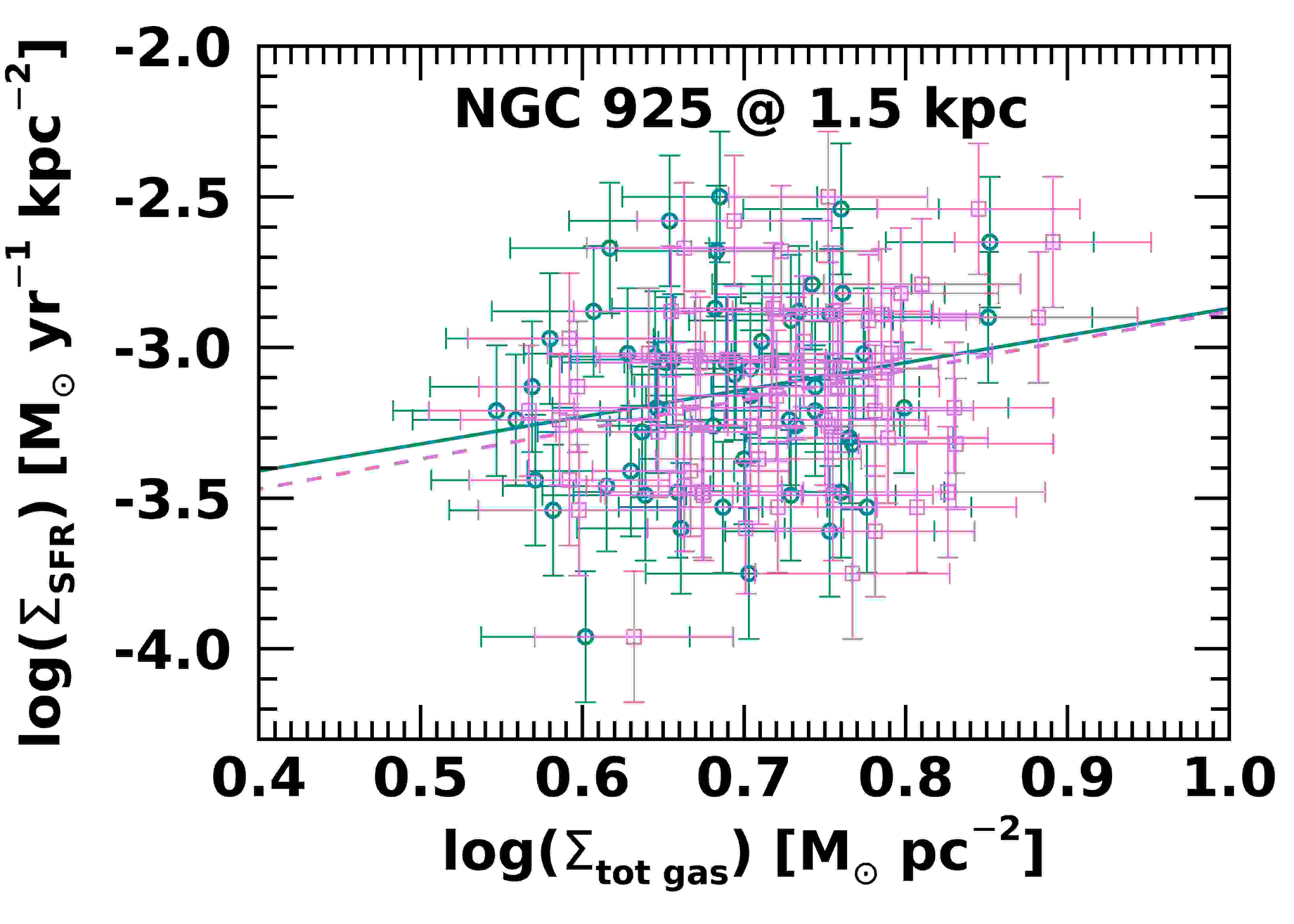}
\includegraphics[width=0.33\textwidth]{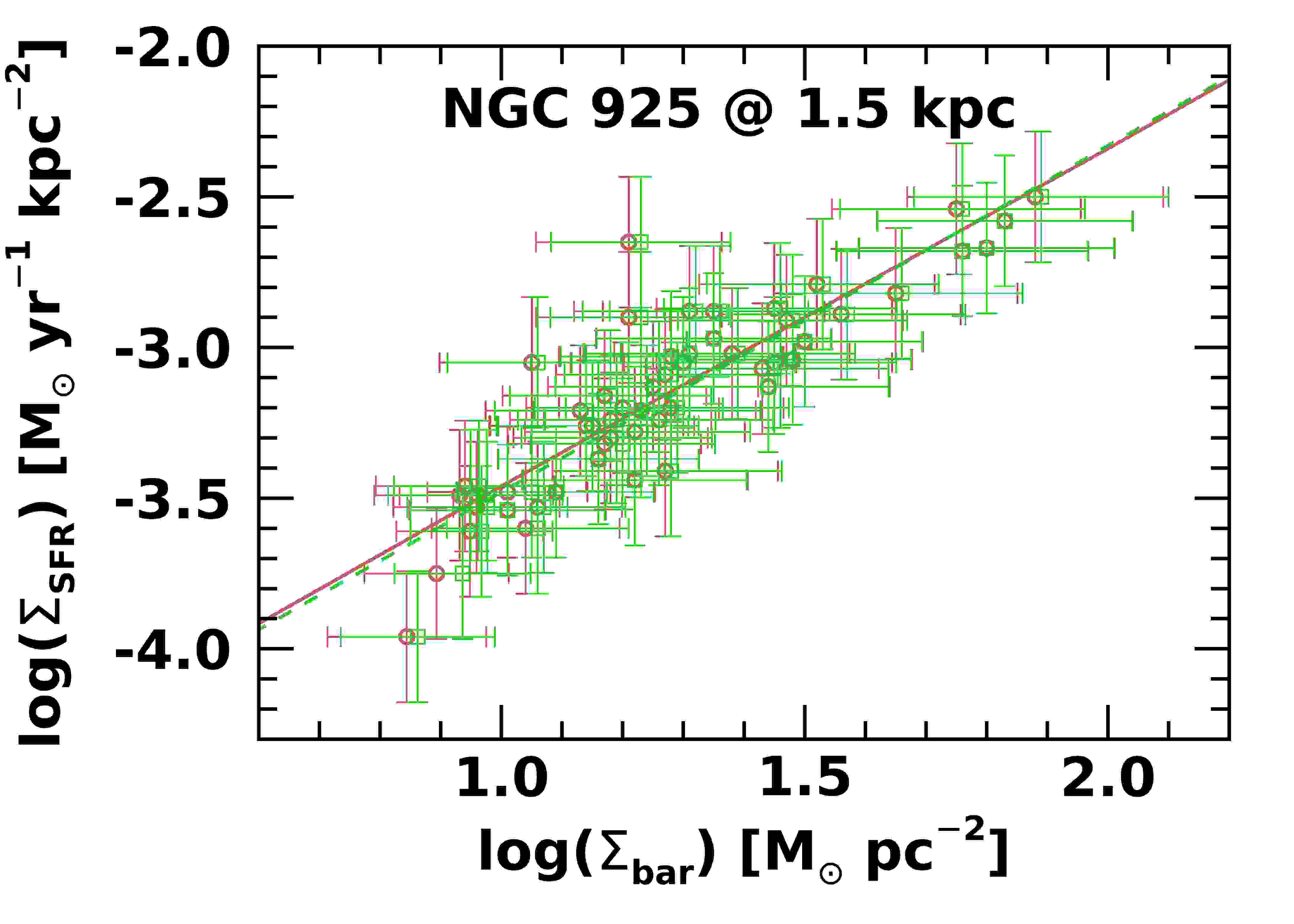}
\includegraphics[width=0.33\textwidth]{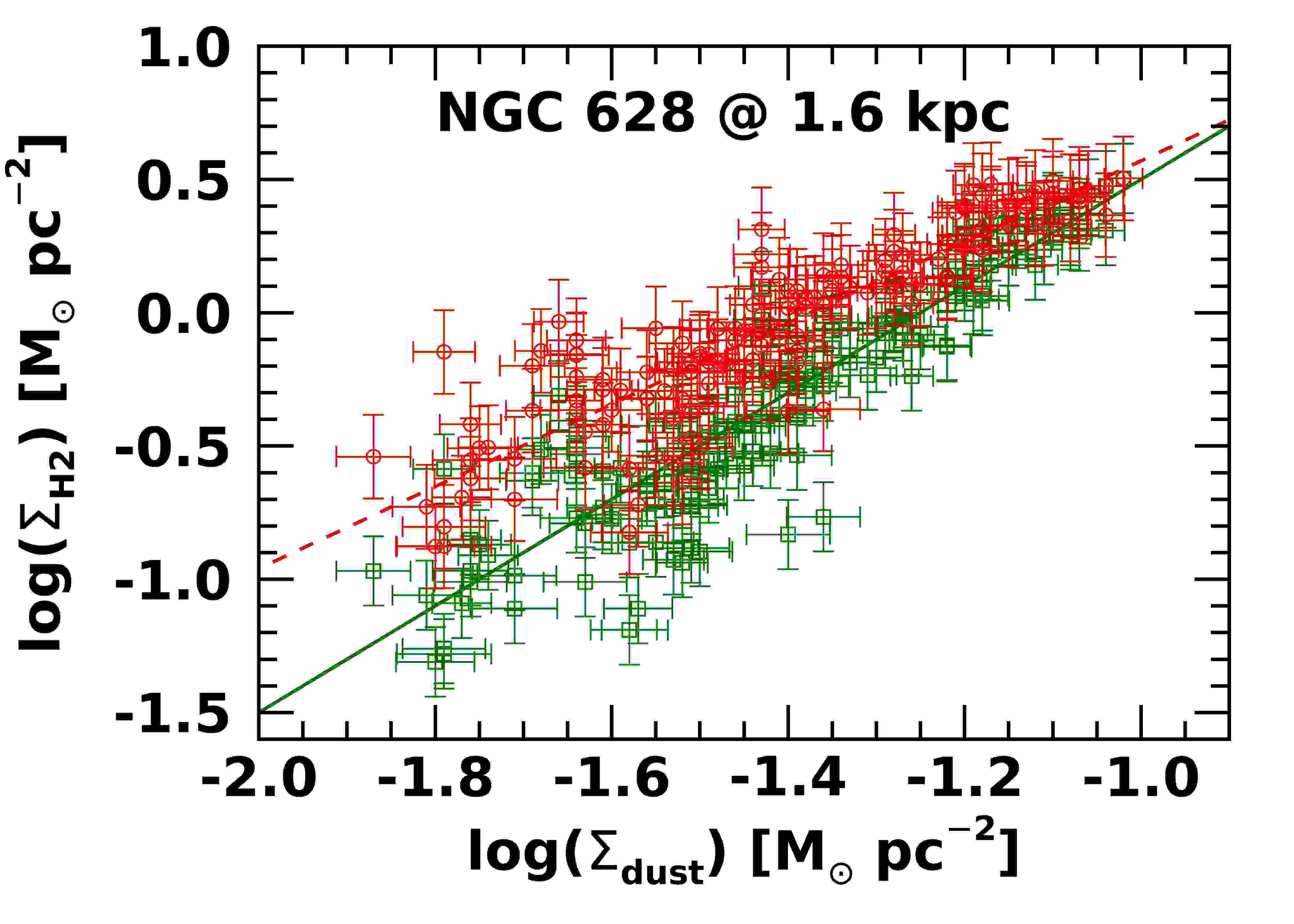}
\includegraphics[width=0.33\textwidth]{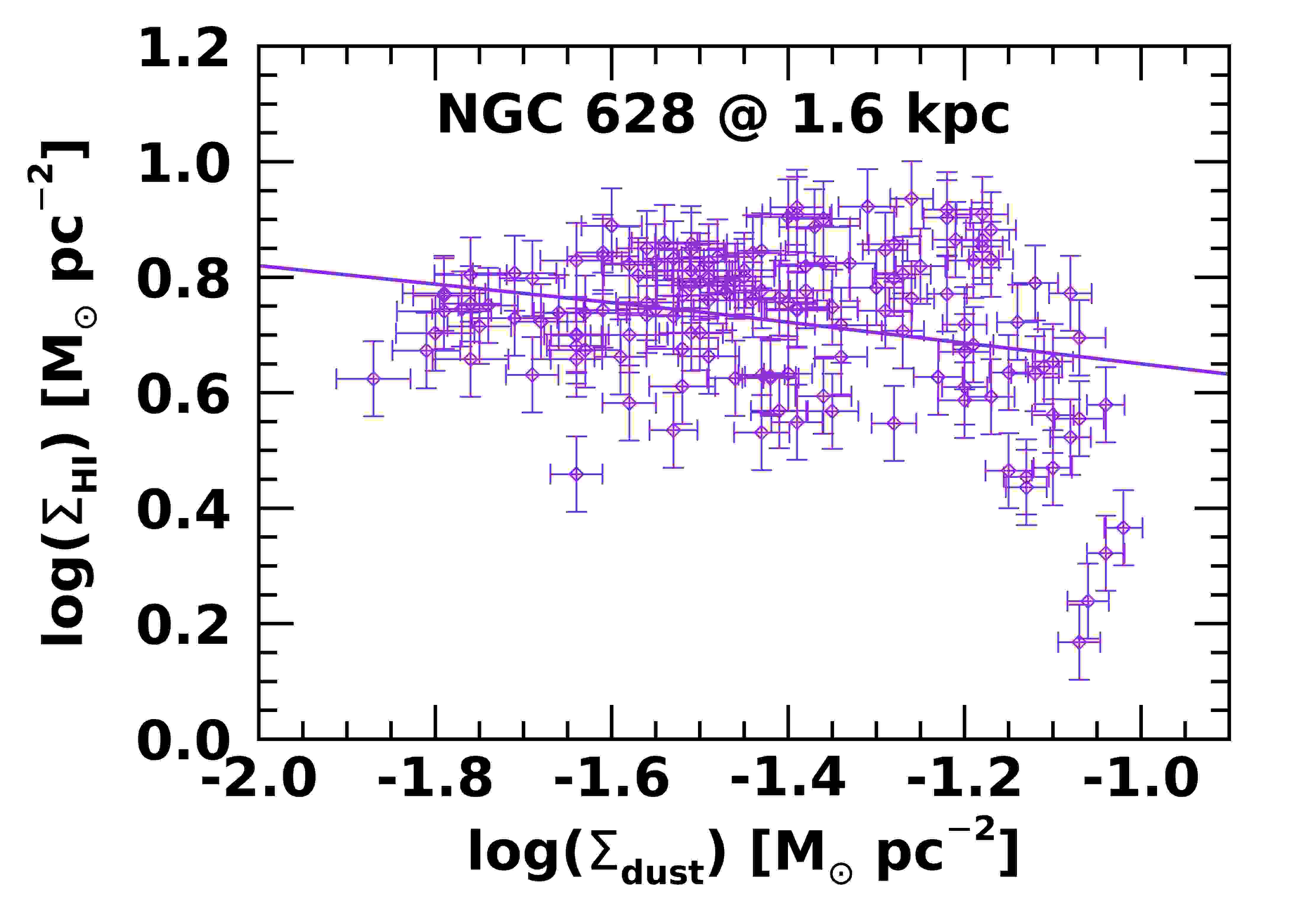}
\includegraphics[width=0.33\textwidth]{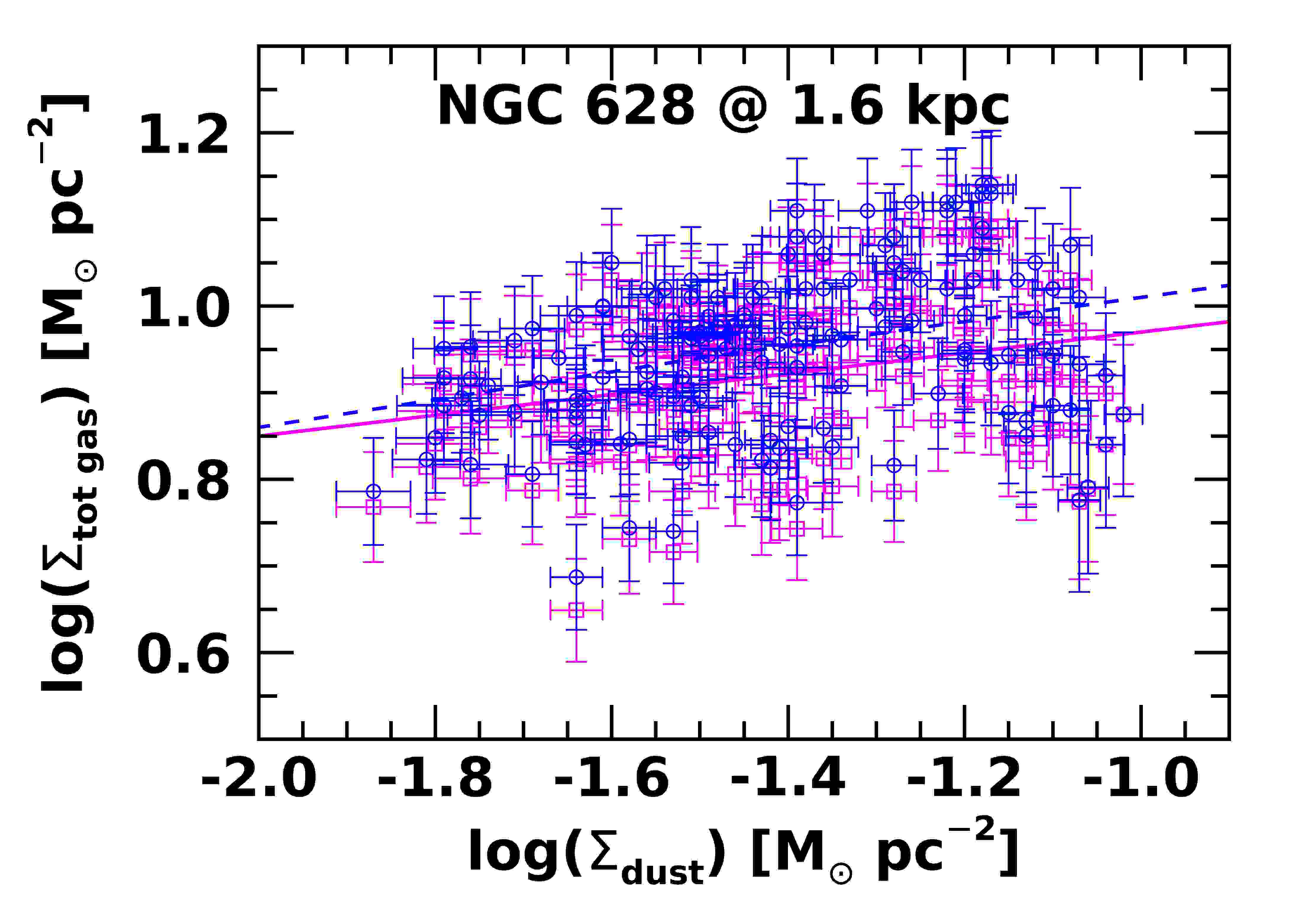}
\includegraphics[width=0.33\textwidth]{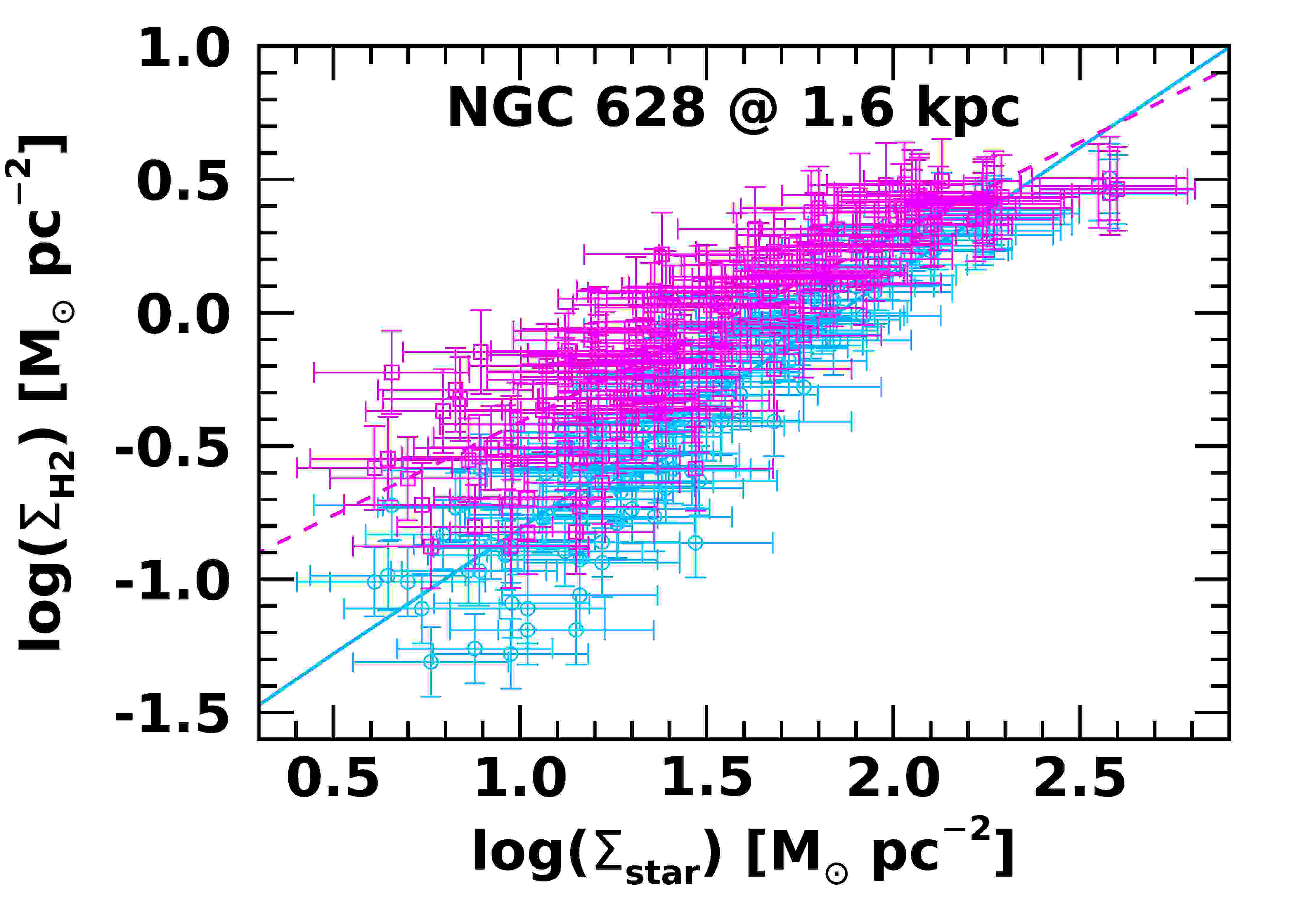}
\includegraphics[width=0.33\textwidth]{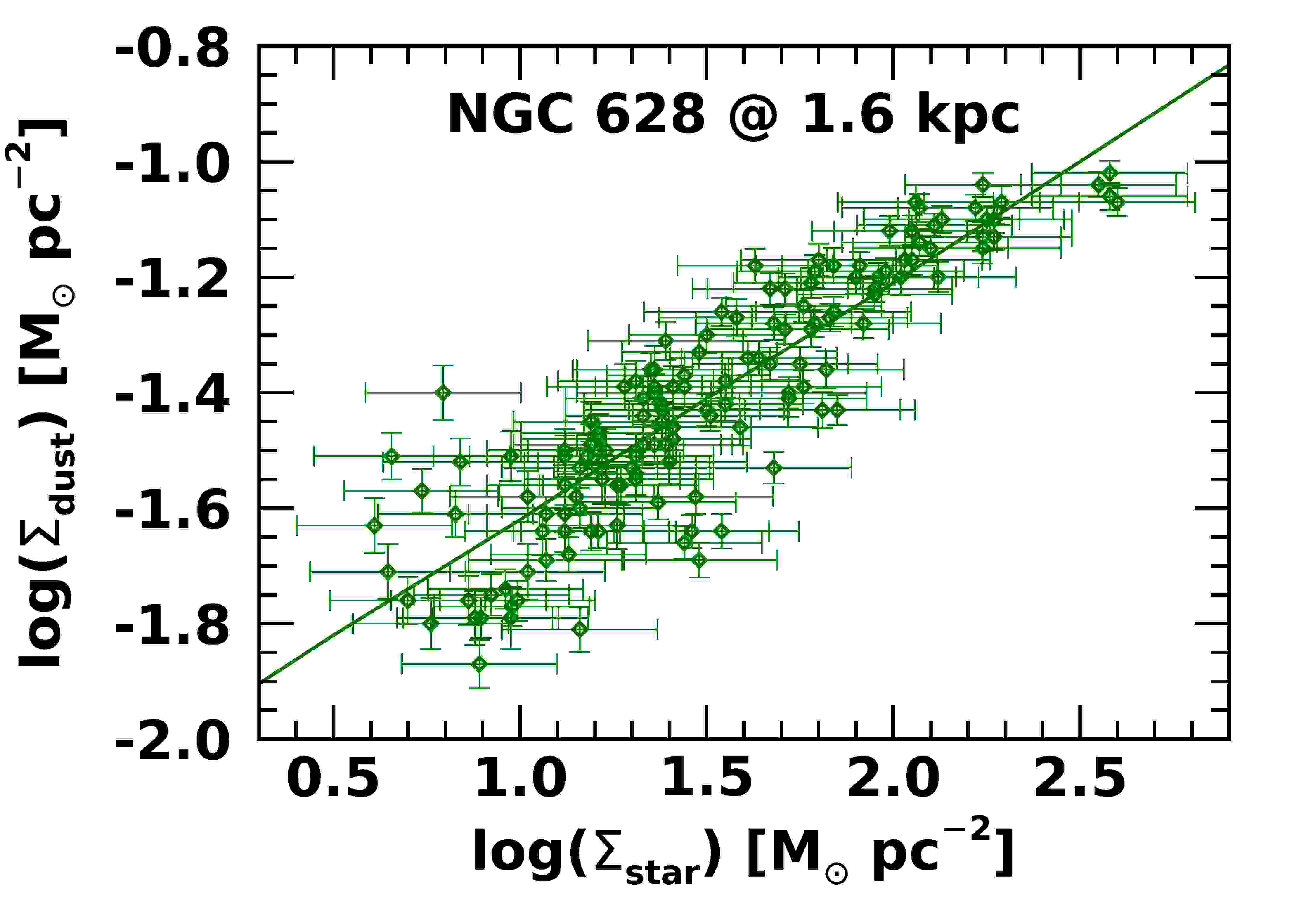}
\includegraphics[width=0.33\textwidth]{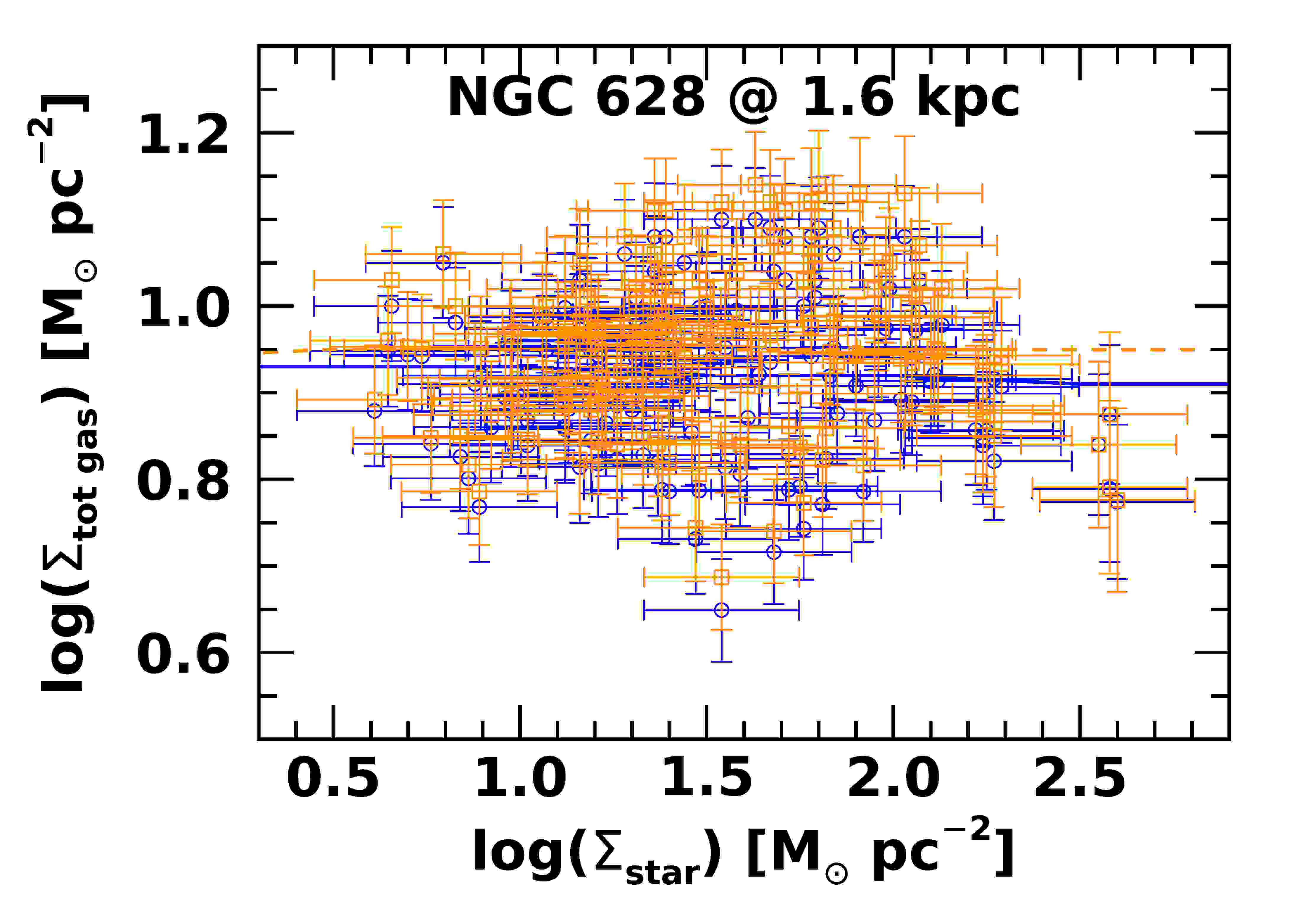}
\includegraphics[width=0.33\textwidth]{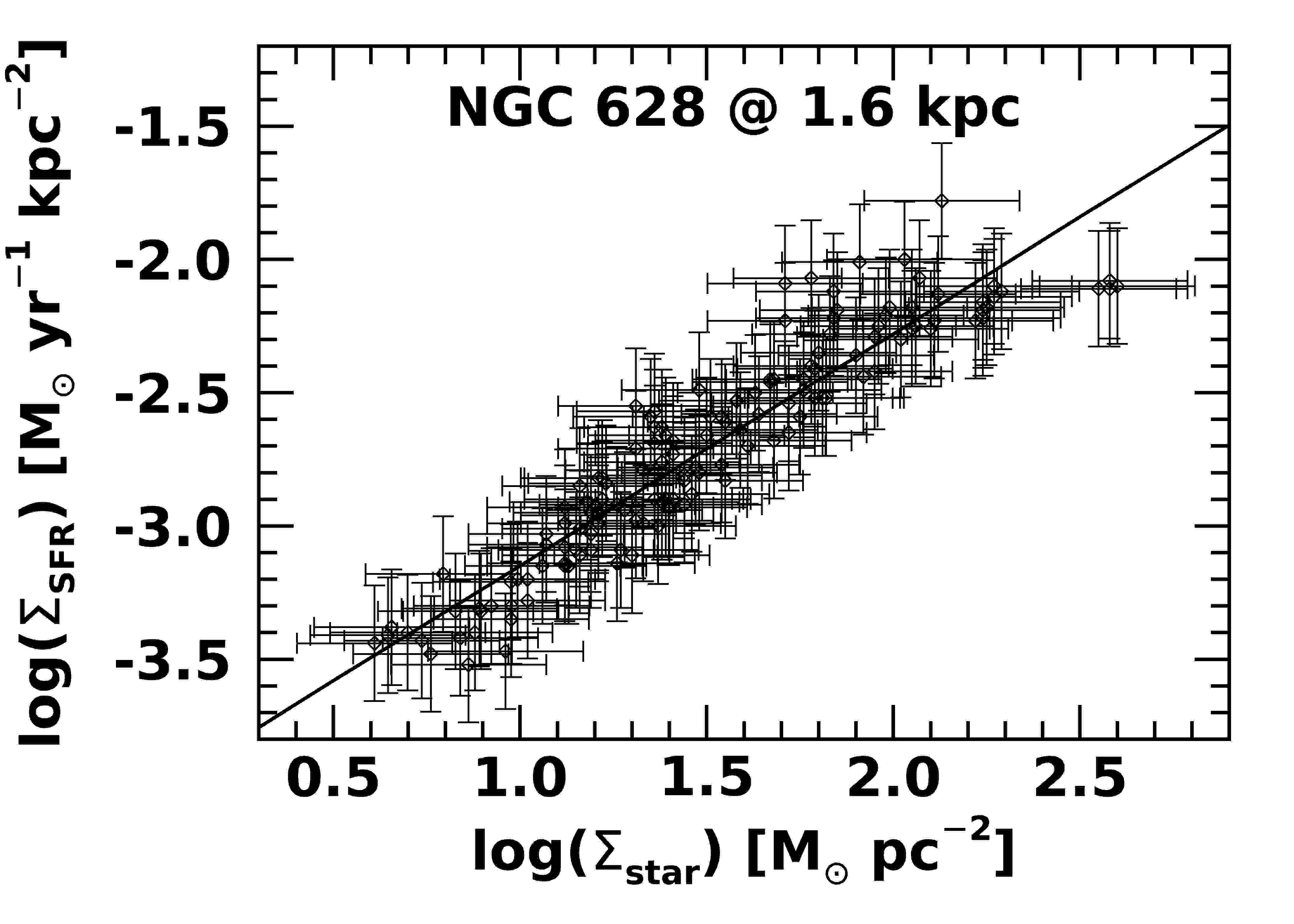}
\includegraphics[width=0.33\textwidth]{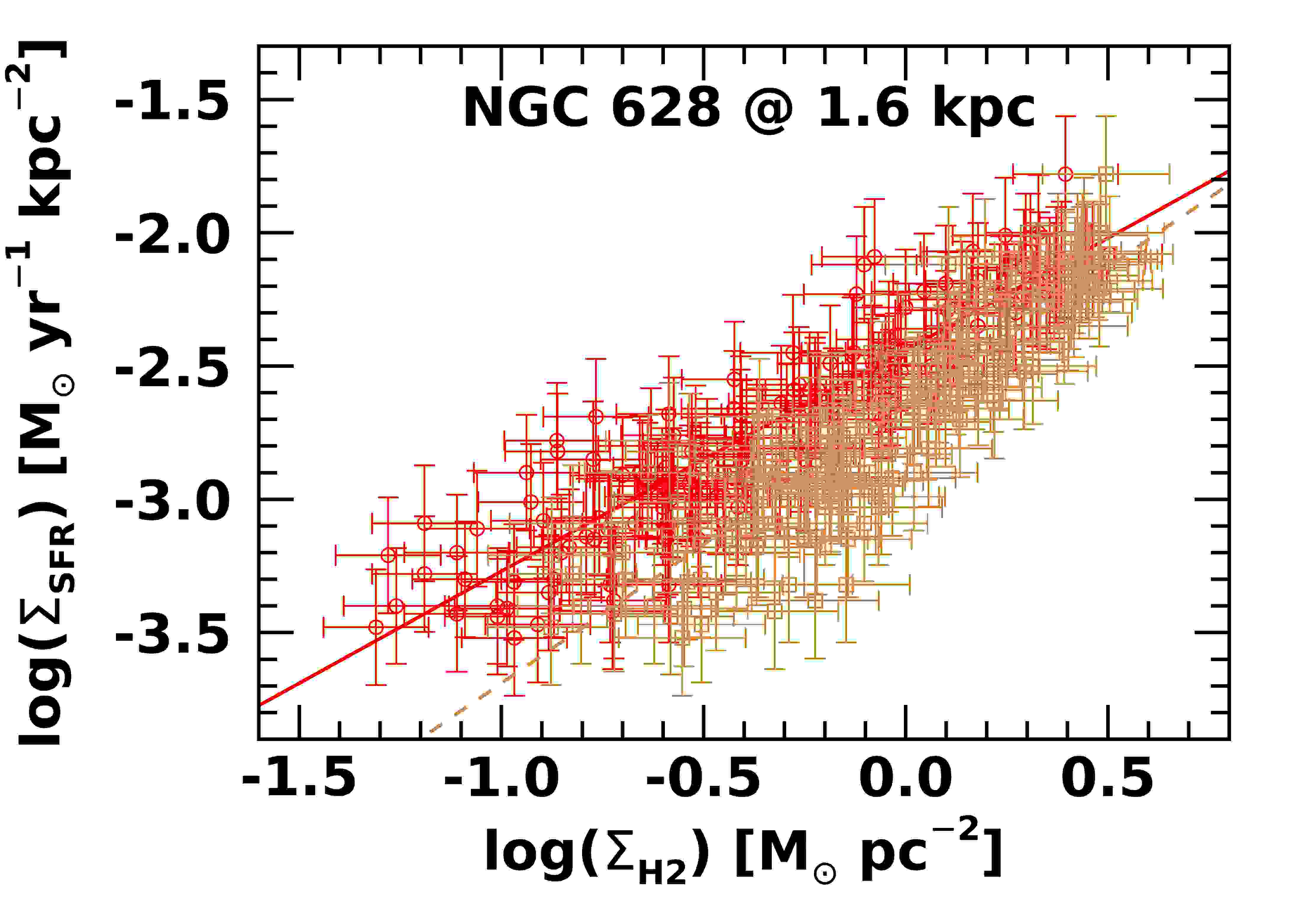}
\caption*{Figure~\ref{fig:add-ism} continued}
\end{figure*}

\begin{figure*}
\centering
\includegraphics[width=0.33\textwidth]{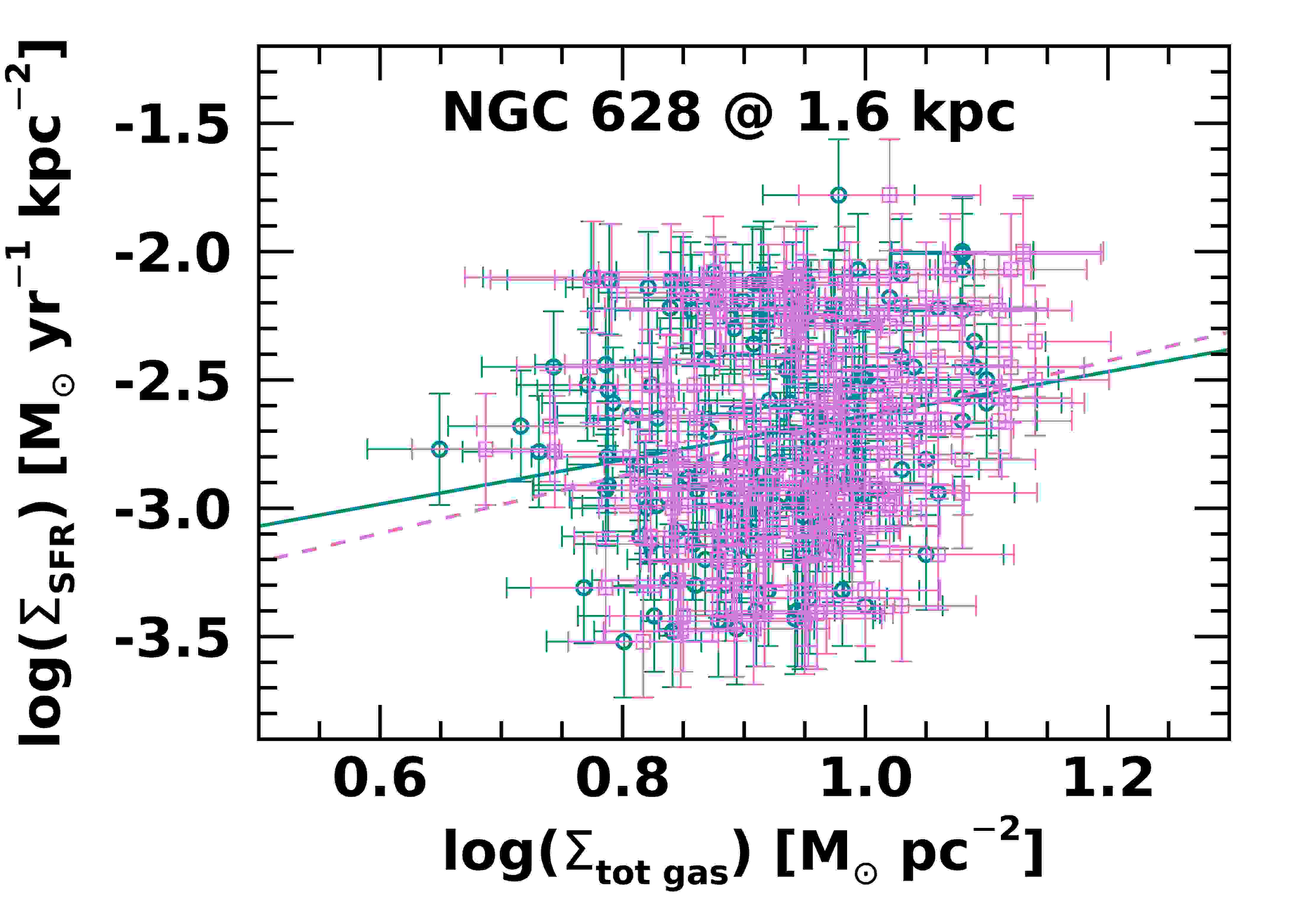}
\includegraphics[width=0.33\textwidth]{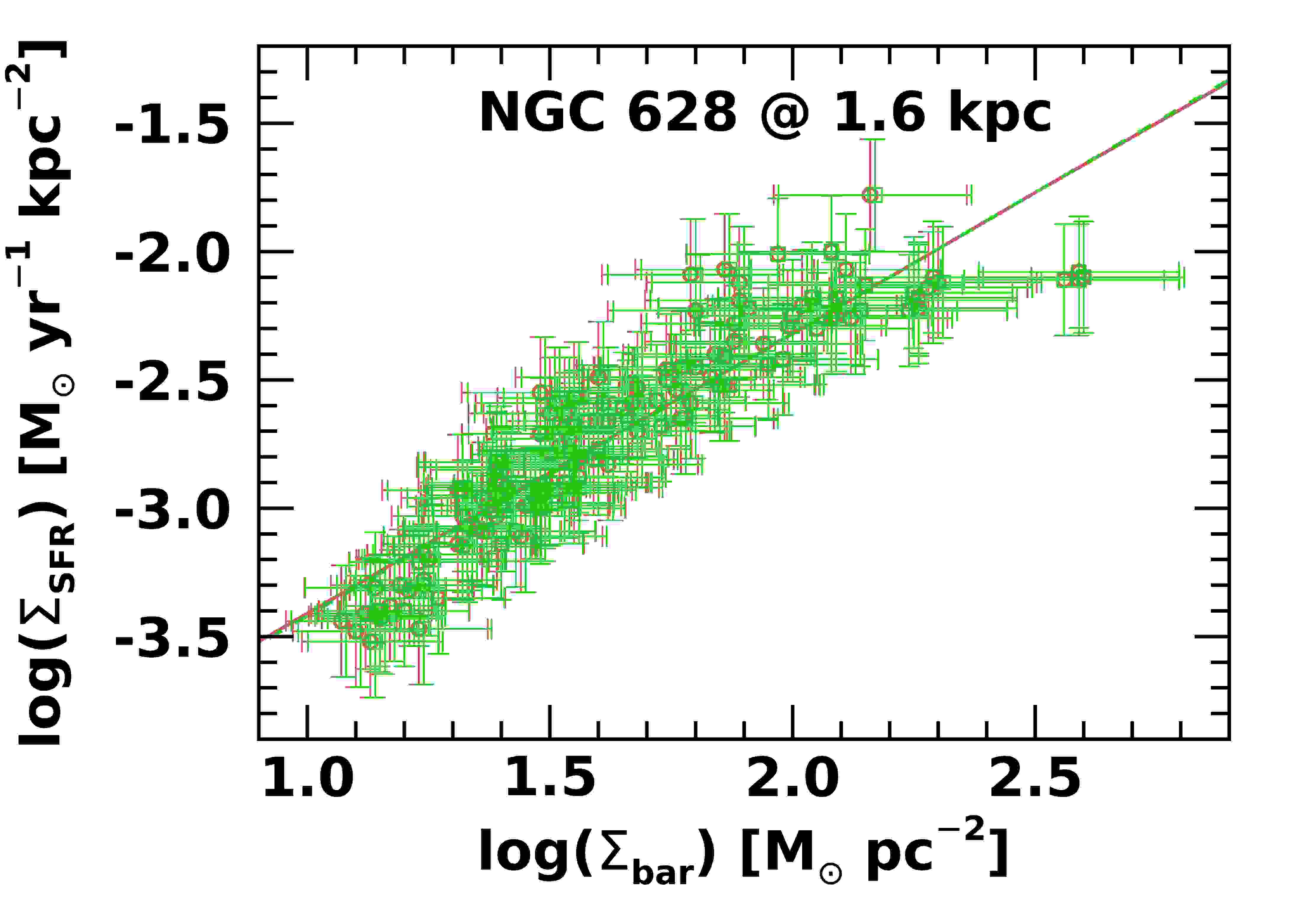}
\includegraphics[width=0.33\textwidth]{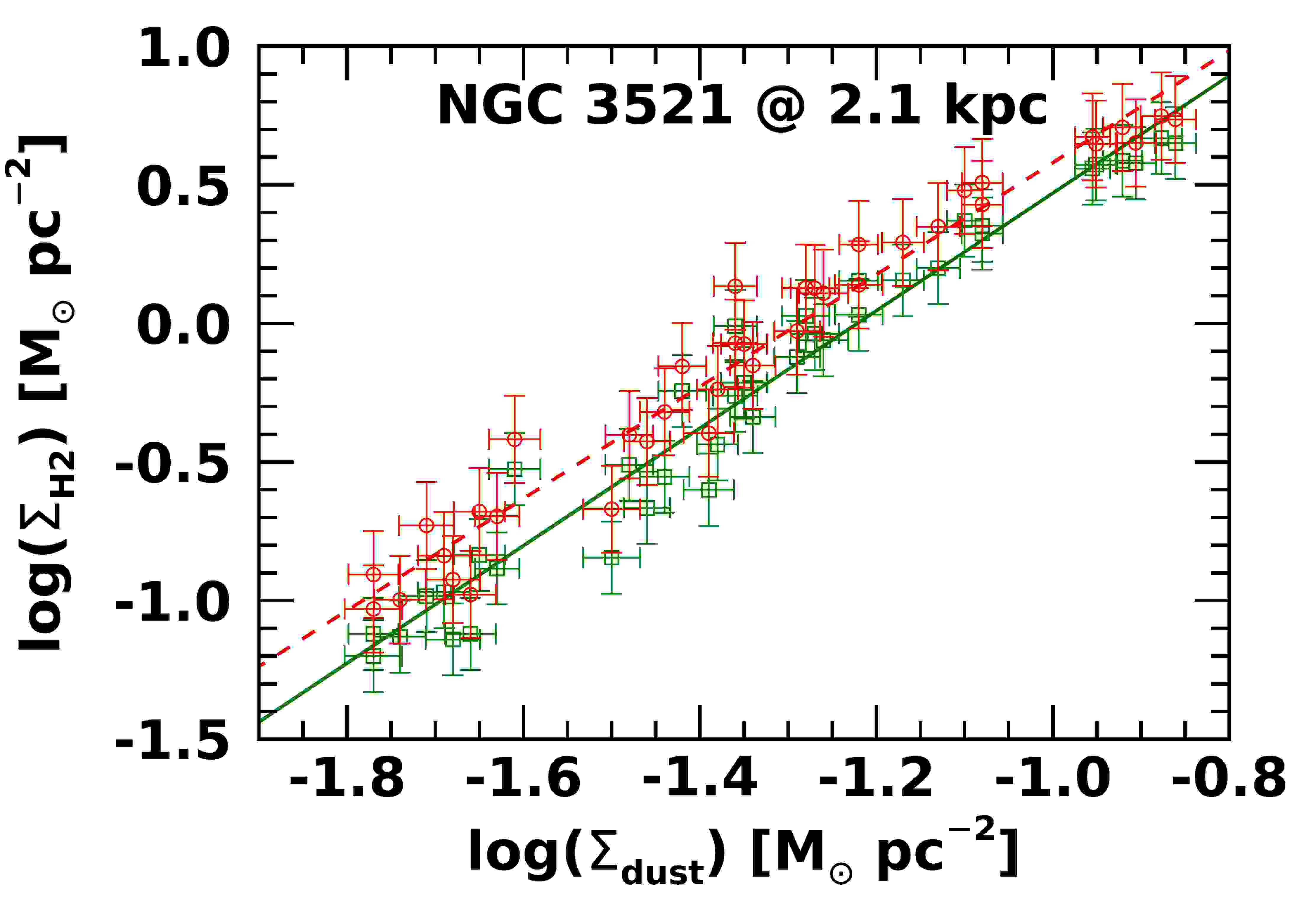}
\includegraphics[width=0.33\textwidth]{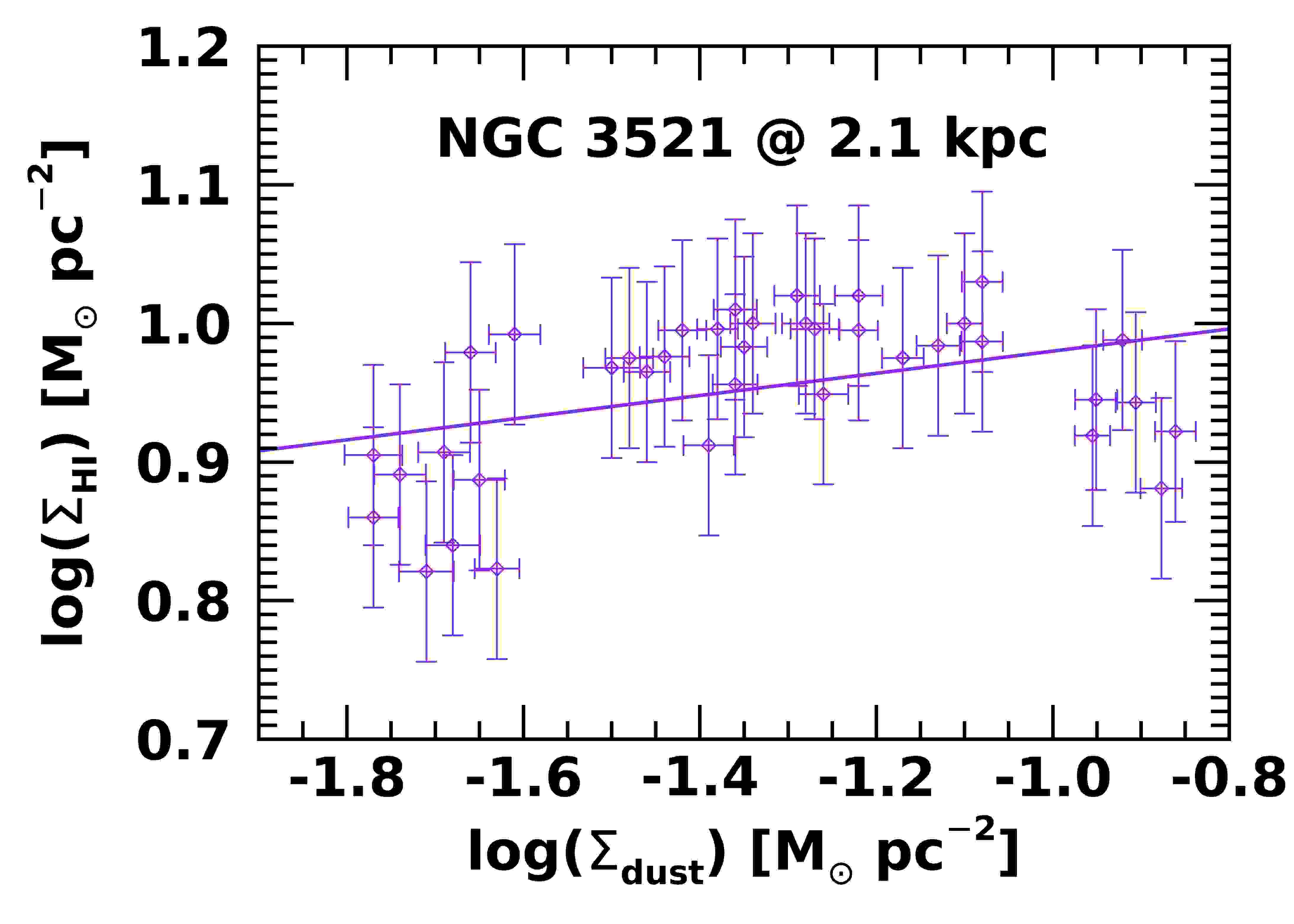}
\includegraphics[width=0.33\textwidth]{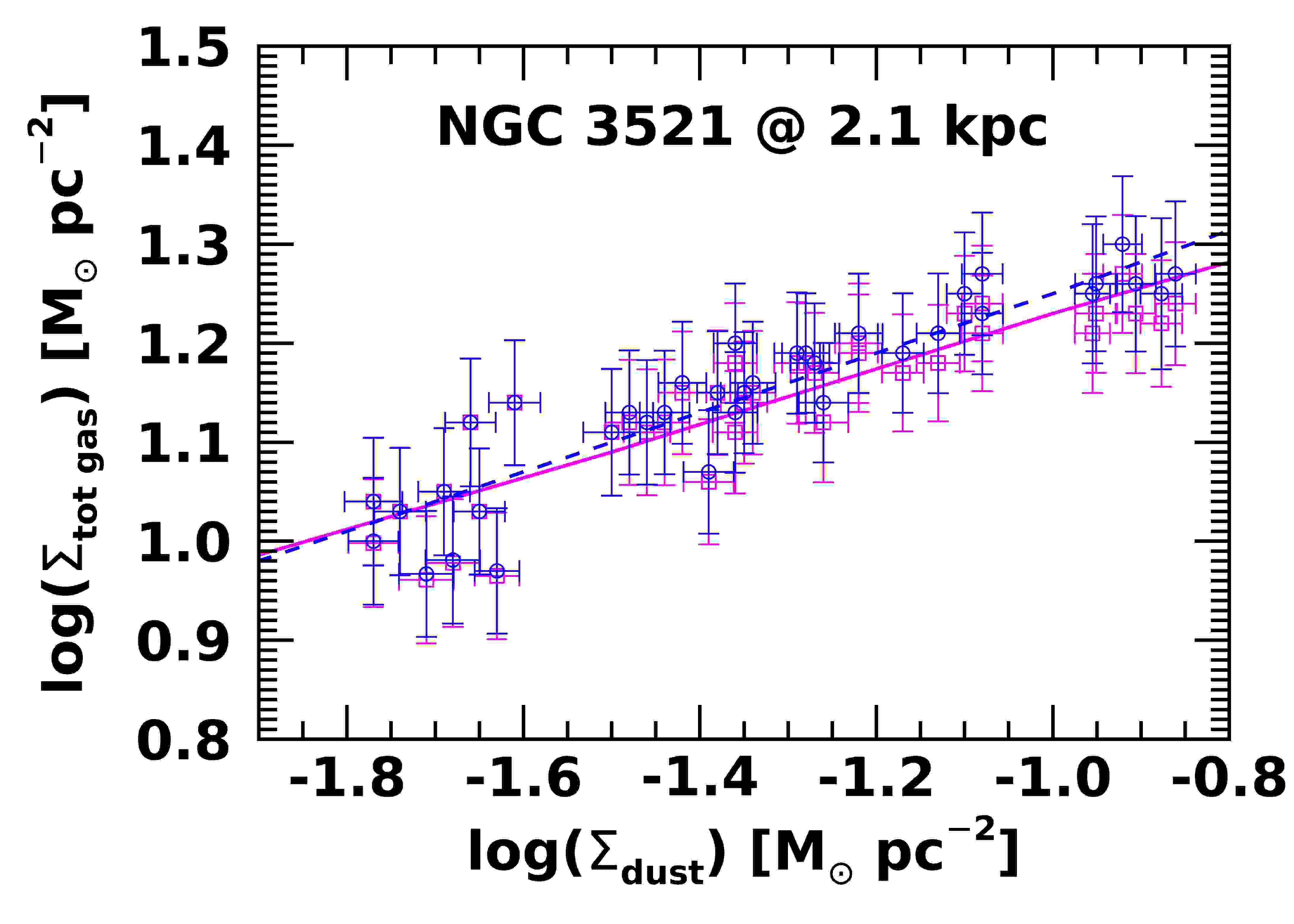}
\includegraphics[width=0.33\textwidth]{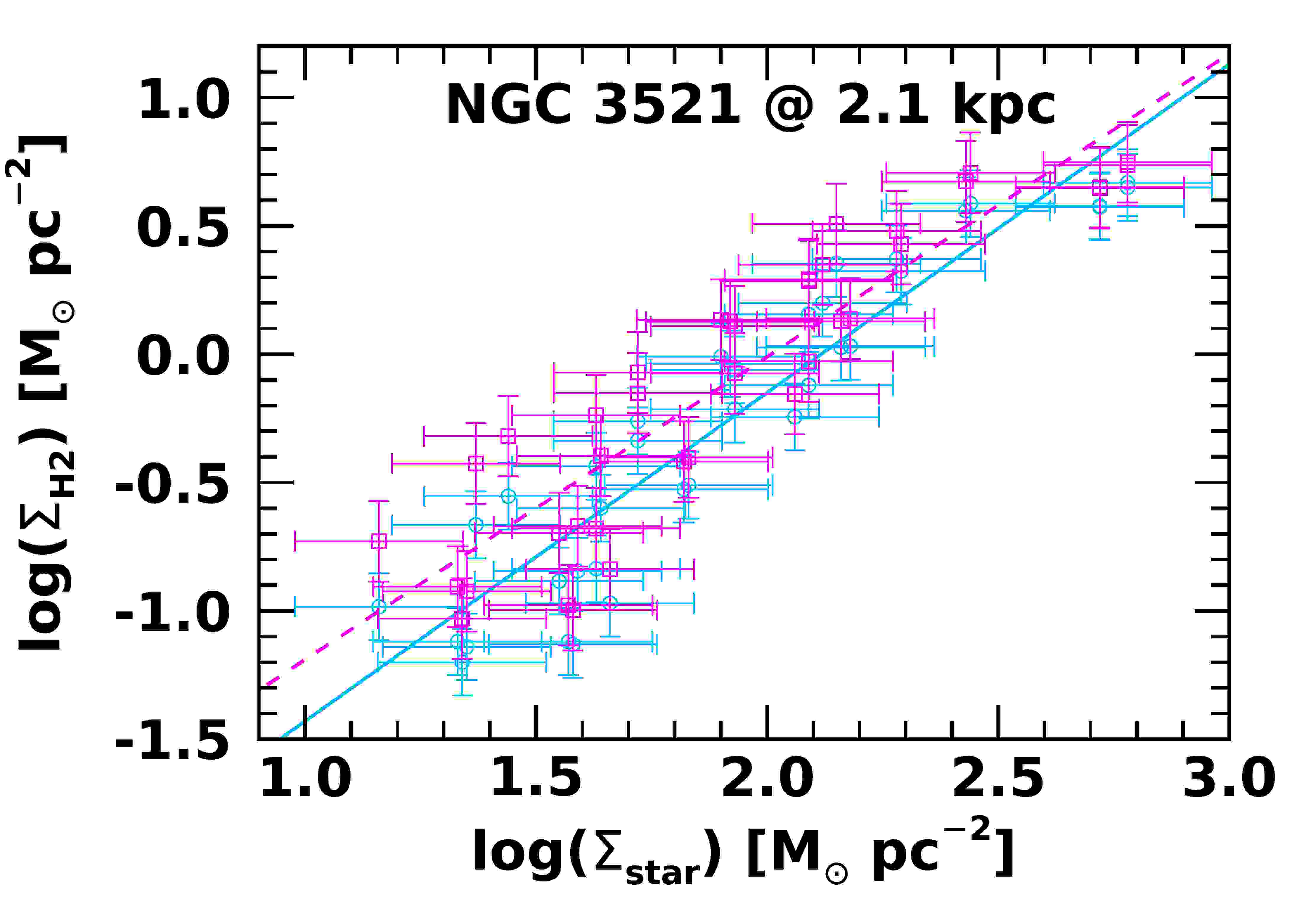}
\includegraphics[width=0.33\textwidth]{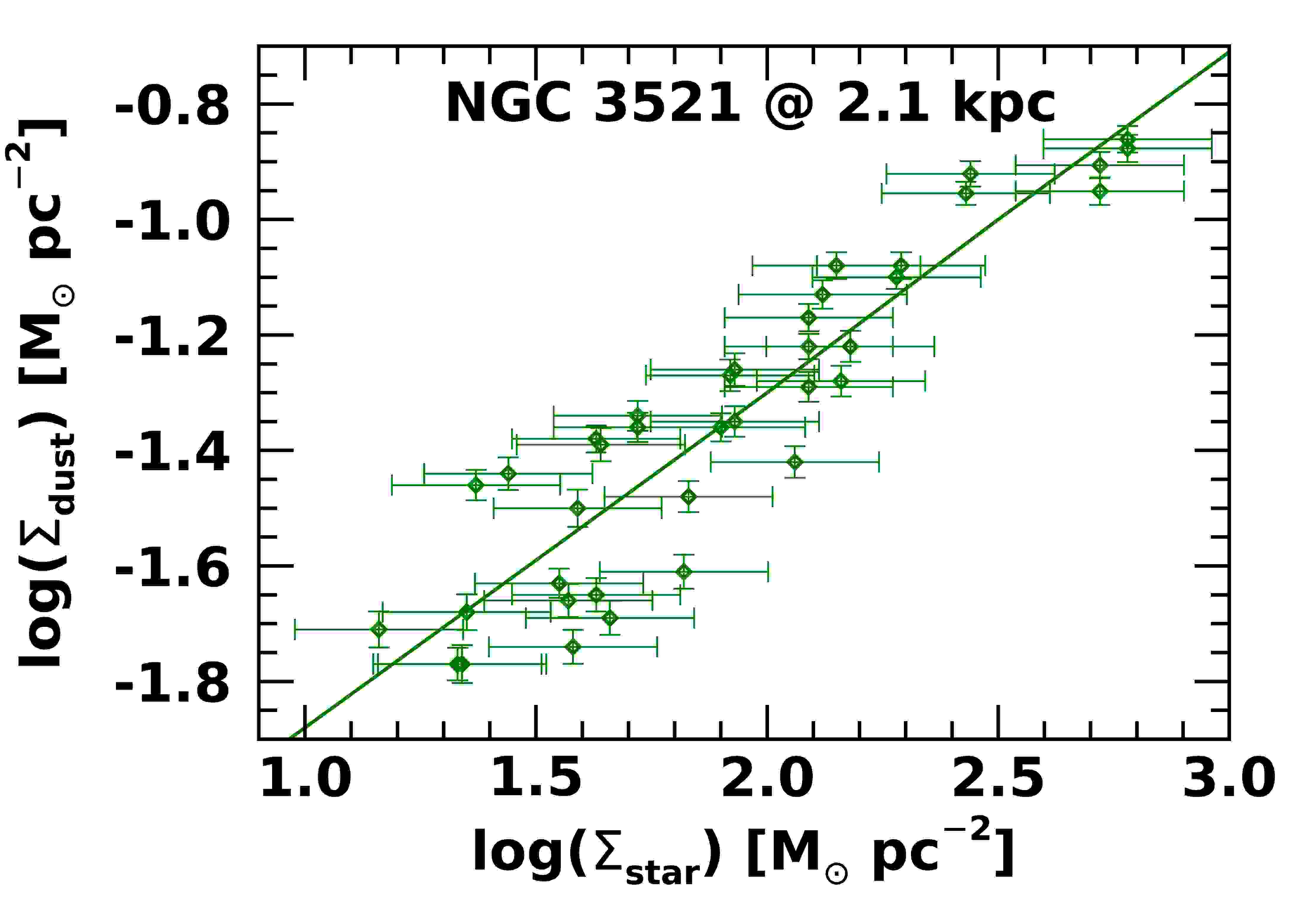}
\includegraphics[width=0.33\textwidth]{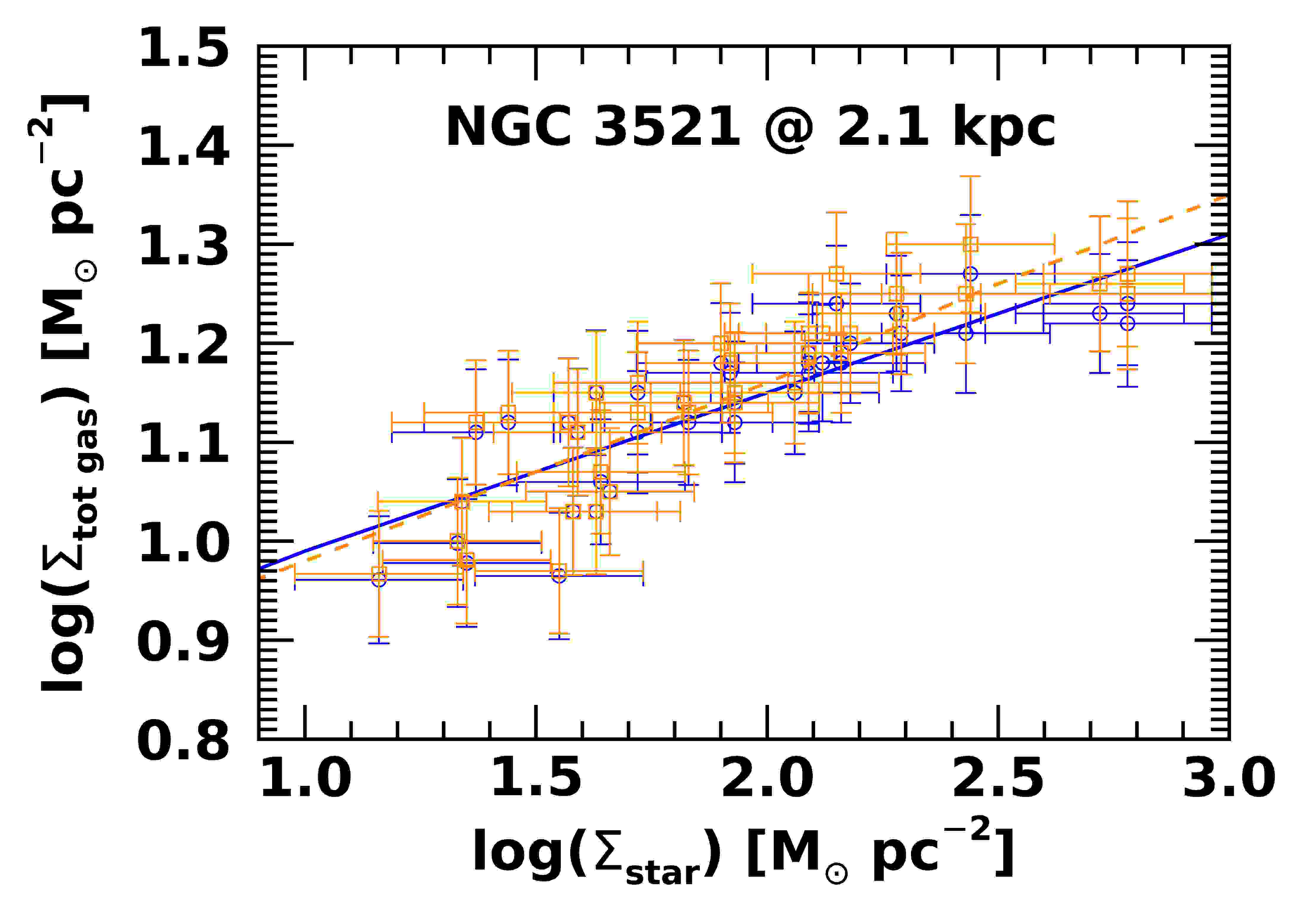}
\includegraphics[width=0.33\textwidth]{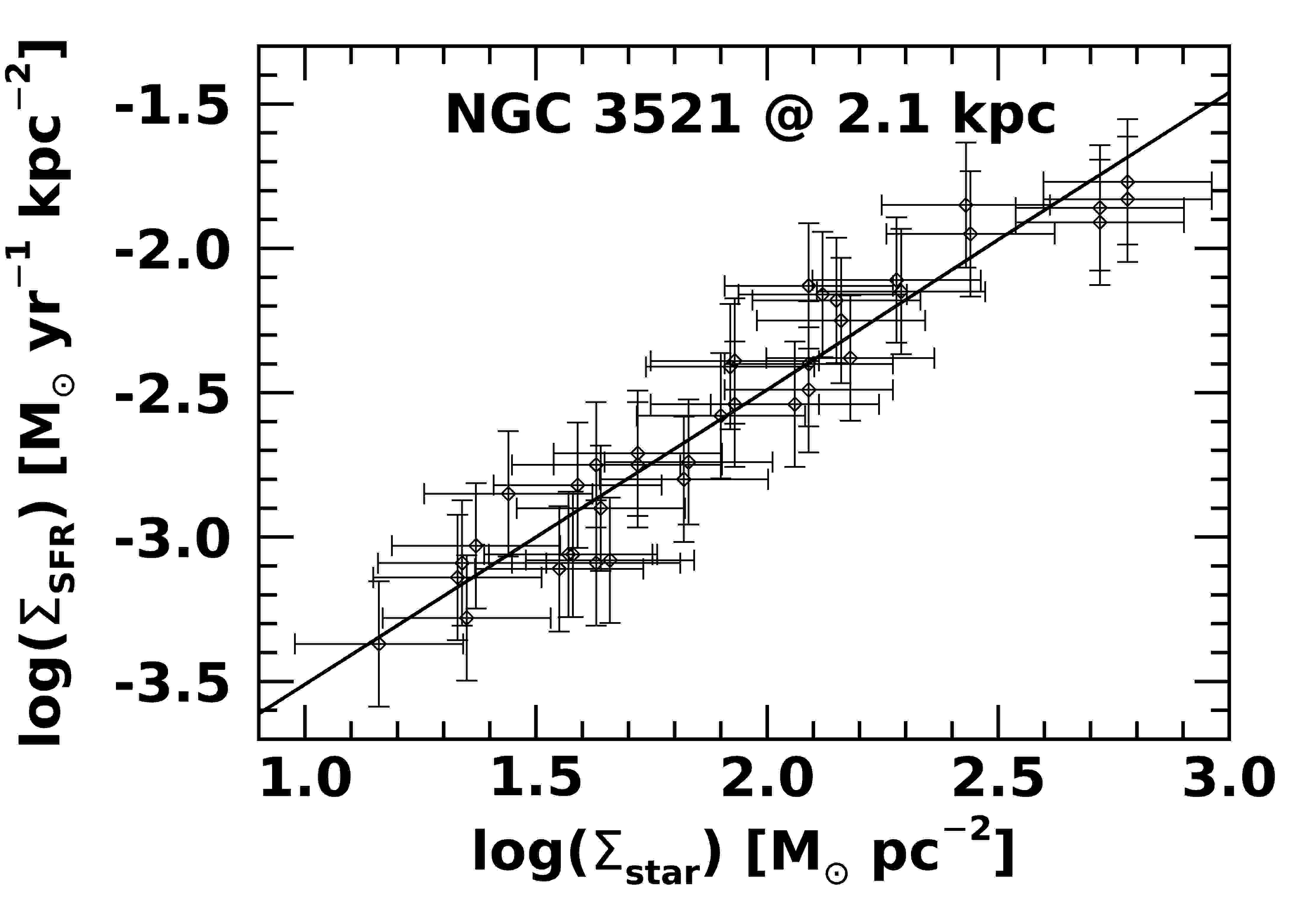}
\includegraphics[width=0.33\textwidth]{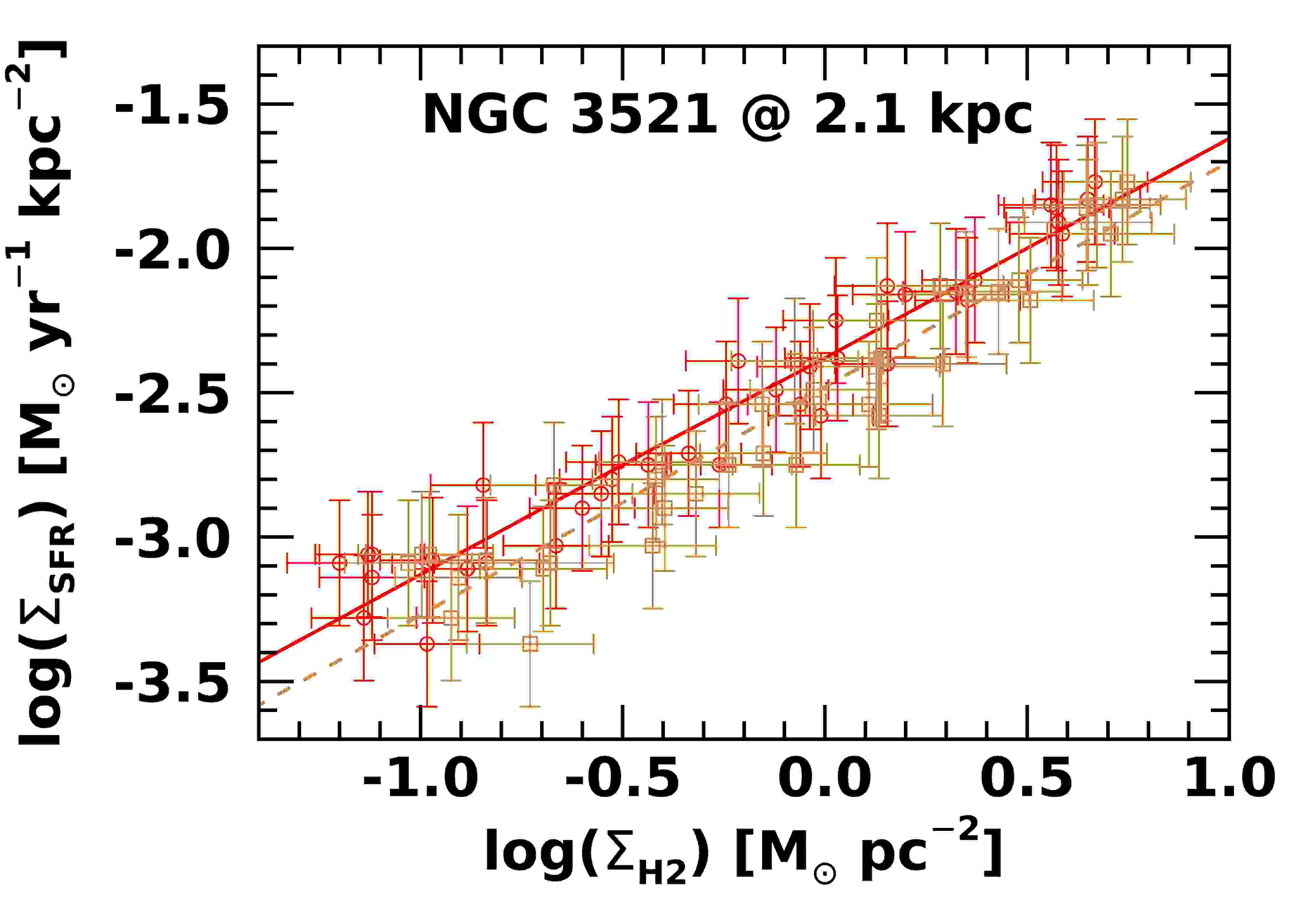}
\includegraphics[width=0.33\textwidth]{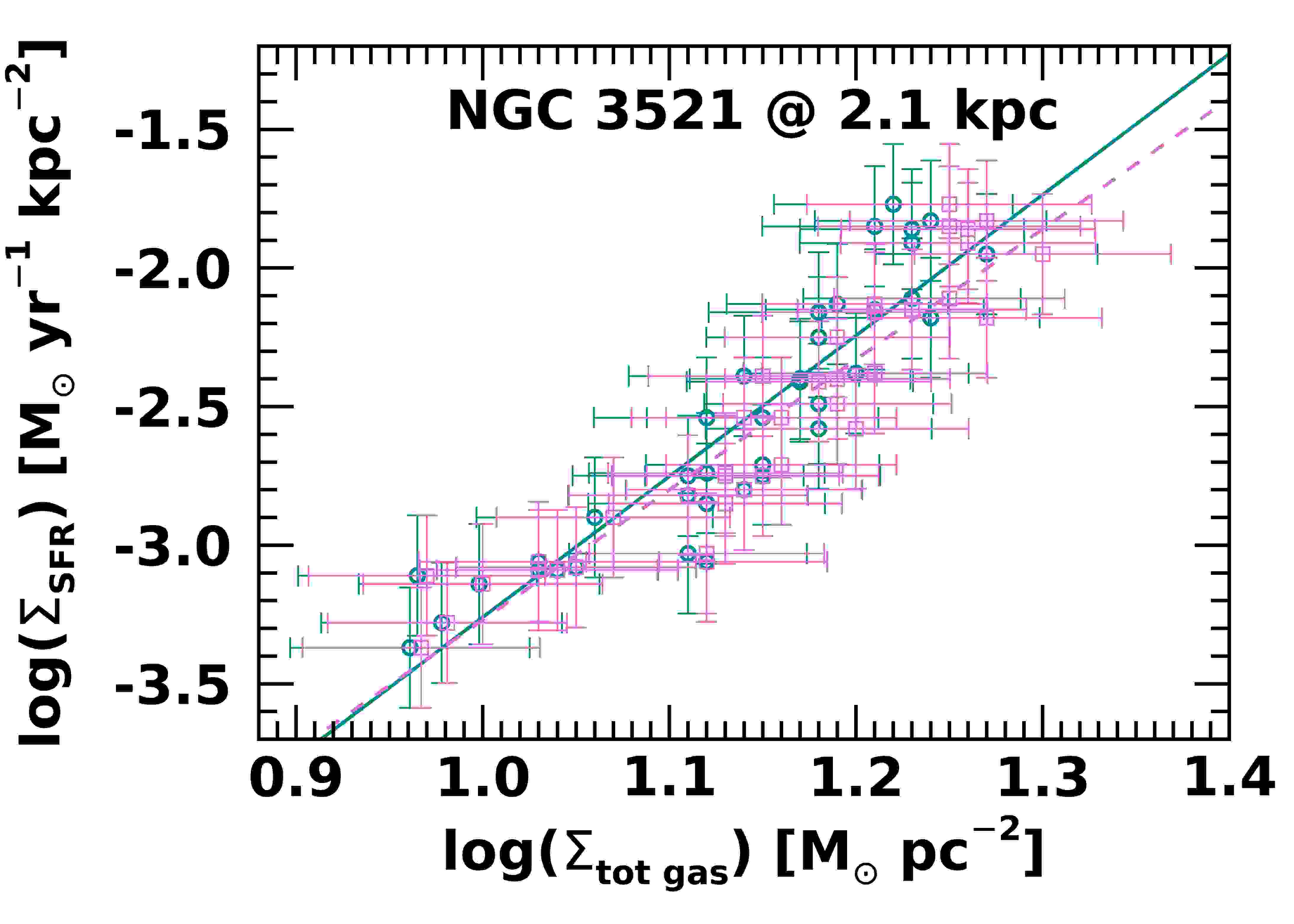}
\includegraphics[width=0.33\textwidth]{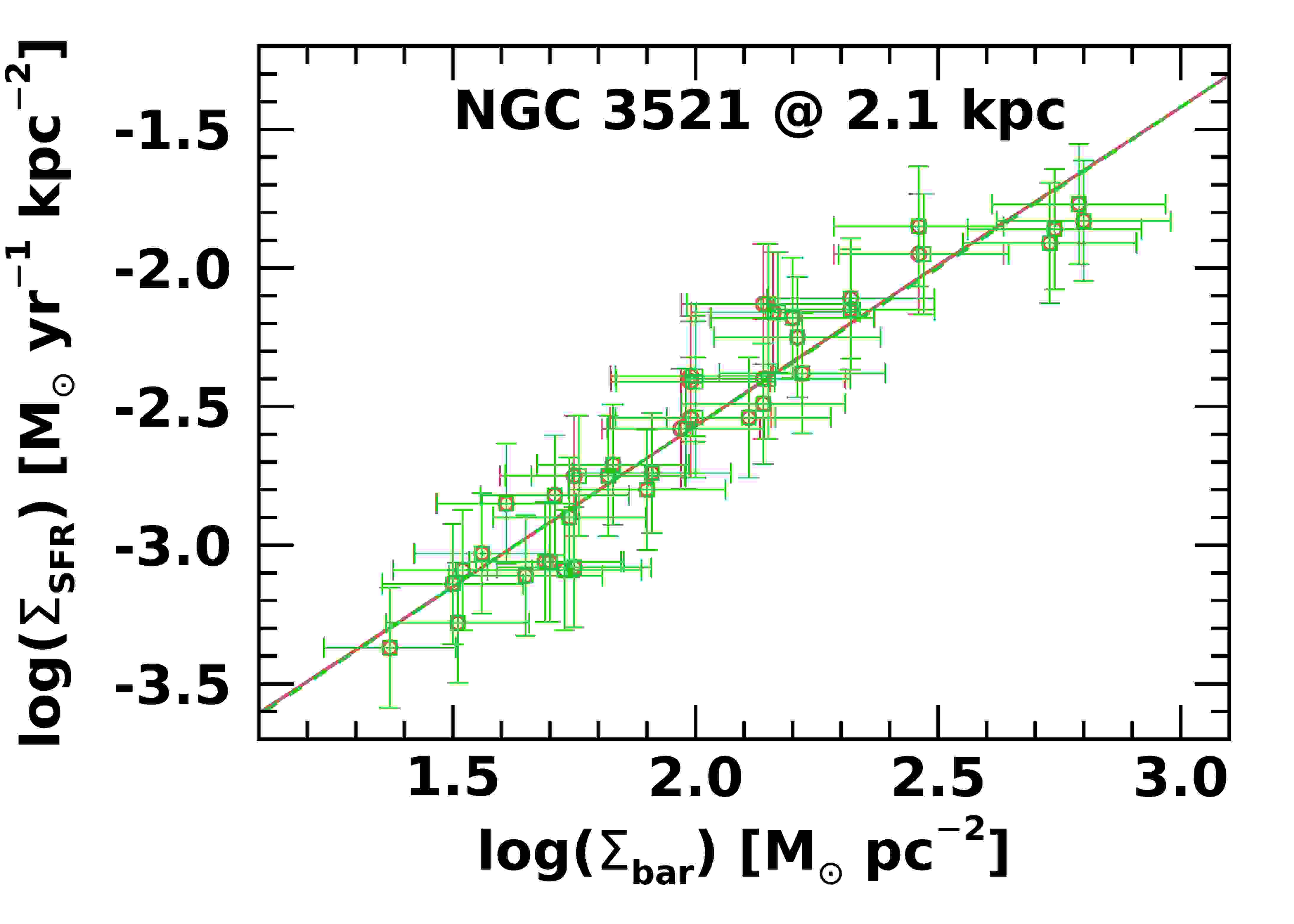}
\includegraphics[width=0.33\textwidth]{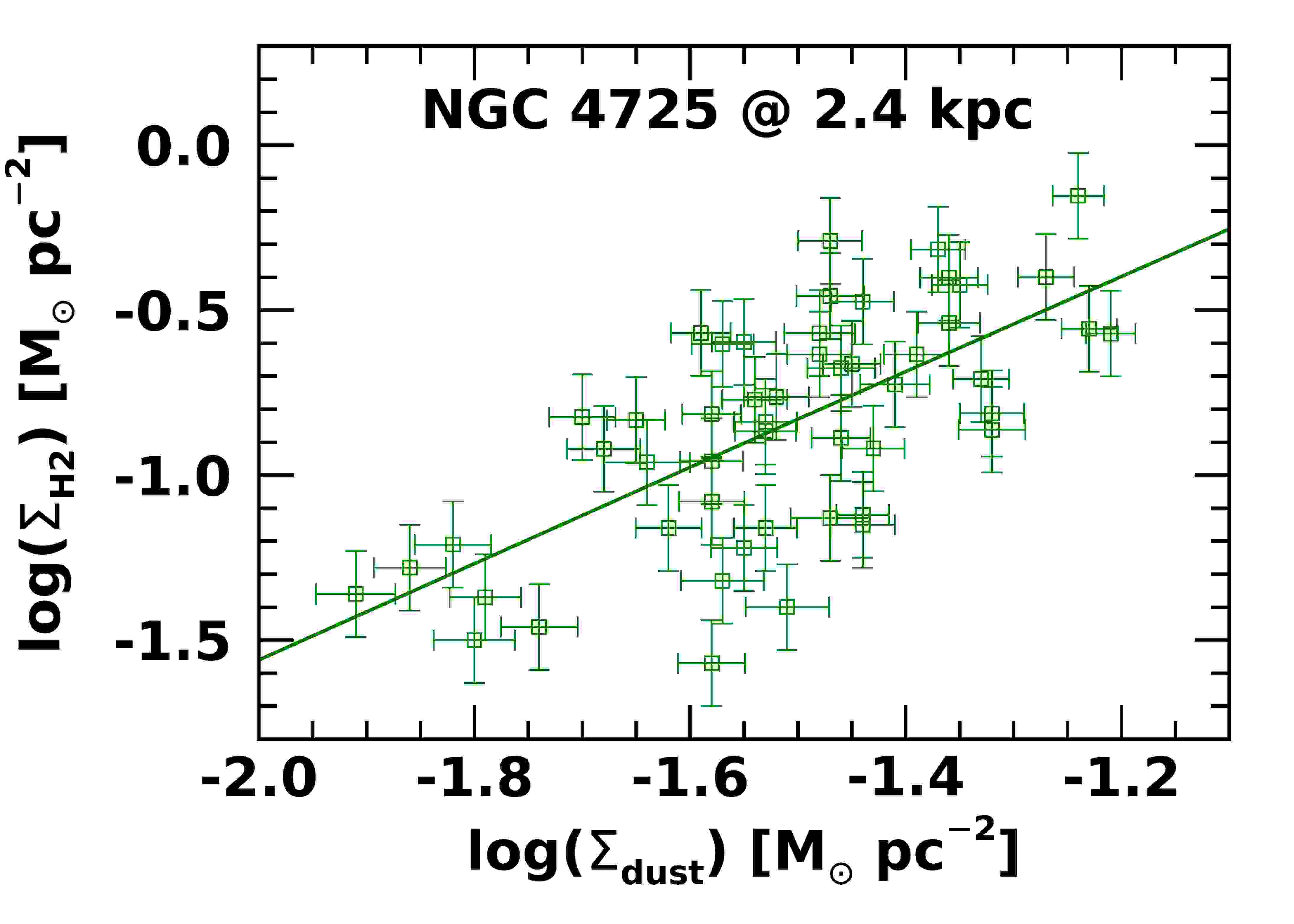}
\includegraphics[width=0.33\textwidth]{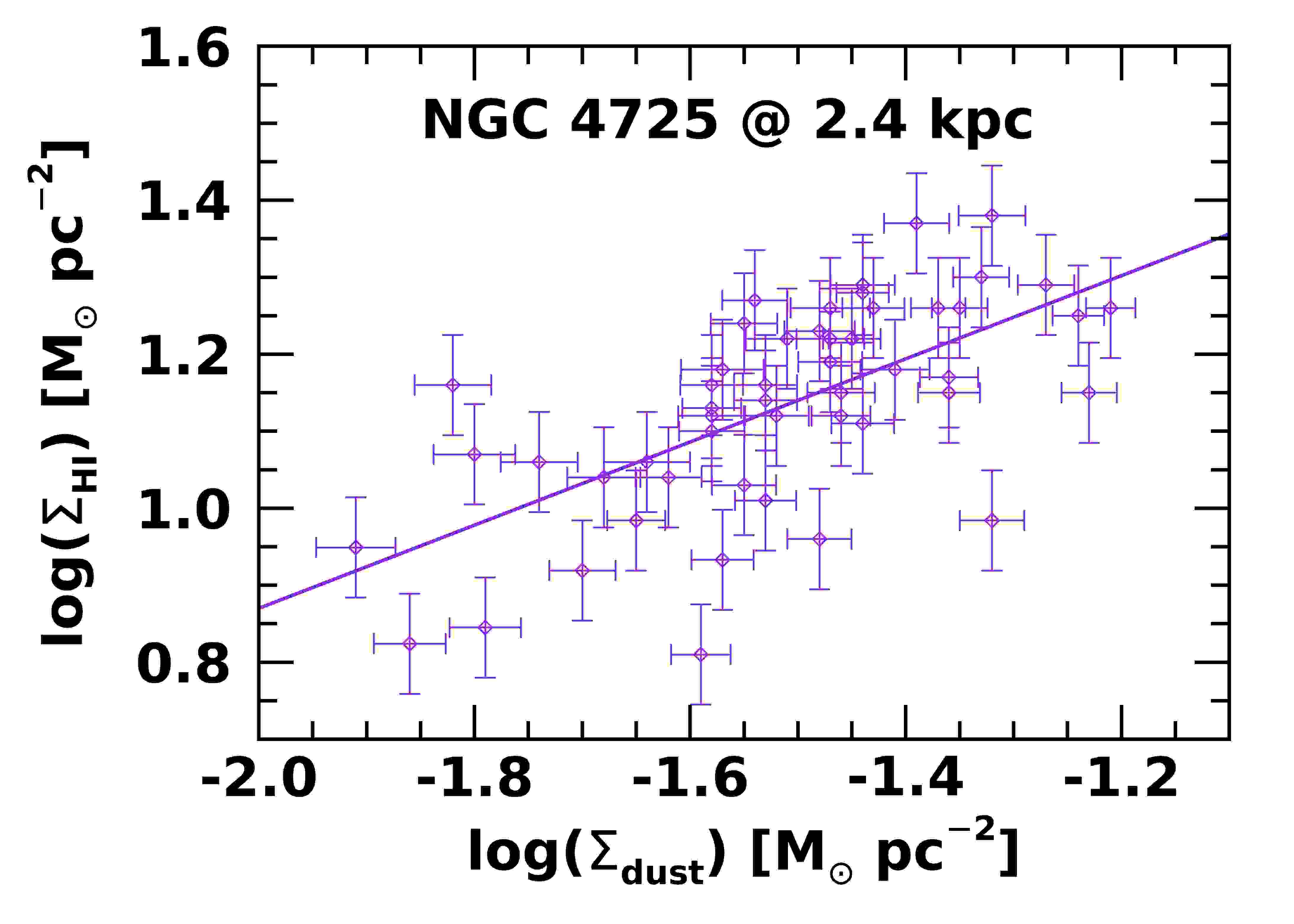}
\includegraphics[width=0.33\textwidth]{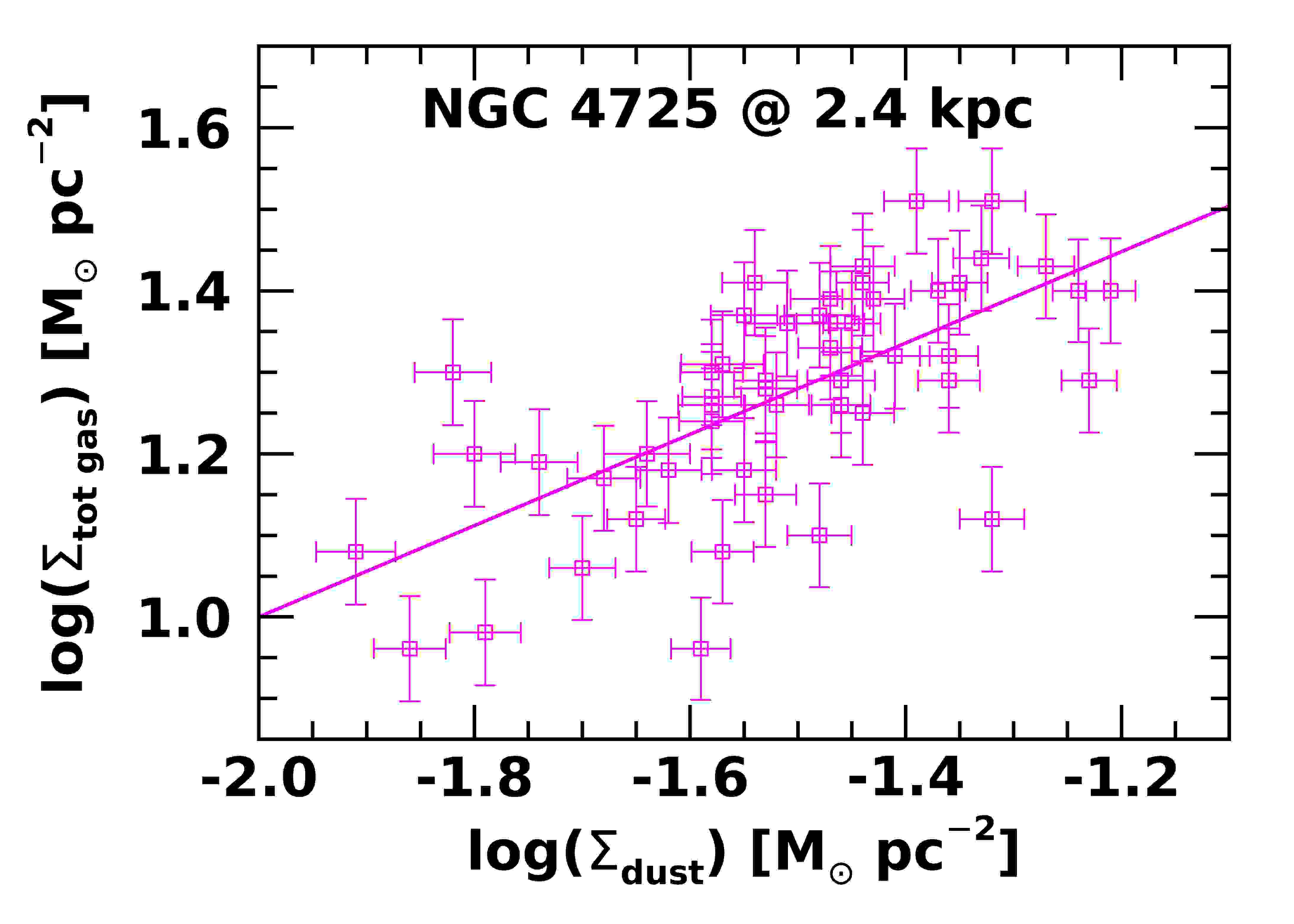}
\caption*{Figure~\ref{fig:add-ism} continued}
\end{figure*}

\begin{figure*}
\centering
\includegraphics[width=0.33\textwidth]{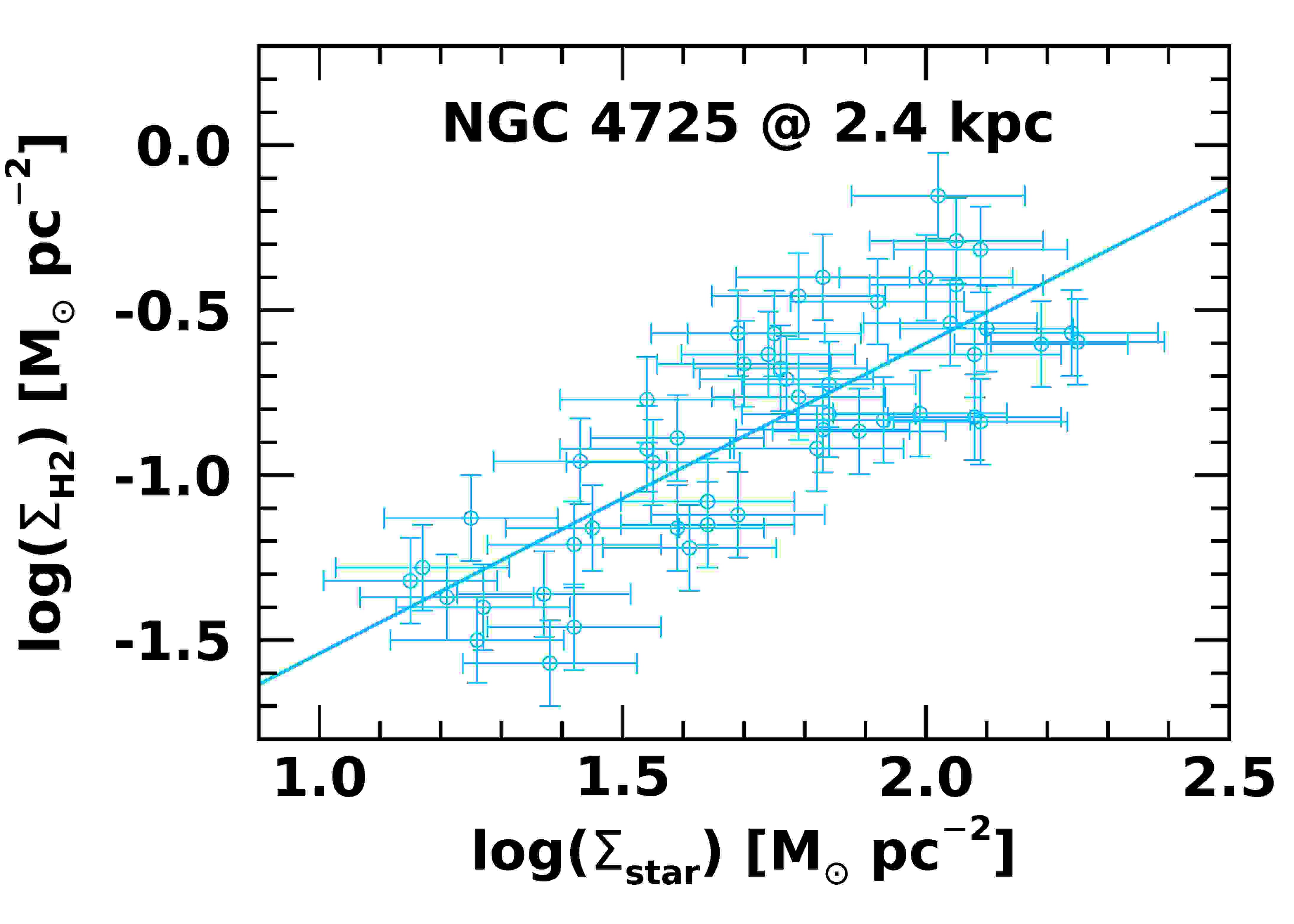}
\includegraphics[width=0.33\textwidth]{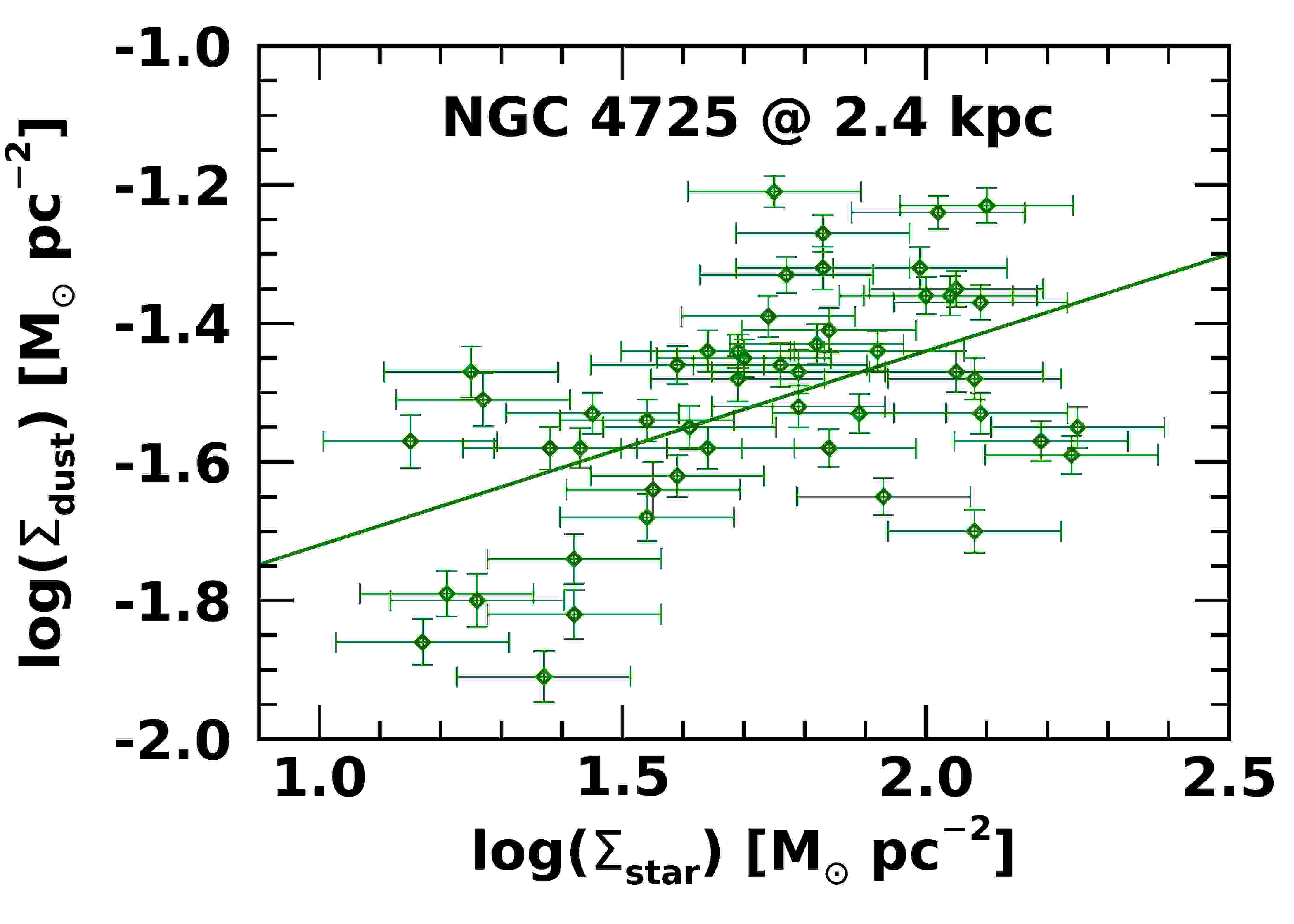}
\includegraphics[width=0.33\textwidth]{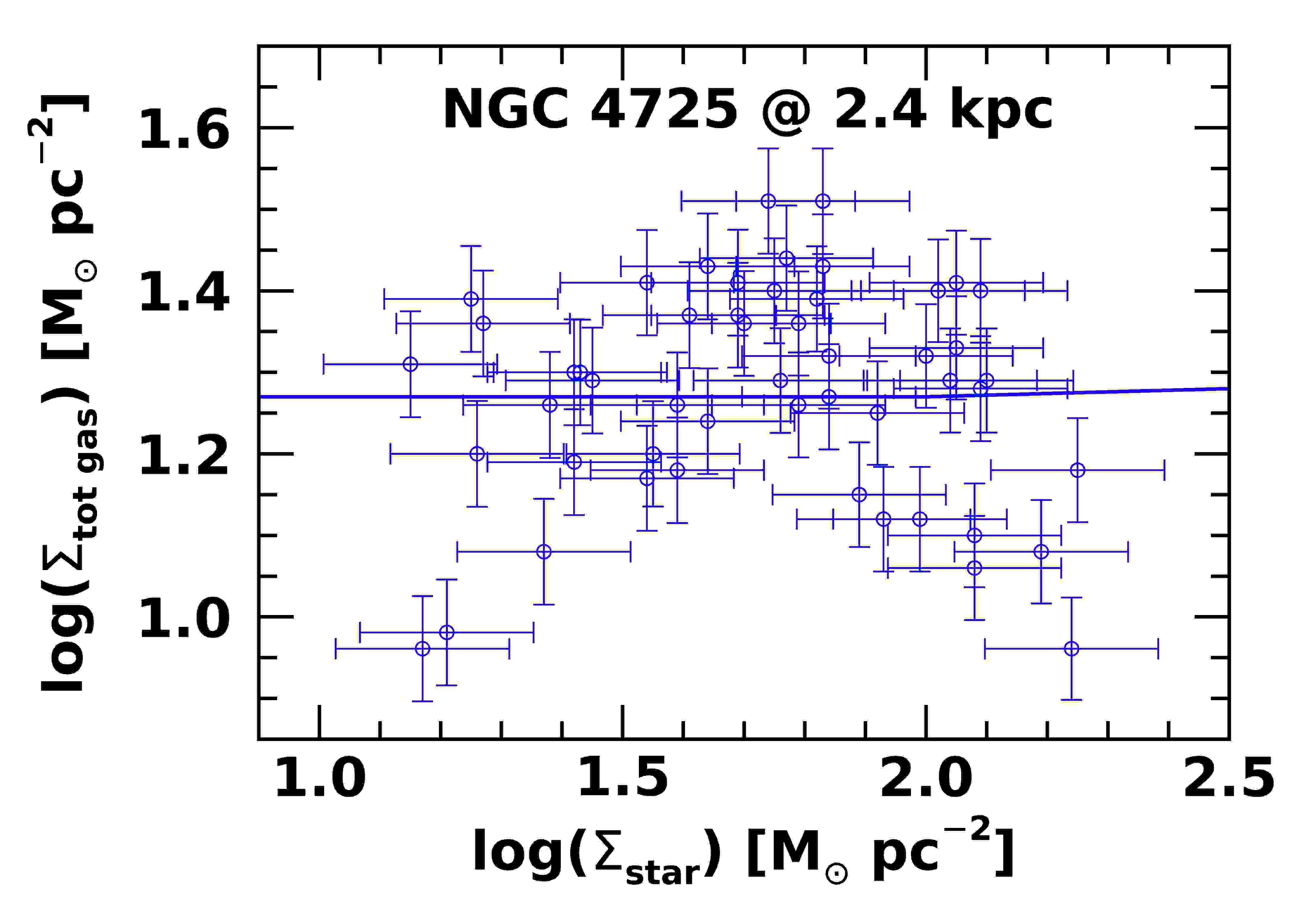}
\includegraphics[width=0.33\textwidth]{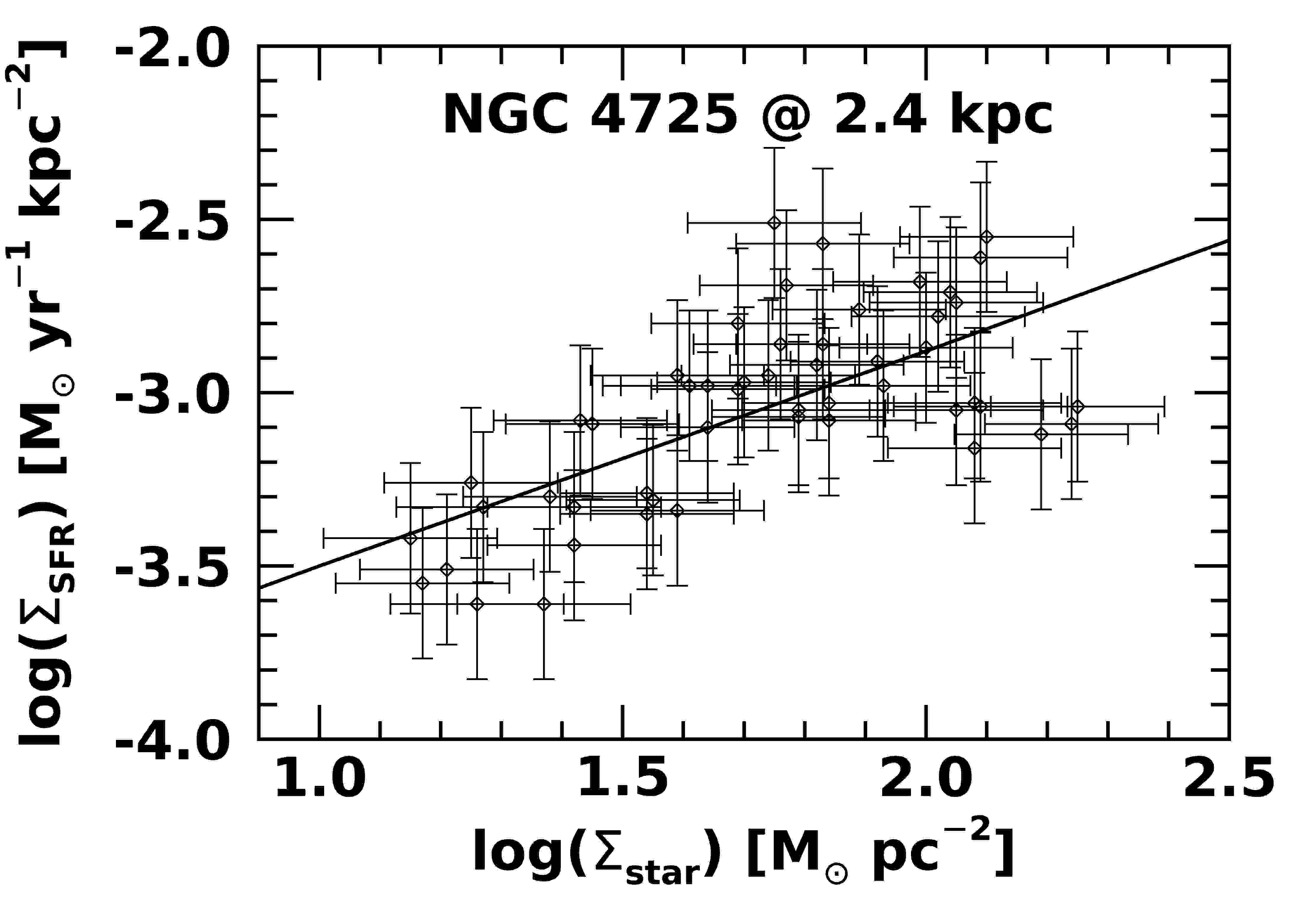}
\includegraphics[width=0.33\textwidth]{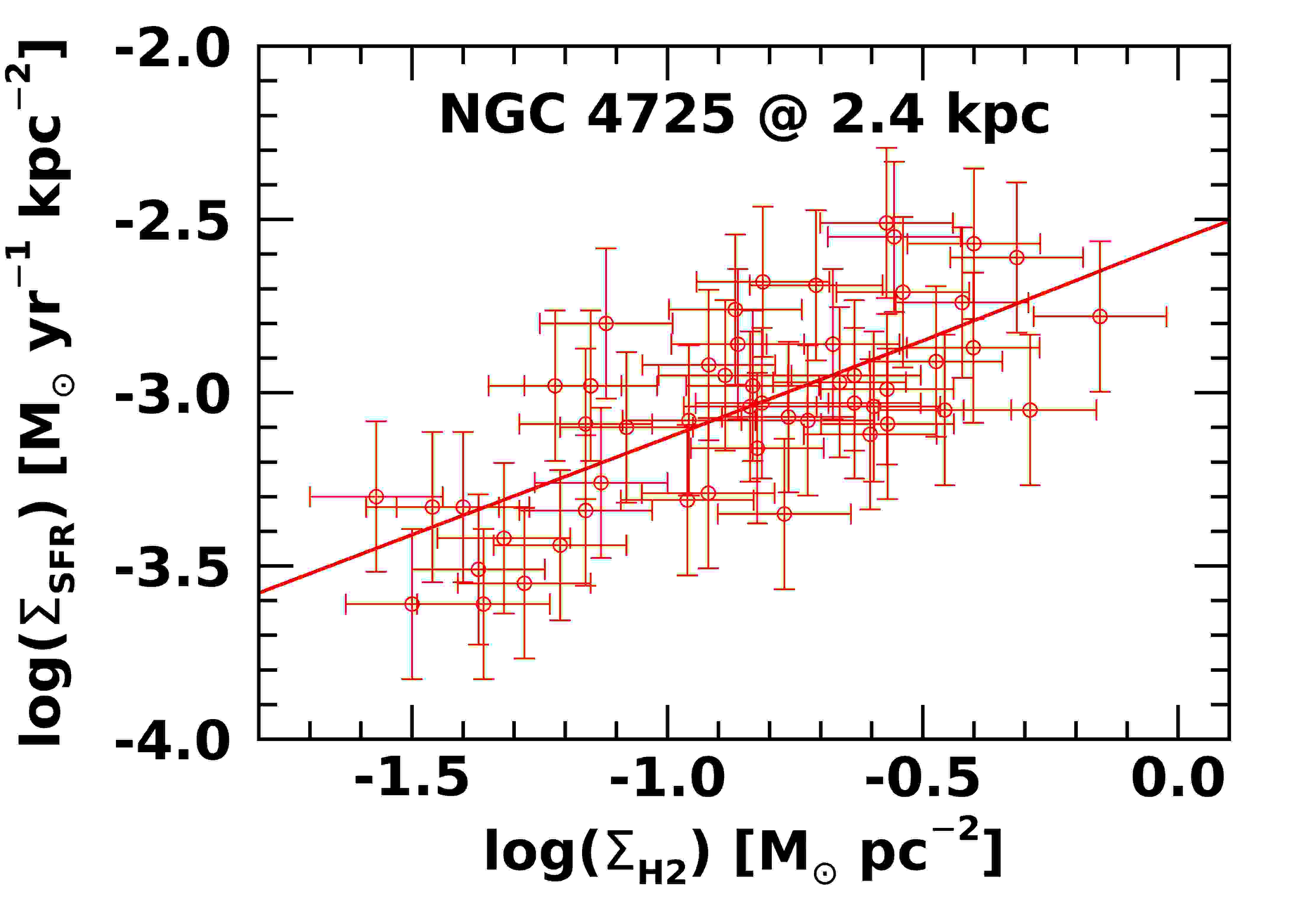}
\includegraphics[width=0.33\textwidth]{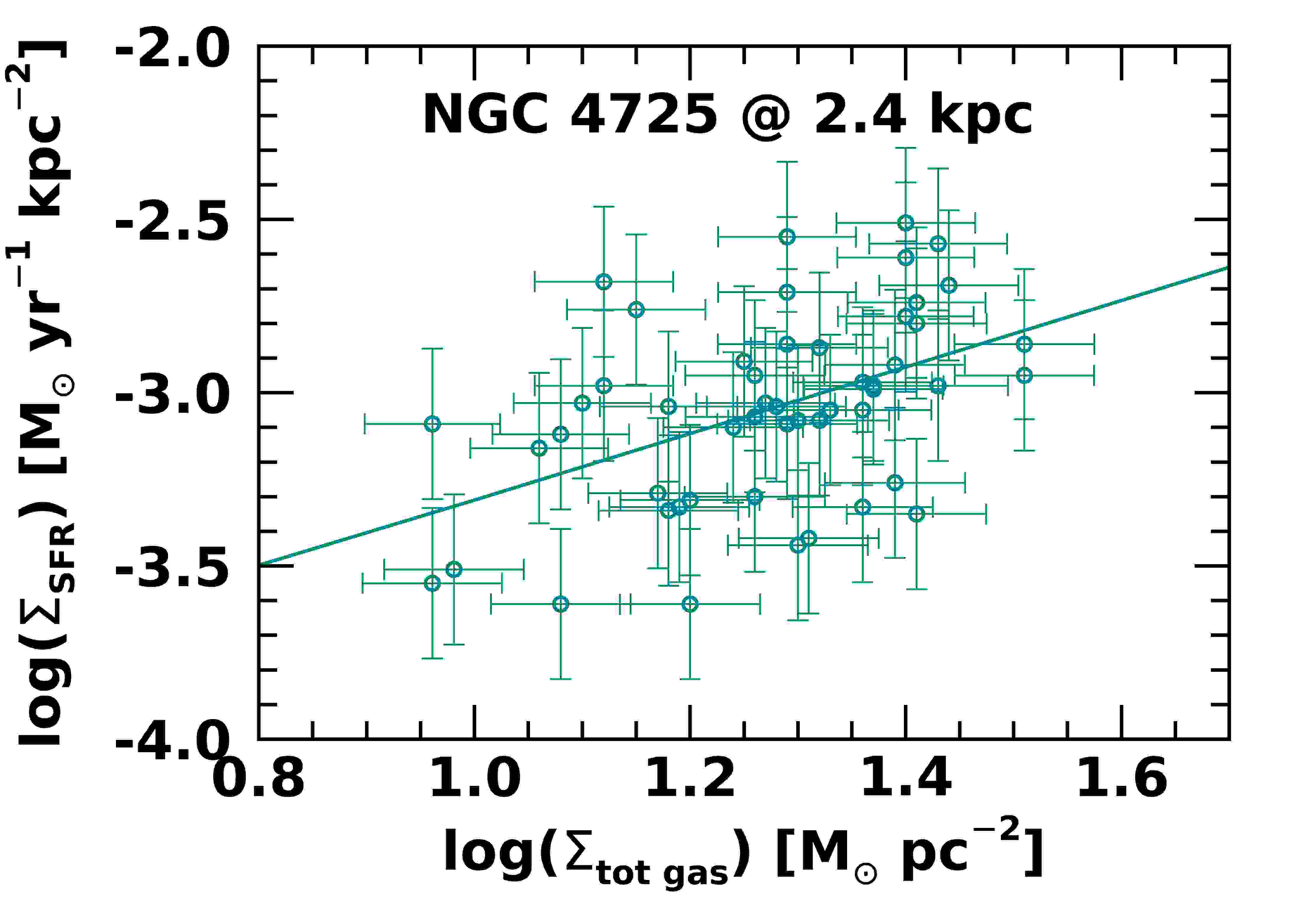}
\includegraphics[width=0.33\textwidth]{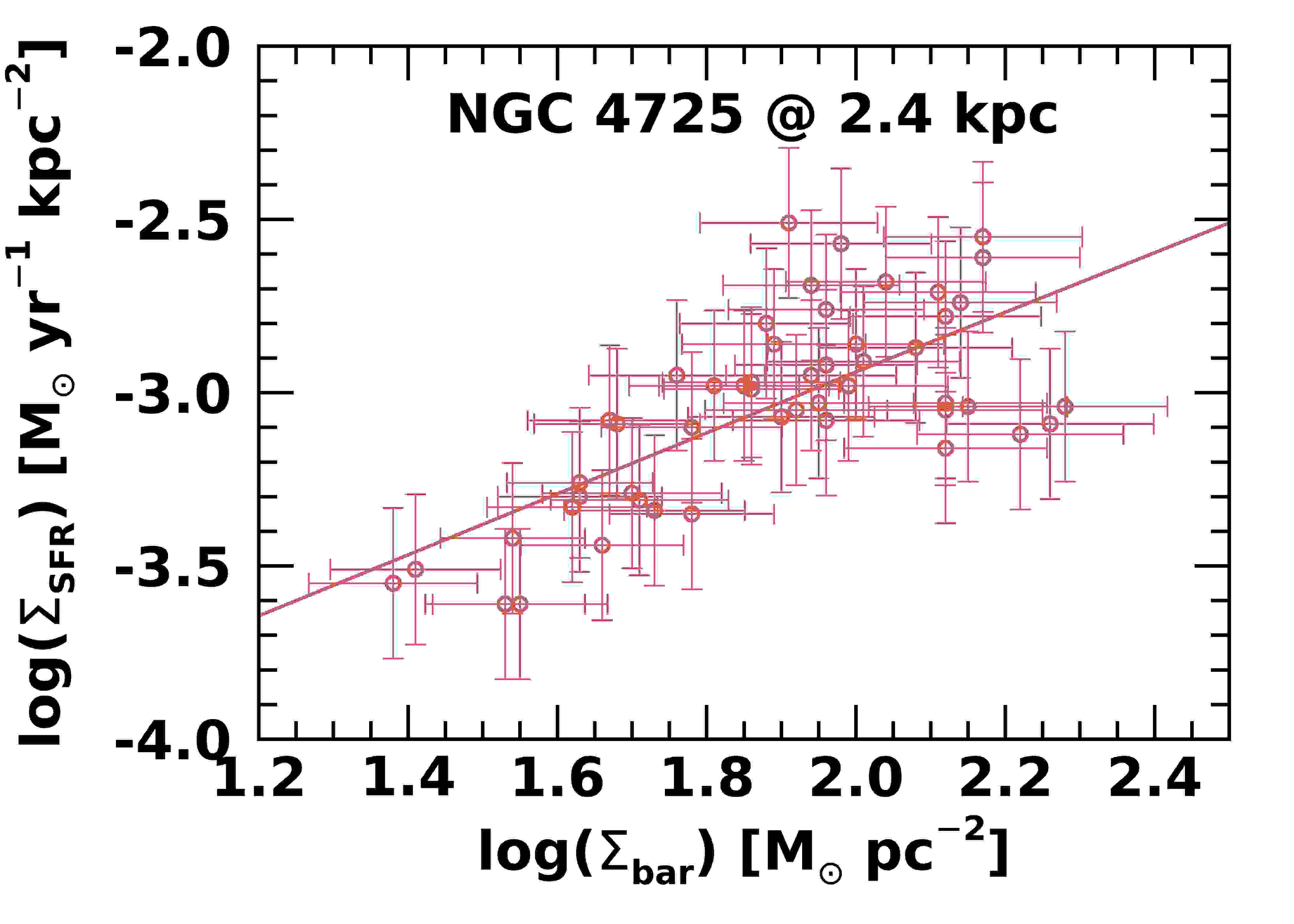}
\includegraphics[width=0.33\textwidth]{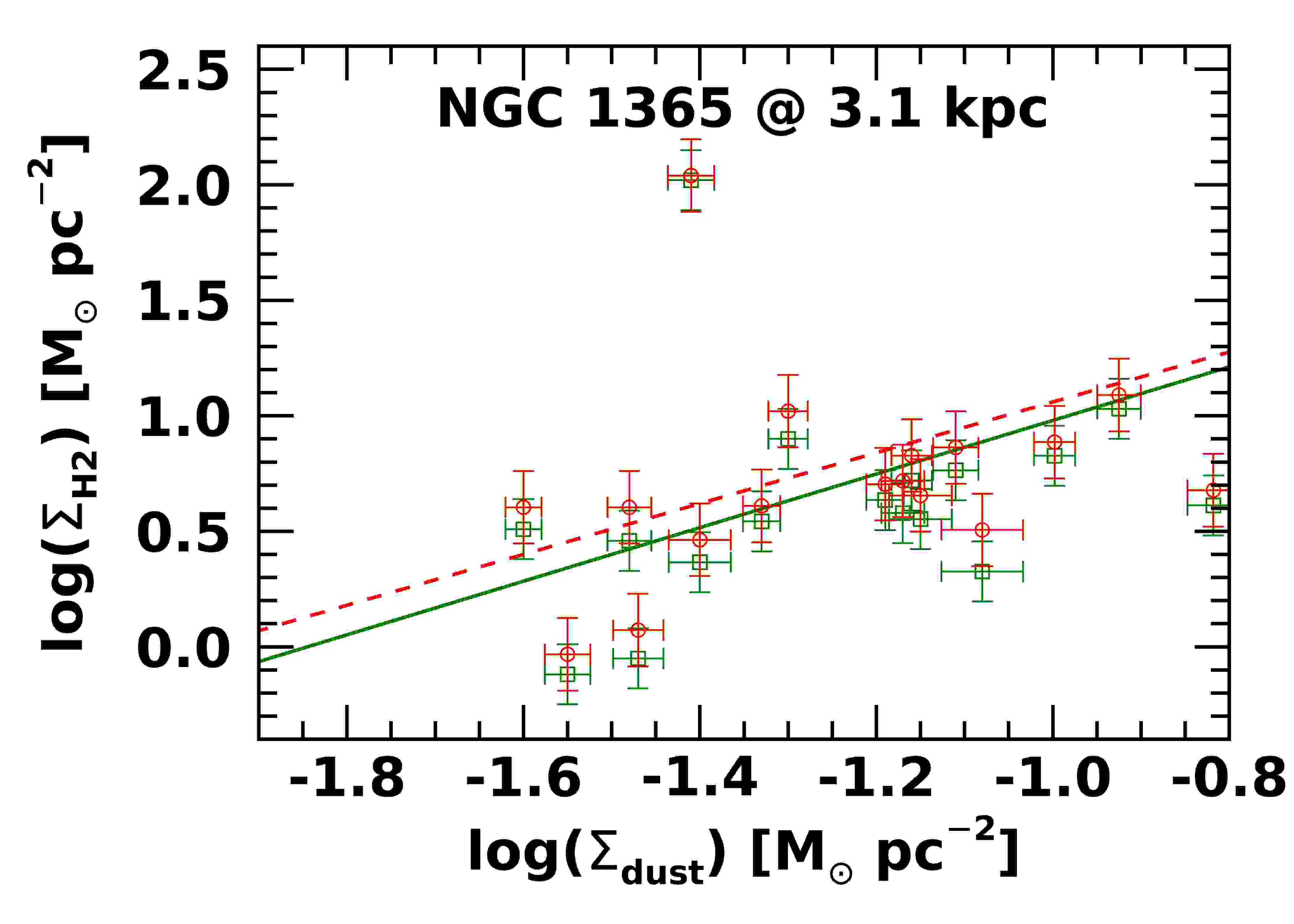}
\includegraphics[width=0.33\textwidth]{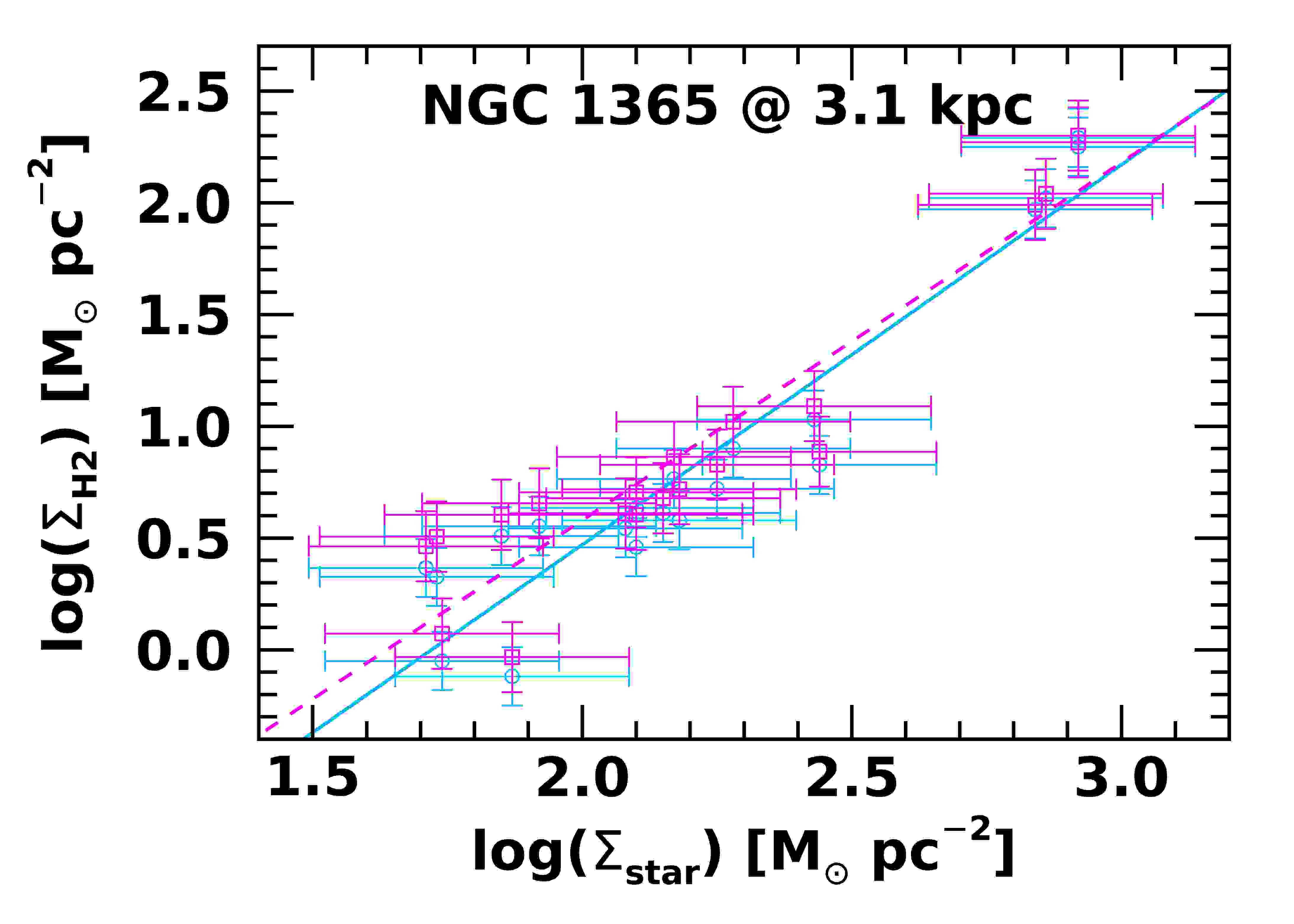}
\includegraphics[width=0.33\textwidth]{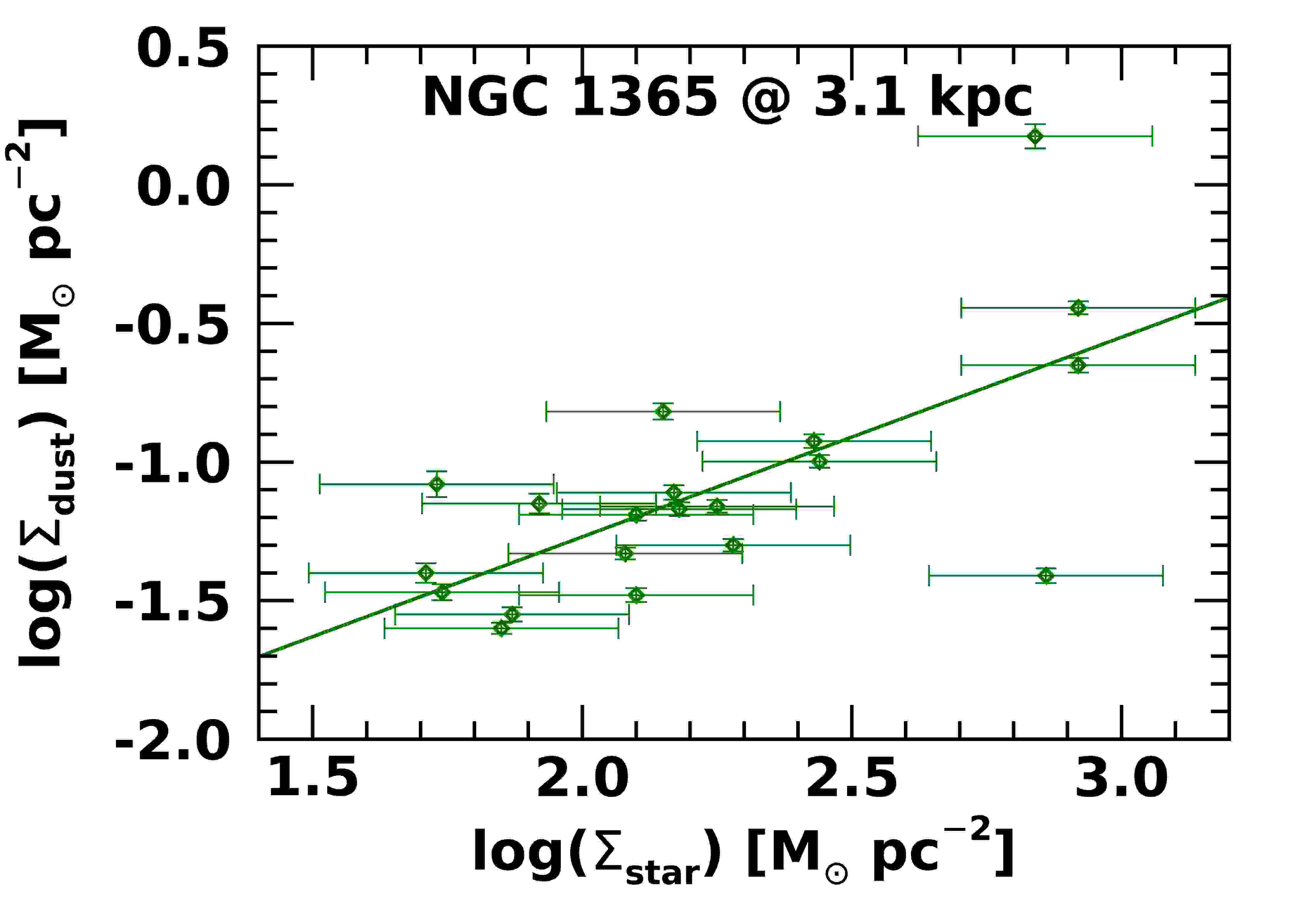}
\includegraphics[width=0.33\textwidth]{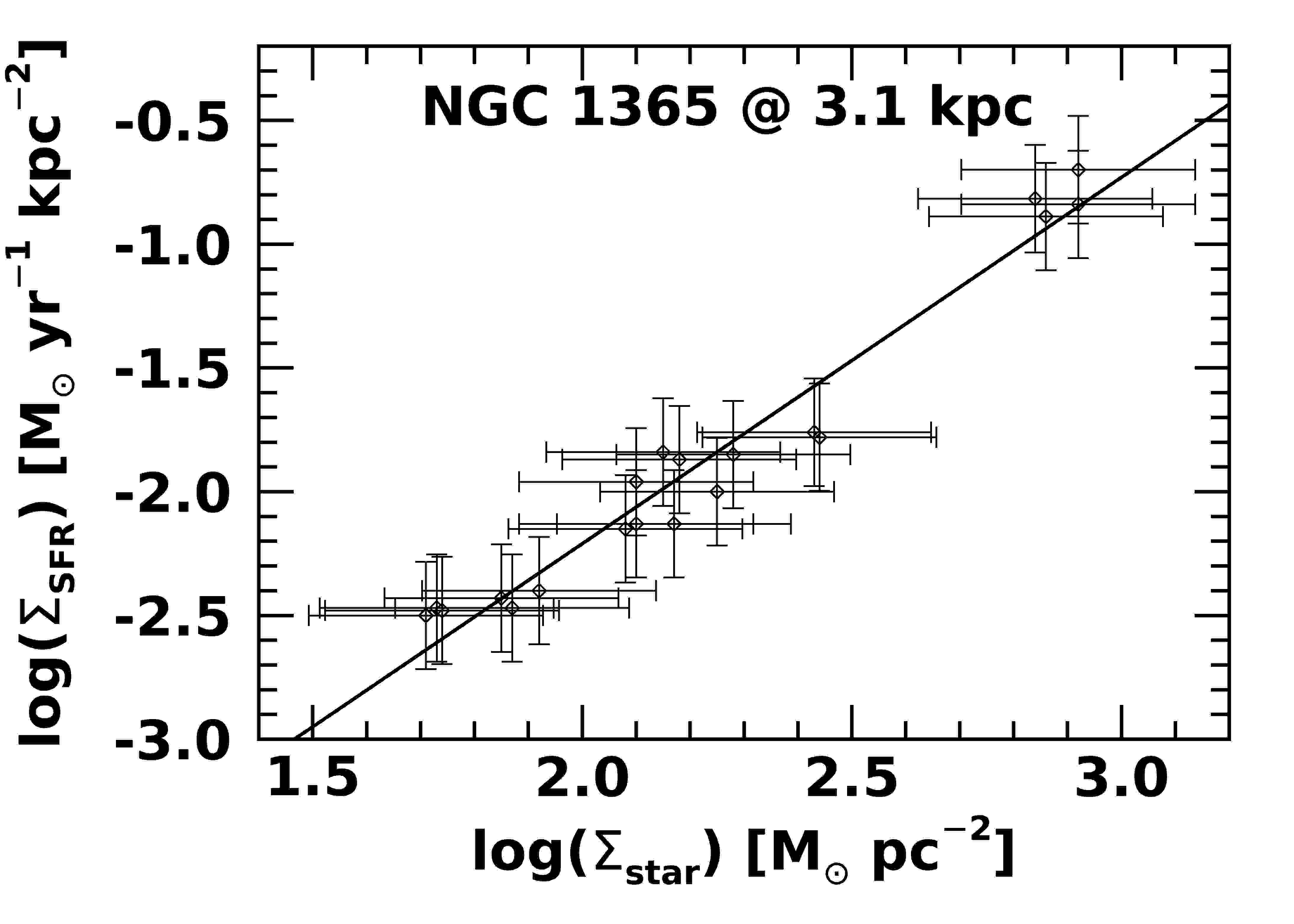}
\includegraphics[width=0.33\textwidth]{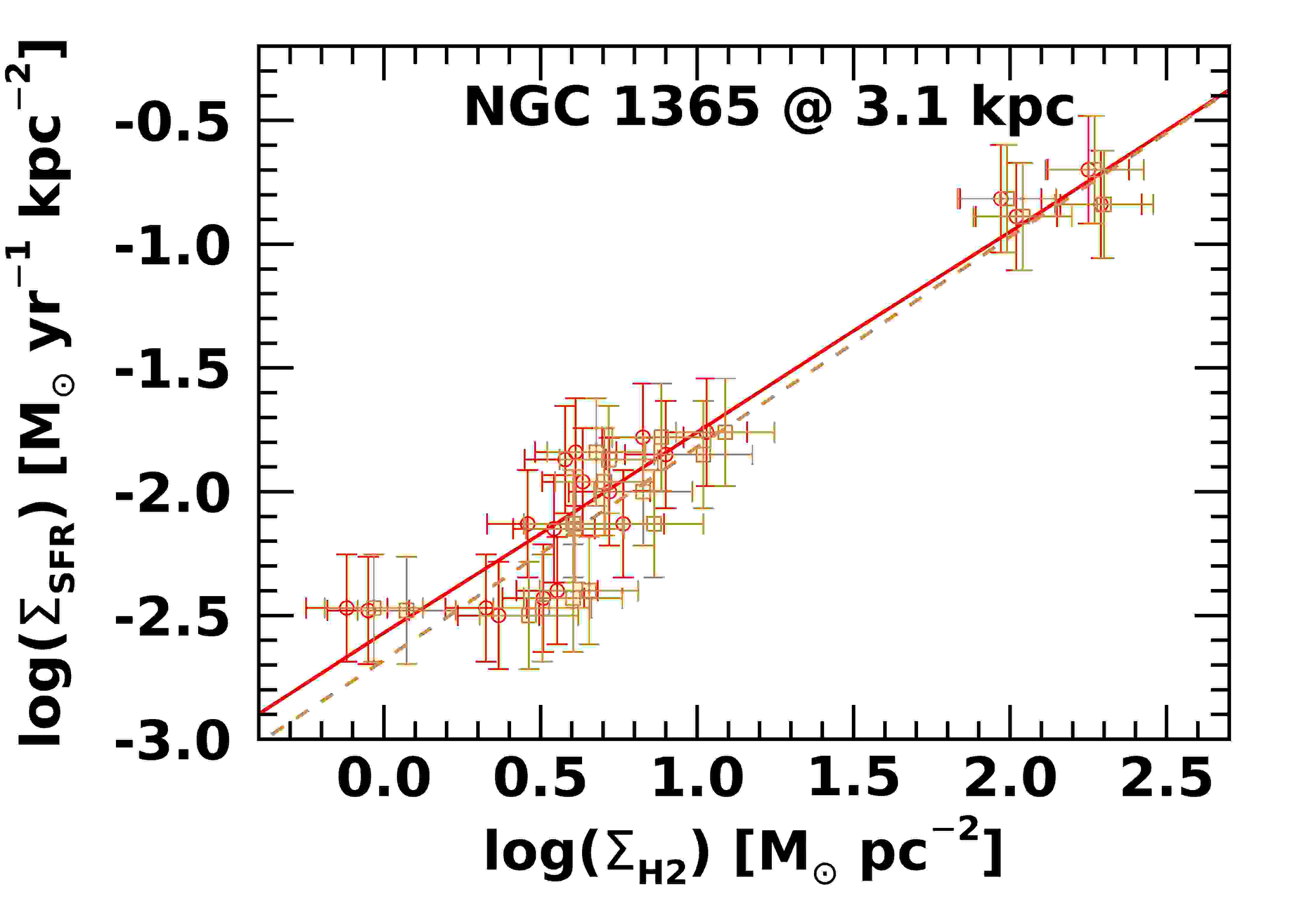}
\includegraphics[width=0.33\textwidth]{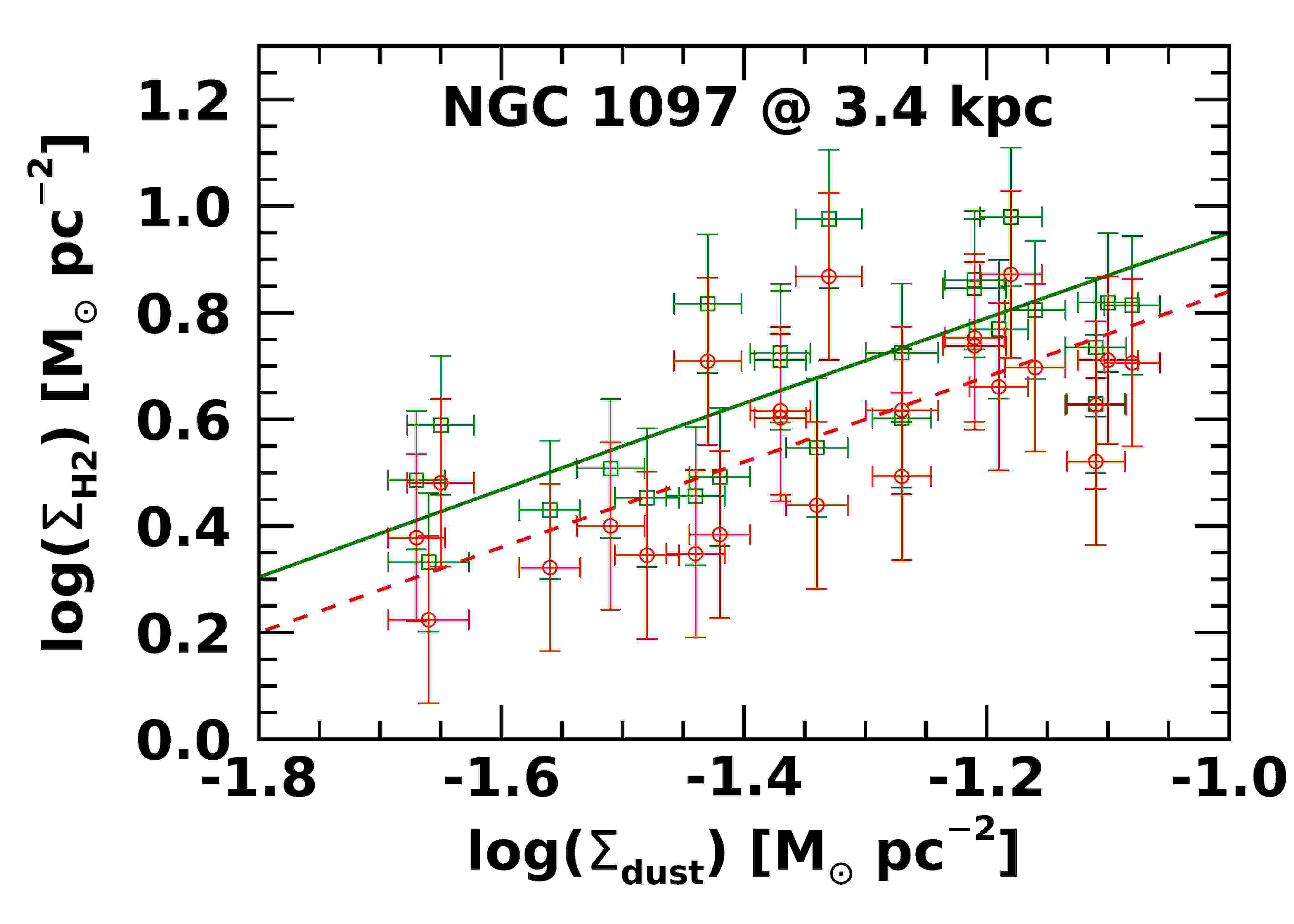}
\includegraphics[width=0.33\textwidth]{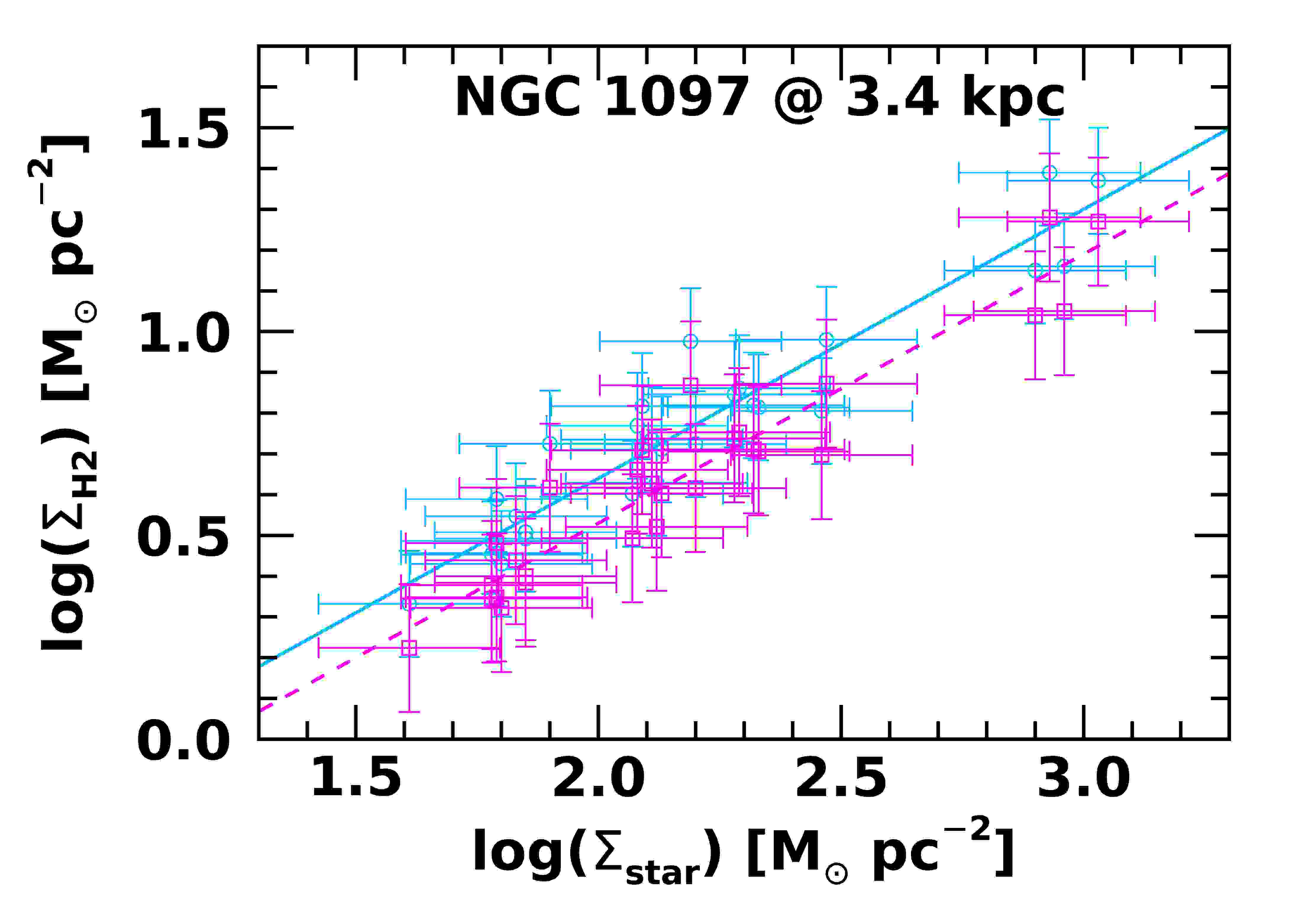}
\includegraphics[width=0.33\textwidth]{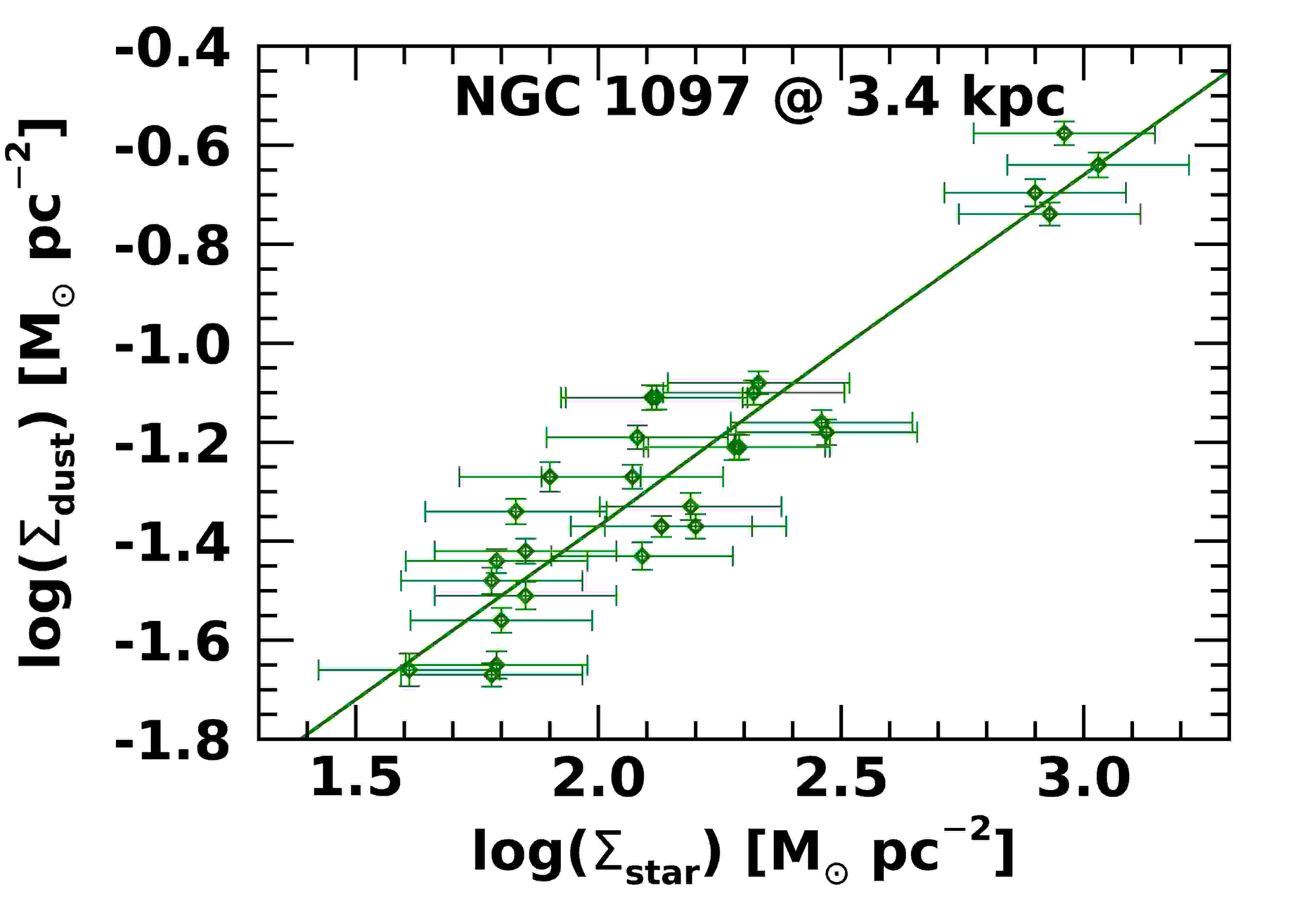}
\caption*{Figure~\ref{fig:add-ism} continued}
\end{figure*}

\begin{figure*}
\centering
\includegraphics[width=0.33\textwidth]{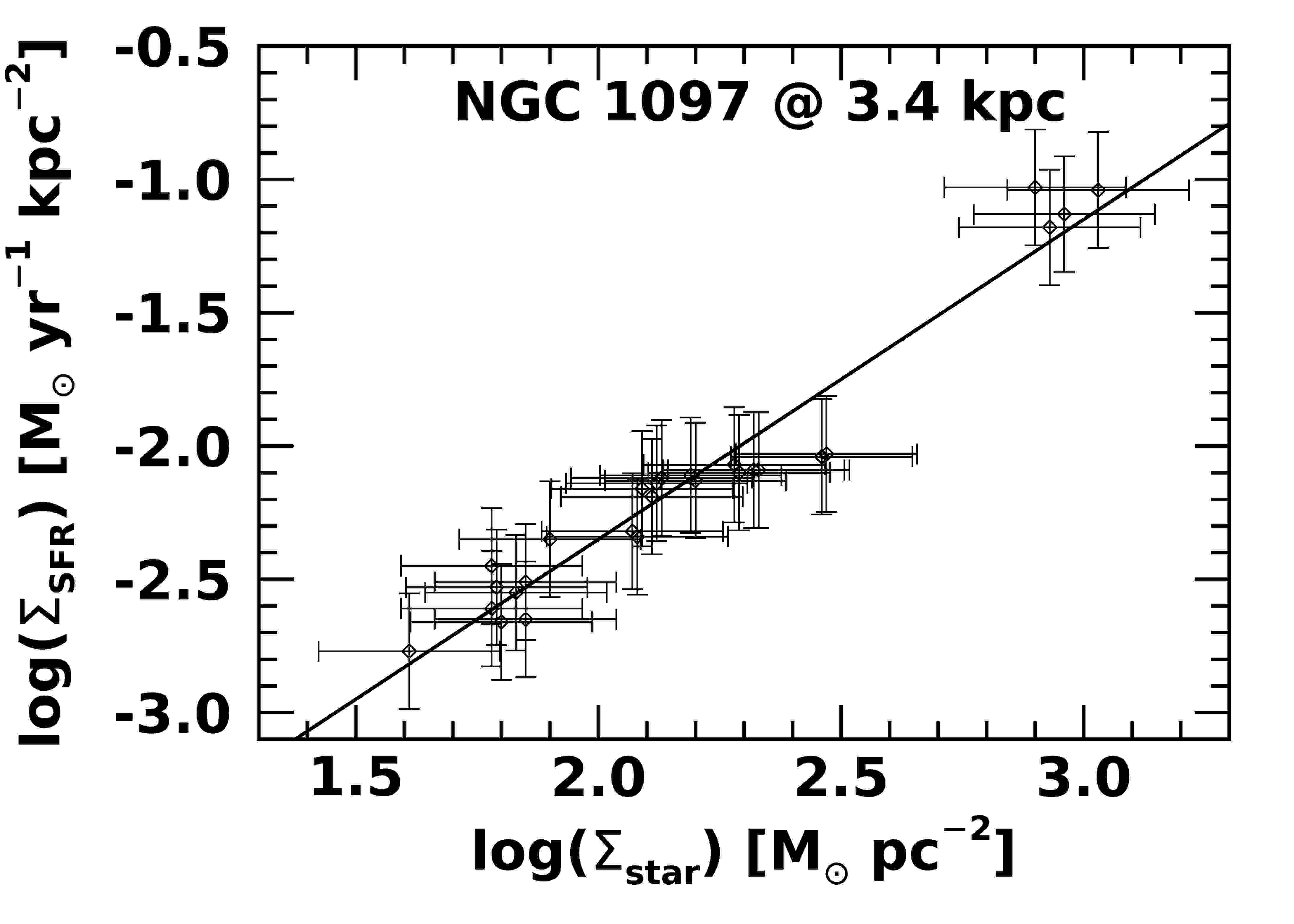}
\includegraphics[width=0.33\textwidth]{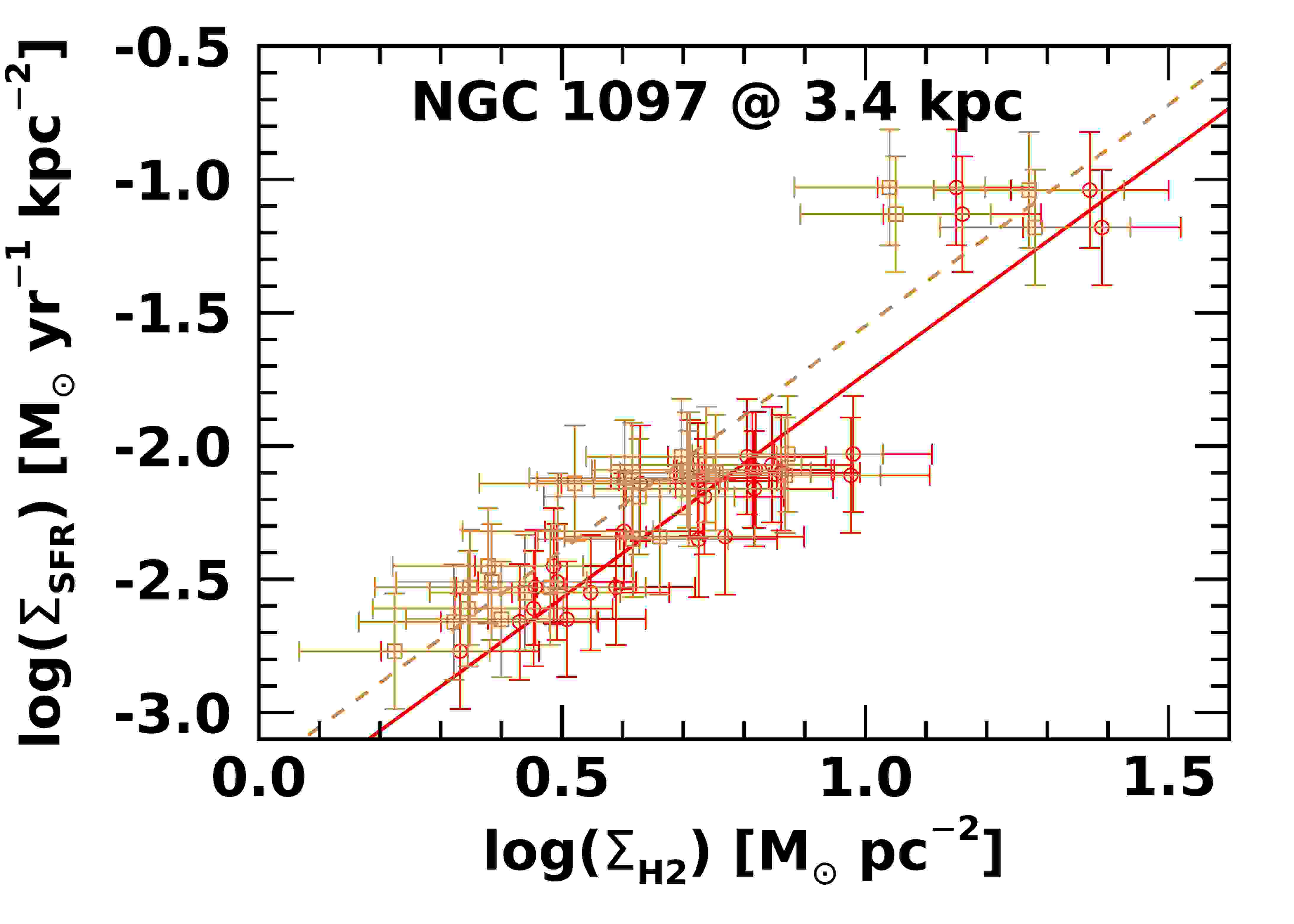}
\caption*{Figure~\ref{fig:add-ism} continued}
\end{figure*}

\onecolumn
\longtab{
\begin{landscape}
\begin{longtable}{lcccccc}
\caption{Parameters of the linear fits applied to the pixel-by-pixel SRs for each sample galaxy at the physical scale imposed by the dust map and at 3.4~kpc,
for all galaxies evaluated together at the common scale of 3.4~kpc, and the median slopes of the SRs at 3.4~kpc presented in Sect.~\ref{sec:sr}.} \label{tab:fit-sr}\\
\hline
\hline
Galaxy	& Scaling relation ($x - y$)$^{(1)}$	& Physical scale $^{(2)}$ & $m$ $^{(3)}$ & $q$ $^{(3)}$ & $R$ ($\sigma$, n. pts) $^{(3)}$ & $X_{\rm CO}$ $^{(4)}$  \\
		&							& [kpc]\\
\hline
\endfirsthead
\caption{continued}\\
\hline
Galaxy	& Scaling relation ($x - y$)$^{(1)}$	& Physical scale $^{(2)}$ & $m$ $^{(3)}$ & $q$ $^{(3)}$ & $R$ ($\sigma$, n. pts) $^{(3)}$ & $X_{\rm CO}$ $^{(4)}$  \\
		&							& [kpc]\\
\hline
\endhead
\hline
\endfoot
\hline
\endlastfoot
NGC~300             		& ${\rm log}(\Sigma{_{\rm star}})-{\rm log}(\Sigma{_{\rm dust}})$	& 0.3		& $0.16\pm0.04$ & $-1.72\pm0.06$ & 0.24 (0.12, 283) 	& -- 	\\
					& ${\rm log}(\Sigma{_{\rm star}})-{\rm log}(\Sigma{_{\rm SFR}})$ 	& 0.3		& $1.20\pm0.06$ & $-4.91\pm0.11$ & 0.74 (0.20, 283) 	& -- \\
					& ${\rm log}(\Sigma{_{\rm star}})-{\rm log}(\Sigma{_{\rm dust}})$	& 3.4		& $0.17\pm0.06$ & $-1.58\pm0.08$ & 0.63 (0.07, 16) 	& -- 	\\
					& ${\rm log}(\Sigma{_{\rm star}})-{\rm log}(\Sigma{_{\rm SFR}})$ 	& 3.4		& $1.08\pm0.33$ & $-4.15\pm0.46$ & 0.66 (0.40, 16) 	& -- \\										
\hline
IC~342              		& ${\rm log}(\Sigma{_{\rm dust}})-{\rm log}(\Sigma{_{\rm H2}})$ 	& 0.5 	& $1.43\pm0.04$ & $2.24\pm0.05$ & 0.86 (0.15, 353) 	& Const.   \\
					& ${\rm log}(\Sigma{_{\rm dust}})-{\rm log}(\Sigma{_{\rm H2}})$		& 0.5 	& $1.20\pm0.04$ & $2.16\pm0.04$ & 0.85 (0.14, 353) 	& $Z$--dep.\\
					& ${\rm log}(\Sigma{_{\rm dust}})-{\rm log}(\Sigma{_{\rm HI}})$ 		& 0.5 	& $0.09\pm0.07$ & $0.63\pm0.07$ & 0.07 (0.23, 353) 	& -- \\
					& ${\rm log}(\Sigma{_{\rm dust}})-{\rm log}(\Sigma{_{\rm tot gas}})$ 	& 0.5 	& $0.99\pm0.03$ & $2.17\pm0.03$ & 0.89 (0.09, 353) 	& Const. \\
\end{longtable}
\tablefoot{
$^{(1)}$ Explored scaling relation.
$^{(2)}$ Physical scale imposed by the angular resolution of the surface density dust mass map (36\arcsec) 
at the galaxy distance or at the common scale of 3.4~kpc (see Sect.~\ref{sec:treatment}).
$^{(3)}$ Slope $m$, intercept $q$, Pearson correlation coefficient $R$, dispersion $\sigma$, and number of pixels corresponding to Eq.~(\ref{eq:fit}).
$^{(4)}$ The two assumptions on $X_{\rm CO}$: ``Const.'' corresponds to constant $X_{\rm CO}$, ``$Z$--dep.'' to metallicity-dependent $X_{\rm CO}$ (see Sect.~\ref{sec:gas}).
}
\end{landscape}
}
\twocolumn

\section{Peculiar galaxies in scaling relations}
\label{sec:peculiar}
In this section we provide details on galaxies showing different behaviors from most galaxies in terms of SRs.
We present these galaxies based on the studied SRs.

\subsection{$\Sigma_{\rm dust}$--$\Sigma_{\rm H2}$}
NGC~925 and NGC~1097 present the $\Sigma_{\rm dust}$--$\Sigma_{\rm H2}$ SR with 
linear and sublinear slope, respectively.
NGC~925 has a dust-to-H$_2$ mass ratio within $r_{25}$ approximately consistent within 1$\sigma$ with the mean value 
computed for galaxies of the same morphological stage (T~=~7, C20).
The dust-to-H$_2$ mass ratio of NGC~925 is not therefore able to explain the lower (linear) slope of the 
$\Sigma_{\rm dust}$--$\Sigma_{\rm H2}$ SR, that is close to the lower limit of the slopes of the other sample galaxies.   
NGC~1097 has instead a very much lower dust-to-H$_2$ mass ratio than the mean ratio of galaxies of the same morphological stage 
(T~=~3, C20), and this could explain the sublinear slope in the $\Sigma_{\rm dust}$--$\Sigma_{\rm H2}$ SR of NGC~1097.

\subsection{$\Sigma_{\rm dust}$--$\Sigma_{\rm HI}$}
The $\Sigma_{\rm dust}$--$\Sigma_{\rm HI}$ SR is strong for NGC~7793 and NGC~4736 and moderate for NGC~5194, NGC~5055, 
NGC~3521, and NGC~4725.
These galaxies have dust-to-\hi\ mass ratios a factor $\sim$0.3--0.6 times lower than mean ratios for galaxies of the 
same morphological stages (C20).
This could indicate a peculiar overabundance of \hi, with respect to dust, that correlates well with the dust itself,
within $r_{25}$.
The pixel-by-pixel SR between dust and \hi\ in NGC~7793 has been recently studied by \citet{saikia20}
finding a weaker correlation than ours. 
This discrepancy could be due to a mixture of different factors. 
\citet{saikia20} studied the dust-\hi\ correlation at the angular resolution imposed by the \hi\ map of 15\farcs6~$\times$10\farcs8 
($\sim$0.3~kpc~$\times$~0.2~kpc), higher than ours, and they used different, individually treated, IR tracers of dust 
(emissions at 8, 24, 70, 100, 160~$\mu$m) and not an estimation of the dust mass.
They found a trend in the variation of dust-\hi\ SRs with individual IR bands, and this is particularly evident
for NGC~7793 where the values of the correlation coefficient increase drastically with an increase 
in IR emission wavelength.  

\subsection{$\Sigma_{\rm star}$--$\Sigma_{\rm H2}$}
NGC~925 and NGC 1365 are two outlier cases in the $\Sigma_{\rm star}$--$\Sigma_{\rm H2}$ SR with  slope sublinear and superlinear, respectively. 
For NGC~925 this SR breaks down under the assumption of metallicity-dependent $X_{\rm CO}$.
We note that NGC~925 is an outlier case also in the $\Sigma_{\rm dust}$--$\Sigma_{\rm H2}$ SR.
This galaxy has a star-to-H$_2$ mass ratio within $r_{25}$ a factor $1.5$ times higher than (although consistent within 1$\sigma$ with) 
the corresponding mean value for galaxies of the same morphological stage (T~=~7, data from DustPedia database).  
This could explain the particularly low slope of the $\Sigma_{\rm star}$--$\Sigma_{\rm H2}$ SR in NGC~925. 
Also, the particularly high slope of NGC~1365 is likely explainable by its star-to-H$_2$ mass ratio, an order of magnitude 
lower than the corresponding mean value for galaxies of the same morphological stage (T~=~3, data from DustPedia database).

\subsection{$\Sigma_{\rm star}$--$\Sigma_{\rm dust}$}
For NGC~300 the $\Sigma_{\rm star}$--$\Sigma_{\rm dust}$ SR could break down for different reasons such as
$i)$ the physical scale under analysis, $ii)$ an anomalous distribution of dust mass, $iii)$ a low dust-to-star mass ratio.
The physical scale of 0.3~kpc (the smallest of the sample) at which NGC~300 is studied could be too small to observe a spatial correlation
between $\Sigma_{\rm star}$ and $\Sigma_{\rm dust}$.    
However, \citet{viaene14} found a strong correlation between star and dust, expressed in terms
of the relationship between M$_{\rm dust}$/M$_{\rm star}$ and the stellar mass surface density (indicated with $\mu_{\rm star}$ in that paper),
for the very nearby ($D = 785$~kpc) Andromeda galaxy (M~31) at the 0.1~kpc scale. 
It is also well known that dust is generally distributed in a disk with an exponential decline as a function of radius (see references in Sect.~\ref{sec:discussion}), 
however in C17 we found that $\Sigma_{\rm dust}$ in NGC~300 is flat and it can not be fitted with an exponential curve
(this happens only for NGC~300 in the studied sample, see Table~6 and Fig.~A.2 in C17).  
Finally, the dust-to-star mass ratio of NGC~300 is a factor 2 times higher than the corresponding mean value for galaxies 
of the same morphological stage (T~=~7, data from DustPedia database and Table~7 of C20).
The reason of a low dust-to-star mass ratio is not therefore viable to explain the breaking down of the $\Sigma_{\rm star}$--$\Sigma_{\rm dust}$ SR in NGC~300.
All these considerations seem to suggest that $\Sigma_{\rm star}$ and $\Sigma_{\rm dust}$ are not correlated in NGC~300
because of the anomalous dust mass distribution.

For NGC~3031 studied at 0.6~kpc the weak $\Sigma_{\rm star}$--$\Sigma_{\rm dust}$ SR could be explained 
by its dust-to-star mass ratio, a factor 2 times lower than the corresponding mean value for galaxies 
of the same morphological stage (T~=~2, data from DustPedia database and Table~7 of C20).
We recall that NGC~3031 is a kind of prototype of CO-poor galaxies 
\citep[e.g.,][]{solomon75,combes77,brouillet91,sage91,sakamoto01,helfer03,knapen06,casasola07,sanchez-gallego11}.
Since the formation of H$_2$ molecules takes place on the surface of dust grains \citep[][]{gould63}, 
a connection between dust and molecular gas is expected and observed. 
Therefore, the poor content of molecular gas characterizing NGC~3031 can also explain, at least partially, its low dust content.

\subsection{$\Sigma_{\rm star}$--$\Sigma_{\rm tot\,gas}$}
For NGC~2403 and NGC~925 the $\Sigma_{\rm star}$--$\Sigma_{\rm tot\,gas}$ SR is weak.
Since $\Sigma_{\rm star}$ and $\Sigma_{\rm H2}$ are well correlated in NGC~2403, it is H{\sc i} gas that produces the breaking down of the
$\Sigma_{\rm star}$--$\Sigma_{\rm tot\,gas}$ SR.
NGC~2403 is indeed defined as an H{\sc i}-dominated galaxy \citep[see][]{bigiel08} with a H$_2$-to-H{\sc i} mass ratio
a factor $\sim$7 times lower than the corresponding mean value for galaxies 
of the same morphological stage (T~=~6, data from DustPedia database and Table~4 of C20).
This overabundance of H{\sc i} could be due to environmental effects.
NGC~2403 is an outlying member of the M~81 Group and it is classified as an interacting galaxy \citep[][]{casasola04}.    
It is known that interactions between galaxies and environment play an important role in determining galaxy
structure and properties, and the distribution of gas is expected to reflect the effects of the interaction more strongly than that of stars
\citep[e.g.,][]{combes94,casasola04,thorp22}.
Also NGC~925 is an H{\sc i}-dominated galaxy \citep[see][]{bigiel08} with H$_2$-to-H{\sc i} mass ratio 
a factor $\sim$20 times lower than the corresponding mean value for galaxies of the same morphological stage 
(T~=~7, data from DustPedia database and Table~4 of C20).
This should explain why the profile of $\Sigma_{\rm tot\,gas}$ as a function of radius 
in NGC~925 is flat and it cannot be fitted with an exponential curve 
(see Table~6 and Fig.~A.2 in C17).
In addition, NGC~925 has, together with NGC~2403, the lowest mean metallicity in the sample (see Table~\ref{tab:sample}). 
These peculiar properties characterizing NGC~925 cannot be attributed to the environment since it is an isolated galaxy
\citep[][]{bettoni03}.
Maybe the presence of the bar in the centre of NGC~925 could have played a role in defining of properties of this galaxy. 
NGC~925 has been the target of many multiwavelength observational campaigns, with a particular focus on its bar
\citep[e.g.,][]{elmegreen98,pisano98,pisano00}.
These observations showed that NGC~925 is a galaxy fraught with asymmetries and that the centre of its bar is offset from 
the dynamical centre of the galaxy by $\sim$1~kpc. 
\citet{pisano00} suggested that the asymmetries observed in NGC~925 could be related to the presence of H{\sc i} clouds interacting 
with the main galaxy.  
An off-centre bar and asymmetries are typical properties of barred Magellanic galaxies 
\citep[][see in particular the Large Magellanic Cloud]{devaucouleurs72} that do not suddenly appear
but are long-lasting processes occurring during the life of a galaxy \citep[e.g.,][]{levine98,noordermeer01,kruk17}.  

\subsection{$\Sigma_{\rm tot\,gas}$--$\Sigma_{\rm SFR}$}
The moderate $\Sigma_{\rm tot\,gas}$--$\Sigma_{\rm SFR}$ SR found for NGC~2403 is likely due to its H{\sc i}-dominated nature,
while the high slope found for NGC~628 and NGC~3521 does not seem linked to peculiar galaxy properties.

\section{Other calibrations of $X_{\rm CO}$}
\label{app:othercal}
In this section we show the results obtained by adopting other two calibrations of $X_{\rm CO}$.
One calibration is proposed by \citet{bolatto13} and is based on models and observations.
It provides a metallicity and surface density dependent $X_{\rm CO}$ (see their Eq.~(31)) which can be written as:
\begin{eqnarray}
X_{\rm CO} = \frac{X_{\rm CO}^{\rm MW}}{e^{0.4}} \times\, {\rm exp}\left(\frac{0.4}{Z/Z_{\odot}}\right) \times\, \left(\frac{\Sigma_{\rm {star+gas}}}{100\,{\rm M_{\odot}\,pc^{-2}}}\right)^{-\gamma},
\label{eq:otherbolatto13}
\end{eqnarray}

\noindent
where $X_{\rm CO}^{\rm MW}$ is the conversion factor for the Milky Way ($2.0 \times 10^{20}$ cm$^{-2}$ (K~km~s$^{-1}$)$^{-1}$), 
$\Sigma_{\rm {star+gas}}$ is the total (star+gas) surface density in units of M$_{\odot}$~pc$^{-2}$, and 
$\gamma \approx 0.5$ for $\Sigma_{\rm {star+gas}} > 100$~M$_{\odot}$~pc$^{-2}$ and
$\gamma = 0$ otherwise \citep[see][for details]{bolatto13}.
The other calibration of $X_{\rm CO}$ is presented from \citet{madden20} and is based on Cloudy models 
\citep[][]{ferland17} and observations. 
It provides a CO-to-H$_2$ conversion factor 
including both the CO-dark and the CO-bright gas and as a function of metallicity (see their Eq.~(6)).
It can be written as:
\begin{eqnarray}
X_{\rm CO} = 2.4 \times\,10^{20} \times\,\left(Z/Z_{\odot}\right)^{-3.39}.
\label{eq:othermadden20}
\end{eqnarray}

\noindent
This calibration is based on star-forming low-metallicity galaxies of the \textit{Herschel} Dwarf Galaxy Survey 
\citep[][]{madden13} consisting of 50 galaxies ranging from very low metallicity (12~+~log(O/H)~$\sim$~7.1) 
to moderate metallicity ($\sim$8.4).

\begin{figure}
\centering
\includegraphics[width=0.5\textwidth]{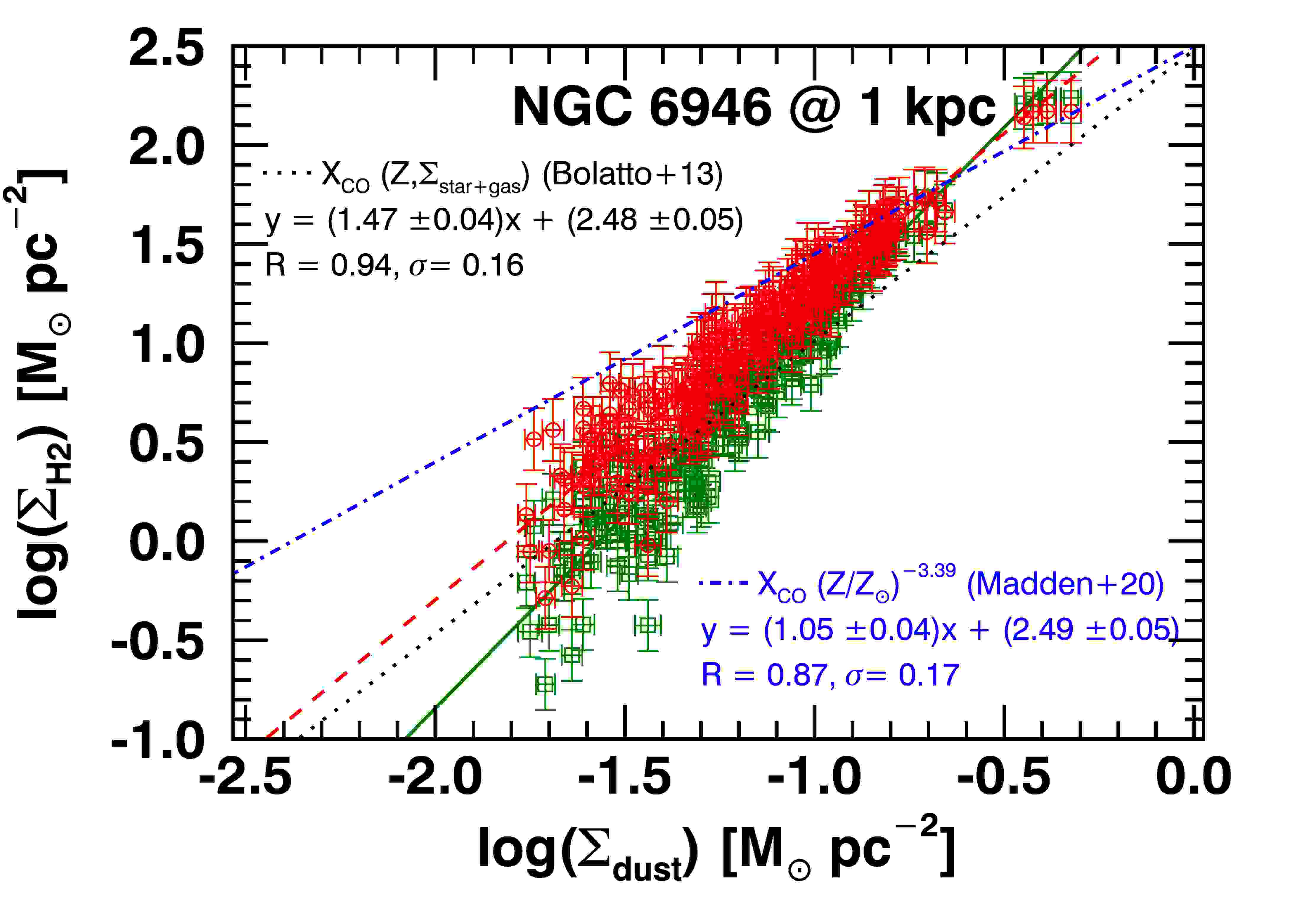}  
\caption{
$\Sigma_{\rm dust}$--$\Sigma_{\rm H2}$ SR for NGC~6946 shown in Fig.~\ref{fig:n6946} with fits to the data obtained adopting 
the prescription of a $\Sigma_{\rm {star+gas}}$ and metallicity-dependent $X_{\rm CO}$ by \citet{bolatto13} (dotted black line)
and the prescription of a metallicity-dependent $X_{\rm CO}$ by \citet{madden20} (dash-dot blue line). 
Equations of these fits are also given in figure.
}
\label{fig:n6946-othercal}
\end{figure}

Figure~\ref{fig:n6946-othercal} shows the $\Sigma_{\rm dust}$--$\Sigma_{\rm H2}$ SR for NGC~6946 with
fits to the data obtained adopting these other two calibrations of $X_{\rm CO}$.
The prescription of \citet{bolatto13} depending on $\Sigma_{\rm {star+gas}}$ and metallicity  
provides similar slopes to (or slightly lower than) those found with the assumption of \citet{amorin16}, 
while that of \citet{madden20} always gives lower slopes maybe due to the lower metallicities 
characterizing their sample (only partially overlap with the metallicity range of our sample).
These other two assumptions of $X_{\rm CO}$ provide slightly lower $R$ (and higher $\sigma$)
than those obtained under the adopted prescriptions. 
Since these other two recipes of $X_{\rm CO}$  do no show extremely different results, 
we have decided not to show them for the all SRs involving molecular gas.

\end{appendix}

\end{document}